\documentstyle[12pt]{article}

\newcounter{lico}

\newcommand{\noi}{\noindent}
\newcommand{\hsp}{\hspace{5.7mm}}
\newcommand{\ssp}{\hspace*{4mm}}
\newcommand{\fsp}{\hspace*{1.8mm}}
\newcommand{\mssp}{\hspace*{-6mm}}
\newcommand{\sspp}{\hspace*{11.3mm}}
\newcommand{\mini}{\hspace{.3mm}}      
\newcommand{\mmini}{\hspace{-.4mm}}    
\newcommand{\mimini}{\hspace{-.3mm}}
\newcommand{\mA}{\hspace*{-1mm}}
\newcommand{\mE}{\hspace{-1.7mm}}

\let\ssection=\section
\renewcommand{\section}{\setcounter{equation}{0}\ssection}

\newfont{\BBB}{msbm10 scaled\magstephalf}
\newcommand{\BB}[1]{\mbox{\BBB #1}}
\newcommand{\BBn}[2]{\mbox{{\BBB #1}$^{#2}$}}

\newfont{\bbviii}{msbm8}
\newcommand{\bb}[1]{\mbox{\bbviii #1}}

\newfont{\oox}{cmsy10}
\newfont{\EEx}{cmex10}
\newcommand{\bgrec}[1]{\mbox{\boldmath $#1$}}
\newcommand{\otens}{\: \mbox{\oox \symbol{'12}}\:}
\newcommand{\bxi}{\xi \hspace{-1.9mm} \xi \hspace{-1.9mm} \xi
   \hspace{-1.9mm} \xi \hspace{-1.9mm} \xi \hspace{-1.9mm} \xi}
\newcommand{\El}{\mbox{\raisebox{3mm}{\EEx\symbol{'12}}}}
\newcommand{\Er}{\mbox{\raisebox{3mm}{\EEx\symbol{'13}}}}

\newcommand{\rank}{\mbox{rank\,}}
\newcommand{\ranks}{\mbox{\scriptsize rank\mini}}
\newcommand{\lcm}{\mbox{lcm}}
\newcommand{\emb}{\mbox{\raisebox{.8mm}{$\hspace{1.8mm}\scriptstyle
   \subset\mA$}\raisebox{-.4mm}{$\longrightarrow\hspace{1.8mm}$}}}
\newcommand{\bfh}{\mbox{$ {\bf h} \hspace{-2.5mm} {\bf h}
   \hspace{-2.5mm} {\bf h}$}}
\newcommand{\bfn}{\mbox{$ {\bf n} \hspace{-2.5mm} {\bf n}
   \hspace{-2.5mm} {\bf n}$}}
\newcommand{\bfs}{\mbox{$ {\bf s} \hspace{-1.8mm} {\bf s}
   \hspace{-1.8mm} {\bf s}$}}
\newcommand{\bfw}{\mbox{$ {\bf w} \hspace{-3.4mm} {\bf w}
   \hspace{-3.4mm} {\bf w}$}}
\newcommand{\bfsig}{\mbox{$ {\bf \Sigma} \hspace{-3.35mm}
   {\bf \Sigma} \hspace{-3.35mm} {\bf \Sigma}$}}
\newcommand{\bfom}{{\vec{\omega}}}

\newcommand{\maxsym}{
  \begin{picture}(0,0)(.025,0)\circle{.31}\end{picture}}
\newcommand{\lowdim}{
  \begin{picture}(0,0)(.12,.1)\framebox(.19,.19){}\end{picture}}
\newcommand{\chacon}{
  \begin{picture}(0,0)(.185,.16)\dashbox{.04}(.32,.32){}\end{picture}}
\newcommand{\noLmin}{\raisebox{-.6mm}
  {\hspace*{-.1mm}$\scriptstyle\times$}}
\newcommand{\nolodiLmin}{\raisebox{-.6mm}
  {\hspace*{-.7mm}$\scriptstyle(\!\times\!)$}}
\newcommand{\dimms}[1]{\makebox(0.16,-0.21)[t]{$\scriptstyle #1$}}
\newcommand{\dimg}[1]{\makebox(0.16,0.35)[t]{$\scriptstyle #1$}}

\newcommand{\la}{<\!}
\newcommand{\ra}{\!>\:}

\newcommand{\be}{\begin{equation}}
\newcommand{\ee}{\end{equation} \vspace{0mm}}
\newcommand{\ben}{$$}       
\newcommand{\een}{$$ \vspace{0mm}}
\newcommand{\bea}{\begin{eqnarray}}
\newcommand{\eea}{\end{eqnarray} \vspace{0mm}}
\newcommand{\ba}{\begin{array}{c}}
\newcommand{\ea}{\end{array}}

\newcommand{\cfr}[1]{(\ref{#1})}
\newcommand{\cfrs}[2]{(\ref{#1}) and~(\ref{#2})}

\newcommand{\hah}{\hsp\mbox{and}\hsp}

\newcommand{\hoh}{\hsp\mbox{or}\hsp}

\newcommand{\hwh}{\hsp\mbox{with}\hsp}

\newcommand{\hfa}{\hsp\mbox{for all }\,}
\newcommand{\hfs}{\hsp\mbox{for some }\,}

\newcommand{\vav}{\vspace{-4.5mm}\noi and \vspace{-3mm}}
\newcommand{\vwv}{\vspace{-4.5mm}\noi where \vspace{-3mm}}
\newcommand{\vwiv}{\vspace{-4.5mm}\noi with \vspace{-3mm}}
\newcommand{\mam}{\vspace{-4.5mm} \noi and \vspace{-3mm}}  

\newcommand{\miem}{\vspace{-4.5mm} \noi i.e., \vspace{-3mm}}

\newcommand{\mhm}{\vspace{-4.5mm} \noi hence \vspace{-3mm}}
\newcommand{\mwm}{\vspace{-4.5mm} \noi where \vspace{-3mm}}
\newcommand{\mwim}{\vspace{-4.5mm} \noi with \vspace{-3mm}}

\newcommand{\US}{\,U(1)_{em}\mmini \times SU(2)_{spin}}
\newcommand{\USo}{\,U(1) \times SU(2)\mini}
\newcommand{\U}{$\,U(1)$}

\newcommand{\uh}{$\,\hat{u}(1)$}
\newcommand{\Ue}{$\,U(1)_{em}$}
\newcommand{\SU}{$\,SU(2)$}
\newcommand{\suh}{$\,\hat{su}(2)$}

\newcommand{\SUs}{$\,SU(2)_{spin}$}
\newcommand{\gh}{$\,\hat{g}$}
\newcommand{\suhN}{$\,\widehat{su}(N)$}

\newcommand{\ui}{$u(1)$}

\newcommand{\da}{\mbox{d}}

\newcommand{\dd}[1]{\frac{\partial}{\partial #1}}
\newcommand{\ddd}[2]{\frac{\partial #1}{\partial #2}}
\newcommand{\Es}{E^{(s)}}
\newcommand{\psis}{\psi^{(s)}}
\newcommand{\psiss}{\psi^{(s)*}}
\newcommand{\psisss}{\psi^{(s)\sharp}}
\newcommand{\psisx}{\psi^{(s)}(\bx)}
\newcommand{\Us}{U^{(s)}}
\newcommand{\D}[1]{{\cal D}_{#1}}
\newcommand{\Dh}[1]{\hat{\cal D}_{#1}}
\newcommand{\am}[1]{a_{#1}(x)}
\newcommand{\wms}[1]{w^{(s)}_{#1}(x)}
\newcommand{\oms}[1]{\omega^{(s)}_{#1}(x)}
\newcommand{\rms}[1]{\rho^{(s)}_{#1}(x)}
\newcommand{\wso}{w^{(s)}}
\newcommand{\rso}{\rho^{(s)}}
\newcommand{\oso}{\omega^{(s)}}

\newcommand{\ahm}[1]{\hat{a}_{#1}\ty}
\newcommand{\psih}{\hat{\psi}\ty}
\newcommand{\ghik}[1]{\hat{g}^{#1}\ty}
\newcommand{\whms}[1]{\hat{w}^{(s)}_{#1}\ty}
\newcommand{\fhj}[1]{\hat{f}^{#1}\ty}
\newcommand{\phity}{\phi\ty}
\newcommand{\Omhab}{\hat{\Omega}^{A}_{\hspace{.4em} B}\ty}
\newcommand{\Usty}{U^{(s)}\ty}

\newcommand{\tphi}{(t,\mini \phi\ty)}
\newcommand{\fhky}[1]{\hat{f}_{#1}(y)}
\newcommand{\Ahky}[1]{\hat{A}_{#1}(y)}
\newcommand{\psihy}{\hat{\psi}(y)}
\newcommand{\psihsy}{\hat{\psi}^\ast(y)}

\newcommand{\bx}{{\bf x}}
\newcommand{\by}{{\bf y}}
\newcommand{\bbf}{\,{\bf f}\,}
\newcommand{\bM}{{\bf M}}
\newcommand{\bBc}{\,{\bf B}_c\,}
\newcommand{\Bcpara}{{\bf B}_c^{\mini\parallel}}

\newcommand{\bA}{{\bf A}}
\newcommand{\bJ}{{\bf J}}
\newcommand{\es}{\mbox{\boldmath$\varepsilon$}}

\newcommand{\tx}{(t,\mini \bx)}
\newcommand{\ty}{(t,\mini \by)}

\newcommand{\eai}[2]{e^{#1}_{\hspace{.4em} #2}({\bf x})}
\newcommand{\eaix}[2]{e^{#1}_{\hspace{.4em} #2}(x)}
\newcommand{\eaio}[2]{e^{#1}_{\hspace{.4em} #2}}
\newcommand{\Eai}[2]{{\cal E}_{#1}^{\hspace{.4em} #2}({\bf x})}
\newcommand{\Eaix}[2]{{\cal E}_{#1}^{\hspace{.4em} #2}(x)}
\newcommand{\Eaio}[2]{{\cal E}_{#1}^{\hspace{.4em} #2}}
\newcommand{\om}[1]{\omega^{A}_{\hspace{.4em} #1}({\bf x})}
\newcommand{\omx}[1]{\omega^{A}_{\hspace{.4em} #1}(x)}
\newcommand{\lam}[1]{\lambda^{A}_{\hspace{.4em} #1}({\bf x})}
\newcommand{\ka}[1]{\kappa^{A}_{\hspace{.5em} #1}({\bf x})}
\newcommand{\eps}[1]{\varepsilon_{#1}}
\newcommand{\ops}[1]{\varepsilon^{#1}}
\newcommand{\epps}[2]{\varepsilon_{#1}^{\hspace{.4em} #2}}

\newcommand{\halb}{{1 \over 2}}
\newcommand{\sH}{\sigma_{\!\scriptscriptstyle H}}
\newcommand{\nH}{n_{\mimini\scriptscriptstyle H}}
\newcommand{\dH}{d_{\mimini\scriptscriptstyle H}}
\newcommand{\sHs}{\sigma_H^{\mbox{\scriptsize\hspace{.1mm}spin}}}
\newcommand{\ndH}{{{n_H\rule[-.8mm]{0mm}{2mm}} \over d_H}}
\newcommand{\RH}{R_{\scriptscriptstyle H}}
\newcommand{\RL}{R_{\scriptscriptstyle L}}
\newcommand{\rH}{\rho_{\scriptscriptstyle H}}
\newcommand{\rx}{\rho_{xx}}
\newcommand{\ry}{\rho_{yy}}

\newcommand{\Seff}{S^{\mbox{\scriptsize\hspace{.15mm}eff}}}

\newcommand{\SLeff}{S_\Lambda^{\mbox{\scriptsize\hspace{.15mm}eff}}}
\newcommand{\SL}{S_\Lambda}

\newcommand{\StLeff}{S^{\mbox{\scriptsize \hspace{-.85mm}
   eff}}_{\tha\Lambda_o}(a^\kt,\mini w^\kt)}
\newcommand{\Ssaw}{S^{\mini\ast}_{\Lambda_o}(\tilde{a}\mini ,\mini
   \tilde{w})}
\newcommand{\Piast}{\Pi^\ast}
\newcommand{\Sast}{S^{\mini\ast}}
\newcommand{\Sti}{\tilde{S}}
\newcommand{\Stieff}{\tilde{S}^{\mbox{\scriptsize\hspace{.15mm}eff}}}
\newcommand{\Stito}{\tilde{S}_{\mbox{\scriptsize tot}}}
\newcommand{\Stitoeff}{\tilde{S}_{\mbox{\scriptsize
   tot}}^{\mbox{\scriptsize\hspace{.15mm}eff}}}
\newcommand{\Iper}{I_{\mbox{\scriptsize pert}}}
\newcommand{\Ssa}{S^{\mini\ast}_{\Lambda_o}(\tilde{a})}
\newcommand{\Ssw}{S^{\mini\ast}_{\Lambda_o}(w)}
\newcommand{\SWZ}{{S_{WZNW}}}
\newcommand{\iL}{\int_\Lambda}

\newcommand{\iLo}{\int_{\Lambda_o}}
\newcommand{\idLo}{\int_{\partial \Lambda_o}}
\newcommand{\dLo}{{\partial \Lambda_o}}
\newcommand{\BT}{\mbox{B.T.}}
\newcommand{\GI}{\mbox{G.I.}}
\newcommand{\ih}{{i \over \hbar}}
\newcommand{\hi}{{1 \over \hbar}}
\newcommand{\jc}{j_c^{\mu}}
\newcommand{\mm}{m_3^{\mu}}
\newcommand{\at}[1]{\tilde{a}_{#1}}
\newcommand{\wt}[1]{\tilde{w}_{#1}}

\newcommand{\ti}[1]{\tau_1^{#1}(\xi)}
\newcommand{\tii}[1]{\tau_2^{#1}(\xi)}
\newcommand{\wcc}{w_{c,03}^\kt(\xi)}
\newcommand{\JJ}{{\cal J}^{\!i}\xx}
\newcommand{\JJc}{{\cal J}_{\!c}^{\!i}\xx}

\newcommand{\Bt}{\tilde{B}_3\xx}
\newcommand{\Etv}{\vec{\tilde{E}}\xx}
\newcommand{\Etvc}{\vec{\tilde{E}}_c\xx}
\newcommand{\Et}{\tilde{E}}
\newcommand{\bt}{\tilde{b}}
\newcommand{\ft}{\tilde{f}}
\newcommand{\Ot}{\tilde{\Omega}\xx}
\newcommand{\R}{{\cal R}\xx}

\newcommand{\gp}[1]{g_\Vert^{(#1)}}
\newcommand{\go}{g_\bot}
\newcommand{\lp}[1]{l_\Vert^{(#1)}}
\newcommand{\lpp}{l_\Vert}
\newcommand{\lo}{l_\bot}
\newcommand{\fft}[1]{\tilde{f}_{#1}}
\newcommand{\kk}[1]{h_{#1}}
\newcommand{\LsA}[2]{L^{(#1)}_{#2}}

\newcommand{\const}{\mbox{const.}}
\newcommand{\sgi}{{1 \over \sqrt {g(x)}}}
\newcommand{\sggi}{{1 \over \sqrt {\gamma(\xi)}}}
\newcommand{\sg}{\sqrt {g(x)}}
\newcommand{\sgg}{\sqrt {\gamma}}
\newcommand{\sggx}{\sqrt {\gamma(\xi)}}
\newcommand{\tha}{\theta}
\newcommand{\kt}{{(\theta)}}
\newcommand{\xx}{(\xi)}
\newcommand{\vxx}{\vec{\xi}\,}
\newcommand{\vxxo}{\vec{\xi}}

\newcommand{\tlr}{_{L/R}}
\newcommand{\Ag}{A|_{\Gamma_{\!o}}}

\newcommand{\cdL}{\chi|_{\partial\Lambda_o}}
\newcommand{\dir}{D(\Ag)}
\newcommand{\slashi}{\raisebox{.3mm}{\hspace{-2.2mm}/}}
\newcommand{\GLR}{\hi \Gamma\!_{L/R}(\Ag)}
\newcommand{\tGLR}{\Gamma\!_{L/R}(\Ag)}

\newcommand{\ehc}{\left( {e \over \hbar c} \right)^2}
\newcommand{\alp}{\alpha_+}

\newcommand{\alm}{\alpha_-}

\newcommand{\dpp}{\partial_+}
\newcommand{\dmm}{\partial_-}
\newcommand{\du}{\:d^{\mini 2}u}

\newcommand{\Ki}{K^{-1}}

\newcommand{\qel}{q_{el}}
\newcommand{\Q}{{\bf Q}}
\newcommand{\G}{\Gamma}
\newcommand{\Gp}{\Gamma_{\!phys}}
\newcommand{\Gs}{\Gamma^\ast}
\newcommand{\Gw}{\Gamma_W}
\newcommand{\Gwi}[1]{\Gamma_{W_#1}}
\newcommand{\Gws}{\Gamma_W^\ast}

\newcommand{\QHL}{\mbox{$(\G,\Q)$}}
\newcommand{\veb}{{\bf e}}
\newcommand{\ve}[1]{{\bf e}_#1}
\newcommand{\veed}{\mbox{\boldmath $\varepsilon$}}
\newcommand{\ved}[1]{\mbox{\boldmath $\varepsilon$}^#1}
\newcommand{\vom}{\mbox{\boldmath $\omega$}}
\newcommand{\omv}{\mbox{\raisebox{-2.2mm}{$\stackrel{\displaystyle
   \omega}{\scriptstyle \rightarrow}$}}}
\newcommand{\omvi}[1]{\mbox{\raisebox{-2.2mm}{$\stackrel{\displaystyle
   \omega}{\scriptstyle \rightarrow}$}
   \raisebox{-1.2mm}{$\scriptstyle\hspace{-1.5mm}#1$}}}
\newcommand{\hw}{h_{\vom}}
\newcommand{\hws}{{h_{\mbox{$\scriptstyle\omega\hspace{-1.9mm}
   \omega\hspace{-1.9mm}\omega\hspace{-1.9mm}\omega$}}}}
\newcommand{\vq}{{\bf q}}
\newcommand{\vv}{{\bf v}}
\newcommand{\vn}{{\bf n}}
\newcommand{\Qv}{\mbox{\raisebox{-2.2mm}{$\stackrel{\displaystyle
   Q}{\scriptstyle \rightarrow}$}}}
\newcommand{\Qvp}{\mbox{\raisebox{-2.2mm}{$\stackrel{\displaystyle
   Q^\prime}{\scriptstyle \rightarrow}$}}}
\newcommand{\lb}{\lambda}

\newcommand{\sub}[1]{\raisebox{-1.6mm}{$\scriptstyle #1$}}
\newcommand{\subi}[1]{\raisebox{-1.2mm}{$\scriptstyle #1$}}
\newcommand{\thesymbol}{\raisebox{-3.6mm}{$\scriptstyle N$} 
   {\left( \ndH \right)}^g_\lambda\;[\mini\lm,\lM\mini]}
\newcommand{\thesymbolpr}{\raisebox{-3.6mm}{$\scriptstyle
   {N^\prime}$}{\left( \ndH \right)}^{g^\prime}_{\lambda^\prime}\;
   [\mini\lm^{\mini\prime},\lM^{\mini\prime}\mini]}
\newcommand{\QHLsymbol}[4]{\raisebox{-3.6mm}{$\scriptstyle #1$} 
   {\left( #2 \right)}^{#3}\hspace{-1,6mm}
   \raisebox{-3.6mm}{$\scriptstyle #4$}}

\newcommand{\Lmini}{$L\mini$-minimal}
\newcommand{\Lminiy}{$L\mini$-minimality}
\newcommand{\LMiniy}{$L\mini$-Minimality}

\newcommand{\sca}[2]{\la #1 , #2 \ra}
\newcommand{\avecd}[1]{#1_{1} , \ldots , #1_{N} }
\newcommand{\avecu}[1]{#1^{1} , \ldots , #1^{N} }
\newcommand{\diix}{d^{\mini 2}\mmini x}
\newcommand{\BQ}{{\cal B}_{\!\Q}}
\newcommand{\BQs}{\raisebox{-.5mm}{\mbox{${\textstyle\cal
   B}_{\scriptscriptstyle\bf Q}$}}}

\newcommand{\Sp}{{\cal S}_p}
\newcommand{\Sigp}{\Sigma_p}
\newcommand{\Sigi}{\Sigma_1}
\newcommand{\Z}{\BB{Z}}
\newcommand{\q}{{\bf q}}
\newcommand{\e}{{\bf e}}

\newcommand{\n}{{\bf n}}

\newcommand{\Hp}{{\cal H}_p}
\newcommand{\Hi}{{\cal H}_1}

\newcommand{\Del}{\Delta}
\newcommand{\LM}{L_{max}}
\newcommand{\Lm}{L_{min}}
\newcommand{\Ls}{\ell_{*}}
\newcommand{\Ns}{N_{*}}
\newcommand{\lM}{\ell_{max}}
\newcommand{\lm}{\ell_{min}}

\newcommand{\OO}{{\cal O}\mini}
\newcommand{\GG}{{\cal G}\mini}
\newcommand{\GGh}{\hat{\cal G}}

\newcommand{\PRL}{Phys.\ Rev.\ Lett.\ }
\newcommand{\PRB}{Phys.\ Rev.\ B }
\newcommand{\PRD}{Phys.\ Rev.\ D }
\newcommand{\PR}{Phys.\ Rev.\ }
\newcommand{\PL}{Phys.\ Lett.\ }
\newcommand{\NPB}{Nucl.\ Phys.\ B }
\newcommand{\AP}{Ann.\ Phys.\ (N.Y.) }
\newcommand{\CMP}{Commun.\ Math.\ Phys.\ }
\newcommand{\SurS}{Surf.\ Sci.\ }
\newcommand{\HPA}{Helv.\ Phys.\ Acta }
\newcommand{\IJMP}{Int.\ J.\ Mod.\ Phys.\ }
\newcommand{\FP}{Found.\ Phys.\ }
\newcommand{\JSP}{J.\ Stat.\ Phys.\ }

\begin{document}

\baselineskip=16pt plus 2pt minus 1pt


\thispagestyle{empty}



\begin{center}
{\large\bf QUANTUM THEORY OF LARGE SYSTEMS}

{\large\bf OF NON-RELATIVISTIC MATTER}

\vspace{20mm}

Research Group in Mathematical Physics$^*$

Theoretical Physics

ETH-H\"onggerberg

CH-8093 \ Z\"urich

\end{center}

\vspace{14cm}

\hrulefill\hspace{9cm}
\medskip

$^*$ J. Fr\"ohlich, U.M.~Studer$^\dagger$, E. Thiran

$^\dagger$ Work supported in part by the Swiss National Foundation


\newpage
\pagenumbering{roman}
\setcounter{page}{1}
\tableofcontents

\newpage
\pagenumbering{arabic}


\baselineskip=16pt plus 2pt minus 1pt


\section{Introduction}
\label{sI}

\noindent
These are some notes to the course taught by J. Fr\"ohlich at
the 1994 summer school at Les Houches. While teaching and being taught
at that school -- and the hiking, early in the morning and, sometimes,
during week ends -- were extremely pleasant and rewarding, the
preparation of publishable notes superceeding those more than three
hundred pages of handwritten lecture notes prepared and distributed
during the school turned out to be a very painful process. It
constantly collided with more urgent duties (which ETH pays us for). 

The reasons why we finally managed to produce some notes are, first of
all, the infinite patience of Fran\c cois David and Paul Ginsparg,
and, second, the circumstance that, in these times of computers and of
\TeX, we could adapt and modify \TeX files of papers written for other
purposes.


\subsection{Sources, and acknowledgements}
Here are the main sources for this
first part of our notes:

\begin{description}
\item[] J. Fr\"ohlich, R. G\"otschmann, and P.-A. Marchetti,
  J. Phys. A {\bf 28}, 1169 (1995): for Chapter 5.
\item[] J. Fr\"ohlich, R. G\"otschmann, and P.-A. Marchetti, ``The
  Effective Gauge Field Action of a System of Non-Relativistic
  Electrons'', to appear in Commun. Math. Phys. (1995): for Chapter
  5. 
\item[] J. Fr\"ohlich, T. Kerler, U.M. Studer, and E. Thiran,
  ``Structuring the Set of Incompressible Quantum Hall Fluids'',
  submitted to Nucl. Phys. B: for Chapter 8.
\item[] J. Fr\"ohlich and U.M. Studer, in: ``New Symmetry Principles
  in Quantum Field Theory'', J. Fr\"ohlich, G. t'Hooft, A. Jaffe,
  G. Mack, P.K. Mitter, and R. Stora (eds), New York: Plenum Press
  1992 (page 195): for Chapters 2, 4 and 6. 
\item[] J. Fr\"ohlich and U.M. Studer, Rev. Mod. Phys. {\bf 65}, 733
  (1993): for Chapters 2, 3, 4, 6 and 7.
\item[] J. Fr\"ohlich, U.M. Studer, and E. Thiran, ``A Classification
  of Quantum Hall Fluids'', to appear in J. Stat. Phys. (1995): for
  Chapter 8. 
\end{description}

\noindent R. G\"otschmann, T. Kerler and P.-A. Marchetti deserve our
  thanks for having contributed their ideas to some of the work
  underlying these notes. We are also indebted to L. Michel for having
  guided us through high-dimensional lattices, a mathematical theory
  that is important for the material in Chapter 8. We thank J. Avron
  for his encouraging interest in our work. J. Fr\"ohlich is very
  grateful to U.M. Studer, who spent many hours in front of a computer
  screen putting these notes in shape. We also thank A. Schultze for
  her help in adding final touches to these notes. J. Fr\"ohlich
  sincerely thanks the organizers, Fran\c cois and Paul, for having
  invited him to present these lectures, for having organized such a
  marvellous school, and for their patience.


\subsection{Topics not treated in these notes}
The ``{\it standard model}''
of the physics of atoms, molecules and condensed matter, at energies
per particle below, say, a tenth of its relativistic rest energy and
usually much smaller, describes non-relativistic, quantum-mechanical
electrons and nuclei coupled to the quantized radiation field (cutoff
in the ultraviolet) and to additional, external electromagnetic
fields. It was planned to include a chapter about fundamental
properties of the standard model, including stability, non-existence
of the ultraviolet limit of quantum electrodynamics (QED) with
non-relativistic matter, its ``cutoff independence'', etc.; 
(for a recent paper on some aspects of QED with non-relativistic
matter see: V. Bach, J. Fr\"ohlich, and I.M. Sigal, ``Mathematical
Theory of Non-Relativistic Matter and Radiation'', to appear in
Lett. Math. Phys. (1995), and refs. given there). We planned to sketch
the answer to the question why a condensed matter physicist studying
transport properties of three-dimensional electron gases can usually
forget the radiation field and the fact that electrons are coupled to
it. We also planned to explain why if the world were two-dimensional
the radiation field would most likely have a drastic effect on the
low-energy properties of electron gases; (emergence of a marginal
Fermi liquid). A physical system that appears to mimic two-dimensional
QED is a quantum Hall fluid at a filling factor of $\frac 1 2$ \ (or
$\frac 1 4$). It would have been desirable to include an analysis of
this system in Chapter 6. 

Unfortunately, time did not permit us to include any of this basic
material in our notes; (some of it was included in the handwritten
notes available at the school). We strongly encourage the reader to
study E.H. Lieb's selecta, ``The Stability of Matter: From Atoms to
Stars'' (Berlin, Heidelberg, New York: Springer-Verlag 1991), for
fundamental results on non-relativistic quantum theory. 

Another topic, originally planned to appear in these notes, concerns
the {\it magnetic properties} of non-relativistic matter. There was a
chapter on magnetism in the handwritten notes; but we were unable to
include it here. Various forms of magnetism are intimately related to
electron spin and the $SU(2)$-gauge invariance of non-relativistic
quantum theory. The $U(1)\times SU(2)$-gauge invariance of
non-relativistic quantum theory and some of its consequences {\it are}
discussed in our notes at some length, but, for example, the extension
of the approach sketched in Chapter 5 for the electric ($U(1)$)
current density to the {\it spin} ($SU(2)$) {\it current density} had
to be omitted. In this connection, we recommend that the reader
consult the papers by Cattaneo et al. and by Leutwyler and the book by
Fradkin quoted in the references. 

Of course, a study of magnetic properties of various tight-binding
models and Heisenberg type models had to be omitted, too. Rigorous
results in this area are scarce; but there is steady progress, at the
heuristic and at the mathematically rigorous level, that we would have
liked to review, had time permitted us to do so.

We hope there will be another opportunity to make up for these
omissions. They leave the picture of non-relativistic quantum theory
drawn in these notes rather incomplete. Nevertheless we hope that, in
these notes, we are laying out, with adequate precision, sufficiently
many pieces of the puzzle to enable the reader to guess what would fit
into some of the holes we leave. We also hope that we are able to
convey to the reader some of the excitement we felt when working on
the problems reviewed in our notes. We believe that our analysis of
the $U(1)\times SU(2)$-gauge invariance of non-relativistic quantum
theory, of the quantum-mechanical Larmor theorem and of their general
consequences in condensed matter physics, of the generating functions
of (connected) current Green functions (effective actions), and our
approach towards classifying states of non-relativistic matter
(``gauge theory of thermodynamic phases'') and, as a special example,
of incompressible quantum Hall fluids will turn out to be of some
general usefulness. These topics {\it are} treated fairly carefully in
these notes.

We hope that, during less hectic times, we shall be able to iron out
various imperfections and complete the puzzle.




\section{The Pauli Equation and its Symmetries}
\label{sPE}

In this section we describe non-relativistic electrons and other
non-relativistic  particles with spin in an external electromagnetic
field. In a one-particle language, the wave functions of such
particles satisfy the Pauli equation. We show that the Pauli
equation exhibits a basic $\US$\ gauge symmetry. This symmetry is,
for example, a corner stone of our analysis of incompressible quantum
fluids presented in Sects.\,\ref{sSL}--\ref{sClQHFs}. In order to
provide a first illustration of the {\em general$\,$} usefulness of
this symmetry in quantum theory, we include, at the end of this
section, discussions of the Aharonov-Bohm effect and its
$\,SU(2)_{spin}$-cousin, the Aharonov-Casher effect. Further
applications of gauge invariance to  quantum-mechanical effects will
be discussed in Sect.~\ref{sKE}.

\subsection{Gauge-Invariant Form of the Pauli Equation}
\label{ssGI}

In order to describe non-relativistic particles of arbitrary spin
$\,s$, mass $\,m\,$ and charge $\,q$, we have to find the correct
Zeeman and spin-orbit terms in their Hamiltonian. We recall that
Bargmann, Michel and 
Telegdi~(1959) have found a relativistic description of the motion of
a classical spin $\,{\bf S}\,$ in a (slowly varying) external
electromagnetic field ({\bf E}, \mini{\bf B}). Expanding their result
in powers of $\,{v\over c}$, one obtains the equation of motion
already found by Thomas (1927; see also Jackson, 1975)

\be
{d {\bf S} \over dt} \:=\: {q \over mc}\, {\bf S} \times \biggl[\, 
{g \over 2}\,{\bf B} - \left( {g \over 2}-{1 \over 2} \right){{\bf
v} \over c} \times {\bf E}\, \biggr] + {\cal O} \left( ( \mbox{$v
\over c$} )^2 \right) \ ,
                            \label{SP}
\ee

\noi
where $\,{\bf v}\,$ is the velocity of the spin (with
respect to the laboratory frame), and $\,g\,$ is its gyromagnetic ratio.
The spin-orbit term (second term in (\ref{SP}))
consists of two contributions: The terms proportional to $\,g/2\,$
describe the precession of the spin (or {\em magnetic moment$\,$}) in
the magnetic field in the rest frame of the spin. The remaining term
describes the purely kinematical effect of the {\em Thomas
precession$\,$} which is a consequence of the acceleration a
{\em charged$\,$} particle experiences in an electric
field. Recalling the Poisson bracket relations, $\,\{S_i,\mini
S_j\} = \eps{ijk}\,S_k$, for a classical spin $\,{\bf S}$, we find
that Eq.~(\ref{SP}) is a hamiltonian equation of motion
corresponding to the Hamilton function

\be
H_{cl}\:=\:-{\bf S}\cdot \biggl[\, {q \over mc}{g \over 2}\,{\bf B} -
{q \over mc}\left( {g \over 2}-{1 \over 2} \right){{\bf v} \over c}
\times {\bf E}\, \biggr] \ .
\ee

If we accept (\ref{SP}) as the appropriate non-relativistic
Heisenberg equation of motion for the (quantum-mechanical) spin
operator $\,{\bf S}\,$ (in the spin-$s\,$ representation) then the
wave function $\,\psis\,$ of the particle, a $\,(2s+1)$-component
complex spinor, satisfies the following {\em Pauli equation}

\bea
i\hbar \dd{t} \psis & = & q\mini \Phi\, \psis
- \bgrec{\mu}_{spin} \cdot {\bf B}\, \psis + {1 \over 2\mini m}\,
{\bf\Pi}^2 \mini\psis \nonumber \\ 
& & \hspace{-2.7cm} \mbox{} - {1 \over 2\mini mc} \biggl[ {\bf\Pi}
\cdot \left( ( \bgrec{\mu}_{spin} - {q \over 2mc}\,{\bf S} ) \times
{\bf E} \right) +  \left( ( \bgrec{\mu}_{spin} - {q \over
2\mini mc}\,{\bf S} ) \times  {\bf E} \right) \cdot {\bf\Pi} \biggr]
\psis \,,
                             \label{gen}
\eea

\noi
where the (intrinsic) magnetic moment of the particle is defined by

\be
\bgrec{\mu}_{spin} \::=\: {g\mini\mu \over \hbar}\,{\bf S} \ ,
                          \label{mamo}
\ee

\noi
and, in the spin-$s\,$ representation, the spin operator $\,{\bf
S}\,$ is given by

\be
{\bf S} \::=\: {\hbar \over 2}\,{\bf L}^{(s)} \:=\: {\hbar \over
2}\,(\LsA{s}{1},\,\LsA{s}{2},\,\LsA{s}{3}) \ . 
                          \label{spin}
\ee

\noi
Here $\,(L_A^{(s)})_{A=1}^{3}\,$ are hermitian generators of the
Lie algebra $\,su(2)\,$ in the spin-$s\,$ representation normalized
such that $\,L_A^{(1/2)} = \sigma_A\,, \mbox{ where } \,\sigma_1,\,
\sigma_2 \,\mbox{ and } \,\sigma_3\,$ are the usual Pauli matrices.
Furthermore, for charged particles, we have that $\,\mu = {q\hbar
\over 2mc}$. In particular, for an {\em electron,$\,$} $-\mu =
{e\hbar \over 2 m_o c} =: \mu_B = 5.79 \times 10^{-9}\,$eV/Gauss,
the Bohr magneton, and $\,g=2$. Other examples are the {\em
proton$\,$} and the {\em neutron,$\,$} where $\,\mu = {e\hbar \over
2mc}$, with $\,m\,$ the proton mass, and the $g$-factors are given
by $\,g=5.59\,$ and $\,g=-3.83$, respectively.

Next, we show that, by "completing the square" in the Pauli equation
(\ref{gen}), we obtain an equation with an astonishingly rich
symmetry, namely with a {\em local} $\US$ {\em symmetry$\mini$}.
Since the modification needed is a term of order $\,{\cal
O}(1/m^3)$, this symmetry should really be viewed as a fundamental
property of non-relativistic quantum mechanics.

Let $x^{0} := c\,t$, and $x := (x^{\mu}) = (x^{0},\, {\bf x})$, where
${\bf x} := (x^1,\,x^2,\,x^3) \in \BBn{E\mini}{3}$ (the
three-dimensional euclidean space). We introduce the {\em covariant 
derivative$\,$} in the $\mu$-direction by setting

\be
D_{\mu} \::=\: \dd{x^{\mu}} + i\mini a_{\mu}(x) + \rho_{\mu}(x)\ ,
\hspace{1cm} \mu = 0, \ldots ,3 \ ,
                          \label{D1}
\ee

\noi
where the real-valued functions $\,a_{\mu}\,$  are given by

\be
a_0(x) \::=\: {q \over \hbar c} \Phi(x) \ , \hspace{.5cm} \mbox{and}
\hspace{.5cm} a_k(x) \::=\: - {q \over \hbar c}A_k(x) \
,\hspace{.5cm} k=1,2,3\ ,               \label{D2}
\ee

\noi
and where the $\,su(2)$-valued functions $\,\rho_{\mu}\,$ are
defined by

\be
\rho_{\mu}(x) \::=\: i \sum_{A=1}^{3} \rho_{\mu A}(x)\, L_A^{(s)} \ ,
\hspace{1cm} \mu = 0,\ldots ,3 \ .
                          \label{D3}
\ee

\noi
The coefficients $\,\rho_{\mu A}\,$ are given by

\be
\rho_{0A}(x) \::=\: - {g\mini\mu \over 2 \hbar c}\, B_A(x) \ ,
\hspace{1cm} A=1,2,3 \ ,
                            \label{D4}
\ee

\mam

\be
\rho_{kA}(x) \::=\: \left( -{g\mini\mu \over 2 \hbar c} + {q \over
4\mini mc^2} \right) \sum_{B=1}^{3} \eps{kAB}\, E_B(x) \ ,
\hspace{1cm} A,\, k=1,2,3 \ ,
                            \label{D5}
\ee

\noi
where $\,\eps{kAB}\,$ is the sign of the permutation $(k\,A\,B)$
of $(1\,2\,3)$.

With the help of the covariant derivatives $\,D_{\mu}$ we are able
to write the Pauli equation (\ref{gen}) in the compact form

\be
i\hbar c\, D_{0}\, \psis(x) \:=\: - {\hbar^2 \over 2m} \,
\sum_{k=1}^{3} D_{k}D_{k}\, \psis(x) \ ,
                       \label{cov}
\ee

\noi
where a term of order ${\cal O}(\rho^{2})$ has been added. For an
electron it can be seen to be equal to half of $\,{e^2 \hbar^2 \over
8\mini m_o^3 c^4}\,{\bf E}^2\,\psi\,$, which can be absorbed into a
one-body potential acting on $\,\psi$.

The form (\ref{cov}) of the Pauli equation shows that
non-relativistic quantum mechanics has a $\US\ ${\em gauge
symmetry$\mini$}. The {\em gauge transformations} are defined as
follows:

\be
\begin{array}{ll}
U(1)_{em}:\ 
  & a_{\mu}(x) \:\mapsto\: \mbox{}^{\chi}a_{\mu}(x) \::=\: a_{\mu}(x)
+ (\partial_{\mu}\,\chi)(x)
\rule[-4mm]{0mm}{5mm}  \\
  & \psis(x) \:\mapsto\: \mbox{}^{\chi}\psis(x) \::=\: e^{-i\chi
(x)}\,\psis(x) \ , \end{array}
                            \label{ug}
\ee

\noi
where $\,\chi\,$ is an arbitrary, real-valued function on
space-time $\,\mbox{\BBB{R}} \times \BBn{E\mini}{3}$, and

\be
\begin{array}{ll}
SU(2)_{spin}:\ 
  & \rho_{\mu}(x) \:\mapsto\: \mbox{}^{g}\rho_{\mu}(x) \::=\: g(x)\,
\rho_{\mu}(x)\, g^{-1}(x) + g(x)\, (\partial_{\mu}\,g^{-1})(x)
\rule[-4mm]{0mm}{5mm}  \\
  & \psis(x) \:\mapsto\: \mbox{}^{g}\psis(x) \::=\: g(x)\,\psis(x) 
\ , \end{array}
                           \label{sg}
\ee

\noi
where $\,g\,$ is (the spin-$s\,$ representation of) an arbitrary
$\,SU(2)$-valued function on $\,\mbox{\BBB{R}} \times \BBn{E\mini}{3}$.
Note that, for constant gauge transformations $\,g$, $\,\rho_\mu\,$
transforms according to the adjoint action of $\,SU(2)\,$ (on its
Lie algebra $\,su(2)\mini$) which, by (\ref{D4}) and (\ref{D5}),
(in an active interpretation) corresponds to global rotations of
the fields $\,${\bf E}$\,$ and $\,${\bf B}$\,$ in physical space.
For space-time dependent gauge transformations, there appears an
additional inhomogeneous term in (\ref{sg}). A full geometrical
interpretation of the \SU\ gauge symmetry will be given in the next
section. See also the work by Anandan~(1989, 1990) for related
observations.

We note that Eq.~(\ref{cov}) can be thought of as the
{\em Euler-Lagrange equation} corresponding to the following
$\USo\ ${\em gauge-invariant$\,$} action functional

\bea
S(\psiss,\, \psis;\, a,\, \rho) & = & \int dt\, d^{\,3}{\bf x}\,
\biggl[\, i\hbar c\, \psiss (x) (D_{0}\, \psis)(x)  Ê\nonumber  \\
&  & \mbox{}- {\hbar^2 \over 2\mini m} \, \sum_{k=1}^{3} (D_{k}
\psis)^* (x)\, (D_{k} \psis) (x) \biggr] \ .
                         \label{act}
\eea

\noi
This action functional and generalizations thereof provide a
convenient starting point for a functional integral formulation of
non-relativistic many-body theory. 

We illustrate the formalism described so far by reviewing some basic
effects in non-relativistic quantum mechanics from the point of view
of its $\USo$\ gauge symmetry.


\subsection{Aharonov-Bohm Effect}
\label{ssAB}

A key effect reflecting Weyl's \Ue\ gauge principle realized
in quantum theory is the {\em Aharonov-Bohm effect$\,$}~(Aharonov
and Bohm, 1959): Consider the scattering of quantum-mechanical
particles at a magnetic solenoid. [The wave functions of the
particles are required to vanish inside the solenoid.] Then the
diffraction pattern seen on a screen depends non-trivially on the
magnetic flux, $\mini\Phi$, through the solenoid. The dependence is
periodic with period $\,hc \over q$, where $\,q\,$ is the charge of
the particles. This is a consequence of the fact that the vector
potential $\mini{\bf A}\mini$ outside the solenoid cannot be gauged
away {\em globally$\mini$}, in spite of the fact that there is no
electromagnetic field, thus leading to {\em non-integrable
\mbox{$\,U(1)$}\ phases$\,$} of quantum-mechanical wave functions
which change interference patterns. 

In formulae, we have that $\,F_{\mu \nu} := \partial_\mu a_\nu
- \partial_\nu a_\mu = 0\,$ outside the solenoid. Thus,
locally, $\,a_\mu = \partial_\mu \chi$, with $\,\chi({\bf x}) =
{-q\over \hbar c} \int_{\ast}^{\bf x} {\bf A} \cdot d
\mbox{\boldmath $l$} \,$, where $\,d \mbox{\boldmath $l$} \,$
denotes the line element along some path of integration from an
arbitrary point $\ast$ to {\bf x}. The phase factors affecting the
interference patterns are then given by $\,\exp\!\left[ 2\pi i{q
\over h c} \oint_{\, \Gamma} {\bf A} \cdot d \mbox{\boldmath
$l$} \right] = \exp\!\left[ 2\pi i {q\mini \Phi \over h c} \right]$,
where $\,\Gamma\,$ is a closed path enclosing the solenoid.

The Aharonov-Bohm effect explains the possibility of
{\em fractional$\,$} (or $\theta$- or {\em abelian braid-})
{\em statistics$\,$} of anyons~(Leinaas and Myrheim, 1977; Goldin,
Menikoff, and Sharp, 1980, 1981, 1983; Wilczek, 1982a, 1982b; for
a review, see Fr\"ohlich, 1990) in two-dimensional systems: Anyons
are particles carrying both electric charge $\,q\,$ and magnetic flux
$\Phi$ ($= \sH^{-1} q\,$ where $\,\sH\,$ is a ``Hall conductivity'')
and hence give rise to Aharonov-Bohm phases which one can interpret
as statistical phases; see Subsect.~\ref{ssQP}.


\subsection{Aharonov-Casher Effect}
\label{ssAC}

One might wonder whether there is a similar interference effect
due to the $SU(2)_{spin}$ gauge symmetry of non-relativistic
quantum mechanics. The answer is yes: It is the {\em Aharonov-Casher
effect$\,$}~(Aharonov and Casher, 1984). Consider a system of
quantum-mechanical particles with spin $\mini s$, electric charge
$\,0$, but with a magnetic moment  $\,\mu_{spin} \neq 0$, moving in
a plane or in three-dimensional space. [The particles could be
neutrons, or neutral atoms, $\ldots\;$.] Following Aharonov and
Casher, we study the influence of a (static) external
electric field on the dynamics of such particles. As a consequence
of relativistic effects, the moving particles will, in their rest
frame, feel a magnetic field that interacts with their magnetic
moment. Up to order ${\cal O}(v/c)$, this is taken into account by
the {\em spin-orbit term} in the Pauli equation (\ref{gen}); see also
(\ref{SP}).

In the formalism developed above, this effect should be described as
follows: The $\,SU(2)$\ gauge potential $\,\rho_\mu\,$ is defined in
Eqs.~(\ref{D3})--(\ref{D5}), and we find that

\be
\rho_{0A}({\bf x}) \:=\: 0\, , \hspace{.5cm} \mbox{and} \hspace{.5cm}
\rho_{kA}({\bf x}) \:=\:  -{g\mini\mu \over 2 \hbar c} \sum_{B=1}^{3}
\eps{kAB}\, E_B({\bf x}) \ , \hspace{.5cm} A,\, k=1,2,3 \ .
                        \label{AC1}
\ee

\noi
For general electric fields, the $\,SU(2)$\ curvature, defined by

\ben
G_{\mu \nu}^A \::=\: \partial_\mu \rho_{\nu A} - \partial_\nu
\rho_{\mu A} - 2\! \sum_{B,C = 1}^3 \eps{ABC}\,\rho_{\mu B}\mini
\rho_{\nu C} \, , \hspace{.5cm} A = 1,2,3, \  \mbox{and}\  \mu, \nu
= 0, \ldots ,3 \:, 
\een

\noi
does not vanish on full-measure sets of space, and so we are not
surprised to find that the electric field $\,{\bf E}\,$ causes
non-trivial spin-orbit interactions. However, if we consider a
system of particles confined to the $\mini(x,\mini y)$-plane in
$\,\BBn{E\mini}{3}\,$ which move in the electric field of a charged wire
placed along the $\,z$-axis, with constant charge $\,Q\,$ per unit of
length, we encounter an \SU\ {\em version of the Aharonov-Bohm
effect$\,$}: Here, the electric field $\,{\bf E}\,$ is given by
$\,{\bf E}({\bf x}) = {Q \over 2\pi r^2}(x,\, y,\, 0)$, where $\,r =
\sqrt{x^2 + y^2}$, and, with (\ref{AC1}), we find

\be
\bgrec{\rho}({\bf x})\::=\: ( \rho_{13}({\bf x}),\,
\rho_{23}({\bf x}) )\: =\: {g\mini \mu\, Q \over 4\pi \hbar c\,
r^2}\, (y,\, -x) \ ,
                        \label{AC2}
\ee

\noi
and $\,\rho_{i1} = \rho_{i2}=0$, for $i = 1,2$. Note that
$\rho_{3A}$ -- which does {\em not} vanish for $A=1,2$ -- does not
enter the dynamics of a system confined to the $\mini(x,\mini
y)$-plane. One then easily checks that, for a two-dimensional system
confined to the $\mini(x,\mini y)$-plane, the only component of the
$\,SU(2)$\ curvature which does not vanish identically is given by

\ben
G_{12}^{\mini 3}({\bf x}) \:=\: - {g\mini \mu \over 2 \hbar c}\,Q \,
\delta ({\bf x}) \ . 
\een

\noi
The distribution $\,G_{12}^{\mini 3}\,$ is supported at the
origin, i.e., the $\,SU(2)$\ connection $\,\rho\,$ is ``flat''
outside the wire. Thus, locally, it is possible to write $\,\rho\,$
as a pure gauge, i.e., $\,\rho_k = g\, \partial_k g^{-1}$, with $\,g
= \exp\!\left[ -i \int_\ast^{\bf x} \bgrec{\rho} \cdot d
\mbox{\boldmath $l$} \; L_3^{(s)} \right]$, where $\,d
\mbox{\boldmath $l$}\,$ is as above. However, the scattering of the
particles at the charged wire depends on its charge per unit length,
$\,Q$, because, although $\,\rho\,$ is flat except at the origin, it
cannot be gauged away {\em globally$\mini$}. Therefore, $\,\rho\,$
gives rise to ``{\em non-integrable \mbox{$SU(2)$}\ phase
factors$\,$}'' in the wave functions of the particles which affect
their interference patterns. These phase factors are given by $\:
\exp\Bigl[ i \oint_{\,\Gamma} \bgrec{\rho} \cdot d \mbox{\boldmath
$l$} \Bigr] = \exp\!\left[ 2\pi i {g\mini \mu \over 2hc} Q \right]$,
where $\,\Gamma\,$ is a path enclosing the wire, and the patterns are
periodic in $\,Q\,$ with a period given by $\,2hc \over g\mini \mu$.

This effect was first described by Aharonov and Casher (1984) in a
somewhat more classical language. A general discussion
of this effect, much along the lines of thought sketched above,
can be found in Anandan~(1989, 1990).

We recall that the Aharonov-Bohm effect explains why two-dimensional
quantum theory can describe anyons with fractional statistics,
namely particles carrying charge and flux. It is natural to ask
whether the Aharonov-Casher effect also has something to do with
exotic statistics in two-dimensional quantum theory. The answer is
yes! The Aharonov-Casher effect is closely related to the existence
of particles in two-dimensional quantum theory with {\em
non-abelian braid statistics$\,$} (Fr\"ohlich, 1988; Fredenhagen, Rehren,
and Schroer, 1989; Fr\"ohlich and Gabbiani, 1990; Fr\"ohlich,
Gabbiani, and Marchetti, 1990; Fr\"ohlich and Marchetti, 1991). Such
particles have topological interactions that can be described by an 
$\,SU(2)$\ Knizhnik-Zamolodchikov connection (Knizhnik and
Zamolodchikov, 1984; Tsuchiya and Kanie, 1987). Consider, for
example, a two-dimensional chiral spin liquid made of particles with
spin $\,s_o \geq 1\,$ (if such systems exist). An incompressible
chiral spin liquid of this type will most likely exhibit excitations
of arbitrary spin $\,s = \halb, \ldots, s_o$. The claim is that an
excitation of non-zero spin $\,s < s_o\,$ exhibits non-abelian braid
statistics, (as pointed out in Fr\"ohlich, Kerler, and Marchetti
(1992)). This will be discussed further in Subsect.~\ref{ssQP}.




\section{Gauge Invariance in Non-Relativistic Quantum Many-Particle
Systems}
\label{sGI}

In this section, we build upon and generalize the formalism
outlined in the previous section. It is our aim to describe
non-relativistic quantum-mechanical systems in (one), two, and
three space dimensions, composed of particles with arbitrary spin
coupled to external electromagnetic fields, variable background
metrics, and affine spin connections on spaces of non-vanishing
curvature and torsion.

The energy-mass
distribution of the background gives rise to curvature which describes
gravitational fields; (torsion is
assumed to vanish in gravity). Torsion and curvature can also
provide an effective description of crystalline backgrounds with
dislocations and disclinations. Such a geometric description of the
background is reasonable, provided the energy of a quantum-mechanical
particle is so small that the lattice structure of the background
cannot be resolved, and the background may be treated as a
``smooth'' manifold, i.e., provided the typical wave length of 
particles is much larger than the crystal lattice spacing.
Furthermore, non-trivial background metrics can account for
off-diagonal disorder in the systems and for a variable effective
mass.

We begin this section by reviewing a geometrical framework which is
well suited to describe all these phenomena; (for more mathematical
background, see, e.g., Eguchi, Gilkey, and Hanson, 1980, Bleecker,
1981, and de Rham, 1984; for a brief summary of basic notions in
differential geometry, see also Sect.~2 in Alvarez-Gaum\'e and
Ginsparg, 1985). Apart from describing possible physical effects
related to curvature and torsion, the purpose of the general
formalism developed here is to elucidate the geometrical meaning and
origin of the $\US$\ gauge invariance of non-relativistic quantum
mechanics.

Since we are interested in time-dependent many-body systems, it will
be convenient to work in a Feynman-Berezin path integral formalism
(see, e.g., Negele and Orland, 1987, Fradkin, 1991, and Feldman,
Kn\"orrer, and Trubowitz, 1992). We show that the action functionals
governing such systems are $\USo$\ gauge-invariant. This gives rise
to powerful Ward identities which are central in a general treatment
of incompressible quantum fluids in two dimensions and their
generalized Hall effects; see Sect.\,\ref{sSL} below, and
Fr\"ohlich and Studer (1993b). At the end of this section, we present
a quantum-mechanical version of Larmor's theorem, including spin
degrees of freedom. Further applications of the general formalism
developed in this section to basic effects in quantum many-body
theory will be given in Sect.~\ref{sKE}.


\subsection{Differential Geometry of the Background}
\label{ssDG}

As stated above, the
main reasons for introducing the following geometrical framework are
an elucidation of the geometrical meaning of the $\US$\ gauge
invariance of non-relativistic quantum mechanics and a preparation
for treating such systems in ``moving coordinates''; see
Subsect.~\ref{ssMC}.

Under the condition of low energy described above, physical space is
a $(d = 2\ \mbox{or}\ \,3)$-dimensional manifold, $\,M$, with a
possibly time-dependent riemannian metric, and space-time is given
by $\,N := \BB{R} \times M$. The system is confined to the interior
of a space-time cylinder $\,\Lambda \subset N$. The intersection of
$\,\Lambda\,$ with a fixed-time slice is denoted by
$\,\Omega_{\mini t}\,$, where $\,t\,$ is time. In local coordinates,
points in $\,M\,$ are denoted by $\,{\bf x},\, {\bf y}, \ldots\ $,
points in $\,N\,$ by $\,x:=(t,{\bf x}),\ y:=(t,{\bf y}), \ldots\;$.
The riemannian metric on $\,M\,$ is denoted by $\,g_{ij}(t,{\bf
x})$, and space-time $\,N\,$ carries the ``lorentzian'' metric
$\,\eta_{\mu \nu}(x)$, where $\,\eta_{00}(x)=1,\ \eta_{0i}(x) =
\eta_{i0}(x)=0,\  \eta_{ij}(x)=-g_{ij}(t,{\bf x})$, where the
indices range over $\,i,j = 1,\ldots ,d\,$, and $\,\mu ,\nu = 
0,\ldots ,d\,$. In the tangent space at a point $\,{\bf x} \in M\,$
we also have the flat cartesian metric, $\,\delta_{AB}\,$, with
$\,A,B = 1, \ldots ,d\,$. [Similarly, in the tangent space at a
space-time point $\,x \in N\,$ we have the usual ``minkowskian''
metric $\,\eta^0_{\alpha \beta}\,$, with $\,\alpha ,\beta = 0,\ldots
,d\,$.]

If the dimension of $\,M\,$ is two, we imagine that $\,M\,$ is a surface
embedded in a three-dimensional riemannian manifold $\,L\,$ with metric
also denoted by $\,g_{ij}(t,{\bf x})$, and the metric on $\,M\,$ is
the induced one.

Since, in {\em non-relativistic$\,$} quantum mechanics, time is
merely a {\em parameter$\,$} we temporarily omit it from our notations
and focus our attention on the description of the ``spatial''
geometry of $\,M\,$ or $\,L\,$, respectively. In order to be able to
describe particles of {\em arbitrary spin} $\,s = 0,\, \halb,\, 1,
\ldots\,$ moving in $\,M$, it is necessary to make use of the {\em
(co)tangent frame}~- or {\em dreibein formalism$\mini$}. This
formalism, involving {\em local$\,$} bases in the (co)tangent spaces
(orthonormal frames), naturally incorporates two {\em local$\,$}
symmetry groups: the group of coordinate reparametrizations of the
manifold (diffeomorphisms) and the group of local frame rotations
($SO(3)$\ gauge transformations). Wave functions of particles of
half-integral spin will transform under spinor representations of
the frame rotation group.

In the cotangent bundle to $\,L$, $\,T^{\ast}(L)$, we choose
(smooth) sections of 1-forms, $\,(\eaio{A}{})_{A=1}^3\,$, with the
property that they form an orthonormal basis (or {\em orthonormal
frame$\mini$}) in the cotangent spaces $\,T_{\bf x}^{\ast}(L),\
\,{\bf x}\in L$. The components of the orthonormal frame
$\,(\eai{A}{})_{A=1}^3\,$ in the coordinate basis $\,(dx^j)_{j=1}^3
\,$ of $\,T_{\bf x}^{\ast}(L)\,$ are denoted by $\,\eai{A}{i}\,$ and
are called {\em dreibein (fields)}. If $\,$dim\mini$M = 2\,$ we
choose $\,(\eai{A}{})_{A = 1}^3\,$ such that, for $\,{\bf x} \in M
\subset L, \  \eai{3}{}\,$ is orthogonal to $\,T_{\bf
x}^{\ast}(M)\,$ in the metric of $\,T_{\bf x}^{\ast}(L)$. The metric
on $L$ can be expressed in terms of the dreibein as
follows:$\mini$\footnote{Throughout this work, if not stated
explicitly, the Einstein summation convention over repeated indices
is understood.}

\be
g_{ij}({\bf x}) \:=\: \delta_{AB}\, \eai{A}{i}\, \eai{B}{j}\ .
                      \label{M1}
\ee

\noi
If $\,$dim\mini$\,M = 2$, we choose local coordinates on $\,L\,$ in
a neighbourhood of $\,M\,$ such that the metric on $\,M\,$ at a point
$\,\bx\,$ is given by

\be
g_{ij}({\bf x}) \:=\: \sum_{A,B=1}^2 \delta_{AB} \, \eai{A}{i}\,
\eai{B}{j} \ , \hspace{1cm} i,j=1,2\, ,
\ee

\noi
i.e., the coordinate $\,x^3\,$ is transversal to $\,M$. In the
following we focus on the geometry of $\,L$, thinking of the
``background manifold'' $\,M\,$ as being identified with $\,L\,$ (for
$\,d=3$), or as being a proper submanifold of $\,L\,$ (for
$\,d=2$)$\,$ embedded in $\,L\,$ in the way just explained.

The inverse of the dreibein $\,\eai{A}{i}\,$ is given by

\be
\Eai{A}{i} \::=\: \delta_{AB}\, g^{ij}(\bx)\, \eai{B}{j} \ ,
\ee

\noi
where $\,(g^{ij}(\bx))\,$ is the inverse matrix of $\,(g_{ij}(\bx))
\,$. Clearly,

\be
\Eai{A}{i}\, \eai{B}{i} \:=\: \delta_A^B\ , \hspace{.5cm} \mbox{and}
\hspace{.5cm} g^{ij}(\bx) \:=\: \delta^{AB}\, \Eai{A}{i}\, \Eai{B}{j}\ .
                       \label{M0}
\ee

To summarize, the dreibein $\,\eai{A}{i}\,$ is the matrix which
transforms the coordinate basis $\,(dx^i)\,$ of $\,1$-forms in
$\,T^\ast_\bx(L)\,$ to an orthonormal basis of $\,1$-forms
(orthonormal frame), $(\eai{A}{})$, in $\,T^\ast_\bx(L)$, i.e.,

\be
\eai{A}{} \:=\: \eai{A}{i}\, dx^i \ , \hspace{.5cm} \mbox{and}
\hspace{.5cm} g^{ij}(\bx)\, \eai{A}{i}\, \eai{B}{j} \:=\: \delta^{AB}\ .
                       \label{M2}
\ee

\noi
Similarly, $\Eai{A}{i}\,$ transforms the basis $\,(\dd{x^i})\,$ of
vector fields in $\,T_\bx(L)\,$ to an orthonormal basis of vector
fields, $(\Eai{A}{})$, in $\,T_\bx(L)\,$, i.e.,

\be
\Eai{A}{} \:=\: \Eai{A}{i}\, \dd{x^i} \:=\:
\Eai{A}{i}\,\partial_i \ , \hspace{.5cm} \mbox{and} \hspace{.5cm}
g_{ij}(\bx)\, \Eai{A}{i}\, \Eai{B}{j} \:=\: \delta_{AB} \ .
                       \label{M3} 
\ee

{}From Eq.~(\ref{M1}) it follows that the dreibien $\,\eaio{A}{i}\,$
is a ``square root'' of the metric $\,(g_{ij})$. This ``square
root'', however, is not unique. It is only defined up to {\em local
frame rotations$\mini$}. [Note that the dreibein $\,\eaio{A}{i}\,$
has 9 independent components while the metric $\,(g_{ij})\,$ has
only 6. It is the group of local frame rotations which accounts for
the mismatch: dim$\mini SO(3) = 3$.] Thus every cotangent space
$\,T^\ast_\bx(L)\,$, $\bx \in L$, carries a three-dimensional
(spin-1) representation, $R(\bx) \in SO(3)$, of the rotation group.
The rotations $\,R(\bx)\,$ act on the dreibein $\,\eai{A}{i}\,$ as
``{\em gauge transformations$\,$}'' in the following way:

\bea
\eai{A}{i} \:\mapsto\: \mbox{}^R \eai{A}{i} & := &
R(\bx)^A_{\hspace{.4em} B}\, \eai{B}{i} \ , \hspace{.5cm} \mbox{or}
\nonumber \\ e(\bx) \:\mapsto\: \mbox{}^R e(\bx) & := & R(\bx)\,
e(\bx) \ .          
                        \label{eGT1}
\eea

In order to define {\em parallel transport$\,$} on $\,L\,$ in the
(co)tangent frame formalism, one introduces the notion of an {\em
affine spin connection$\mini$},  $\,\omega^{A}_{\hspace{.4em} B}$. This
connection is an $\,so(3)$-valued $\,1$-form on $\mini L\mini$. It can
be expanded 
in the coordinate basis $\,(dx^i)$, or in the orthonormal frame
$\,(\eai{A}{})\,$ of $\,T^\ast_\bx(L)\mini$:

\be
\om{B}\: =\: \om{Bi}\mini dx^i \: =\: \om{BC}\, \eai{C}{i}\mini
dx^i \: =\: \om{BC}\mini\eai{C}{}   \ .
                             \label{M7}
\ee

\noi
Notice that, with the help of the dreibein and its inverse
(see (\ref{M2}) and (\ref{M3})), the indices of any tensor can be
changed at will from coordinate indices, $i,j, \ldots\:$, to frame
indices, $A,B, \ldots\;$.

\noi
Geometrically, the connection $\,\omega^{A}_{\hspace{.4em} B}\,(\xi;x)$
determines an isomorphism from $T_x^*(L)$ to $T_{x+\xi}^* (L)$, where
$\bxi = (\xi^i)$ is an infinitesimal vector, and 

\be
\omega^{A}_{\hspace{.4em} B} (\bxi\,;\, x)\,:=\,
\omega^{A}_{\hspace{.4em} Bi} (x) \xi^i (x)\; . 
                        \label{pt}
\ee

A tensor important to characterize the affine spin connection
$\,\omega^{A}_{\hspace{.4em} B}\,$ is the {\em torsion
2-form$\mini$}, ${\cal T}^A$, associated to $\,\eaio{A}{}\,$ and
$\,\omega^{A}_{\hspace{.4em} B}$. It is defined through {\em
Cartan's first structure equation}

\bea
{\cal T}^A (\bx) & = & {\cal T}^A_{\hspace{.7em} ij} (\bx)\, dx^i
\wedge\! dx^j  \nonumber   \\
 & = & \da \eai{A}{} + \om{B} \wedge\! \eai{B}{} \ ,
                        \label{M4}
\eea

\noi
where $\,\da\,$ denotes exterior differentiation (given in local
coordinates by $\,\da = dx^i\, \dd{x^i}\,$) and $\,\wedge$
stands for the (totally antisymmetric) exterior product; see, e.g.,
Eguchi, Gilkey, and Hanson (1980), Bleecker (1981), and de Rham
(1984).

It is customary to decompose the affine spin connection into two
parts:

\be
\om{B} \:=\: \lam{B} + \ka{B} \ ,
                        \label{M5}
\ee

\noi
where $\,\lambda^{A}_{\hspace{.4em} B}\,$ is the {\em
Levi-Civita connection$\,$} and $\,\kappa^{A}_{\hspace{.4em}
B}\,$ is the so-called {\em contorsion field$\mini$}. The
Levi-Civita connection plays a prominent role in general
relativity, and hence it is important if we wish to study quantum
mechanics in a curved space-time. On any riemannian manifold
$\,(L,\, g_{ij})$, it is uniquely determined by requiring that its
torsion vanishes, i.e., if one replaces $\,\omega^{A}_{\hspace{.4em}
B}\,$  by $\,\lambda^{A}_{\hspace{.4em} B}\,$ in Eq.~(\ref{M4}) the
resulting expression has to vanish. The components
$\,\lambda^{A}_{\hspace{.4em} Bi}\,$ can be expressed purely in
terms of the dreibein $\,e^{A}_{\hspace{.4em} i}\mini$, its
derivatives, and its inverse $\,{\cal E}_{A}^{\hspace{.4em}i}\,$
(Eguchi, Gilkey, and Hanson, 1980; Alvarez-Gaum\'e and Ginsparg,
1985).

\bea
\lam{Bi} & = & {1 \over 2} \Bigl[\, \Eai{B}{k}\, \Bigl( \partial_k
\eaio{A}{i} - \partial_i \eaio{A}{k} \Bigr)(\bx)
+ \delta^{AC} \delta_{BD}\, \Eai{C}{k}\, \Bigl( \partial_i
\eaio{D}{k} - \partial_k \eaio{D}{i} \Bigr)(\bx)  \nonumber  \\
 & & + \mbox{}\, \delta^{AC} \delta_{DE}\, \eai{D}{i} \Eai{B}{k}
\Eai{C}{l}\, \Bigl( \partial_k \eaio{E}{l} - \partial_l \eaio{E}{k}
\Bigr)(\bx)\, \Bigr] \ .
                          \label{M8}
\eea

\noi 
The contorsion field $\,\kappa^{A}_{\hspace{.4em} B}\,$ contains
additional information about the affine geometry of $\,L$. This information
is relevant if one considers the motion of {\em
spinning$\,$} particles in $\,L\mini$; see Subsect.~\ref{ssSS}.

As a last geometrical notion we introduce the {\em curvature 2-form}
$\,{\cal R}^A_{\hspace{.4em} B}(\bx)\,$ of the connection
$\,\omega^{A}_{\hspace{.4em} B}\,$ on $\,L$. It is defined through
{\em Cartan's second structure equation}

\bea
{\cal R}^A_{\hspace{.4em} B}(\bx) & = & {\cal R}^A_{\hspace{.4em}
Bij}(\bx)\, dx^i \wedge\! dx^j \nonumber \\
 & = & \da \om{B} + \om{C} \wedge\! \omega^{C}_{\hspace{.4em}
B}(\bx)  \ .
                        \label{M6}
\eea

It is easy to deduce from Eqs.~(\ref{M4}) and (\ref{M6}) how
$\,\omega\,$ and $\,{\cal R}\,$ transform under the gauge
transformations (\ref{eGT1}) of the dreibein:

\bea
\omega(\bx) \:\mapsto\: \mbox{}^R \omega(\bx) & := & R(\bx)\,
\omega(\bx)\, R^T(\bx) + R(\bx)\, \da R^T(\bx) \ ,  \nonumber  \\
{\cal R}(\bx) \:\mapsto\: \mbox{}^R {\cal R}(\bx) & := & R(\bx)\,
{\cal R}(\bx)\, R^T(\bx) \ ,
                        \label{eGT2}
\eea

\noi
where the superscript $\,\mbox{}^T\mini$ denotes transposition of a
matrix. Furthermore, the following transformation properties follow
from Eq.~(\ref{eGT2}) and from the decomposition of $\,\omega\,$
into the two parts $\,\lambda\,$ and $\,\kappa$, as given
in~(\ref{M5}):

\bea
\lambda(\bx) \:\mapsto\: \mbox{}^R \lambda(\bx) & := & R(\bx)\,
\lambda(\bx)\, R^T(\bx) + R(\bx)\, \da R^T(\bx) \ ,  \nonumber  \\
\kappa(\bx) \:\mapsto\: \mbox{}^R \kappa(\bx) & := & R(\bx)\,
\kappa(\bx)\, R^T(\bx) \ .
                        \label{eGT3}
\eea

\noi
The contorsion field $\,\kappa\,$ transforms
{\em homogeneously$\,$} under gauge transformations, i.e., according
to the adjoint action of the gauge group.

\bigskip
We end this subsection with some remarks about the physical
relevance of the geometrical notions introduced above in connection
with crystalline backgrounds exhibiting dislocations and
disclinations. We summarize results contained, e.g., in
Kleinert~(1989; see also Katanaev and Volovich, 1992) where more
details can be found.

Let $\,\by_n \in \BBn{E\mini}{3}$ denote the lattice sites of a perfect
crystalline background. If the crystal suffers some distortion,
the original lattice sites get shifted to $\,\bx_n$, where
$\,\bx_{\mini n} = \by_n + {\bf u}(\bx_{\mini n})$, and {\em
defects$\,$} may form. In order to study these defects in the
framework of differential geometry, one assumes that the crystalline
background can be treated as a continuous (isotropic) medium. Then
$\,{\bf u}(\bx)\,$ is called the {\em total distortion
field$\mini$}. It describes a {\em singular$\,$} infinitesimal
transformation of euclidean space $\BBn{E\mini}{3}$ (with metric
$\,\delta_{ij}$) into a space $\,L\,$ containing {\em defects$\,$}
or, from a geometrical point of view, into a manifold with
non-vanishing {\em curvature$\,$} and {\em torsion$\mini$}.

Densities of different types of defects in $\,L\,$ can be expressed
in terms of (derivatives of) the distortion field $\,u_i(\bx)\,$ as
follows:

\noi
The {\em density of dislocations$\,$} (or translational defects) is
given by

\be
\alpha_{ij}(\bx) \::=\: \epps{i}{hk}\, \partial_h \partial_k
u_j(\bx)  \ ,
                             \label{de1}
\ee

\noi
the {\em density of disclinations$\,$} (or rotational defects) by

\be
\Theta_{ij}(\bx) \::=\: {1 \over 2}\, \epps{i}{hk} \epps{j}{mn}\,
\partial_h \partial_k \partial_m u_n(\bx) \ , 
\ee

\noi
and the general {\em defect density$\,$} by

\be
\eta_{ij}(\bx) \::=\: {1 \over 2}\, \epps{i}{hk} \epps{j}{mn}\,
\partial_h \partial_m \Bigl( \partial_k u_n + \partial_n
u_k \Bigr)(\bx) \ .
                           \label{de2}
\ee

\noi
These expressions are non-vanishing, in general, because
the distortion field $\,u_i(\bx)\,$ is {\em singular$\mini$}! In the
presence of a single defect line $\,\Gamma$, the dislocation and
disclination densities are both proportional to a $\delta$-function
along the line $\,\Gamma\mini$; see Kleinert~(1989). 

The geometric properties of the manifold $\,L\,$ are coded into its
metric $\,g_{ij}(\bx)\,$ and contorsion field $\,\kappa_{jhk}(\bx)$.
In terms of the distortion field $\,u_i(\bx)\,$ they are given, in
linear approximation, by

\be
g_{ij}(\bx) \:=\: \delta_{ij} - \Bigl( \partial_i u_j + \partial_j
u_i \Bigr)(\bx) \ ,
                             \label{de3}
\ee

\mam

\be
\kappa_{jhk}(\bx) \:=\: {1 \over 2} \Bigl[\,
\partial_k \Bigl( \partial_j u_h - \partial_h u_j \Bigr)\mmini(\bx)
- \partial_j \Bigl( \partial_h u_k + \partial_k u_h
\Bigr)\mmini(\bx) + \partial_h \Bigl( \partial_j u_k + \partial_k
u_j \Bigr)\mmini(\bx)\, \Bigr] \ ,
                             \label{de4}
\ee

\noi
where $\,\kappa_{jhk} = \delta_{AC}\, e^{C}_{\hspace{.4em} j}\,
e^{B}_{\hspace{.4em} h}\, \kappa^A_{\hspace{.4em} Bk}\mini$; see
(\ref{M1}) and (\ref{M5}).

This allows for a comparison of the defect densities of $\,L$, given in
(\ref{de1}) to (\ref{de2}), with the expressions for torsion and
curvature of the manifold $\,L$; see (\ref{M4})--(\ref{M6}). The
following relations hold:

\bea
\alpha_{ij}(\bx) & = & \epps{i}{hk} \kappa_{jhk}(\bx) \,  \\
\Theta_{ij}(\bx) & = & {\cal R}_{ij}(\bx) - {1 \over 2}\,
g_{ij}(\bx) \, {\cal R}(\bx) \ , \hspace{.5cm} \mbox{and}  \\
\eta_{ij}(\bx) & = & \stackrel{\rm LC}{\cal R}_{ij}(\bx) - {1
\over 2}\, g_{ij}(\bx) \, \stackrel{\rm LC}{\cal R}(\bx) \ ,
\eea

\noi
where $\,{\cal R}_{ij} := {\cal R}^A_{\hspace{.4em} ijA} = 
{\cal E}_{A}^{\hspace{.4em} k}\, e^B_{\hspace{.4em} i}\, {\cal
R}^A_{\hspace{.4em} Bjk}\,$ is the {\em Ricci tensor$\mini$}, and
$\,{\cal R} := g^{ij}\, {\cal R}_{ij}\,$ is the {\em scalar
curvature$\,$} of the affine spin connection
$\,\omega^{A}_{\hspace{.4em} Bj}$. Similarly, $\,\stackrel{\rm
LC}{\cal R}_{ij}\,$ and $\, \stackrel{\rm LC}{\cal R}\,$ denote the
Ricci tensor and scalar curvature, respectively, of the
Levi-Civita connection $\,\lambda^{A}_{\hspace{.4em} Bj}$, the
torsion-free part of the affine spin connection; see Eq.~(\ref{M5}).

For quantum-mechanical particles moving in a crystalline
background with defects, the following assumption appears to be
reasonable: If the energy of the particles is so small that the
lattice structure of the background cannot be resolved, then a
(metrically non-trivial) riemannian manifold provides an effective
description of the background when studying the {\em orbital$\,$}
motion of the particles. [Technically, the Laplacian in the
Schr\"odinger-Pauli equation should be replaced by the
Laplace-Beltrami operator associated with the riemannian metric on
the manifold.] Moreover, the formalism presented above is well
adapted to describing the orbital motion of particles confined
(e.g., by some potential) to a curved surface in $\,\BBn{E\mini}{3}$.

The question of the motion of the spin degrees of freedom, however,
is more subtle and will be addressed in the next subsection.


\subsection{Systems of Spinning Particles Coupled to External
Electromagnetic and Geometric Fields}
\label{ssSS}

We start this section by showing how to describe systems of
{\em spinning$\,$} particles moving in a geometrically non-trivial
background and coupled to an external electromagnetic field.

We assume that the manifold $\,L\,$ admits a {\em spin
structure$\mini$}. Then we may introduce {\em spinor bundles$\,$}
over $\,L\,$ (associated to the cotangent bundle $\,T^\ast(L)\,$
over $\,L$). Let $\,s=0,\, \halb,\, 1, \ldots\,$ denote the spin of
the particles, i.e., $\,2s+1\,$ is the dimension of an irreducible
representation of $\,SU(2) = \widetilde{SO(3)}\,$ with spin $\mini
s$. The fiber of the spin-$s\,$ spinor bundle, $\,\Es\mmini (L)$,
over $\,L\,$ is isomorphic to the $\,(2s+1)$-dimensional Hilbert
space, $\,{\cal D}^{(s)}$, carrying the spin-$s\,$ representation of
$\,SU(2)$. {\em Sections$\,$} of the spin-$s\,$ spinor bundle are
denoted by $\,\psisx$. {}From now on, we choose the gauge
transformations $\,R(\bx)\,$ to be $\,SU(2)$-valued. The action of
these gauge transformations on the cotangent bundle $\,T^\ast(L)\,$
is given by their adjoint (spin-1) representation, also denoted by
$\,R(\bx)$; see (\ref{eGT1}).\footnote{There is little danger of
confusion.} Under a gauge transformation $\,R(\bx)$, a section
$\,\psisx\,$ of $\,\Es\mmini (L)\,$ transforms as follows:

\be
\psisx \:\mapsto\: \mbox{}^R\psisx \::=\: \Us(R(\bx))\, \psisx \ ,
                          \label{psigt1}
\ee

\noi
where $\,\Us\,$ is the spin-$s\,$ representation of $\,SU(2)$. The
transition functions of the spin-$s\,$ spinor bundle
$\,\Es\mmini (L)$ -- which must be specified if the topology of the
base manifold $\,L\,$ is non-trivial -- are inherited from the
transition functions of the cotangent bundle $\,T^\ast(L)\,$ by
lifting them to the spin-$s\,$ representation of $\,SU(2)$. [Since
we have assumed that $\,L\,$ admits a spin structure, this is possible
even if $s$ is half-integer!]
 
Physically, what is meant by ``spin up'' or ``spin down'' is now a
{\em local notion$\mini$}, depending on the point $\,\bx \in L\,$
at which the spin is located and determined by the local frames
$\,(\eai{A}{})_{A = 1}^3\,$; see Subsect.~\ref{ssDG}.

The spaces of wave functions in non-relativistic, one-particle quantum
mechanics are Hilbert spaces of sections of these spinor bundles. In
non-relativistic quantum mechanics, wave functions are
complex-valued. We therefore tensor the fiber space $\,{\cal
D}^{(s)}\,$ -- real when $\,s\,$ is an integer -- by \BB{C}$\,$. The
structure group of the resulting complexified bundle, denoted by
$\,\Es_{\mbox{\bbviii C}}\mmini (L)$, is then $\,U(1) \times SU(2)$.
The factor $\,U(1)\,$ (phase transformations of $\,\psis$) is
connected to electromagnetism, as recognized by Weyl more
than sixty years ago (Weyl, 1928; see also Weyl, 1918).

In order to keep our notation simple, it is advantageous to
formulate quantum mechanics of many-particle systems in the
language of second quantization. The sections $\,\psisx\,$ of
$\,\Es_{\mbox{\bbviii C}}\mmini (L)\,$ over $\,L\,$ are then
interpreted as {\em operator-valued distributions$\,$} acting on
Fock space and subject to {\em canonical equal-time commutation
or anti-commutation relations} 

\bea
\biggl[\psis_\alpha (\bx),\: \psiss_\beta (\by) \biggr]_\pm & = & {1
\over \sqrt{g(\bx)}}\, \delta_{\alpha\beta}\, \delta(\bx - \by) \ , 
\hspace{.5cm} \mbox{and}   \nonumber  \\
\biggl[\psisss_\alpha (\bx),\: \psisss_\beta (\by)
\biggr]_\pm & = & 0 \ , \hspace{1cm} \alpha,\, \beta = 1, \ldots ,\,
2\mini s+1 \ ,
                           \label{CR}
\eea

\noi
where $\,[\ ,\ ]_+\,$ denotes the anticommutator and $\,[\ ,\ ]_-\,$
the usual commutator, $\,\psisss = \psis \ \mbox{or}\ \psiss\, ;\
\psiss$, the {\em creation operator$\mini$}, is the adjoint (on Fock
space) of $\,\psis$, the {\em annihilation operator}$\, ;\ g(\bx)\,$
denotes the determinant of the metric $\,(g_{ij}(\bx))\,$ on $\,L$. The
usual connection between spin and statistics is to choose
anticommutators in (\ref{CR}), corresponding to Fermi statistics,
when $\,s\,$ is half-integer, and commutators, corresponding to Bose
statistics, when $\,s\,$ is integer.

Our purpose is to specify some non-relativistic dynamical laws
governing the time-evolution of the operators $\,\psisss\,$ in the {\em
Heisenberg picture$\mini$}. Let $\,\psisss(x) = \psisss(t,\mini
\bx)\,$ denote the Heisenberg picture creation~- and annihilation
operators with initial conditions $\,\psisss(0,\mini \bx) = \psisss
(\bx)\,$. Geometrically, these operators are sections of a trivially
extended spin-$s\,$ spinor bundle, $\,\Es_{\mbox{\bbviii
C}}\mmini({\BB R} \times L)$, over the space-time manifold, $\,{\BB
R} \times L\mini$. In order to formulate local dynamical laws for
$\,\psisss(x)$, we need to be able to differentiate these fields in
$\,t\,$ and $\,\bx$. This necessitates introducing a notion of {\em
parallel transport$\,$} in $\,\Es_{\mbox{\bbviii C}}\mini({\BB R}
\times L)$. Parallel transport in the spinor bundle
$\,\Es_{\mbox{\bbviii C}}\mmini({\BB R} \times L)\,$ is defined with
the help of a $\,U(1) \times SU(2)$ {\em connection$\mini$}, i.e., by
a {\em vector potential$\,$} with values in $\,{\BB R}\,
\oplus\,su(2)$. Once
such a connection is fixed, derivatives of sections $\,\psisss(x)\,$
are defined as {\em covariant derivatives$\mini$}. Setting $\,x^{0}
:= c\,t$, and $\,x := (x^{\mu}) = (x^{0},\, {\bf x})$, where $\,{\bf
x} \in L$, the covariant derivative in the $\mu$-direction is
defined by

\be
\D{\mu} \::=\: \dd{x^\mu} + i\mini \am{\mu} + \wms{\mu} \ ,
\hspace{1cm} \mu=0, \ldots ,\, 3 \ ,
                          \label{DD1}
\ee

\noi
where the real-valued $\mini 1$-form $\,a = \am{\mu}\mini dx^\mu\,$
is the $\,U(1)$ connection, and the $\,su(2)$-valued $\mini 1$-form
$\,w^{(s)} = \wms{\mu}\mini dx^\mu\,$ is the $\,SU(2)$ connection in
the spin-$s\,$ representation of $\,su(2)$, i.e.,

\be
\wms{\mu} \:=\: i \sum_{A=1}^3 w_{\mu A}(x)\, L^{(s)}_A \ ,
                          \label{DD2}
\ee

\noi
where we have adopted the same notation as in Eq.~(\ref{D3}):
$(L^{(s)}_A)_{A=1}^3\,$ are hermitian generators of $\,su(2)\,$ in
the spin-$s\,$ representation, normalized such that $\,L_A^{(1/2)} =
\sigma_A, \mbox{ where }\, \sigma_1,\, \sigma_2\, \mbox{ and }\,
\sigma_3\,$ are the standard Pauli matrices. 

We shall argue shortly that we can identify $\,a\,$ with
the electromagnetic vector potential (up to multiplication by
physical constants). This is no surprise, given the observations
in Subsect.~\ref{ssGI} (see (\ref{D1}) and (\ref{D2})). What about
$\,\wso$? First, from a geometrical point of view, it is clear that
the affine spin connection $\,\omega^{A}_{\hspace{.4em} B}$,
introduced in (\ref{M7}), enters the definition of $\,\wso\,$
since the spinor bundle  $\,\Es_{\mbox{\bbviii C}}\mmini ({\BB R}
\times L)\,$ is associated to the cotangent bundle $\,T^\ast({\BB R}
\times L)$, i.e., it inherits the geometrical structure of
$\,T^\ast({\BB R} \times L)$. Second, based on the observations in
Subsect.~\ref{ssGI} (see (\ref{D1}) and (\ref{D3}) to (\ref{D5})),
we expect the interaction of the external electromagnetic field with
the magnetic moment carried by the particles (Zeeman and spin-orbit
couplings) to be described by an additional term, $\,\rho^{(s)} =
\rms{\mu}\mini dx^\mu$, in the $\,SU(2)$ connection $\,\wso$. Since
the sum of $\,\omega\,$ and $\,\rho^{(s)}\,$ must be an
\SU\ connection, $\,\rso\,$ has to transform under \SU\ gauge
transformations according to the adjoint action of the gauge group.
We use the following notations:

\be
\wms{\mu} \:=\: \oms{\mu} + \rms{\mu} \ ,
                            \label{w1}
\ee

\mwm

\be
\oms{\mu} \:=\: {i \over 2} \sum_{A,B,C=1}^{3} \epps{A}{\, BC}\,
\omx{B\mu}\, L^{(s)}_C \ ,
                            \label{w2}
\ee

\noi
$\epps{A}{\, BC}= \eps{ABC}\,$ is the sign of the permutation
$\,(A\, B\, C)\,$ of $\,(1\, 2\, 3)\,$, and

\be
\rms{\mu} \:=\: i \sum_{A=1}^{3} \rho_{\mu A}(x)\, L^{(s)}_A \ .
                            \label{w3}
\ee

\noi
All notions introduced in Subsect.~\ref{ssDG} -- defined over space
$\,L\,$ and its cotangent bundle $\,T^\ast(L)\,$ -- can easily be
extended to space-time, $\, {\BB R} \times L$, and its cotangent
bundle, $\,T^\ast({\BB R} \times L) \simeq {\BB R} \times T^\ast(L)$.
In a non-relativistic setting, the space-time metric $\,\eta_{\mu
\nu}(x)\,$ has the property that $\,\eta_{0i}(x)=0=\eta_{i0}(x)$,
(see the beginning of Subsect.~\ref{ssDG}), and, as a consequence,
most ``temporal'' components of the different geometrical
fields introduced in Subsect.~\ref{ssDG} vanish. In (\ref{w2}),
$\,\omx{Bi},\ i=1,\,2,\,3$, is given by (\ref{M7}), (\ref{M5}) and
(\ref{M8}), and

\be
\omx{B0} \:=\:  {1 \over 2} \Bigl[\, \delta^{AC} \delta_{BD}\,
\Eaio{C}{k}(x)\, \partial_0 \eaio{D}{k}(x) - \Eaio{B}{k}(x)\,
\partial_0 \eaio{A}{k}(x)\, \Bigr] \ .
                         \label{w4}
\ee

Under an $\,SU(2)$ {\em gauge transformation} $\,R(x)$, i.e., under
{\em local frame rotations$\,$} in the cotangent bundle
$\,T^\ast({\BB R} \times L)$ (see (\ref{eGT1})), the different terms
of the $\,SU(2)$ connection $\,\wso\,$ transform as follows:

\be
\oms{\mu} \:\mapsto\: {}^R\oms{\mu} \::=\: \Us(R(x))\, \oms{\mu}\,
\Us(R(x))^\ast + \Us(R(x))\, \partial_\mu \Us(R(x))^\ast \ ,
                         \label{wgt1}
\ee

\noi
which can be inferred from (\ref{eGT2}), and

\be
\rms{\mu} \:\mapsto\: {}^R\rms{\mu} \::=\: \Us(R(x))\, \rms{\mu}\,
\Us(R(x))^\ast \ ,
                         \label{wgt2}
\ee

\noi
where ${}^\ast$ denotes the adjoint of a matrix.

If the metric on $\,L\,$ is time-independent $\oms{0}$ vanishes; see
(\ref{w2}) and (\ref{w4}). After a time-dependent \SU\ gauge
transformation, however, it may be different from zero.
Furthermore, the $\rso$-part of the \SU\ connection $\,\wso\,$ will,
in general, be different from zero. Geometrically, it corresponds to
an additional contorsion field yielding non-vanishing torsion; see
(\ref{M4}). The physical interpretation of the $\oso$-part of the
\SU\ connection $\,\wso$, as well as the precise identifications with
physical quantities of the $\rso$-part of $\,\wso\,$ and of the
\U\ connection $\,a\,$ will be given below. (The material in
Subsect.~\ref{ssGI} will serve us as a guide.)

Having introduced a $\,U(1) \times SU(2)$ connection and defined
covariant differentiation of the sections $\,\psisss$, we are now in
a position to formulate {\em local dynamical laws$\mini$}. It is
convenient to use the {\em lagrangian formalism$\mini$}, but we
could also work in the hamiltonian formalism; see Fr\"ohlich and
Kerler (1991). Let us consider a system of non-relativistic
particles of fixed spin $s$, and, to simplify our notations, we drop
the superscript ${}^{(s)}$ from the field operators $\,\psisss\,$
and the \SU\ connection $\,\wso=\oso+\rso$. Our ansatz for the
action of the system is an obvious generalization of the action
(\ref{act}) found in Subsect.~\ref{ssGI}. It reads (with
$\,x:=(ct,\, \bx)$)

\bea
S_\Lambda(\psi^\ast,\, \psi;\, g,\, a,\, w) & := & \int_\Lambda
\sqrt{g\tx}\, dt\, d^{\,3}{\bf x}\, \biggl[\, i\hbar c\, \psi^\ast(x)
(\D{0} \psi)(x)   \nonumber  \\   
 & & \hspace{-1.7cm} \mbox{} - {\hbar^2 \over 2m} \, g^{kl}(t,\mini
\bx)\, (\D{k} \psi)^\ast(x)\, (\D{l} \psi)(x) -
U(\psi^\ast\!,\,\mini\psi)(x)\, \biggr] \ ,
                         \label{Cact1}
\eea

\noi
where the covariant derivatives are given in (\ref{DD1}), $\,m\,$
is the {\em effective mass$\,$} of the particles (sometimes also
denoted by $\,m^\ast$; in common situations of solid state physics,
it can be considerably smaller than $\,m_o$, the mass of the
particles in the vacuum), and $\,U(\psi^\ast\!,\mini \psi)(x)\,$ is a
$\,U(1) \times SU(2)${\em -invariant$\,$} functional of $\,\psi\,$
and $\,\psi^\ast$, e.g., ($y:=(c\mini t,\, \by)$)

\bea
U(\psi^\ast\!,\mini \psi)(x) & := & v(x)\, \psi^\ast(x) \psi(x)
\nonumber     \\
 & & \hspace{-3.2cm} \mbox{} + {1 \over 2} \int_\Lambda
\!\sqrt{g\ty}\, d^{\,3}\by\, :\! \Bigl( \psi^\ast(x) \psi(x) - n
\Bigr)\, V(t,\mini\bx,\mini \by)\,\Bigl( \psi^\ast(y) \psi(y) -
n \Bigr)\!:\ .
                         \label{Cact2}
\eea

\noi
The double colons indicate Wick ordering, $\,v\,$ is a possibly
time-dependent one-body background potential (depending on the
background and on the scalar curvature $\,\cal R\,$ of $\,M\mini$),
$\,V\,$ is some two-body potential (e.g., for charged particles a
(possibly screened) Coulomb potential, or for neutral atoms or
molecules a van der Waals potential), and $\,n\,$ is approximately
equal to the background density of the system (related to its
chemical potential). We recall that $\,\Lambda \subset \BB{R} \times
M\,$ is a cylindrical region in space-time. At fixed time $\,t$, we
impose Dirichlet boundary conditions at the boundary, $\,\partial
\Omega_{\mini t}\,$, of the region $\,\Omega_{\mini t}\,$ to which
the system is confined.

The {\em field equation$\,$} (or {\em Euler-Lagrange equation}) for
$\,\psi\,$ or $\,\psi^\ast\,$ follows by setting the variation of
$\,S_\Lambda\,$ with respect to $\,\psi^\ast\,$ or $\,\psi$, equal
to zero. The resulting equation is a {\em generalization$\,$} of the
Pauli equation (\ref{cov}) to a system of spinning particles in a
geometrically non-trivial background. 

In order to illustrate these matters, let us consider a simple
situation: We choose space $\,M\,$ to be given by
$\,$\BBn{E\mini}{2} (the $(x,\mini y)$-plane in $\,L =
\BBn{E\mini}{3}$) or by $\,$\BBn{E\mini}{3}; $\,g_{ij}\tx =
\delta_{ij}$, for all times $\,t\,$ and all $\,\bx \in M, \ \Lambda
= \BB{R} \times \Omega\,$ where $\,\Omega\,$ is some
time-independent open set in $\,M$. The field equation for
$\,\psi\,$, obtained by varying the action $\,S_\Lambda\,$ defined
in (\ref{Cact1}) with respect to $\,\psi^\ast$, then reduces to the
Pauli equation given in (\ref{cov}). This equation coincides with
the usual form given in (\ref{gen}) (up to a modification of order
$\,{\cal O}(1/ m\mini m_o^2)$, see Subsect.~\ref{ssGI}) provided we
make the same identifications as in Eqs.~(\ref{D2})--(\ref{D5}): The
components of the \U\ connection $\,a\,$ (with respect to the
coordinate basis $\,(dx^\mu)\,$ of the cotangent space
$\,T_x^\ast(\BB{R} \times L)\,$) are given by the electromagnetic
potentials $\,\Phi\,$ and $\mini\bf A\mini$,

\be
a_0(x) \:=\: {q \over \hbar c} \Phi(x) \ , \hah a_k(x) \:=\: - {q
\over \hbar c}A_k(x) \ ,
                         \label{a}
\ee

\noi
where $\,q\,$ is the charge of the particles. Furthermore, the
components of the $\rho$-part of the \SU\ connection $\,w\,$ 
are expressed in terms of the electromagnetic field $\,({\bf
E},\mini {\bf B})\,$ as follows:

\be
\rho_{\mini 0A}(x) \:=\: - {g\mini\mu \over 2 \hbar c}\, B_A(x) \ ,
                            \label{rho1}
\ee

\noi
where $\,B_A(x)\,$ is the $\mini A$-component of the magnetic field
$\,{\bf B}(x)\,$ in the basis $\,(\eaix{A}{})\,$ of
$\,T_x^\ast(\BB{R} \times L)$, and

\be
\rho_{kA}(x) \:=\: \left( -{g\mini\mu \over 2 \hbar c} + {q \over
4\mini mc^2} \right) \sum_{B=1}^{3} \eps{kAB}(x)\, E_B(x) \ ,
                            \label{rho2}
\ee

\noi
where $\,E_B(x)\,$ is the $\mini B$-component of the electric field
$\,{\bf E}(x)\,$, and the symbol $\,\eps{kAB}(x)\,$ is defined by

\be
\eps{kAB}(x) \::=\: \eaix{C}{k}\, \eps{CAB} \ ,
                            \label{rho3}
\ee

\noi
where $\,\eps{CAB}\,$ is the sign of the permutation $\,(C\mini
A\mini B)\,$ of $\,(1\,2\,3)$. In Eqs.~(\ref{rho1})
and~(\ref{rho2}), the magnetic moment of the particles enters via
$\,g\mu\mini$; see (\ref{mamo}). Although the orthonormal frames
$\,\eaix{A}{}\,$ could be chosen to vary, it is simplest, in the
present situation, to choose them as $\,\eaix{A}{\mu}= \delta^A_{\,
\mu}\,$. Then the ``geometrical'' part $\,\omega\,$ of the
\SU\ connection $\,w\,$ clearly vanishes.

In a general situation, when the background of the system has the
structure of an {\em arbitrary riemannian spin manifold} $\,M$, the
physical interpretation of the connections $\,a\,$ and $\,w\,$ is
straightforward: The \Ue\ {\em connection} $\,a\,$ is still
expressed in terms of the {\em electromagnetic potentials$\mini$},
as in Eq.~(\ref{a}). The \SUs\ {\em connection} $\,w\,$ has been
given in Eq.~(\ref{w1}) with the ``geometric'' part $\,\omega\,$
being specified in terms of the {\em affine spin connection}
$\,\omega^{A}_{\hspace{.4em} B}\,$ on $\,\BB{R} \times L\, $; see
(\ref{w2}). Its $\rho$-part (describing {\em Zeeman$\,$} and {\em
spin-orbit couplings$\,$} of the magnetic moment of particles to the
external electromagnetic field) always contains the terms in 
(\ref{rho1})--(\ref{rho3}). The only difference is that, on a
general riemannian manifold, it is not possible to choose the
dreibein $\,\eaix{A}{\mu}\,$ to be {\em constant$\,$} on all of
$\,\BB{R} \times L$.

\bigskip
{\bf Remark}$\mini$.$\;$ We should comment on the physical
status of the $\,\omega$-part in the \SU\ connection $\,w$. In the
study of {\em gravitational fields$\mini$}, the affine spin connection
$\,\omega\,$ is torsion-free. It is given by the Levi-Civita
connection $\,\lambda\,$ which is canonically associated to the
gravitational metric field $\,g\,$; see~(\ref{M8}). Hence, if we
consider a quantum-mechanical system in a (strong) external
gravitational field then $\,\omega\,$ enters into the description of
the motion of the spin of the particles as a {\em fundamental$\,$}
physical field. 

At the end of Subsect.~\ref{ssDG}, we have also argued that the
geometrical framework of riemannian manifolds provides an effective
description for the {\em orbital$\,$} motion of low-energy particles
in a {\em crystalline background with defects$\,$} (or of particles
confined to a curved surface in $\,\BBn{E\mini}{3}$). Given the
Levi-Civita connection, $\,\lambda$, we must ask whether the
corresponding affine spin 
connection, $\,\omega=\lambda +\kappa\,$ (see~(\ref{M5})), might
provide an effective description of the interaction of the {\em
spin$\,$} of the particles with the crystalline background through
which they are moving. In general, this is not likely to be so! For
example, let us consider a spinning particle with vanishing magnetic
moment which moves in a crystalline background. Then, from the point
of view of basic one-body quantum mechanics (see
Subsect.~\ref{ssGI}), we do not expect that the dynamics of the spin
of the particle is coupled to the effective metric associated with
the background. (In this situation, the spin can be viewed as an
internal degree of freedom.) More generally, in one-body quantum
mechanics (in the absence of gravitational fields), the dynamics of
the spin of a particle (moving in some background or constrained to a
surface in $\,\BBn{E\mini}{3}$) is completely determined by the Zeeman
effect and by spin-orbit coupling, including the kinematical effect
of the Thomas precession. The $\,\rho$-part of the \SU\ connection
$\,w\,$ fully accounts for these effects, and $\,\omega\,$ can be
transformed to $\,0\,$ in a suitable \SU\ gauge.

We emphasize that the main reasons for introducing the
geometrical framework have been to elucidate, from a geometrical
point of view, the meaning and origin of the \SU\ gauge invariance
(i.e., the introduction of an \SU\ connection and of {\em local$\,$}
frame rotations), and to prepare for the description of
quantum-mechanical systems in ``moving coordinates'' which will be
the subject matter of the following subsection.

\bigskip
Finally, we recall that with the help of the dreibein
$\,\eaio{A}{i}\,$ and its inverse $\,\Eaio{A}{i}\,$ the components
of the electromagnetic field and its vector potential can easily be
changed from the form they take in orthonormal frames to the form
they take in local coordinates (see~(\ref{M2}) and~(\ref{M3})), e.g.,

\be
B_{A}(x) \:=\: \Eaix{A}{k}\, B_{k}(x) \ , \hah A_{k}(x) \:=\:
\eaix{C}{k}\, A_{C}(x) \ .
                            \label{rho4}
\ee 
 
Note that the expressions~(\ref{rho1}) and~(\ref{rho2}) are
consistent with the transformation law (\ref{wgt2}) of
$\,\rho^{(s)}_{\mu}\,$ under \SUs\ gauge transformations.
Moreover, defining \Ue\ gauge transformations as in
(\ref{ug}) (with $\,\chi\,$ an arbitrary, real-valued function on
$\,\BB{R} \times L$), the discussion above (see, in particular,
Eq.~(\ref{Cact1}) for the action $S_\Lambda$) proves that:

{\em Non-relativistic quantum mechanics of charged, spinning
particles moving in an external electromagnetic field and in a
geometrically non-trivial background is $\US$ gauge-invariant.}


\subsection{Moving Coordinates and Quantum-Mechanical Larmor
Theorem}
\label{ssMC}

We now imagine that the background of the system is {\em moving\mini}
on the manifold $M$ according to some {\em classical flow}
$\,\phi(t,\mini\cdot)$. Here $\,\phity\,$ is the position in $M$ of
a point particle at time $\,t\,$ starting at position $\,\by\,$ at
time 0. Then, in the $x$-coordinates (``laboratory coordinates'')
fixed to $\,\BB{R} \times M$, the one-body potential $\,v(x)\,$ and
the electric and magnetic fields $\,{\bf E}(x)\,$ and $\,{\bf
B}(x)\,$ created by the background are {\em time-dependent$\mini$}.
This implies that, in the (time-independent) $\,x$-coordinates on
$\BB{R} \times M$, the Hamiltonian of the system is {\em
time-dependent$\,$} which complicates the mathematical analysis of
the system. In particular, it complicates the analysis of its {\em
thermal equilibrium properties$\mini$}. It is quite clear,
physically, that approximate thermal equilibrium in such a system
will be reached {\em locally$\,$} in regions moving {\em with$\,$}
the background according to the flow $\,\phi(t,\mini\cdot)$. Thus,
we ought to formulate quantum mechanics in ``{\em moving
coordinates$\,$}'', $(y^1,\mini y^2,\mini y^3)$, where

\be
\bx \:=\: \phity \ , \hspace{.5cm} \mbox{i.e.,} \hspace{.5cm} \by \:=\:
\phi^{-1}\tx \ .
                           \label{T1}
\ee

\noi
In accordance with our non-relativistic treatment of quantum
theory, time will not be transformed, and in our calculations only
terms to order $\,{\cal O}(f/c)\,$ are taken into account, where
$\mini f\mini$ is the modulus of the velocity field of the moving
background, (see below). We shall see that the geometrical formalism
introduced in the first part of this section allows for a natural
description of the transformation to ``moving coordinates'',
since, from the outset, it incorporates the local symmetry group of
coordinate reparametrizations of the manifold, i.e.,
diffeomorphisms. For a different account of quantum mechanics in
moving coordinates (or in non-inertial reference frames),
see Schmutzer and Pleba\'nski (1977).

In the new coordinates $\,(y^1,\mini y^2,\mini y^3)$, the one-body
potential $\,v\ty\,$ and the background fields $\,{\bf E}\ty\,$
and $\,{\bf B}\ty\,$ may be expected to be (approximately) {\em
time-independent$\mini$}. In this situation, the Hamiltonian for
spinless particles $\,(s=0)\,$ will be (approximately)
time-independent, and we can apply the rules of gibbsian statistical
mechanics to study thermal equilibrium.

Unfortunately, for spinning particles $\,(s={\halb,\, 1, \ldots})$,
the situation is not quite as neat, because, in the
$y$-coordinates, the dreibein $\,\tilde{e}^{A}_{\hspace{.4em}
i}(y)\,$ is {\em time-dependent$\mini$},

\be
\tilde{e}^{A}_{\hspace{.4em} k}(y) \::=\:
e^{A}_{\hspace{.4em} j}(t,\mini \phity)\, \dd{y^k} \phi^{\, j}\ty
\:=\: e^{A}_{\hspace{.4em} j}(t,\mini \phity)\,
\ddd{x^j}{y^k} \ .
                           \label{T2}
\ee

\noi
In order to eliminate as much of this undesirable time-dependence
as possible, we attempt to find a suitable \SU\ {\em gauge
transformation$\,$} of the new dreibein
$\,\tilde{e}^{A}_{\hspace{.4em} k}(y)\mini$; see (\ref{eGT1}). What
is the optimal choice? The answer is, perhaps, somewhat ambiguous, in
general. But the following choice tends to be quite optimal. Let
$\,(f^j\tx)\,$ be the {\em velocity field$\,$} generating the
flow $\,\phi(t,\mini\cdot)$, i.e.,

\be
c\, \ddd{x^j}{y^0} \:=\: \dd{t}\phi^{\, j}\ty \:=\: f^j(t,\mini
\phi\ty) \ ,\hspace{1cm} j=1,\mini 2,\mini 3\ .
                           \label{T22}
\ee

\noi
Then the infinitesimal rotation of an orthonormal frame carried
along by the flow $\,\phi(t,\mini\cdot)$, at the point $\,\bx \in
M\,$ and at time $\,t$, is given by

\be
\Omega^{A}_{\hspace{.4em} B}\tx \::=\: \halb \Bigl[\, (\partial_B
f^A)\tx - \delta^{AC} \delta_{BD}\, (\partial_C f^D)\tx\, \Bigr] \ ,
                           \label{Rot0} 
\ee

\noi
where $\,\partial_A := \Eaix{A}{i}\dd{x^i}\, $, and $\,f^A\tx :=
\eaix{A}{j}f^j\tx\, $; see (\ref{M3}) and the remark after
(\ref{M7}). The vector $\, {\bf\Omega}\tx\,$ dual to the antisymmetric
matrix  $\,(\Omega^{A}_{\hspace{.4em} B}\tx)\,$ is called the {\em
vorticity$\,$} or {\em circulation$\,$} of the vector field $\,{\bf
f}\tx\,$ and is the local angular velocity of the rotation
induced by $\,\phi(t,\mini\cdot)\,$ of a frame at the point $\,\bx$,
at time $\,t$.

We define a rotation matrix $\,(R^{A}_{\hspace{.4em} B}\tx)\,$ by
setting

\be
R^{A}_{\hspace{.4em} B}\tx \::=\: T \exp \Bigl[\, \gamma \int_0^t
dt'\, \Omega(t',\, \bx) \Bigr]^A_{\hspace{.4em} B} \ ,
                           \label{Rot} 
\ee

\noi
where ``$\mini T\exp\mini$'' denotes a time-ordered exponential, and
$\,\gamma\,$ is a real constant to be chosen later. (Its physical
meaning will become clear at the end of this subsection; see also
Subsect.~\ref{ssBEH}). The r.h.s.\ of (\ref{Rot}) can be defined,
for example, by a convergent Dyson series if $\,{\bf\Omega}\tx\,$ is
uniformly bounded in $\,t$. We now define (see (\ref{eGT1}))

\be
\hat{e}^{A}_{\hspace{.4em} k}(y) \;:= \;
{}^R\tilde{e}^{A}_{\hspace{.4em} k}(y)\; =\; R^{A}_{\hspace{.4em}
B}(t,\mini \phi\ty)\: \tilde{e}^{B}_{\hspace{.4em} k}(y) \ ,
                          \label{T3}
\ee

\noi
where $\,\tilde{e}^{B}_{\hspace{.4em} i}(y)\,$ is given by (\ref{T2}).
We also define the following transformed quantities:

\bea
\psih & := & \Usty\, \psi\tphi \ ,
                          \label{T4}
\eea

\vspace{-1.6cm}
\bea
\ghik{kl} & := & \ddd{y^k}{x^i}\ddd{y^l}{x^j}\, g^{ij}\tphi \ ,
                          \label{T5}
\eea

\mam

\bea
\ahm{0} & := & a_0\tphi + \ddd{x^j}{y^0}\, a_j\tphi \ , \nonumber  \\
\ahm{k} & := & \ddd{x^j}{y^k}\, a_j\tphi \ .
                          \label{T6}
\eea

\noi
Furthermore,

\bea
\whms{0} & := & \Usty\, \biggl[\, w_0^{(s)}\tphi + \ddd{x^j}{y^0}\,
w_j^{(s)}\tphi\, \biggr]\, \Usty^\ast       \nonumber  \\ 
 & & \mbox{} + \Usty\, \dd{y^0} \Usty^\ast \ , \hspace{1.5cm}
\mbox{and} \nonumber  \\ 
\whms{k} & := & \Usty\, \biggl[\, \ddd{x^j}{y^k}\, w_j^{(s)}\tphi\,
\biggr]\, \Usty^\ast       \nonumber  \\ 
 & & \mbox{} + \Usty\, \dd{y^k} \Usty^\ast \ , 
                          \label{T7}
\eea

\mwm

\be
\Usty  \::=\:  \Us\Bigl( R\tphi \Bigr) \ .
                          \label{T8}    
\ee

Our aim is to rewrite the action $\,\SL\,$ introduced in
(\ref{Cact1}) and (\ref{Cact2}) in moving coordinates, $y$,
using the transformations (\ref{T3})--(\ref{T8}). By (\ref{T4})
and (\ref{T8}),

\bea
\psi\tx & = & \Us(t,\mini \phi^{-1}\tx)^\ast\, \hat{\psi}(t,\mini
\phi^{-1}\tx)  \nonumber  \\
 & = & \Us\Bigl( R\tx \Bigr)^\ast\, \hat{\psi}(t,\mini \phi^{-1}\tx)
\ .   
\eea

\noi
Hence, (with $\,\dd{x^0}= {1 \over c}\dd{t} = \dd{y^0}\mini$)

\bea
\lefteqn{\Us\Bigl( R\tx \Bigr)\, \dd{x^0}\psi\tx\:
\rule[-3mm]{.1mm}{8.7mm}_{\: \bx\mini=\mini\phi\ty}  = }  \nonumber 
\\
 & = & \dd{y^0}\psih + \Usty\, \dd{y^0}\Usty^\ast\, \psih  
\nonumber  \\
 & & \mbox{} - {1 \over c} \fhj{k} \biggl[\,
\dd{y^k}\psih + \Usty\, \dd{y^k}\Usty^\ast\, \psih\, \biggr] 
\nonumber \\
 & = & \dd{y^0}\psih - {1 \over c} \fhj{k}\,
\dd{y^k}\psih \nonumber  \\
 & & \mbox{} - {i\mini\gamma \over 4c} \sum_{A,B,C = 1}^3
\epps{A}{\,BC}\, \Omhab\, L_C^{(s)}\, \psih \ ,
                             \label{T9}
\eea

\noi
where $\,-\fhj{k}:=-\ddd{y^k}{x^j}\,f^j\mmini\tphi\,$ is the
$k^{\rm\mini th}$ component of the vector field generating
$\,\phi^{-1}(t,\mini \cdot)\,$ in the $y$-coordinates, and 
$\,(\Omhab)\,$ is the vorticity of the generating vector field in
the $y$-coordinates with respect to the orthonormal frame
$\,(\hat{e}^A \ty)$, given in (\ref{T3}). By comparing (\ref{T9})
with (\ref{T6})--(\ref{T8}) we find that

\bea
\lefteqn{\Us\Bigl( R(x) \Bigr) \left\lgroup \dd{x^0} + i\mini\am{0}
+  \wms{0} \right\rgroup \psi(x)\: \rule[-3mm]{.1mm}{8.7mm}_{\:
x\mini=\mini(c\mini t,\,\phi\ty)} = }   \nonumber \\
 & = & \left\lgroup \dd{y^0} + i\mini\hat{a}_{0}(y) +
\hat{w}^{(s)}_{0}(y) \right\rgroup \hat{\psi}(y) - {1 \over c}
\hat{f}^k(y) \left\lgroup \dd{y^k} + i\mini \hat{a}_{k}(y) +
\hat{w}^{(s)}_{k}(y) \right\rgroup \hat{\psi}(y) \ .  \nonumber \\
 & &
                             \label{T10}
\eea

\vspace{-1cm}
\noi
We define the transformed covariant derivatives

\bea
\Dh{0} & := & \dd{y^0} + i\mini\hat{a}_{0}(y) + \hat{w}^{(s)}_{0}(y) 
\ , \nonumber  \\
\Dh{k} & := & \dd{y^k} + i\mini\hat{a}_{k}(y) - i\mini{m \over \hbar}
\hat{f}_k(y) + \hat{w}^{(s)}_{k}(y) \ , 
                             \label{T11}
\eea

\noi
where $\,\hat{f}_k(y) = \hat{g}_{kl}(y) \hat{f}^l(y)$, and the
transformed one-body potential

\bea
\hat{v}\ty & := & v\tphi - {m \over 2}\hat{f}_k(y)\hat{f}^k(y)
- i\mini{\hbar \over 2} {1 \over \sqrt{\hat{g}(y)}} \dd{y^k}
\left(\sqrt{\hat{g}(y)}\, \hat{f}^k(y) \right) \ ,\hspace{.7cm}
                           \label{T12}
\eea

\noi
as well as the transformed two-body potential

\bea
\hat{V}(t,\mini \by,\mini \by') & := & V(t,\mini\phi(t,\mini
\by),\mini \phi(t,\mini \by')) \ .
                           \label{T13}
\eea

\noi
After these preparations, one easily verifies the following
\rule[-4mm]{0mm}{5mm}

{\bf Theorem~\ref{ssMC}.1.} {\em In moving coordinates, the action
of a non-relativistic system of charged, spinning particles moving
in an external electromagnetic field and on a geometrically
non-trivial manifold is given by}

\bea
\lefteqn{\SL(\psi^\ast\!,\mini \psi\mini ;\mini g,\mini a,\mini w)\; 
= \; \hat{S}_{\hat{\Lambda}}(\hat{\psi}^\ast\!,\mini \hat{\psi}\mini
;\mini \hat{g},\mini \hat{a},\mini \hat{f},\mini \hat{w})} 
\nonumber  \\
 & := & \int_{\hat{\Lambda}} \sqrt{\hat{g}(y)}\, dt\mini d^{\,3}{\bf
y}\, \biggl[\, i\hbar c\, \hat{\psi}^\ast(y) (\Dh{0} \hat{\psi})(y)
\nonumber  \\
& & \mbox{} - {\hbar^2 \over 2\mini m} \,
\hat{g}^{kl}(y)\, (\Dh{k} \hat{\psi})^\ast(y)\, (\Dh{l}
\hat{\psi})(y) - \hat{U}(\hat{\psi}^\ast\!,\mini \hat{\psi})(y)\,
\biggr] \ , 
                           \label{mCact}
\eea

\noi
{\em where in the definition of $\:\hat{U}(\hat{\psi}^\ast\!,\mini
\hat{\psi})\,$ (see~(\ref{Cact2})) the potentials $\,\hat{v}\,$ and
$\,\hat{V}\,$ of (\ref{T12}) and (\ref{T13}) are used, and
}$\,\hat{\Lambda}:=\{\ty\:|\:(t,\mini \bx=\phity) \in \Lambda\}$.
\rule[-4mm]{0mm}{5mm}

To prove (\ref{mCact}), one expands the r.h.s.\ of (\ref{mCact}) in
powers of $\,\hat{f}^k$, integrates by part, and compares the
resulting expression to (\ref{T10}), (\ref{Cact1}) and
(\ref{Cact2}), using (\ref{T11}) through (\ref{T13}) and the fact
that $\,(\Us\psi)^\ast(\Us\psi)=\psi^\ast\psi$.

We pause to interpret the result (\ref{mCact}). By (\ref{T11}),
$\,-{m \over \hbar}\fhky{k}$ enters the action
$\,\hat{S}_{\hat{\Lambda}}\,$ as a contribution to the
\U\ connection. By (\ref{a}), $\,m\fhky{k}\,$ and $\,{q
\over c} \Ahky{k}\,$ play analogous roles, i.e., (in $x$-coordinates)

\be
m\,{\bf f}(x)\; \leftrightarrow \; {q \over c}\, {\bf A}(x) \ .
                           \label{cori}
\ee

\noi
The vector potential $\,{\bf A}(x)\,$ gives rise to the {\em
Lorentz force$\,$} in the classical limit. The Lorentz force has the
same form as the {\em Coriolis force$\,$} if one replaces $\,{q \over
c}\, {\bf B}(x)\,$ by $\,2\mini m\,{\bf\Omega}(x)$, where $\,
{\bf\Omega}(x)\,$ is the local angular velocity which is precisely
half the curl of the vector field $\,{\bf f}(x)$; see
Eq.~(\ref{Rot0}). Thus $\,{\bf f}(x)\,$ is the vector potential
that gives rise to the Coriolis force in the classical limit. 

By (\ref{T9}) and (\ref{T10}), the action $\,\hat{S}_{\hat{\Lambda}}
\,$ contains a term

\be
\gamma\, \psihsy \left( \sum_{A=1}^3 \hat{\Omega}_A(y)\, {\hbar
\over 2} L_A^{(s)} \right) \psihy \ , 
                            \label{wh1}
\ee

\noi
where $\,\hat{\bf\Omega}(y) = \halb\, curl\:\hat{\bf f}(y)$, in
$\,y$-coordinates. It has the same form as the {\em Zeeman term}

\be
{g\mini \mu \over \hbar}\, \psihsy \left( \sum_{A=1}^3 \hat{B}_A(y)\,
{\hbar \over 2} L_A^{(s)} \right) \psihy \ ,
                            \label{wh2}
\ee

\noi
which, by (\ref{T7}), (\ref{rho1}), (\ref{w3}), and (\ref{w1}), also
appears in $\,\hat{S}_{\hat{\Lambda}}$. Recall that the magnetic
moment of a particle with spin $\,s\,$ has been defined by
$\bgrec{\mu}_{spin}:= {g\mini\mu \over 2}\, {\bf L}^{(s)}$; see
(\ref{mamo}). Thus $\,{g\mini\mu \over \hbar} \,\hat{\bf B}\,$ is
the angular velocity of {\em spin precession$\,$} in the
magnetic field $\,\hat{\bf B}$.

Next, we analyze the one-body potential $\,\hat{v}\,$ in moving
coordinates. By (\ref{T12}), $\,\hat{v}\,$ is {\em complex-valued,
unless}

\be
div_{\hat{g}}\,\hat{\bf f}(y) \::=\: {1 \over
\sqrt{\hat{g}(y)}}\, \dd{y^k}\!\left(\sqrt{\hat{g}(y)}\,
\hat{f}^k(y) \right) \:=\: 0 \ , 
                                          \label{cori2}
\ee

\noi
i.e., {\em unless the vector field $\,\hat{\bf f}\mini$ is
divergence-free$\mini$}. A divergence-free vector field generates a
{\em volume-preserving$\,$} flow $\,\phi(t,\mini \cdot)$, hence

\be
\hat{g}(y) \::=\: \mbox{det}\,(\hat{g}_{kl}\ty) \:=\: g\tphi \ .
                                          \label{ghut}
\ee

\noi
Thus, for volume-preserving (i.e., {\em incompressible}) flows, and
{\em only$\,$} for such flows, $\,\hat{v}\,$ is again {\em
real-valued$\mini$}. [This is, because if volume is preserved by
$\,\phi(t,\mini \cdot)\,$ then, by (\ref{ghut}), the
quantum-mechanical time-evolution in the moving coordinate system
preserves probabilities with respect to the volume element
$\,\sqrt{g\tphi}\:d^{\,3}\by\,$ and hence is generated ba a {\em
hermitian (selfadjoint)$\,$} Hamiltonian!$\,$] But $\,\hat{v}\,$
contains an additional term, $- {m \over 2}\hat{f}_k\hat{f}^k$, that
was not present in the original one-body potential, see (\ref{T12}).
What does it correspond to physically? It is the potential of the
{\em centrifugal force$\mini$}, because $\,{m \over
2}\dd{y^l}(\hat{f}_k\hat{f}^k)\,$ is precisely the $l^{\rm\mini th}$
component of the centrifugal force at the point $\,\by$, at time
$\,t$. [Note, incidentally, that $\,{m \over 2}\,
\hat{f}_k\hat{f}^k\,$ is the classical kinetic energy of the
particle in the time-independent frame which must be subtracted in
the $\,y$-coordinates.]

In conclusion, we find that quantum mechanics in moving coordinates
is hamiltonian, with a {\em hermitian$\,$} (but possibly still
time-dependent) Hamilton operator, if, and only if, the flow
$\,\phi(t,\mini \cdot)\,$ defining the moving coordinate system is
{\em volume-preserving$\mini$}, or {\em incompressible$\mini$}.
Henceforth this property is usually assumed. It is worthwhile
recalling that in {\em two$\,$} space dimensions, incompressible
flows are automatically symplectic (hamiltonian) flows, because the
vector fields generating them are divergence-free and hence are dual
to the gradient of some (scalar) Hamilton function. (This is the
basis of a mathematical analysis of the two-dimensional Euler
equations.)

\bigskip
In order to provide a concrete application illustrating the general
formulae given in this subsection, we rewrite the {\em one-particle
Pauli equation in moving coordinates}. The flow $\,\phi(t,\mini
\cdot)\,$ defining the moving coordinate system (the moving
background) is assumed to be volume-preserving; see (\ref{cori2}).
Varying the action (\ref{mCact}) (without two-body potential term)
with respect to $\hat{\psi}^\ast$, the one-particle Pauli
equation reads 

\be
i\hbar c\,(\Dh{0}\hat{\psi})(y) \:=\:
- {\hbar^2 \over 2\mini m} \,{1\over \sqrt{\hat{g}(y)}}\,
\hat{g}^{kl}(y)\,\Dh{k}(\sqrt{\hat{g}}\,\Dh{l}\hat{\psi})(y)
+\hat{v}(y)\hat{\psi}(y)\ ,
                           \label{opPEmC}
\ee

\vwv

\be
\Dh{\mu} \::=\: \dd{y^\mu} +i\mini \at{\mu}(y) + i \sum_{A=1}^3
\hat{w}_{\mu A}(y)\, L^{(s)}_A\ , \hsp \mu=0,\ldots,3\ ,
\ee

\vwiv

\bea
\at{0}(y) &:=&
\hat{a}_0(y) \:=\: {q\over \hbar c}\,\Phi\tphi - {q\over \hbar c}\,
({\bf f}\cdot{\bf A})\tphi \nonumber\ , \\
\at{k}(y) &:=&
\hat{a}_k(y)-{m\over \hbar}\hat{f}_k(y) \:=\: -{q\over \hbar c}
\hat{A}_k(y) - {m\over \hbar}\hat{f}_k(y)\ ,\nonumber \\
\hat{w}_{0 A}(y) &:=&
-{g\mini \mu\over 2\hbar c}\, \hat{B}_A(y)-{1\over c} \left(
-{g\mini\mu \over 2 \hbar c} + {q \over 4\mini mc^2} \right)
\sum_{B,C=1}^{3} \eps{ABC}\,\hat{f}_B(y)\hat{E}_C(y)\nonumber \\
& & - {\gamma\over 2c}\,\hat{\Omega}_A(y) + \; {\rm terms \ } \propto \;
\omega \; ,  
\eea

\noi
and $\,\hat{w}_{kA}(y)\,$ is given by~(\ref{T7}) which contains terms
proportional to first, $\dd{y^k}\phi^j$, and second derivatives,
$\dd{y^k}\dd{y^l}\phi^j$, of the flow $\,\phi(t,\mini\cdot)$.
Furthermore, in~(\ref{opPEmC}), we have the terms

\be
\hat{v}(y) \::=\: v\tphi - {m \over 2}\hat{f}_k(y)\hat{f}^k(y)\ ,
\ee

\vav

\be
\hat{g}^{kl}(y) \::=\: \ddd{y^k}{x^i}\ddd{y^l}{x^j}\, g^{ij}\tphi \ ,
\ee

\noi
where $\,g_{ij}(t,\mini\bx)\,$ is the metric on the manifold $M$ (in
the time-independent $x$-co\-ordi\-nates) on which the background of
the one-particle system is moving according to the flow
$\,\phi(t,\mini\cdot)\mini$; $\,\hat{g}(y)$ is the determinant of
the metric $\hat{g}^{kl}$, as defined in~(\ref{ghut}).

\bigskip
An interesting consequence of formulating quantum mechanics in
moving coordinates is a quantum-mechanical version of {\em Larmor's
theorem$\,$} in which also the spin degrees of freedom of particles
moving in a (variable) external magnetic field are taken into
account. 

For definiteness, let us consider a system of particles with
effective mass $\mini m$, charge $\mini q$, spin $\mini s$, and
magnetic moment $\,\mu_{spin}$, (where $\,\bgrec{\mu}_{spin} :=
{g\mini \mu \over \hbar} {\bf S}$, with $\,\mu = {q\hbar \over
2\mini m_o c}$, and $\,m_o\,$ is the mass of the particles in the
vacuum; see (\ref{mamo})). Furthermore, for simplicity, we choose
the background manifold $\,M\,$ to be euclidean space
$\,\BBn{E\mini}{2}$ or $\,\BBn{E\mini}{3}$, i.e., the metric takes
the form $\,g_{ij}(x) = \delta_{ij}$, and all geometrical
contributions to the \SU\ connection $\,w(x)\,$ are absent; see
(\ref{w1})--(\ref{w3}). We now suppose that the system is under the
influence of a (variable) external magnetic field, ${\bf B}(x)$, and
assume that there is no external electric field; $\,{\bf
E}(x)=0\mini$. As we shall see shortly, it is convenient to work in
a \U\ gauge where $\,div\,{\bf A}(x)=0\,$ (Coulomb gauge). We then
have the following result.
\rule[-4mm]{0mm}{5mm}

{\bf Theorem~\ref{ssMC}.2 (Quantum-Mechanical Larmor Theorem).}
{\em For the system just described, the effect of an external
magnetic field $\,{\bf B}(x)$ can be eliminated to first order by
choosing to work in moving coordinates generated by the velocity
field $\: {\bf f}(x) = -{q \over mc} {\bf A}(x)$, and by performing
an \SU\ gauge transformation $\,\Us(R(x))$ where the local frame
rotation $\,R(x)$ is given by (\ref{Rot}) with $\,\gamma = g{m \over
m_o}\mini$}. \rule[-4mm]{0mm}{5mm}

Note that the vorticity field $\,{\bf \Omega}(x) := \halb\,
curl\, {\bf f}(x) = - {q \over 2\mini mc}\mini {\bf B}(x)\,$ of the
velocity field $\,{\bf f}(x)\,$ is precisely the so-called {\em
Larmor angular velocity$\mini$}. Choosing the vector potential 
$\,{\bf A}(x)$ in the Coulomb gauge renders the flow generated by
$\,{\bf f}(x)$ divergence free (i.e., volume preserving).

The proof of this theorem is straightforward. We use that $\,\wms{j}
=0= \am{0}\,$ by (\ref{w1})--(\ref{w3}), and
(\ref{a})--(\ref{rho2}). Then, adopting the identifications given in
the theorem, it follows from (\ref{T11}) and (\ref{T12}), using
(\ref{T6}) and (\ref{T7}), that in moving coordinates:

\bea
\Dh{0} & = & \dd{y^0} + {\cal O}\Bigl( \mbox{max$\, (\, \left\vert
\hat{\bf f} \right\vert^2\!(y),\, \left\vert \dd{y^k}\hat{\bf B}
\right\vert\!(y)\, )$} \Bigr) \ , \nonumber  \\
\Dh{k} & = & \dd{y^k} + {\cal O}\Bigl( \mbox{$ \left\vert
\dd{y^k}\hat{\bf B} \right\vert\!(y) )$} \Bigr)  \ , \hspace{1cm}
\mbox{and} \nonumber  \\ 
\hat{v}(y) & = & v\tphi + {\cal O}\Bigl( \left\vert \hat{\bf f}
\right\vert^2\!(y) \Bigr) \ ,
                            \label{new}
\eea

\noi
see also (\ref{wh1}) and (\ref{wh2}). Note that if the external
magnetic field is constant (in the moving coordinates), $\,\hat{\bf
B}(y)=\hat{\bf B}_o\mini$, then the ``tidal vector potential''
$\,\hat{\bf f}(y)\,$ may be written as

\be
\hat{\bf f}(y)\; =\; \hat{\bf \Omega} \times \by\; =\; {q \over
2\mini mc}\, \by \times  \hat{\bf B}_o \ .
                            \label{fhc}
\ee 

\noi
Hence, in this situation, the terms proportional to
$ \dd{y^k}\; \hat{\bf B}(y) $  are absent
in (\ref{new}), and $\,{\cal O}\Bigl( \left\vert \hat{\bf f}
\right\vert^2\!(y) \Bigr) = {\cal O}\Bigl( \left\vert \hat{\bf B}_o
\right\vert^2\mini \Bigr)$.

\bigskip

The formalism developed in this section applies
equally well to (one-), two- and three-dimensional systems.

It often happens in solid state physics, e.g. in two-dimensional
heterostructures used in measurements of the quantum Hall effect,
that the system exhibits an approximate internal symmetry described
by some compact group $\,G$. The spinors $\,\psisss\,$ then
transform according to some non-trivial representation, $\pi$, of
$\,G$. A breaking of this symmetry by an external background might be described as the effect of
coupling $\,\psisss\,$ to an external gauge field in the
representation $\,\pi_\ast\,$ of the Lie algebra of $\,G$. Let us
denote this gauge field by $\,z$. By modifying the covariant
derivatives given in (\ref{DD1}),

\be
\D{\mu}\: \mapsto \:
\D{\mu}^{\mini\prime}\::=\:\D{\mu}+ z_\mu(x) \ ,
\ee

\noi
we may easily extend the entire formalism developed in this section
to systems with {\em gauged internal symmetries$\mini$}. This can be
important in applications.

Note that, in this situation, the action $\,\SL\,$ introduced in
Eq.~(\ref{Cact1}) is $\,U(1) \times SU(2) \times G$ {\em gauge
invariant$\mini$}, i.e., it does not change if, for an arbitrary
real-valued fucntion $\,\chi$, an \SU-valued function $\,R$, and a
$\,G$-valued function $\,g$, all defined over {\em space-time}
$\,\BB{R} \times M$, the following substitutions are made:

\bea
\psis(x) 
& \hspace{-1.5mm} \:\mapsto\: \hspace{-1.5mm} &
e^{-i\chi(x)}\,\Us(R(x))\: \mbox{\oox \symbol{'12}}\: \pi(g(x))\,
\psis(x) \ ,  \nonumber  \\
\am{\mu}+{m \over \hbar}f_\mu(x) 
& \hspace{-1.5mm} \:\mapsto\: \hspace{-1.5mm} &
\am{\mu}+{m \over \hbar}f_\mu(x)+\partial_\mu \chi(x) \ ,   
\nonumber  \\  
\wms{\mu} 
& \hspace{-1.5mm} \:\mapsto\: \hspace{-1.5mm} & 
\Us(R(x))\,\wms{\mu}\,\Us(R(x))^\ast + \Us(R(x))\, \partial_\mu
\Us(R(x))^\ast \ ,   \nonumber  \\
& \hspace{-65.5mm} \mbox{and}  &    \nonumber  \\
z_\mu(x) 
& \hspace{-1.5mm} \:\mapsto\: \hspace{-1.5mm} & 
\pi(g(x))\, z_\mu(x)\, \pi(g(x))^\ast + \pi(g(x))\, \partial_\mu
\pi(g(x))^\ast \ .
                         \label{gauget}
\eea

\noi
Thus, barring gauge anomalies, which actually {\em cannot$\,$}
appear in systems of {\em finitely$\,$} many non-relativistic
particles, the non-relativistic quantum mechanics of such systems is
$\,U(1) \times SU(2) \times G\mini$ gauge invariant. Ward
identities expressing this gauge invariance turn out to play an
important role in establishing certain universal properties of such
systems; see Fr\"ohlich and Studer~(1992a\mini--1992d) and
Sect.~\ref{sSL} below.




\section{Some Key Effects Related to the $U(1) \times
SU(2)\mini$ Gauge Invariance of Non$\,$-Relativistic Quantum
Mechanics}
\label{sKE}

In this section, we describe some effects in quantum mechanics from
the point of view of its $\,U(1)_{em} \times
SU(2)_{spin}$ gauge-invariance. We continue and expand the
discussion started at the end of Sect.~\ref{sPE}. Most of the
material reviewed here is well-known, but our perspective,
emphasizing gauge-invariance, may be somewhat novel.


\subsection{``Tidal''$\mmini$ Aharonov$\,$-$\,$Bohm and
``Geometric''$\mmini$ Aharonov$\,$-Casher Effects}
\label{ssTG}

After what we have learned in Subsect.~\ref{ssMC} on the
$U(1)$ vector potential of the Coriolis force, present in moving
coordinates (see (\ref{cori})), it is clear that there must exist a
``{\em tidal$\,$}'' {\em Aharonov-Bohm effect}$\,$: Consider a
mass-current conducting superfluid in a large container penetrated
by some straight cylindrical tube that excludes the quantum fluid.
Now, set the fluid into circular motion around the axis of the tube
with velocity field $\bbf\!$, where $\,\left\vert {\bf f}(r)
\right\vert = {\kappa \over 2\pi\mini r}\,$ at a distance $\,r\,$
from the axis of the tube, and $\,\kappa\,$ is a quantity of
dimension $\,cm^2/sec$, the total {\em vorticity$\,$} or {\em
circulation}$\,$, $\kappa := \oint{\bf f}\cdot d{\bf l}\,$. [Note
that $\,\kappa={2\pi\mini \overline{L_z} \over NM}$, where $\,M\,$
is the mass of the particles constituting the quantum fluid,
$\,\overline{L_z}\,$ is the expectation value of the component of
the total angular momentum operator parallel to the tube in the
given state of the system, and $\,N\,$ is the number of paricles in
the system.] Small mass currents excited in this system, scattered
at the tube, will exhibit an Aharonov-Bohm effect depending
periodically on $\,\kappa$, with period $\,{h \over m}$, where
$\,m\,$ is the mass of the particles in the scattering currents;
compare to Subsect.~\ref{ssAB}.

While this effect may be somewhat difficult to test experimentally,
it is important theoretically: Consider a superfluid film with
manifestly (e.g., through rotation) or spontaneously broken
time-reversal and reflections-in-lines (i.e., two-dimensional
parity) invariance. The formation of such films at a particular
superfluid $\mbox{}^3\mmini He$-A/B interface has been discussed by
Salomaa and Volovik (1989). Such a two-dimensional superfluid will,
in general, exhibit {\em vortex excitations$\,$} of vorticity
$\,\kappa=n{h \over M}\, ,\; n \in \BB{Z}\,$, where $\,M\,$ is the
mass of the constituent particles in the superfluid, and {\em
fractional mass$\,$} (rather than fractional charge) $\,\sH\kappa\,
$, where $\,\sH=\sigma\,{M^2 \over h}\,$ is a ``tidal Hall
conductivity''. Such excitations give rise to Aharonov-Bohm phases
and hence are anyons if $\,\sigma\,$ is not an integer, i.e., if
the superfluid shows a {\em fractional$\,$} ``tidal'' quantum Hall
effect; a detailed discussion of this effect can be founed in
Subsect.~VII.B in Fr\"ohlich and Studer (1993b). Such excitations
may be observed experimentally by measuring fluctuations in the
longitudinal resistance of superfluid current conduction. See
Simmons {\em et al.}~(1989) and Hwang {\em et al.}~(1992) for an
analogous experiment in two-dimensional electronic systems
exhibiting the fractional quantum Hall effect.

A remarkable experimental observation of a ``tidal'' Aharonov-Bohm
effect has been provided by Werner, Staudenmann, and
Colella~(1979). Using a neu\-tron interferometer they have detected
a quantum-mechanical interference effect due to the rotation of the
Earth. For a brief theoretical comment on this interference
experiment that is close to our discussion in Subsect.~\ref{ssMC},
see Sakurai~(1980).

Other interesting systems where mixed ``tidal'' and electromagnetic
effects play a role are, for example, rotating superconductors.
Following an analysis by Semon~(1982; see also Schmutzer and
Pleba\'nski, 1977), an experiment performed by Zimmerman and
Mercereau~(1965) can be interpreted as the realization of a thought
experiment proposed by Aharonov and Carmi~(1973): Given a
(uniformly) rotating sample that is not simply connected, the
``tidal'' forces (Coriolis - and centrifugal force) felt by the
particles in the moving system (all of the same charge to mass
ratio) can be cancelled by an electromagnetic field whose vector
potential does not cancel the ``tidal'' vector potential completely
everywhere; see (\ref{cori}). This uncancelled ``tidal'' vector
potential then leads to a quantum-mechanical interference
effect.

\bigskip
In Subsect.~\ref{ssAC}, we have discussed the Aharonov-Casher effect
as an \SU\ version of the Aharonov-Bohm effect. Furthermore, in
Subsect.~\ref{ssSS}, systems in geometrically non-trivial
backgrounds have been described by including a ``geometrical'' term
in the \SU\ connection $\,w\mini$; see (\ref{w1})--(\ref{w4}) and
(\ref{M5}). Here we combine these findings and discuss a ``{\em
geometrical version$\mini$}'' {\em of the Aharonov-Casher
effect$\,$}: We consider a {\em two-dimensional$\,$} system of
particles with non-zero spin on a {\em cone$\,$} with tip at
$\,\bx=0$. Then, although $\,\rho=0\,$ if there are no
electromagnetic fields, the \SU\ connection $\,w\,$
determined by $\,\omega^{A}_{\hspace{.4em} B}$, the affine spin
connection on the cone, cannot be gauged away globally, although
$\,\omega^{A}_{\hspace{.4em} B}\,$ is {\em flat$\,$} for $\,\bx\neq
0$. The \SU\ connection $\,w\,$ has the same form as the
electromagnetic part $\,\rho\,$ given in Eq. (\ref{AC2}), but
$\,Q\,$ now denotes the defect angle of the cone. Scattering of
particles at the tip of the cone will yield interference patterns
depending on the defect angle $\,Q$. This effect is perhaps better
known than its electromagnetic cousin. It has attracted attention,
for example, in connection with quantum mechanics and quantum field
theory in the presence of cosmic strings (Deser and Jackiw, 1988;
t'Hooft, 1988; Kay and Studer, 1991).

Do spinless particles ``see'' the tip of the cone, or is spin
important? The answer depends on our choice of a quantum-mechanical
state space: We must impose some ``boundary conditions'' on the
wave functions, i.e., $\,\psi(r,\,\varphi+2\pi-Q)=e^{i\theta}
\psi(r,\, \varphi)$, where $\,\varphi\,$ is the polar angle, and
$\,\theta\,$ is some phase to be specified; plus some boundary
condition at $\,r=0$. But no matter how we choose $\,\theta$, we can
make the tip of the cone ``invisible'' to spinless particles by
threading a magnetic flux through $\,\bx=0$. If the particles have
spin {\em and$\,$} a non-zero magnetic moment then, in addition, we
would have to put a charged wire through $\,\bx=0$, in order to make
the tip ``invisible''.

\bigskip
Finally, we remark that there is an analogue of the
Aharonov-Casher effect where $\,SU(2)\,$ is replaced by a
gauged internal symmetry group $\,G$. This effect can, perhaps, be
tested in inhomogeneous heterostructures. It is related, both
physically and mathematically, to the existence of particles in
two-dimensional quantum theory with topological pair interactions
described by a $\,G$-Knizhnik-Zamolodchikov connection that, just
as in the case of $\,SU(2)$, may give rise to non-abelian
braid statistics; see Subsect.~\ref{ssQP}.


\subsection{Flux Quantization}
\label{ssFQ}

A superconductor exhibits the {\em Meissner-Ochsenfeld effect$\,$}:
A magnetic field cannot penetrate into the bulk of a superconducting
material. However, in a type II$\,$ superconductor, thin magnetic
field tubes can thread through the bulk. They have the property
that they carry a magnetic flux $\,\Phi\,$ which is an integer
multiple of $\,{hc \over q}$, where $\,q\,$ is the charge of the
particles in the condensate, (e.g., $\,q=-2e$, for BCS pairs of
electrons). These tubes are called {\em Abrikosov
vortices$\mini$}. The quantization of $\,\Phi\,$ is explained by
requiring that, outside an Abrikosov vortex, the superconducting
state of the system remain undisturbed. From what we have said
about the Aharonov-Bohm effect in Subsect.~\ref{ssAB}, it follows
that this requirement is fulfilled precisely if $\,\Phi\,$ is an
integer multiple of $\,{hc \over q}$. Experimentally, this effect
has been established first in Deaver and Fairbank~(1961) and Doll
and N\"abauer~(1961; for a review, see, e.g., Chap.~6 in Tilley and
Tilley, 1986).

The discussion of the ``tidal'' Aharonov-Bohm effect above makes
it clear that the Meissner-Ochsenfeld effect and flux quantization
in Abrikosov vortices have their counterparts in the theory of
superfluidity: Consider a superfluid  in some container. Now
set the container into uniform rotation. The superfluid inside the
container abhors angular velocity which could destroy its
superfluidity and does, therefore, {\em not$\,$} follow the rotation
of the container's walls. However, just like there can be Abrikosov
vortices in a type II$\,$ superconductor, the superfluid can
eventually be set into motion, and the motion is generated by a
velocity field $\,\bf f$, whose circulation (or vorticity),
$\,{\bf\Omega}=\halb\, curl\mini \bbf$ (see (\ref{Rot0})), is
localized along thin tubes. The ``tidal'' Aharonov-Bohm effect then
predicts that the total circulation in such a tube is quantized in
integer multiples of $\,{h \over M}$, where $\,M\,$ is the mass of
the particles (e.g., $\mbox{}^3\mmini He$-pairs) constituting the
superfluid. [This can also be understood by appealing to the
quantization of orbital angular momentum.] If, in such a fluid, one
can excite mass currents of quantum-mechanical particles,
``dopants'', of mass $\,m < M\,$ one may be able to test the
``tidal'' Aharonov-Bohm effect.

Our conclusions agree with another theoretical analysis given by
Pines and No\-zi\-\`eres~(1989). Experimentally, the first
verification of the quantization of circulation (in superfluid
$\,He\,I\!I\mini$) has been given by Vinen~(1961; for a review, see,
e.g., Chap.~6 in Tilley and Tilley, 1986). The phenomena described
here may be relevant in the astrophysics of rotating neutron stars
(pulsars) which appear to be superfluid (see, e.g., Tsakadze and
Tsakadze, 1980).


\subsection{Barnett and Einstein$\,$-$\,$de$\:$Haas Effects}
\label{ssBEH}

The Barnett and Einstein$\mini$-$\mini$de$\:$Haas effects (see,
e.g., Landau and Lifshitz, 1960) find a very natural explanation in
the light of the quantum-mechanical Larmor theorem discussed at the
end of Subsect.~\ref{ssMC}. Consider a cylinder of iron or some
other ferromagnetic material suspended at a wire in such a way that
it can freely rotate around its axis. Let us suppose that,
initially, the cylinder is at rest and demagnetized. Now, imagine
that the cylinder is set into rapid rotation around its axis. As
explained in Subsect.~\ref{ssMC}, the quantum mechanics of the
electrons in this material should now be described in a {\em
uniformly rotating$\,$} coordinate system fixed to the cylindrical
background. In this coordinate system the electronic Hamiltonian
will be {\em time-independent$\mini$}, but it now contains a Zeeman
term

\be
\hat{\psi}^\ast\: \hat{\bf\Omega} \cdot {\hbar \over 2}\,
\bgrec{\sigma}\: \hat{\psi} \ ,
                            \label{Zem}
\ee

\noi
where $\,\hat{\bf\Omega}=\const\,$ is the angular velocity of the
rotation; (see (\ref{wh1}) where we have set $\,\gamma=1$, i.e., by
(\ref{Rot0})--(\ref{T3}), the orthonormal frame in the
(co)tangent space (``spin space'') rotates with the {\em same$\,$}
angular velocity as the rotating coordinate system). Note that, in
(\ref{Zem}), $\,\hat{\bf\Omega}\,$ plays the role of the magnetic
field $\,\bf B\,$ in an inertial frame. Furthermore, the Hamiltonian
contains a tidal vector potential $\,\hat{\bf f} = \hat{\bf
\Omega} \times \by\,$ in the covariant derivatives $\,\Dh{k}\,$,
and a potential $\,- {m \over 2} \left\vert\, \hat{\bf \Omega}
\times \by \right\vert^2\,$ of the centrifugal forces; see
(\ref{T11}) and (\ref{fhc}), and (\ref{T12}), respectively. All the
additional terms can be combined into the term

\be
\hat{\psi}^\ast\: \hat{\bf \Omega}\cdot \hat{\bf J}\:\hat{\psi} \ ,
\ee

\noi
where $\,\hat{\bf J}= \hat{\bf S}+\hat{\bf L}\,$ is the total
angular momentum operator (see also Kerman and Onishi, 1981). The
effect of centrifugal forces will be balanced by the chemical
potential of the background. Thus the electronic Hamiltonian is
essentially equivalent to the one for a cylinder at rest but in a
magnetic field $\, {\bf B}= \left( {g\mini\mu_B \over \hbar}
\right)^{\! -1}\!{\bf \Omega}\, $. The result is, in both
situations, that the cylinder is {\em magnetized$\mini$}, because
the spins of the electrons will align with $\, {\bf \Omega}\, $ or
$\,{\bf B}$, respectively. This is the {\em Barnett effect$\mini$}.

\bigskip
Conversely, in the {\em Einstein$\mini$-$\mini$de$\:$Haas effect$\mini$},
one turns on a magnetic field, $\,{\bf B}$, antiparallel to the
spontaneous magnetization of a {\em magnetized$\,$} piece of iron at
rest, thereby {\em increasing$\,$} the {\em free energy$\,$} of the
system. The system reacts to this perturbation by starting to {\em
rotate$\,$} around the axis of the external magnetic field so as to
offset the effect of $\, {\bf B}\, $ on the electrons by rotation.
It thereby returns to a state corresponding to a local minimum of
the free energy. By (\ref{wh1}) and (\ref{wh2}), the angular
velocity of this rotation, $\,{\bf \Omega}$, is given by $\, {\bf
\Omega} = {g\mini\mu_B \over \hbar}\, {\bf B}\, $ which is precisely
the angular velocity of spin precession in the magnetic field $\,
{\bf B}$. A similar effect is observed when one tries to magnetize a
paramagnet. It would appear interesting to test a {\em local
version$\,$} of this effect in a ``ferro-fluid''. If the magnetic
field acting on a highly mobile ferro-fluid, locally in thermal
equilibrium, is modified locally the fluid reacts by starting to
flow with a velocity field that optimally offsets the change in the
magnetic field so as to restore local equilibrium. The particle~-
and magnetic current densities induced are given by $\,n\bbf$ and
$\,\bM\, \raisebox{.3mm}{\mbox{\oox \symbol{'12}}}\, {\bf f}$,
respectively, where $\bbf$ is the velocity field, $\,n\,$ the
particle density, and $\,\bM\,$ the magnetization density. A
somewhat analogous effect for quantum Hall fluids will be discussed
in Subsect.~\ref{ssLR}.

\bigskip
There is another variant (Bell and Leinaas, 1983, 1987) of the
Barnett effect: Consider a beam of non-relativistic particles, e.g.
heavy ions, with spin, moving in a storage ring with some mean
angular velocity $\, {\bf \Omega}$. Then they experience a
``tidal'' Zeeman effect, given by (\ref{wh1}), in addition to the
usual magnetic Zeeman effect, given by (\ref{wh2}). After
relaxation to a steady state, the ``tidal'' Zeeman energy obviously
affects the ratio of ``spin-up'' to ``spin-down'' ions in the beam!

Similar considerations are important e.g. in the study of
electronic spectra of rotating molecules in the Born-Oppenheimer
approximation (Kerman and Onishi, 1981).


\subsection{Meissner$\,$-$\,$Ochsenfeld Effect and London Theory
of Superconductivity}
\label{ssSAH}

Consider a superconducting condensate of charged bosons (e.g.,
electron pairs) of charge $\,q\,$ and mass $\,M\mmini$, in
equilibrium. Imagine that a (weak) magnetic field, $\bBc\!$, is
turned on inside the bulk of this system, resulting in an increase
of the free energy of the system (weak form of the
Meissner-Ochsenfeld effect). Then the condensate {\em reacts$\,$} to
the field $\bBc$ by starting to {\em flow$\,$} according to a
velocity field $\bbf$ in such a way as to cancel $\bBc$ in the
moving coordinate system. For, in this way, the free energy
in regions of the system moving with the flow is minimal.
Neglecting the centrifugal potential, $\,-{m \over 2}\, {\bf
f}\cdot{\bf f}$, and (in a first step) the magnetic field created by
the resulting current, it follows from Eqs.~(\ref{mCact}),
(\ref{cori}) and (\ref{cori2}) that the optimal velocity field
$\bbf$ is given by

\ben
\bbf \:=\:-{q \over Mc}\, {\bf A}_c^T\ ,
\een

\noi
where $\,{\bf A}_c^T$ is the vector potential of $\bBc$ in the
Coulomb gauge (i.e., $\,div\,{\bf A}_c^T=0$). Thus, the system
exhibits a supercurrent density, ${\bf J}_s$, given in our
approximation by

\be
{\bf J}_s \:\approx\: q\mini n_s\bbf\:=\:-{q^2\mini n_s \over Mc}\,
{\bf A}_c^T\ ,
                        \label{LONo}
\ee

\noi
where $\,n_s\,$ is the density of the condensate. Of course, this
supercurrent $\,{\bf J}_s$ will give rise to an additional vector
potential, $\,{\bf A}_s^T$, determined by Maxwell's equation:
$\,\Delta\,{\bf A}_s^T=-{1 \over c}\,{\bf J}_s$. Adding this
potential to the r.h.s.\ of (\ref{LONo}) we obtain the equation

\be
{\bf J}_s \:=\: -{q^2\mini n_s \over Mc}\,{\bf A}^T\ ,
                        \label{LON}
\ee

\noi
where $\,{\bf A}^T:={\bf A}_c^T+{\bf A}_s^T\,$ is the total
vector potential of the external magnetic field, $\bBc$, and the
magnetic field created by the supercurrent. This is the {\em London
equation$\,$} characterizing the superconducting state! Relating the
external magnetic field $\bBc$ to an external current density,
$\,{\bf J}_c$, via Maxwell's equation $\,\Delta\,{\bf A}_c^T=-{1
\over c}\,{\bf J}_c$, and assuming that the external current
$\,{\bf J}_c\,$ does not enter into the bulk of the superconductor,
i.e., $\,{\bf J}_c=0\,$ inside the superconducting region, the
London equation~(\ref{LON}) immediately implies the equation

\be
\Delta\,{\bf A}^T\:=\: {q^2\mini n_s \over Mc^2}\,{\bf
A}^T\ ,
                        \label{MEI}
\ee

\noi
which shows that, in a stationary state, currents and
magnetic fields in superconductors can exist only within a surface
layer of thickness $\,\Lambda := \left({Mc^2 \over q^2\mini n_s}
\right)^{\! 1/2}$, the so-called London penetration depth (see,
e.g., de Gennes, 1966). This is the {\em Meissner-Ochsenfeld
effect$\mini$}. Note that by Eq.~(\ref{LON}) a {\em supercurrent}
$\,\bJ_s$ is really a sign for the presence of a {\em vector
potential$\mini$}, $\,{\bf A}^T$, and thus can be used for
experimental tests of the Aharonov-Bohm effect. However, if
$\,\oint_{\,\Gamma} {\bf A}^T \cdot d \mbox{\boldmath $l$} =
n\mini{hc\over q}$, $\,n \in \BB{Z}\,$, for any closed curve
$\,\Gamma\,$ contained in the superconducting phase, then the
Aharonov-Bohm phase factors of the charged bosons are trivial (see
Subsect.~\ref{ssAB}), and no supercurrent results. This is the
phenomenon of flux quantization.

The expulsion of a magnetic field from the interior of a
superconductor is related to the fact that, inside a superconductor,
``photons are massive'', i.e., one observes the phenomenon of the
{\em Anderson-Higgs mechanism$\mini$}. We now show that, in a
superconductor, the presence of a ``{\em mass term for the
photons$\,$}'' can also be inferred directly from the London
equation (\ref{LON}): Since the electric current density, $\,\bJ$,
and the {\em free energy$\mini$}, $F(\bA)$, of a system in the
presence of a (static) electromagnetic field with vector potential
$\,\bA:=(A_1,\mini A_2,\mini A_3)$, are related by

\be
\bJ(\bx)\:=\:-\mini c\,{\delta F(\bA) \over \delta\bA(\bx)} \ ,
                           \label{bJ}
\ee

\noi
it follows from Eq.~(\ref{LON}) that, in the bulk of a
superconductor,

\be
F(\bA) \:=\: {1 \over 2 \Lambda^2} \int d^{\,3}{\bx}\; \bA^T\!(\bx)
\cdot \bA^T\!(\bx)\: +\: \cdots \ ,
                           \label{FE}
\ee

\noi
where $\,A^T_{\mini i}(\bx) := \left[\,\delta^j_i - \partial_i\,
\Delta^{-1} \partial^j\,\right]\!A_j(\bx)$, with $\,\Delta\,$ the
three-dimensional Laplacian, and the dots stand for higher
derivative terms. Notice that (\ref{FE}) is a {\em non-local$\,$}
functional of $\,\bf A\,$ which is manifestly {\em invariant$\,$}
under \U\ gauge transformations, $\bA \mapsto \bA + \mbox{\boldmath
$\nabla$} \chi$. It provides a ``mass term for the photons'' in the
bulk of a superconductor.


\subsection{Quantum Hall Effect}
\label{ssQH}

Just as the Aharonov-Bohm effect reflects the \Ue\ {\em gauge
invariance$\,$} of quantum theory, so does the {\em quantum Hall
(QH) effect for the electric current$\mini$}, as emphasized by
Laughlin~(1981; see also Halperin, 1982). In the same vein, the
Aharonov-Casher effect and the {\em quantum Hall effect for the
spin current$\,$} reflect the \SUs\ {\em gauge invariance$\,$}
of non-relativistic quantum theory, as emphasized by Fr\"ohlich and
Studer~(1992a, 1992c). In this subsection, we review some basic
facts concerning the integer (von Klitzing, Dorda, and Pepper,
1980) and fractional (Tsui, Stormer, and Gossard, 1982) quantum Hall
effect; for comprehensive reviews see Chakraborty and
Pietil\"ainen~(1988), Morandi~(1988), Prange and Gervin~(1990),
Wilczek~(1990), and Stone~(1992). The purpose of this review is to
set the stage for Sects.~\ref{sSL}--\ref{sClQHFs}, where we attempt
to unravel the {\em universal$\,$} aspects of the quantum Hall
effect in two-dimensional, {\em incom\-press\-i\-ble$\,$} quantum
fluids.

Experimentally, the {\em QH effect$\,$} is observed in
two-dimensional systems of electrons confined to a planar
region $\,\Omega\,$ and subject to a strong, uniform
magnetic field $\mini{\bf B}_c$ transversal to $\,\Omega$. For
definiteness, we choose the region $\,\Omega\,$ to be a rectangle
in the $\mini(x,\mini y)$-plane $\,\Omega\,$ with dimensions
$\,l_x\,$ and $\,l_y\,$ in the $\,x$- and $\,y$-directions,
respectively. By tuning the $y$-component, $I_y$, of the total
electric current to some value and then measuring the voltage drop,
$V_x$, in the $x$-direction\,--\,i.e., the difference in the
chemical potentials of the electrons at the two edges parallel to
the $y$-axis\,--\,one can calculate the {\em Hall resistance}

\be
\RH \::=\: {V_x \over I_y} \ ,
                          \label{RH}
\ee

\noi
and finds that, for a fixed density, $n$, of electrons and at
temperatures close to $\,0\,K$, the value of $\RH$ is independent
of the current $\,I_y$. It depends only on the external magnetic
field $\mini{\bf B}_c$. If the electrons are treated classically
one finds, by equating the electrostatic - to the Lorentz force,
that

\be
\RH \:=\: {B_c \over n\mini ec}\ ,
                             \label{Hallc}
\ee

\noi
where $B_c$ is the $z$-component of $\mini{\bf B}_c$ perpendicular
to the plane of the system, $e\,$ is the elementary electric charge,
and $\,c\,$ is the velocity of light.

By also measuring the voltage drop, $V_y$, in the $y$-direction,
one can determine the longitudinal resistance, $\RL$, from the
equation

\be
\RL \::=\: {V_y \over I_y} \ .
\ee

\noi
Neither classical, nor quantum theory make simple predictions about
the behaviour of $\RL$, but $\,\RL>0\,$ means that there are
{\em dissipative processes\mini} in the system.

Two-dimensional systems of electrons (and, similarly, of holes) are
realized, in the laboratory, as {\em inversion layers\,} which form
at the interface between an insulator and a semiconductor when an
electric field (gate voltage) perpendicular to the interface, the
plane of the system, is applied. An example of such a material is a
heterojunction (a ``sandwich'') made from $GaAs$ and
$Al_xGa_{1-x}As\mini$. The quantum-mechanical motion of the
electrons in the $z$-direction perpendicular to the interface is
then constrained by a deep potential well with a minimum on the
interface. Quantum theory predicts that electrons of sufficiently
low energy, i.e., at low enough temperatures, remain bound to the
interface and form a very nearly two-dimensional system.

In classical physics, the connection between the electric field
$\,\vec{E}=(E_x,\mini E_y)\,$ in the plane of the system and the
electric current density $\,\vec{J}=(J_x,\mini J_y)\,$ is given by
the {\em Ohm-Hall law}

\be
\vec{E}\:=\:\bgrec{\rho}\,\vec{J} \ , \hwh
\bgrec{\rho}\::=\,\left(
\begin{array}{cc} \rx & -\rH \\
\rH & \ry \end{array} \right) \ ,
                             \label{OH}
\ee

\noi
where the components of the resistivity tensor $\,\bgrec{\rho}\,$
are given as follows: $\,\rx=\RL\,l_y/l_x$, $\,\ry=\RL\,l_x/l_y$,
and $\,\rH = \RH$. This is a phenomenological law valid on
macroscopic distance scales and at low frequencies.


\begin{figure}[p]

\begin{flushleft}
\vspace*{3mm}
\begin{picture}(10,8)(-.4,0)

\put(4.74,0){\makebox(0,0)[l]
  {$\hspace*{.4mm}\cdot\hspace{1.2mm}{\bf 9\over 19}$}}

\put(4.71,.8){\makebox(0,0)[l]
  {$\circ\hspace{.6mm}{\bf 8\over 17}$}}
\put(5.29,.8){\makebox(0,0)[l]
  {$\circ\hspace{.6mm}{\bf 9\over 17}$}}
\put(5.88,.8){\makebox(0,0)[l]
  {$\hspace*{.4mm}\cdot\hspace{1.2mm}{\bf 10\over 17}$}}

\put(2.67,1.6){\makebox(0,0)[l]
  {$\circ\hspace{.7mm}{\bf 4\over 15}$}}
\put(4.67,1.6){\makebox(0,0)[l]
  {$\circ\hspace{.7mm}{\bf 7\over 15}$}}
\put(5.33,1.6){\makebox(0,0)[l]
  {$\circ\hspace{.7mm}{\bf 8\over 15}$}}

\put(2.3,2.4){\makebox(0,0)[l]
  {$\circ\hspace{.7mm}{\bf 3\over 13}$}}
\put(3.07,2.4){\makebox(0,0)[l]
  {$\hspace*{.4mm}\cdot\hspace{1.2mm}{\bf 4\over 13}$}}
\put(4.62,2.4){\makebox(0,0)[l]
  {$\bullet\hspace{.7mm}{\bf 6\over 13}$}}
\put(5.38,2.4){\makebox(0,0)[l]
  {$\bullet\hspace{.7mm}{\bf 7\over 13}$}}
\put(6.15,2.4){\makebox(0,0)[l]
  {$\bullet\hspace{.7mm}{\bf 8\over 13}$}}
\put(6.92,2.4){\makebox(0,0)[l]
  {$\circ\hspace{.7mm}{\bf 9\over 13}$}}

\put(1.82,3.2){\makebox(0,0)[l]
  {$\circ\hspace{.7mm}{\bf 2\over 11}$}}
\put(2.73,3.2){\makebox(0,0)[l]
  {$\bullet\hspace{.7mm}{\bf 3\over 11}$}}
\put(3.64,3.2){\makebox(0,0)[l]
  {$\hspace*{.4mm}\cdot\hspace{1.2mm}{\bf 4\over 11}$}}
\put(4.54,3.2){\makebox(0,0)[l]
  {$\bullet\hspace{.5mm}{\bf 5\over 11}$}}
\put(5.45,3.2){\makebox(0,0)[l]
  {$\bullet\hspace{.7mm}{\bf 6\over 11}$}}
\put(6.36,3.2){\makebox(0,0)[l]
  {$\circ\hspace{.7mm}{\bf 7\over 11}$}}
\put(7.27,3.2){\makebox(0,0)[l]
  {$\bullet\hspace{.7mm}{\bf 8\over 11}$}}

\put(2.22,4){\makebox(0,0)[l]
  {$\bullet\hspace{.7mm}{\bf 2\over 9}$}}
\put(4.44,4){\makebox(0,0)[l]
  {$\bullet\hspace{.7mm}{\bf 4\over 9}$}}
\put(5.56,4){\makebox(0,0)[l]
  {$\bullet\hspace{.7mm}{\bf 5\over 9}$}}

\put(1.43,4.8){\makebox(0,0)[l]
  {$\circ\hspace{.7mm}{\bf 1\over 7}$}}
\put(2.86,4.8){\makebox(0,0)[l]
  {$\bullet\hspace{.7mm}{\bf 2\over 7}$}}
\put(4.29,4.8){\makebox(0,0)[l]
  {$\bullet\hspace{.7mm}{\bf 3\over 7}$}}
\put(5.71,4.8){\makebox(0,0)[l]
  {$\bullet\hspace{.7mm}{\bf 4\over 7}$}}
\put(7.14,4.8){\makebox(0,0)[l]
  {$\bullet\hspace{.7mm}{\bf 5\over 7}\:${\scriptsize\em
  (B-p\hspace{-.2mm}) }}}

\put(2,5.6){\makebox(0,0)[l]
  {$\bullet\hspace{.7mm}{\bf 1\over 5}$}}
\put(4,5.6){\makebox(0,0)[l]
  {$\bullet\hspace{.7mm}{\bf 2\over 5}\:${\scriptsize\em
  (B-p\hspace{-.2mm}) }}}
\put(6,5.6){\makebox(0,0)[l]
  {$\bullet\hspace{.7mm}{\bf 3\over 5}\:${\scriptsize\em B-p}}}
\put(8,5.6){\makebox(0,0)[l]
  {$\bullet\hspace{.7mm}{\bf 4\over 5}$}}

\put(3.33,6.4){\makebox(0,0)[l]
  {$\bullet\hspace{.7mm}{\bf 1\over 3}$}}
\put(6.67,6.4){\makebox(0,0)[l]
  {$\bullet\hspace{.7mm}{\bf 2\over 3}\:${\scriptsize\em B/n-p}}}

\put(10,7.2){\makebox(0,0)[l]
  {$\bullet\hspace{1mm}{\bf 1}$}}

\put(.73,7.2){\makebox(0,0)[r]{\small$\dH=1$}}
\put(.73,6.4){\makebox(0,0)[r]{\small$3$}}
\put(.73,5.6){\makebox(0,0)[r]{\small$5$}}
\put(.73,4.8){\makebox(0,0)[r]{\small$7$}}
\put(.73,4){\makebox(0,0)[r]{\small$9$}}
\put(.73,3.2){\makebox(0,0)[r]{\small$11$}}
\put(.73,2.4){\makebox(0,0)[r]{\small$13$}}
\put(.73,1.6){\makebox(0,0)[r]{\small$15$}}
\put(.73,.8){\makebox(0,0)[r]{\small$17$}}
\put(.73,0){\makebox(0,0)[r]{\small$19$}}

\thinlines
\put(-.5,-.45){\hspace{1mm}\line(1,0){11}}
\put(-.5,7.6){\hspace{1mm}\line(1,0){11}}

\multiput(0,7.6)(0,-0.1){3}{\hspace{1mm}\line(0,-1){0.04}} 
\multiput(0,7)(0,-0.1){74}{\hspace{1mm}\line(0,-1){0.04}}
\put(0,-.4){\hspace{1mm}\line(0,-1){0.1}}

\multiput(1.43,7.6)(0,-0.1){28}{\hspace{1mm}\line(0,-1){0.04}} 
\multiput(1.43,4.7)(0,-0.1){51}{\hspace{1mm}\line(0,-1){0.04}}
\put(1.43,-.4){\hspace{1mm}\line(0,-1){0.1}}

\multiput(1.67,7.6)(0,-0.2){13}{\hspace{1mm}\line(0,-1){0.1}} 
\put(1.67,4.56){\hspace{1mm}\line(0,-1){0.06}}
\multiput(1.67,4.4)(0,-0.2){25}{\hspace{1mm}\line(0,-1){0.1}}

\multiput(2,7.6)(0,-0.1){20}{\hspace{1mm}\line(0,-1){0.04}} 
\multiput(2,5.5)(0,-0.1){21}{\hspace{1mm}\line(0,-1){0.04}}
\multiput(2,2.9)(0,-0.1){33}{\hspace{1mm}\line(0,-1){0.04}}
\put(2,-.4){\hspace{1mm}\line(0,-1){0.1}}

\multiput(2.5,7.6)(0,-0.2){17}{\hspace{1mm}\line(0,-1){0.1}} 
\put(2.5,3.76){\hspace{1mm}\line(0,-1){0.06}}
\multiput(2.5,3.6)(0,-0.2){5}{\hspace{1mm}\line(0,-1){0.1}}
\put(2.5,2.16){\hspace{1mm}\line(0,-1){0.06}}
\multiput(2.5,2)(0,-0.2){13}{\hspace{1mm}\line(0,-1){0.1}}

\multiput(3.33,7.6)(0,-0.1){12}{\hspace{1mm}\line(0,-1){0.04}} 
\multiput(3.33,6.3)(0,-0.1){37}{\hspace{1mm}\line(0,-1){0.04}}
\multiput(3.33,2.1)(0,-0.1){25}{\hspace{1mm}\line(0,-1){0.04}}
\put(3.33,-.4){\hspace{1mm}\line(0,-1){0.1}}

\multiput(5,7.6)(0,-0.2){25}{\hspace{1mm}\line(0,-1){0.1}} 
\put(5,2.16){\hspace{1mm}\line(0,-1){0.06}}
\put(5,2){\hspace{1mm}\line(0,-1){0.1}}
\put(5,1.36){\hspace{1mm}\line(0,-1){0.06}}
\put(5,1.2){\hspace{1mm}\line(0,-1){0.1}}
\put(5,.56){\hspace{1mm}\line(0,-1){0.06}}
\put(5,.4){\hspace{1mm}\line(0,-1){0.1}}
\put(5,-.24){\hspace{1mm}\line(0,-1){0.07}}
\put(5,-.4){\hspace{1mm}\line(0,-1){0.1}}

\multiput(10,7.6)(0,-0.1){4}{\hspace{1mm}\line(0,-1){0.04}} 
\multiput(10,7.1)(0,-0.1){75}{\hspace{1mm}\line(0,-1){0.04}}
\put(10,-.4){\hspace{1mm}\line(0,-1){0.1}}

\thicklines
\put(0,-.6){\hspace{1mm}\makebox(0,0)[t]{$0$}}
\put(.71,-.66){\hspace{1mm}\makebox(0,0)[t]{$\cdots$}}
\put(-.26,-1){\mbox{$\underbrace{\hspace*{19mm}}_{
  \renewcommand{\arraystretch}{.7} \ba \mbox{\small\em Wigner
  crystal} \\ \mbox{\small\em or carrier} \\ \mbox{\small\em
  freeze-out}\ea}$}}
\put(1.43,-.54){\hspace{1mm}\makebox(0,0)[t]{$1\over 7$}}
\put(1.43,-.41){\hspace{1mm}\line(1,0){.24}}
\put(1.67,-.54){\hspace{1mm}\makebox(0,0)[t]{$1\over 6$}}

\put(2,-.54){\hspace{1mm}\makebox(0,0)[t]{$1\over 5$}}
\put(2,-.41){\hspace{1mm}\line(1,0){.5}}
\put(2.5,-.54){\hspace{1mm}\makebox(0,0)[t]{$1\over 4$}}
\put(1.76,-1.5){\mbox{$\renewcommand{\arraystretch}{.7} \ba \uparrow
  \rule[-2mm]{0mm}{5mm} \\ \mbox{\small\em Fermi liquid} \\
  \mbox{\small\em behaviour}\ea$}}

\put(3.33,-.54){\hspace{1mm}\makebox(0,0)[t]{$1\over 3$}}
\put(3.33,-.41){\hspace{1mm}\line(1,0){1.67}}
\put(5,-.54){\hspace{1mm}\makebox(0,0)[t]{$1\over 2$}}
\put(4.26,-1.5){\mbox{$\renewcommand{\arraystretch}{.7} \ba \uparrow
  \rule[-2mm]{0mm}{5mm} \\ \mbox{\small\em Fermi liquid} \\
  \mbox{\small\em behaviour}\ea$}}

\put(10,-.6){\hspace{1mm}\makebox(0,0)[t]{$1$}}
\put(10,-.41){\hspace{1mm}\line(1,0){.5}}
\put(10.4,-.66){\hspace{1mm}\makebox(0,0)[t]{$\sH$}}
\put(8.41,-1){\mbox{$\underbrace{\hspace*{30mm}}_{
  \renewcommand{\arraystretch}{.7} \ba \mbox{\small\em domain of}^^ca\\
  \mbox{\small\em attraction} \\ \mbox{\small\em of $\,\sH=1$}\ea}$}}

\put(.71,7.85){\hspace{1mm}\makebox(0,0){$\cdots$}}
\put(1.43,7.56){\hspace{1mm}\line(1,0){.24}}
\put(1.55,7.75){\hspace{.85mm}\makebox(0,0)[b]{$\Sigma_3^+$}}
\put(1.83,7.75){\hspace{2mm}\makebox(0,0)[b]{$\Sigma_3^-$}}

\put(2,7.56){\hspace{1mm}\line(1,0){.5}}
\put(2.25,7.75){\hspace{1.9mm}\makebox(0,0)[b]{$\Sigma_2^+$}}
\put(2.92,7.75){\hspace{1.4mm}\makebox(0,0)[b]{$\Sigma_2^-$}}

\put(3.33,7.56){\hspace{1mm}\line(1,0){1.67}}
\put(4.17,7.75){\hspace{1.2mm}\makebox(0,0)[b]{$\Sigma_1^+$}}
\put(7.5,7.75){\hspace{1.2mm}\makebox(0,0)[b]{$\Sigma_1^-$}}

\put(10,7.56){\hspace{1mm}\line(1,0){.5}}

\end{picture}
\end{flushleft}

\vspace{31mm}
\noi
{\bf Figure \ref{sKE}.1. }{\em Observed Hall fractions $\,\sH=\nH/\dH\,$
in the interval $\,0<\sH\leq 1$, and their experimental status in
single-layer quantum Hall systems.}
\rule[-4mm]{0mm}{5mm}

\noi
{\small Well established Hall fractions are indicated by
``$\,\bullet\,$''. These are fractions for which an
$R_{xx}\mini$-minimum and a plateau in $R_H$ have been clearly
observed, and the quantization accuracy of $\sH=1/R_H$ is typically
better than $0.5$\%. Fractions for which a minimum in $R_{xx}$ and
typically an inflection in $R_H$ (i.e., a minimum in ${d\mini R_H /
d\mini B_{\!c}^\perp}$, but no well developed plateau in $R_H$) have
been observed are indicated by ``$\,\circ\,$''. If there are only
very weak experimental indications or controversial data for a given
Hall fraction, the symbol ``$\,\cdot\,$'' is used. Finally, ``{\em
B/n-p\,}'' is appended to fractions at which a magnetic field
{\em\mimini(B)} and/or density {\em\mimini(n)} driven phase
transition has been observed.}

\end{figure}

It is convenient to introduce a dimensionless quantity, the
so-called {\em filling factor\,} $\nu$, by setting

\be
\nu\::=\:{n\,{hc \over e} \over B_c}
                           \label{fifa1}
\ee

\noi
where $\,{hc\over e}\,$ is the flux quantum. Then the {\em
classical\,} Hall law~\cfr{Hallc} says that $\RH^{-1}$ rises {\em
linearly\mini} in $\,\nu$, $\,\RH^{-1}={e^2\over h}\,\nu$, the
constant of proportionality being a constant of nature, ${e^2\over
h}$. Since, experimentally, $B_c$ can be varied and $\,n\,$ can be
varied (by varying the gate voltage), this prediction of classical
theory can be put to experimental tests. Experiments at very low
temperatures and for rather pure inversion layers yield the
following very surprizing data some of which are summarized in
Fig.\,\ref{sKE}.1 below:
\vspace*{4mm}

\begin{list}
{\bf (D\arabic{lico})}{\usecounter{lico}\setlength{\labelsep}{.4cm}
\setlength{\labelwidth}{1cm}\setlength{\leftmargin}{1.4cm}
\setlength{\rightmargin}{.7cm}}
\setcounter{lico}{0}

\item  $\sigma_H = {h \over e^2}\,\RH^{-1}\,$
has plateaux at {\em rational heights$\mini$}, i.e., $\,\sigma_H =
\nH/\dH$, with $\,\nH,\, \dH\,$ relatively prime integers (see,
e.g., Tsui, 1990; Stormer, 1992; Du {\em et al.}, 1993). Typically
$\,\dH\,$ is {\em odd$\,$}, but lately plateaux at $\,\sigma_H = {5
\over 2}\,$ (Willett {\em et al.}, 1987; Eisenstein {\em et al.},
1988; Eisenstein {\em et al.}, 1990) and $\,\sigma_H = {1 \over 2}\,$
(Suen {\em et al.}, 1992; Eisenstein {\em et al.}, 1992) have been
observed. The plateaux at integer height occur with an astronomical
accuracy (measurements are precise to one part in $10^8\mini!)$.

\item  When $\,(\nu,\mini\sigma_H)\,$ belongs to a plateau the
longitudinal resistance, $\,\RL$, very nearly {\em
vanishes$\mini$}, i.e., in plateau regions the system is {\em
dissipationless$\mini$}. Inverting the resistivity tensor
$\,\bgrec{\rho}\,$ (see (\ref{OH})) to obtain the conductivity
tensor, $\,\bgrec{\sigma}=\bgrec{\rho}^{-1}$, yields the result
that the diagonal part of $\,\bgrec{\sigma}\,$ {\em vanishes$\,$}
on a plateau.

\item  The precision of the plateau quantization is insensitive to
details of sample preparation and geometry, hence this quantization
is a ``{\em universal}$\,$'' phenomenon.

\item  One has found (Clark {\em et al.}, 1988; Chang and Cunningham,
1989; Simmons {\em et al.}, 1989; Clark {\em et al.}, 1990; Hwang
{\em et al.}, 1992) that, when $\,(\nu,\mini \sigma_H)\,$ belongs to a
pla\-teau at non-integer height, then the system exhibits {\em
fractionally charged$\,$} excitations, (the fractions of $\,e\,$
being related to the value of $\,\dH$).

\item Studies in ``{\em tilted magnetic fields$\,$}'' provide
evidence that, when $\,(\nu,\mini \sigma_H)\,$ belongs to a plateau at
height $\,\sigma_H = {5 \over 2}\,$ (Willett {\em et al.}, 1987;
Eisenstein, Willett {\em et al.}, 1988, 1990), $\,{4 \over 3}\,$
(Clark {\em et al.}, 1989; Clark {\em et al.}, 1990), $\,{8 \over
5}\,$ (Eisenstein, Stormer {\em et al.}, 1989, 1990a), or $\,{2 \over
3}\,$ (Clark {\em et al.}, 1990; Eisenstein, Stormer {\em et al.},
1990b), then the ground state of the system can be {\em
spin-unpolarized$\mini$}. For certain plateaux it might be a {\em
spin-singlet$\,$} state.

\item Magnetic field and density driven {\em phase transitions\mini}
have been reported at $\,\sH=2/3\,$ (Clark {\em et al.}, 1990;
Eisenstein, Stormer {\em et al.}, 1990b; Engel {\em et al.}, 1992).
A magnetic field driven phase transition has been established at
$\,\sH=3/5\,$ (Engel {\em et al.}, 1992), and a possible
observation of such a phase transition at $\,\sH=5/7\,$ has been
discussed in Sajoto {\em et al.}~(1990).
\end{list}


Next, we propose to study what the Ohm-Hall law (\ref{OH}) tells us
about a two-dimensional system of electrons in an external magnetic
field when $\,(\nu,\mini\sigma_H)\,$ belongs to a {\em
plateau$\mini$}. As noted in {\bf (D2)}, experimentally one finds
that, in this situation, the longitudinal resistance $\,\RL\,$
vanishes. This signals the {\em absence of dissipative
processes$\mini$}. The absence of dissipative processes could be
explained if one succeeded in showing that the spectrum of the
many-electron Hamiltonian of the system exhibits an {\em energy
gap$\mini$}, $\,\delta>0$, above the ground-state energy, (or, at
least, that states of very small energy above the ground-state
energy are {\em localized}$\,$). To exhibit a positive energy gap
for certain values of the filling factor $\,\nu$, physically
interpreted as {\em incompressibility$\,$} of the system, poses
difficult analytical problems. Some recent ideas about how to
establish incompressibility at particular filling factors can be
found in the following papers: studies of Laughlin states are given
in Haldane~(1983, 1990), Halperin~(1983, 1984), Laughlin~(1983a,
1983b, 1984, 1990), Arovas, Schrieffer, and Wilczek~(1984), and
Trugman and Kivelson~(1985); off-diagonal long-range order and
Chern-Simons-Landau-Ginzburg theory in fractional QH
fluids have been studied in Girvin and MacDonald~(1987),
Read~(1989), Zhang, Hansson, and Kivelson~(1989), Lee and
Zhang~(1991), Fr\"ohlich~(1992), Fr\"ohlich, Kerler, and
Marchetti~(1992), and Zhang~(1992); finally, for some numerical
studies concerning the question of incompressibility see, e.g.,
Fano, Ortolani, and Colombo~(1986), Yoshioka~(1986), Chakraborty
and Pietil\"ainen~(1987, 1988), Rezayi~(1987), and
d'Am\-bru\-me\-nil and Morf~(1989). What is easier to show is for
{\em which values$\,$} of the parameter $\,\sigma_H = {h \over
e^2}\,\RH^{-1}\,$ a positive energy gap $\,\delta\,$ {\em
cannot$\,$} occur; more precisely, to prove a ``{\em gap labelling
theorem}$\,$''. Such a theorem is stated at the end of
Subsect.\,\ref{ssEE}, and will form the main theme of
Sect.\,\ref{sClQHFs}\mini.

Thus, if the system is incompressible, in the sense that $\,\RL=0$,
then we have the following form for the {\em Hall law} (we use units
such that $e^2/h=1$):

\be
\vec{J}\:=\:\bgrec{\sigma}\,\vec{E} \ , \hspace{.5cm} \mbox{with}
\hspace{.5cm} \bgrec{\sigma}\::=\:\bgrec{\rho}^{-1}\:=\,\left(
\begin{array}{cc} 0 & \sH \\
-\sH & 0 \end{array} \right) \ ,
                             \label{Hall}
\ee

\noi
where $\,\sH=\RH^{-1}$. This is a {\em phenomenological\,} law valid
on macroscopic distance scales and at low frequencies, as mentioned
after~(\ref{OH}). More fundamental are the following two laws: {\em
Charge conservation}

\be
{1 \over c}\dd{t}\,J^0 + \vec{\nabla}\cdot\vec{J}\:=\:0 \ ,
                           \label{ED0}
\ee

\noi
(continuity equation), where $\,J^0\,$ is ($c\,$ times) the
electric charge density, and {\em Faraday's induction law},

\be
{1 \over c}\dd{t}\,B + \vec{\nabla} \times \vec{E}\:=\:0 \ ,
                           \label{ED1}
\ee

\noi
where $\,B\,$ denotes the component of the magnetic field
perpendicular to the plane of the system, and $\,\vec{E}\,$ is the
electric field in the plane of the system. We note that
the dynamics of charged, spinless particles confined to a plane
only depends on the component of the magnetic field perpendicular
to the plane of the system and the components of the electric field
in that plane. Combining Eqs.~(\ref{Hall}) through (\ref{ED1}), we
find that

\be
\dd{t}\,J^0\:=\:\sH\,\dd{t}\,B \ .
                           \label{ED2}
\ee

Eq.~(\ref{ED2}) can be integrated with respect to time $\,t$. By
$\,J^0=J^0_{tot} - n\mini ec\,$ we denote the difference between
the total electric charge density, $\,J^0_{tot}$, and the
uniform background density, $\,n\mini ec$, of a system in a uniform
background magnetic field $\,B_c$. Likewise, $\,B\,$ denotes the
difference between the total magnetic field, $\,B_{tot}$, and the
uniform background field $\,B_c$. Then Eq.~(\ref{ED2}) implies the
{\em charge-flux relation}

\be
J^0\:=\:\sH\,B \ .
                           \label{ED3}
\ee

It is convenient to introduce the electromagnetic field tensor
which is a 2-form, $\,F$, given by

\be
F\:=\:\halb\sum_{\mu,\nu} F_{\mu\nu}\,dx^\mu\!\wedge\!dx^\nu\ ,
\hspace{.5cm} \mbox{with} \hspace{.5cm} (F_{\mu\nu})\:=\:\left(
\begin{array}{ccc}
0 & E_x & E_y \\
-E_x & 0 & -B \\
-E_y & B & 0
\end{array} \right) \ ,
                             \label{ED4}
\ee

\noi
and the 2-form $\,\cal J$ dual to the current density
$\,J^\mu=(J^0,\mini\vec{J}\,)$, i.e.,

\be
{\cal J}=\:\halb \sum_{\mu,\nu} {\cal J}_{\!\mu\nu} \,
dx^\mu\!\wedge\! dx^\nu\ , \hspace{.5cm} \mbox{with} \hspace{.5cm}
{\cal J}_{\!\mu\nu}=\: \eps{\mu\nu\rho}\,J^\rho \ .
\ee

\noi
Then Eqs.~(\ref{Hall}) and (\ref{ED3}) can be combined into one
equation,

\be
{\cal J}=\:-\sH\,F \ ,
                             \label{ED5}
\ee

\noi
while current conservation (\ref{ED0}) is expressed as

\be
\da{\cal J}=\:\halb \sum_{\alpha,\mu,\nu}\, \partial_\alpha\mini
{\cal J}_{\!\mu\nu} \: dx^\alpha\!\wedge\!dx^\mu\!\wedge\!dx^\nu
\:=\: 0 \ ,
                             \label{ED6}
\ee

\noi
and Faraday's induction law (\ref{ED1}) becomes

\be
\da F\:=\:0 \ .
                             \label{ED7}
\ee

\noi
Eqs.~(\ref{ED5}) through (\ref{ED7}) are compatible with each other
if, and only if, $\,\sH\,$ is {\em constant$\mini$}. If the values
of $\,\sH\,$ along the two sides of a curve $\,\Gamma\,$ differ from
each other -- which happens, for example, at the boundary of the
system -- then an {\em additional$\,$} current, $\,\vec{\cal I}$,
{\em not$\,$} described by Eq.~(\ref{Hall}), is observed in the
vicinity of $\,\Gamma$, in order to reconcile charge conservation
with the induction law. [For time-independent fields one finds that
$\,\vec{\nabla}\cdot\vec{\cal I} = (\vec{\nabla}\sH) \times
\vec{E}\,$; see also Halperin~(1982) and Sect.~\ref{sAC} below.]

Note that Eqs.~(\ref{ED5}) through (\ref{ED7}) are {\em generally
covariant$\,$} and independent of metric properties of the system.
Eqs.~(\ref{ED6}) and~(\ref{ED7}) can be integrated by introducing
the 1-forms (or ``vector potentials'') $\,A\,$ and $\,b$, with

\be
{\cal J}=\:\da b\ , \hspace{.5cm}\mbox{i.e.,}\hspace{.5cm}{\cal
J}_{\!\mu\nu}=\:\partial_\mu b_\nu - \partial_\nu b_\mu \ ,
\hspace{.5cm} \mbox{and} \hspace{.5cm} F\:=\:\da A \ .
\ee

\noi
Eq.~(\ref{ED5}) then reads

\be
\da b\:=\:-\sH\,\da A \ .
                             \label{ED8}
\ee

\noi
Eq.~(\ref{ED8}) is the Euler-Lagrange equation derived form an
{\em action principle}$\,$! The corresponding action, $S(b\mini;
A)$, is given by

\bea
S(b\mini; A) & := & {1 \over 2\mini c^2\sH} \iL b\wedge\! \da b\:
+\, {1 \over c^2} \iL  A \wedge\! \da b\: +\:
\BT(A|_{\partial\Lambda},\mini b|_{\partial\Lambda})  \nonumber  \\
 & = & {1 \over 2\mini c^2\sH} \iL (\da^{-1}{\cal J}) \wedge\! {\cal
J} +\, {1 \over c^2} \iL  A \wedge\! {\cal J} +\:
\BT(A|_{\partial\Lambda},\mini {\cal J}\!|_{\partial\Lambda})  \,,
\hspace{1.4cm}
                             \label{ED9}
\eea

\noi
where $\,\Lambda:=\BB{R}\times\Omega\,$ is the space-time domain to
which the system is confined, and ``$\BT$'' stands for {\em boundary
terms$\,$} which only depend on the restrictions
$\,A|_{\partial\Lambda}\,$ and $\,b|_{\partial\Lambda}\,$ of $\,A\,$
and  $\,b\,$ to the space-time boundary $\,\partial\Lambda\mini$,
(see the remark after~(\ref{ED11}) below). Moreover, $\,S(b\mini;
A)\,$ shall be varied with respect to the {\em dynamical$\,$}
variable, that is, with respect to $\,b\,$; (the vector potential
$\,A\,$ of the electromagnetic field is a tunable, external field).

Why is the result (\ref{ED9}) interesting? It is interesting,
because an equation of motion, like (\ref{ED8}), that can be
derived form an action principle can be quantized easily; for
example by using {\em Feynman path integrals$\mini$}. Clearly, the
current density $\,{\cal J}$ of a system of electrons must be
interpreted as a quantum-mechanical operator-valued distribution.
Hence (\ref{ED8}) must be {\em quantized$\mini$}. We note that, in
the present example, Feynman path integral quantization has a
mathematically rigorous interpretation. In
formal, ``physical'' notation, Feynman's path space measure is
given by

\be
dP_A(b)\:=\: Z(A)^{-1}\, \exp\biggl[\mini\ih\mini S(b\mini ;
A) \biggr] \:{\cal D}b\ ,
                             \label{ED10}
\ee

\noi
where the {\em partition~-} or {\em generating function$\,$}
$Z(A)\,$ is chosen such that (formally) $\,\int dP_A(b)=1\,$. This
implies that

\be
Z(A)  \:=\:  Z_o\,\exp\biggl[ -{i\,\sH \over 2\mini\hbar c^2} \iL A
\wedge\! \da A\: + \:\BT(A|_{\partial\Lambda})\, \biggr] \ ,
                             \label{ED11}
\ee

\noi
where $\,Z_o\,$ is a constant {\em independent$\,$} of $\,A$.
[In~(\ref{ED10}), we have omitted a gauge fixing term for the
integration over the field $\,b\mini$.] At this point, we
emphasize that a non-trivial boundary
term, $\,\BT(A|_{\partial\Lambda})$, in Eq.~(\ref{ED11}) is a
necessity forced upon us by the \Ue\ {\em gauge invariance of
quantum mechanics$\,$} (see Sect.\,\ref{sGI}): Under a \U\ gauge
transformation, $\,A\mapsto A+\da\chi$, the Chern-Simons term
in~(\ref{ED11}) transforms according to

\bea
\iL A \wedge\! \da A & \mapsto & \iL A \wedge\! \da A\: - \:
\int_{\partial\Lambda}\da\chi\wedge\! A \ ,
                             \label{ED111}
\eea

\noi
i.e., there is a {\em gauge anomaly$\,$} localized at the boundary
$\,\partial\Lambda\mini$. Hence, in order for the partition function
$\,Z(A)\,$ to be \U\ gauge invariant, the presence of a boundary
term, $\,\BT(A|_{\partial\Lambda})$, exhibiting a gauge anomaly
cancelling the one in~(\ref{ED111}) is indispensable! We shall see
that the anomalous part of $\,\BT(A|_{\partial\Lambda})\,$ turns out
to be the generating functional of the connected Green functions of
chiral current operators generating a \uh\ current (Kac-Moody)
algebra which, physically, describes charged, chiral waves
circulating at the edge of the QH sample. Sect.~\ref{sAC}
will be devoted to an investigation of this current algebra. Our
analysis will lead to a list of the possible (quantized) values of
the response coefficient of the Hall conductivity $\,\sH\mini$; see
also remark {\bf (iii)} at the end of this subsection.

Given the expression~(\ref{ED11}) for the partition function
$\,Z(A)$, we may ask wheth\-er there is a simple way of recovering
the action $\,S(b\mini ; A)\,$ as a functional of the vector
potential $\,b\,$ of the conserved current density $\,{\cal J}\!$,
as given in~(\ref{ED9}). The answer is yes. It is provided by the
following {\em functional Fourier transform}$\,$ identity:

\be
\exp\biggl[\mini\ih\mini S(b\mini; A)\biggr] \:=\: \const\,^^ca\int
{\cal D}\alpha\: \exp \biggl[-{i \over \hbar c^2} \int \alpha
\wedge\! \da b\biggr]\: Z(A+\alpha) \ ,
                             \label{ED12}
\ee

\noi
where we again omit a suitable gauge fixing term for the integral
over $\,\alpha\mini$; see Fr\"ohlich and Kerler~(1991), and the general
discussion in Sect.\,\ref{sSLaC} below.

\bigskip
{\bf Remarks. (i)}$\;$ Defining the {\em effective action$\mini$},
$\,\Seff(A)$, of a two-dimensional electronic system by

\be
\Seff(A) \::=\: {\hbar \over i}\, \ln Z(A) \ ,
                             \label{ED13}
\ee

\noi
our circle of arguments can be closed from Eq.~(\ref{ED11}) back to
the starting point of the Hall law (\ref{Hall}) (and (\ref{ED3})) by
noting that

\be
J^i \:=\: c^2\, {\delta \Seff(A) \over \delta A_i} \:=\: \sH\,
\varepsilon^{ij}\mini E_j \ ,
                             \label{ED14}
\ee

\noi
where $\,E_j=\partial_j A_0 - \partial_0 A_j$, for $\,j=1,\mini 2$.

Eqs.~(\ref{ED14}), (\ref{ED13}) and (\ref{ED12}) make it clear
that one approach leading to an understanding of the QH effect is to
derive, for a QH fluid at particular values of the filling factor
$\nu$, the effective action $\,\Seff(A)\,$ corresponding to
Eq.~(\ref{ED11}) from ``first principles''. In Subsect.\,\ref{ssSL},
we show that {\em gauge invariance$\,$} of non-relativistic quantum
mechanics and the single assumption of {\em incompressibility$\,$}
of QH fluids are sufficient to uniquely determine their effective
action $\,\Seff(A)\,$ in the ``scaling limit'' and thereby to
derive~(\ref{ED11}). Hence the phenomenology of QH systems at low
frequencies and on large distance scales, including the
quantization of the Hall conductivity $\,\sH\,$ (see remark
{\bf (iii)} below), can be derived from gauge invariance and
incompressibility. This shows that a proof of incompressibility of
electronic systems at particular values of the filling factor $\nu$
is really the essential problem in the theory of the QH effect in
need of further investigation.

{\bf (ii)}$\;$ It can be seen directly from the Ohm-Hall law (\ref{OH})
that the incompressibility of QH fluids is a crucial property which
allows for a description of these systems in terms of an effective
action formalism: Only for {\em dissipationless$\,$} systems, i.e.,
for systems with an {\em antisymmetric$\,$} conductivity tensor
$\,\bgrec{\sigma}$, it is possible to functionally integrate
relation~(\ref{ED14}) in order to obtain an effective action. In
other words, for electronic systems with {\em dissipation$\mini$},
one {\em cannot$\,$} formulate the Ohm-Hall law (coming from
transport theory) within an effective action formalism.

{\bf (iii)}$\;$ Clearly, we must require that the quantum theory
with action $\,S(b\mini; A)\,$ given by (\ref{ED9}), defined by
the Feynman path integral (\ref{ED10}), describe {\em localized,
particle-like excitations$\,$} with the quantum numbers of the {\em
electron$\,$} or {\em hole$\mini$}, i.e., with electric charge
$\,\pm e\,$ and Fermi statistics. We shall see in Sects.\,\ref{ssEE}
and~\ref{ssDic} that this requirement implies that, for {\em
consistency$\,$} of the theory, the dimensionless Hall conductivity
({\em Hall fraction\,}) $\,\sigma = {h \over e^2}\,\sH \,$ must be a
{\em rational number$\mini$}.




\section{Scaling Limit of the Effective Action of Fermi Systems, and
  Classification of States of Non$\,$-Relativistic Matter}
\label{sSLaC}

In this section, we generalize the observations made at the end of
the last subsection. We review a conceptual framework (Fr\"ohlich,
G\"otschmann, and Marchetti, 1995a, 1995b), based on the bosonization
of quantum systems with infinitely many degrees of freedom, which
has proven to be useful in attempting to classify states of
non-relativistic matter at very low temperatures. In this section,
we focus our attention on the analysis of electric charge transport
properties. 
Magnetic properties can be studied in a similarly general manner; for
an example see Sect.~\ref{sSL} where two-dimensional, incompressible
quantum fluids are treated, {\it including} their magnetic properties.

The basic ideas underlying our approach are very simple: The
starting point is to study the response of a quantum system of
charged particles to perturbations by external electromagnetic
fields. Thus we couple the electric current density, $J^\mu$, to an
arbitrary, smooth, external electromagnetic vector potential
(1-form), $A$, and then attempt to calculate the partition
function, $Z(A)$, of the system as a functional of $A$.

Of course, this is a very complicated task. However, in order to
{\em classify electric properties of non-relativistic matter}, we are
really only interested in understanding the behaviour of the {\em
effective action} (see also~\cfr{ED13})

\be
\Seff(A) \::=\: {\hbar \over i}\, \ln Z(A) \ ,
                             \label{SLC1}
\ee

\noi
on very large distance scales and at very low frequencies. We thus
study families of systems confined to ever larger cubes,
$\Omega^{(\tha)}\!:=\!\{\vec{x}\,|\,Ê\vec{x}/\tha\in\Omega \}$, in
physical space \BBn{E}{d}, $d=1,2,3$, where $\Omega$ is a fixed
compact subset of $\BBn{E}{d}$, and $\,1\leq \tha < \infty\,$ is a scale
parameter. We keep the particle density, $n$, and the temperature
$\,T\:(\approx 0\,K)$ constant. We then couple the electric current
density of the system confined to $\Omega^{(\tha)}$ to a vector
potential $\,A^{(\tha)}$ given by

\be
A^{(\tha)}(t,\vec{x}\,) \::=\: \tha^{-1}\,A({t \over\tha},{\vec{x}
\over\tha})\ , \hsp {\vec{x}\over\tha}\in\Omega\ ,
\ee

\noi
where $A$ is an arbitrary, but $\tha$-independent vector potential
on $\,\BB{R}\times\Omega$. We then study the behaviour
of $\,\Seff_{\Omega^{(\tha)}}(A^{(\tha)})\,$ when $\tha$ becomes large.

More precisely, we attempt to expand
$\,\Seff_{\Omega^{(\tha)}}(A^{(\tha)})\,$ in powers of $\,\tha^{-1}$
around $\,\tha=\infty\,$ (to a finite order) and define the {\em
scaling limit}, $\Sast(A)$, {\em of the effective acction\mini} to
be the coefficient of the leading power of $\tha$ in that
expansion; more details are given in Subsect.\,\ref{ssSL}.

Remarkably, all many-body systems of non-relativistic electrons
that can be controled analytically have the property that
$\Sast(A)$ is {\em quadratic\mini} in $A\mini$; this holds, in
particular, for insulators, Landau-Fermi (non-interacting electron)
liquids, metals, incompressible quantum Hall fluids (e.g., Laughlin
fluids), and superconductors. Since we do not know of a proof of this
fact for any imagineable system of non-relativistic electrons at
positive density, we call this fact a {\it ``quasi-theorem''} and
summarize below the explicit results of a case-by-case analysis that
has been presented in Fr\"ohlich, G\"otschmann, and Marchetti
(1995a, 1995b). It should be emphasized at this point, that, except
in the case of insulators and incompressible quantum Hall fluids (see
Subsect.~\ref{ssSL}), the proof of the quasi-theorem is never
trivial and does {\em not\,} simply follow from dimensional analysis
(and gauge-invariance).

It is well known that $\Seff(A)$ is the generating functional of
connected Green functions of the electric current density $J^\mu$ and
that it is {\em gauge-invariant}, i.e.,

\be
\Seff(A+\da\chi) \:=\: \Seff(A)\ ,
\ee

\noi
where $\chi$ is an arbitrary, real function on space-time, and
$\da\chi$ is its exterior derivative. Clearly, gauge-invariance
persists upon passing to the scaling limit. Using our quasi-theorem,
we conclude that

\be
\Sast(A) \:=\: \halb(A,\Piast\,A) \:=\: \halb\int d^{\mini d+1}x \int
d^{\mini d+1}y\,A_\mu(x) \Pi^{\ast\mu\nu}(x-y)A_\nu(y)\ ,
                             \label{SLC2} 
\ee

\noi
where, for $\,x\neq y$, $\Pi^{\ast\mu\nu}(x-y)\,$ is given by the
scaling limit of the connected two-current Green function $\,\El\,
T(\mini J^\mu(x)J^\nu(y)\mini)\, \Er^{\mbox{\scriptsize
con}}$.

By gauge-invariance, or, equivalently, current conservation,

\be
\partial_\mu\Pi^{\ast\mu\nu} \:=\: \partial_\nu\Pi^{\ast\mu\nu}
\:=\: 0\ .
                             \label{SLC3} 
\ee

\noi
Furthermore, $\Pi^{\ast\mu\nu}$ inherits all the symmetries of the
system; (in~\cfr{SLC2}, we have assumed translation invariance, in the
limit when $\Omega$ increases to $\BBn{E}{d}$).
Thus classifying electric properties of a system reduces, in the
scaling limit and under the assumption that the quasi-theorem
holds, to a classification of ``vacuum polarisation tensors'',
$\Pi^{\ast\mu\nu}$, satisfying~\cfr{SLC3} and having certain
symmetries.

In principle, essentially all information concerning electric
properties of a system can be retrieved from its effective action
$\Seff(A)$ by fairly straightforward calculations. However, in
order to evoke and then apply analogies with other physical systems,
in particular with gauge theories of elementary particle physics, it
is useful to embark on a detour. The detour chosen here is {\em
bosonization}. The idea (exemplified by our discussion in
Subsect.~\ref{ssQH}) is as follows: Since the electric current
density $\,J^\mu\ (\mu=0,\ldots,d)\,$ is conserved, it can be derived
from a ``potential'',

\be
J^\mu \:=\: \varepsilon^{\mu\mu_1\cdots\mini\mu_d}\,\partial_{\mu_1}
b_{\mu_2\cdots\mini\mu_d}\ ,
                            \label{SLC4} 
\ee

\noi
where $\,\varepsilon\,$ is the totally antisymmetric tensor in $d+1$
dimensions, and $b_{\mu_2\cdots\mini\mu_d}$ is an antisymmetric tensor
field of rank $\,d-1\mini$. In the language of differential forms,
Eq.~\cfr{SLC4} is expressed as

\be
J \:=\: J_\mu\, dx^\mu \:=\: \ast\,\da b\ ,
                            \label{SLC5}
\ee

\noi
where $\ast$ is the Hodge $\ast$-operation and $\da$ denotes
exterior differentiation; see Subsect.\,\ref{ssDG}. The ``potential''
$b\,$ of the current density $J\,$ is determined by~\cfr{SLC5} only
up to the exterior derivative of an  antisymmetric tensor field of
rank $\,d-2$, (a $(d\!-\!2)$-form). Thus $b\,$ is what one calls a
``{\em gauge form\,}''. For  one-dimensional systems, $b\,$ is a
scalar and is determined  by~\cfr{SLC5} up to a constant. In two
dimensions, $b\,$ is a  $1$-form determined by $J\,$ up to the
exterior derivative of an  arbitrary function.

The basic idea is then to derive the effective field theory for the
field $b\,$ from $\Seff(A)$. This field theory is determined by an
action $\Stieff(b)$. Choosing units in which $\,\hbar=c=1$ and using
an {\em imaginary-time (euclidean)} formulation, $\Stieff(b)$ is
obtained from $\Seff(A)$ by functional Fourier transformation

\be
e^{-\Stieff(b)} \:=\: N^{-1} \int{\cal D}A\:e^{-\Seff(A)}\, e^{i\int
A\wedge d b}\ ,
                            \label{SLC6}
\ee

\noi
where $N$ is a (divergent) normalization factor (proportional to
the volume of the Lie algebra of $U(1)$ gauge transformations); see
also~\cfr{ED12} in the previous subsection. Gauge invariance
implies that

\be
\Stieff(b+\da\Lambda) \:=\: \Stieff(b)\ ,
                            \label{SLC7}
\ee

\noi
for an arbitrary $(d\!-\!2)$-form $\Lambda\ (d\geq 2)$. Our
quasi-theorem then implies that the low-wave-vector, low-energy
modes of the {\em bosonic\mini} field $b\,$ are {\em
non-interacting}, i.e., the scaling limit of the system described
in terms of the $b\mini$-field has a {\em quadratic action},
$\Sti^{\mini\ast}$, whose form is constrained by its gauge
invariance, Eq.~\cfr{SLC7}, and by the symmetries of the system.

Many quantities of interest in the original system can
be expressed in terms of quantities referring to the $b\mini$-field.
For example, current Green functions (at imaginary time) are given by
expectations of products of the ``field strength'' $\da b\,$ in the
functional measure

\be
Z^{-1}\,e^{-\Stieff(b)}\,{\cal D}b\ .
\ee

\noi
Green functions of electron creation - and annihilation operators
turn out to be proportional to expectations of {\em disorder
operators\mini} of the ``dual'' theory formulated in terms of the
$b\mini$-field (Fr\"ohlich, G\"otschmann, and Marchetti, 1995a).

If $\Sigma$ is an open subset of physical space, and ${\cal
Q}_\Sigma$ denotes the operator measuring the total electric charge
inside $\Sigma$ then

\be
e^{i\mini\alpha{\cal Q}_\Sigma} \;\propto\;
e^{i\mini\alpha\int_{\partial\Sigma} b}\ ,\hsp \alpha\in\BB{R}\ ,
                            \label{SLC8}
\ee

\noi
i.e., the charge operator can be reconstructed from operators
analogous to the {\em Wilson loops\mini} of gauge theory. The
operators in~\cfr{SLC8} are {\em dual\,} to the disorder operators
describing electron creation and annihilation, in the sense of
Wegner$\,$-$\mini$'t Hooft duality (Wegner, 1971; 't Hooft, 1978).
This suggests that we can carry over the consequences of
Wegner$\,$-$\mini$'t Hooft duality form gauge theory to condensed
matter physics. As a consequence, we mention that if, at zero
temperature, the total electric charge operator, ${\cal
Q}=\lim_{\Sigma\nearrow\bb{R}^d}{\cal Q}_\Sigma$, is well defined
on the Hilbert space of all physical states of the system then
disorder Green functions\,--\,e.g., the Green function of an
electron creation - and an annihilation operator\,--\,exhibit strong
spatial cluster decomposition properties, and conversely. This
applies to {\em insulators, incompressible quantum Hall fluids, and
superconductors with (unscreened) Coulomb repulsion}. In contrast, if
the field $b\,$ couples the ground state of the 
system to a massless quasi-particle then the total charge operator
does {\em not\,} exist, because charge fluctuations are divergent in
the thermodynamic limit, as in metals and massless superconductors.

For {\em two-dimensional\,} systems, both fields, $A\,$ and $b$, are
$1$-forms defined up to gradients of scalar functions. In this
case, Eq.~\cfr{SLC6} enables us to define a notion of {\em
duality\mini}: a system $1$ and a system $2$ are {\em dual\,} to
each other if, and only if,

\be
\Sti_1^{\mini\ast} \;\propto\; \Sast_2\ .
                            \label{SLC9}
\ee

\noi
It turns out that, in the sense of Eq.~\cfr{SLC9}, a
two-dimensional {\em insulator\mini} is {\em dual\,} to a
two-dimensional {\em London superconductor}, a metal to a
``semi-conductor'', and an incompressible quantum Hall (e.g. a
Laughlin) fluid is {\em self-dual\mini}.

Besides the duality expressed in Eqs.~\cfrs{SLC6}{SLC9}, there is
also a notion of Kramers\mini-Wannier duality: in {\em any dimension}
$d$, a $U(1)$ gauge theory of an antisymmetric tensor field $b$ of
rank $d-1$ is ``Kramers\mini-Wannier dual'' to a {\em scalar\mini}
field theory. Thus, for example, a London superconductor, corrected
by dynamical Abrikosov vortices, is Kramers\mini-Wannier dual to a
Landau-Ginzburg superconductor; see Subsect.~IV.D in Fr\"ohlich and
Studer (1993b).

The two notions of duality sketched here are conceptually quite
clarifying and useful in a classification of electric properties
of non-relativistic matter. One of the principal advantages of
reformulating the theory of a system of electrons in terms of the
tensor field $b\,$ (bosonization) is that this formulation is
convenient to explore properties of systems obtained by perturbing a
given one by 
{\em two-body interactions\mini}. A translation-invariant two-body
interaction, $\Iper\mini$, has the form

\be
\Iper \:=\: \halb \int d^{\mini d+1}x\int d^{\mini d+1}y\, J^\mu(x)
V_{\mu\nu}(x-y) J^\nu(y)\ .
\ee

\noi
After bosonization, the effective action of the perturbed system is
given by

\be
\Stitoeff(b) \:=\: \Stieff(b)+\halb \int d^{\mini d+1}x \int
d^{\mini d+1}y\, (\ast\mini\da b)^\mu(x) V_{\mu\nu}(x-y)
(\ast\mini\da b)^\nu(y)\ ,
                            \label{SLC10}
\ee

\noi
where $\Stieff(b)$ is the effective action of the unperturbed system.
Note that, expressed in terms of the field $b\,$, the two-body
interaction $\Iper$ is {\em quadratic\mini} (rather than quartic)! A
conventional two-body interaction described by an instantaneous
two-body potential corresponds to a kernel $V_{\mu\nu}$ given by

\be
V_{\mu\nu}(x-y) \:=\: \delta_{\mu 0}\,\delta_{\nu 0}\,V(\vec{x}-
\vec{y}\,)\,\delta(x^0-y^0)\ ,
\ee

\noi
with $\,x:=(x^0,\vec{x}\,)$ and $\,y:=(y^0,\vec{y}\,)$.

Suppose now that the action of the perturbed system in the scaling
limit (scale parameter $\,\tha\rightarrow\infty$),
$\Stito^{\mini\ast}$, is given by the scaling limit,
$\Sti^{\mini\ast}$, of the action of the unperturbed system, perturbed
by the long-range tail,  $\Iper^{\mini\ast}\mini$, of the 
two-body interaction $\Iper\mini$. {}From our quasi-theorem we infer
that $\Sti^{\mini\ast}$ is quadratic in $b$, and hence, since
$\Iper^{\mini\ast}$ is quadratic in $b$, $\Stito^{\mini\ast}$ is
{\em quadratic\mini} in $b$, too, and, by~\cfr{SLC7}
and~\cfr{SLC10}, gauge-invariant. It is given by

\be
\Stito^{\mini\ast} \:=\: \Sti^{\mini\ast} + \theta^\kappa\,
\Iper^{\mini\ast}\ ,\hsp (\theta\rightarrow\infty)\ ,
\ee

\noi
for some exponent $\kappa$. It is plausible that the
assumption that perturbation by $\Iper$ and passage to the scaling
limit are commuting operations is justified if $V_{\mu\nu}$ is {\em
positive-definite\mini} and of very {\em long range\mini} (e.g.,
Coulomb repulsion). In this case, the analysis sketched above
yields a variant of the ``random phase
approximation'' (RPA). These ideas have been applied in Fr\"ohlich,
G\"otschmann, and Marchetti~(1995a, 1995b) to the following systems:

(i) A metal pertubed by repulsive two-body Coulomb interactions. In
this case, one obtains the exact formula for the {\em plasmon gap}.

(ii) A massless London superconductor perturbed by repulsive
two-body Coulomb interactions. In this example, one recovers a
precise formulation of the {\em Anderson-Higgs mechanism}; (the
$U(1)$-Goldstone boson becomes massive).

(iii) A Landau-Fermi liquid perturbed by repulsive two-body
interactions, in the presence of a static source. For a
two-dimensional system of this type with Coulomb-Amp\`ere
interaction, a possible crossover to a non-Fermi-liquid behaviour
(``{\em Luttinger liquid\,}'') has been observed.

\bigskip

{\bf Remark}: \ An effective field theory similar to the one described
here for the electric current density can also be derived for the {\it
  spin} $(SU(2)-)$ {\it current density} by using tools developed in
the study of the so-called BF-theories (see Cattaneo et
al.~(1995)). It is useful in the analysis of {\it magnetic properties}
of electron gases (and will be described elsewhere). 

\bigskip

{\bf Examples.} In the second part of this section, we provide some
explicit examples illustrating the ideas presented above.
\rule[-4mm]{0mm}{5mm}

{\bf (F)} We start with the {\em free non-relativistic fermion 
gas\mini}. The euclidean action of the system is given by

\be
S_E(\Psi,\Psi^\ast) \:=\: \int d^{\mini d+1}x \biggl[
\Psi^\ast\partial_0\Psi + {1\over 2m} |\vec{\nabla}\Psi|^2 +\mu\,
\Psi^\ast\Psi \biggr](x)\ ,
\ee

\noi
where $m$ is the mass of the fermions and $\mu$ the chemical
potential.

In $\,d=1$, the Fermi surface consists of only two points, $\pm
k_F$. Since physics in the scaling limit is dominated by excitations with
momenta close to the Fermi surface, one expands the momentum-space
fermionic two-point Green function around the points $\pm k_F$ on
the Fermi surface, in order to calculate its scaling limit.

The result of this analysis is that, in the scaling limit, one can
introduce quasi-particle fields, $\Psi_L,\,\Psi^\ast_L$ and
$\Psi_R,\,\Psi^\ast_R$, corresponding to the points $-k_F$ and
$k_F$, respectively. The fields $\Psi_L,\,\Psi^\ast_L$ 
describe left-moving excitations, while $\Psi_R,\,\Psi^\ast_R$
describe right movers. These excitations approximately obey a
``relativistic'' dispersion relation, $\omega\approx |p|$, where
$\,\omega=E-E_F\,$ and $\,p=k-k_F$. The original field variable
$\Psi$ is related to the relativistic field variables $\Psi_L$ and
$\Psi_R$ by

\be
\Psi(x) \:\approx\: e^{-ik_F x^1}\Psi_R(x) + e^{ik_F x^1} 
\Psi_L(x)\ .
                         \label{Seffexp3}
\ee

\noi
The euclidean action of $\Psi_L$ and $\Psi_R$ is given by

\be
S_E(\Psi_R,\Psi^\ast_R,\Psi_L,\Psi^\ast_L) \:=\: \int d^2x
\biggl[ \Psi^\ast_R(\partial_0+iv_F\mini\partial_1)\Psi_R +
\Psi^\ast_L(\partial_0-iv_F\mini\partial_1)\Psi_L \biggr](x)\ .
                         \label{Seffexp4}
\ee

\noi
The action~\cfr{Seffexp4} is the euclidean action of {\it free, massless
Dirac fermions} (in two dimensions), with the Fermi velocity $v_F$
playing the role of the velocity of light. As a consequence, in the
scaling limit, the effective action $\Sast(A)$ obtained by minimal
coupling of $\Psi$ and $\Psi^\ast$ to the gauge
field $A$ is {\it quadratic} in $A$; for more details, see
Subsect.~\ref{ssIQ}. Using the bosonization method sketched above,
one obtains a quadratic action for the potential $b\,$ of the
conserved $u(1)$ current. These features of one-dimensional systems
have been known for many years; see Lieb and Mattis (1965), and
Luther and Peshel (1975).

More recently, it has been realized (see Luther, 1979, for the
original observations) that similar ideas on the scaling limit can
be used in any dimension $d\mini$. The sum over the two points of the
Fermi surface in~\cfr{Seffexp3} must be replaced by an integration
over a $(d\!-1\!)$-dimensional Fermi surface; see Feldman and
Trubowitz (1990), Benfatto and
Gallavotti (1990), Anderson (1990), Shankar (1991), and Feldman,
Magnen {\em et al.}~(1992).

Let $\Omega$ be the $(d\!-1\!)$-dimensional unit sphere in momentum
space. Elements of $\Omega$ are denoted by $\bfom$. The extension
of formula~\cfr{Seffexp3} to higher dimensions is given by

\be
\Psi(x) \:\approx\: \int_\Omega d\bfom\,e^{-ik_F\mini
\vec{x}\mini\cdot\mini\bfom}\,\Psi_\bfom(x)\ ,
\ee

\noi
and the euclidean action for the fields $\Psi_\bfom$ describing
the scaling limit of the free Fermi gas has the form

\be
S_E(\{\Psi_\bfom,\Psi_\bfom^\ast\}) \:=\: {\rm const.} \, (k_F
\alpha)^{d-1}\; \int_\Omega d\bfom \int
d^{\mini d+1}x \biggl[ \Psi^\ast_\bfom(\partial_0+iv_F\mini\bfom
\cdot \vec{\nabla}) e^{-\alpha(\bfom\wedge\!\vec{\nabla})^2}
\Psi_\bfom \biggr](x)\ ,
                         \label{Seffexp6}
\ee

\noi
where $\alpha$ is a large constant. 

Coupling $\Psi$ and $\Psi^\ast$ minimally to the gauge field $A$, one
can show (see Fr\"ohlich, G\"otschmann, and Marchetti (1995b)) that
$\Sast(A)$ is given by an  integral over the set of directions,
$\pm\bfom$, in momentum space of contributions coming from the
degrees of freedom described by the fields $\Psi_\bfom^\#$ and
$\Psi_{-\bfom}^\#$, corresponding to antipodal points, $\pm\bfom$, on
the Fermi surface. Every such contribution corresponds to the one
of a one-dimensional free fermion system and is
quadratic in $\,A_\bfom=(A_0,\mini\bfom\cdot \vec{A}\,)$. Since
$\Sast(A)$ is an integral of the effective  actions
$\Sast(A_\bfom)$ of one-dimensional systems over all pairs of points
$\pm\bfom$ in $\Omega$, 
it, too, is quadratic in $A$. These considerations yield an explicit
expression for $\Sast(A)$, in particular for the polarization tensor
$\Pi^\ast$. 

Let $\Pi^{\mu\nu}(x,y)$ denote the vacuum polarization tensor of a
system of electrons  at zero temperature, and let
$n^{(1)}=n\,dx^0$, where $\,n\,$ is the density of the system. Then
(assuming that the quasi-theorem holds)

\be
\Sast(A) \:=\: \halb (A,\Pi^\ast A) + i(n^{(1)},A)\ .
                         \label{Seffexp7}
\ee

\noi
By invariance under $U(1)$ gauge transformations, translations,
rotations and parity, the expression for $\Pi^{\mu\nu}$ is given,
in momentum space, by

\bea
\Pi^{ij}(k) &=& \Pi_\perp(k)\biggl( \delta^{ij}-{k^i k^j\over
\vec{k}^2} \biggr) + \Pi_\parallel(k) \,{k^i k^j \over\vec{k}^2}\ ,
\nonumber \\
\Pi^{i0}(k) &=& -\Pi_\parallel(k)\, {k^i \over k_0}\ ,
\nonumber \\
\Pi^{00}(k) &=& \Pi_\parallel(k)\, {\vec{k}^2 \over k_0^2}
\:=:\:\Pi_0(k) \ ,
                         \label{Seffexp8}
\eea

\noi
for $i,j=1,\ldots,d$. For non-interacting electrons  the explicit form of
$\Pi^\ast_0$ in the scaling limit,  in $\,d=1$, is given by

\be
\Pi^\ast_0(k) \:=\: \chi_0\, {v_F^2\,k_1^2\over k_0^2+v_F^2k_1^2}\ ,
                         \label{Seffexp9}
\ee

\noi
and, in higher dimensions, to leading order in $| {{v_F\vec{k}\over
k_0}} |$ and $| {k_0\over v_F\vec{k}} |$, we have that

\bea
\Pi^\ast_0(k) &=& (\chi_0+\lambda_0\, | {k_0\over v_F\vec{k}} |)\,
\Theta\biggl(1- | {k_0\over v_F\vec{k}} |\biggr) +\tilde{\chi}_0\, 
| {v_F\vec{k}\over k_0}|^2 \,\Theta\biggl(| {k_0\over v_F\vec{k}} |-1
\biggr)\ ,\nonumber \\
\Pi^\ast_\perp(k) &=& (\chi_\perp \vec{k}^2 + \lambda_\perp
| {k_0\over v_F\vec{k}} |)\, \Theta\biggl(1- | {k_0\over
v_F\vec{k}} |\biggr) +\tilde{\chi}_0\, \Theta\biggl(| {k_0\over
v_F\vec{k}} |-1 \biggr)\ ,
                         \label{Seffexp10}
\eea

\noi
where $\,\chi_0,\, \lambda_0,\, \tilde{\chi}_0,\,\chi_\perp$, and
$\lambda_\perp\,$ are constants depending on $\,d,\, m$,
and $v_F\mini$; see Pines and Nozi\`ere (1989), and Fetter and
Walecka~(1980).

For $d>1$, according to the argument given before, the expression
for $\Pi^\ast$ can be derived from an integration over $\pm \vec\omega$ of
the one-dimensional vacuum polarizations, $\Pi_\bfom$, referring to
the quasi-particle fields $\Psi_\bfom,\,\Psi_{-\bfom}\mini$:

\be
(A,\Pi^\ast\,A) \:=\: \halb \int_\Omega d\bfom\, (A_\bfom,
\Pi_\bfom^\ast\,A_\bfom)\ .
\ee

It is important to note that the quadratic nature of $\Sast(A)$ does
not just follow from a ``small $A$'' approximation and dimensional
analysis, but it is the result of explicit cancellations arising
from the structure of the fermion two-point function in the scaling
limit.

{\bf Remark. } One can bosonize directly the scaling limit action
of the free Fermi gas expressed in terms of the quasi-particle
fields $\{\Psi_\bfom,\,\Psi_{-\bfom}\},\ \bfom\in\!\Omega$, by
introducing real scalar fields $\{\Phi_\bfom\},\ \bfom\in\!\Omega$,
identifying $\Phi_\bfom$ with $\Phi_{-\bfom}$. The result is 
Luther\mini-Haldane bosonization (Luther, 1979; Haldane, 1992) in the
euclidean path-integral formalism; see also Fr\"ohlich, G\"otschmann, and
Marchetti~(1995a, 1995b).
\rule[-4mm]{0mm}{5mm}

{}From now on we omit the trivial term $i(n^{(1)},A)$ in the
effective actions. This corresponds to redefining the density
$J^0(\Psi,\Psi^\ast)$ by subtracting the background density
$\,n\mini$. Furthermore, we set $\,v_F=1\mini$.
\rule[-4mm]{0mm}{5mm}

{\bf (I)} {\em Insulators and incompressible quantum fluids\mini} form
another class of fermionic systems whose (bulk) effective action
in the scaling limit, $\Sast(A)$, is quadratic in $A$.
Here incompressibility means that the connected correlation
functions of the current density $J^\mu(\Psi,\Psi^\ast)$ have
cluster properties better than those encountered in systems whose
large-scale physics is dominated by Goldstone bosons; see
Subsects.\,\ref{ssQH} and~\ref{ssSL}. {}From incompressibility it
follows, as we will review in Subsect.~\ref{ssSL}, that $\Sast$ is
{\em local}, and, for systems with translation, rotation and parity
invariance, it is given by

\be
\Sast(A) \:=\: \halb \int d^{\mini d+1}x\,\biggl[\mini
g_\parallel\,(\da A)^{0i}(\da A)_{0i} + g_\perp (\da A)^{ij}(\da
A)_{ij}\biggr](x)\ ,
                         \label{Seffexp11}
\ee

\noi
where $g_\parallel,\,g_\perp$ are constants.
\rule[-4mm]{0mm}{5mm}

{\bf (H)} The {\em quantum Hall fluids} are parity-breaking,
two-dimensional, incompressible electron systems. For such 
fluids (Laughlin, 1983a, 1983b) with translation and rotation
invariance, it has been shown in Fr\"ohlich and Kerler (1991), and in
Fr\"ohlich and Studer (1992b, 1992c, 1993b)\,--\,see also
Sect.~\ref{sSL} below\,--\,that $\Sast(A)$ is the abelian
Chern-Simons action

\be
\Sast(A) \:=\: {i\sH\over 4\pi} \int \da A\wedge\!A\ ,
                         \label{Seffexp12}
\ee

\noi
where $\sigma_H$ is the dimensionless Hall conductivity.

\rule[-4mm]{0mm}{5mm}

{\bf (S)} London theory and computations based on perturbation
theory suggest that the effective action $\Sast(A)$ of {\em BCS
superconductors\mini} is quadratic and  is given by

\bea
\Sast(A) &\!=\!& \halb\int d^{\mini d+1}x\,\biggl[\, {1\over
\lambda_L^2}\,(\vec{A}^{\mini T})^2(x) \nonumber\\
& &\mssp + \int d^{\mini d+1}y\,
(A_0-\partial_0\Delta_d^{-1}\partial_j A^j)(x) \Pi_s^\ast (x-y)
(A_0-\partial_0\Delta_d^{-1}\partial_j A^j)(y) \biggr]\ ,
                         \label{Seffexp14}
\eea

\noi
where $\lambda_L$ is a constant (the London penetration depth),

\be
A_i^T \::=\: A_i-\partial_i\Delta_d^{-1}\partial_j A^j\ ,
\ee

\noi
with $\,i,j=1,\ldots,d$, $\Delta_d$ denotes the $d$-dimensional
Laplacian, and $\Pi_s^\ast$ is the scaling limit of the scalar component
of the vacuum polarization tensor in a superconductor. To leading order
in $|{\vec{k}\over k_0}|$ and $| {k_0\over\vec{k}} |$, $\Pi_s$ is
given, in $\,d>1$, by

\be
\Pi_s^\ast (k) \:=\: {1\over \lambda_L^2}\, | {\vec{k}\over k_0} |^2\,
\Theta\biggl( |{k_0\over\vec{k}}|-1 \biggr) + \chi_s\,
\Theta\biggl(1-| {k_0\over\vec{k}} |\biggr)\ ,
                         \label{Seffexp15}
\ee

\noi
where $\chi_s$ is a constant depending on $\,d\,$ and $m\mini$; see
Schrieffer (1964).
\rule[-4mm]{0mm}{5mm}

The scaling limits of the effective actions of systems {\bf (F)},
{\bf (I)}, {\bf (H)}, and {\bf (S)}, given by Eqs.~\cfr{Seffexp7}
through~\cfr{Seffexp10}, \cfr{Seffexp11}, \cfr{Seffexp12}, and
\cfr{Seffexp14} through~\cfr{Seffexp15}, respectively, yield b-field
(dual) actions of the form 

\be
\Sti^\ast(\da b) \:=\: {1\over 8\pi^2}\, ({\ast\mini\da b}\mini,
{(\Pi^\ast)^{-1}} {\ast\mini\da b})\ ,
                         \label{Seffexp16}
\ee

\noi
where, in the notation of \cfr{Seffexp8}, $\Pi^\ast$ is given by

\vspace{7mm}
{\bf (F)}\vspace{-6mm}
\be
\mbox{Eqs.~\cfrs{Seffexp9}{Seffexp10}}\ ,
\ee

{\bf (I)}\vspace{-6mm}
\be
\Pi^\ast_0(k) \:=\: g_\parallel\, \vec{k}^2\ ,\hsp 
\Pi^\ast_\perp(k) \:=\: g_\perp \vec{k}^2 + g_\parallel\, k_0^2\ ,
\ee

{\bf (H)}\vspace{-7mm}
\be
\Pi^{\ast\mini\mu\nu}(k) \:=\: {i\over 2\pi\mini\sH}\,
\varepsilon^{\mu\nu\rho}k_\rho\ ,
\ee

{\bf (S)}\vspace{-7mm}
\be
\Pi^\ast_0(k) \:=\: \Pi_s^\ast (k)\ ,\hsp 
\Pi^\ast_\perp(k) \:=\: {1\over \lambda_L^2}\ .
                         \label{Seffexp17}
\ee

\noi
For  $\Pi^\ast$ given by Eq.~\cfr{Seffexp8}, one finds

\bea
\Sti^{\mini\ast}(\da b) &\mA\!=\mA\! & {1\over 8\pi^2}\int d^{\mini
d+1}x\, d^{\mini d+1}y \biggl[ (\ast\da
b)^i(x)\,[(\Pi^\ast_\perp)^{-\halb}
(\delta_{ij}-\partial_i\Delta_d^{-1}\partial_j)
(\Pi^\ast_\perp)^{-\halb}](x,y)\,(\ast\da b)^j(y) \nonumber\\ &
&\hsp + \,(\ast\da b)^0(x)\,(\Pi^\ast_0)^{-1}(x,y)\,(\ast\da b)^0(y)
\biggr]\ .
                         \label{Seffexp18}
\eea

\noi
In particular, in two space dimensions, $b\,$ is a $1$-form,
and~\cfr{Seffexp18} simplifies to 

\bea
\Sti^{\mini\ast}(\da b) &=& {1\over 8\pi^2}\int d^{\mini 2+1}k\,
\biggl[ (\Pi^\ast_\perp)^{-1}(k)\,\vec{k}^2\,
(b_0 - k_0 {\vec{k}\cdot\vec{b}\over\vec{k}^2})^2 \nonumber\\
& & \hsp+\, (\Pi^\ast_0)^{-1}(k)\,\vec{k}^2\,(\vec{b}^{\mini T})^2
\biggr]\ .
                         \label{Seffexp19}
\eea

\noi
For incompressible Hall fluids, the dual action is given by

\be
\Sti^{\mini\ast}(\da b) \:=\: -{i\over 4\pi\,\sH} \int b\wedge\!
\da b\ .
                         \label{Seffexp20}
\ee

{\bf Duality in two-dimensional systems.} Note that, in two space
dimensions, $A\,$ and $b\,$ are $1$-forms, and from formulae
\cfr{Seffexp7}, \cfr{Seffexp10}, \cfr{Seffexp11}, \cfr{Seffexp14},
\cfr{Seffexp16}, \cfr{Seffexp17}, \cfr{Seffexp19}, and
\cfr{Seffexp20} it follows that the actions $\Sast(A)$ and
$\Sti^{\mini\ast}(\da b)$ are related by the remarkable ``duality'':

\bea
\Sast|_I &\propto& \Sti^{\mini\ast}|_S\ ,\nonumber\\
\Sast|_S &\propto& \Sti^{\mini\ast}|_I\ ,\nonumber\\
\Sast|_H &\propto& \Sti^{\mini\ast}|_H\ ,\nonumber\\
 {\rm (etc.)} 
\eea

\noi
with $(g_\parallel,\mini g_\perp)$ corresponding to
$(\lambda_L^2,\mini\chi_s^{-1})$, and $\sH$ to $\sH^{-1}$.

In particular, since $\Sast(A)|_I$ is the Maxwell action, it
follows that $\Sti^{\mini\ast}(\da b)|_S$ describes a massless
mode, the Goldstone boson of the superconducting state with broken
$U(1)$-gauge symmetry, known as the Anderson-Bogoliubov mode. By
Kramers\mini-Wannier duality, this mode can also be described by an
angular variable which is a free field.




\section{Scaling Limit of the Effective Action of a
Two$\,$-$\,$Dimensional, Incompressible Quantum Fluid}
\label{sSL}

In this section, we study the {\em partition~-} or {\em generating
function$\,$} (at $\,T=0\,$ and for {\em real\,} time) of a
two-dimensional, non-relativistic quantum system confined to a
space-time region $\,\Lambda= \BB{R}\times\Omega\,$ and coupled to
external electromagnetic, ``tidal'', and geometric fields:

\be
Z_\Lambda(a\mini,\mini w) \::=\: \int {\cal D}\psi^\ast\mini {\cal
D}\psi\: \exp\biggl[\mini\ih \SL(\psi^\ast,\mini\psi\,;\mini
a\mini ,\mini w)\biggr] \ ,
                             \label{Pa1}
\ee

\noi
where the gauge potentials $\,a\,$ and  $\,w\,$ have been
introduced in (\ref{DD1})--(\ref{w4}), and $\,\SL(\psi^\ast,
\mini\psi\,;\mini a\mini ,\mini w)\,$ is the action of the system
given in~(\ref{Cact1}) and~(\ref{Cact2}); see also
(\ref{T11})--(\ref{mCact}). The integration variables
$\,\psi^\ast\,$ and $\,\psi\,$ are Grassmann variables (i.e.,
anticommuting $\,c\,$-numbers) for Fermi statistics, and complex
$\,c\,$-number fields for Bose statistics.

We have not displayed the metric, $\,g_{ij}$, on the background
space, $\,M$, explicitly, since it will be kept fixed, and usually
$\,M=\BBn{E\mini}{2}$ with $\,g_{ij}=\delta_{ij}$, for simplicity. We
note, however, that for the study of the {\em stress tensor$\mini$},
pressure~- and density fluctuations, and curvature~- and torsion
effects, we would have to choose a variable external metric (or, at
least, a variable conformal factor in $\mini g_{ij}$). This is
important in the study  of density waves, in particular of 
surface density waves (which are interesting in two-dimensional
quantum fluids), and of {\it critical phenomena} in the theory of
phase transitions. We note, however, that
curvature~- and torsion effects can be studied by analyzing the
dependence of $\,Z_\Lambda(a\mini,\mini w)\,$ on $\,w\,$ which
contains the affine spin connection, $\,\omega^{A}_{\hspace{.4em}
B}\,$; see (\ref{w1}) and~(\ref{w2}), and the remarks at the end
of Subsect.~\ref{ssDG} and after~(\ref{rho3}).

Calculating the partition function (\ref{Pa1}) for an arbitrary
(two-dimensional) non-relativistic quantum system is surely a major
task. In the first part of this section, we
show how the calculation can be carried out for {\em
incompressible$\,$} systems, provided one passes to the {\em
scaling limit\mini}. Once we have an explicit expression for the
partition function of a system, many of its physical properties can
be derived. For incompressible systems, this is the topic of the
rest of this section and of Sects.\,\ref{sAC} and~\ref{sClQHFs}
where we review, in partucular, a classification of incompressible
quantum Hall fluids. Since we are working in the scaling limit, we can
only analyze {\em universal$\,$} properties of such systems,
(i.e., properties independent of the small-scale structure of the
system).


\subsection{Scaling Limit of the Effective Action}
\label{ssSL}

In this subsection, we sketch how one calculates, for {\em
incompressible$\,$} systems, the scaling limit of the effective
action (see (\ref{Pa2A}) below) associated with the partition
function (\ref{Pa1}).

One of our main motivations for studying two-dimensional,
incompressible quantum fluids comes from the phenomenology of the
quantum Hall effect; see Subsect.~\ref{ssQH}. In discussing quantum
Hall fluids, it is often assumed that the magnetic field
transversal to the samples is so strong that the Zeeman energies
are large enough for the systems to be totally spin-polarized.
Moreover, spin-orbit interaction terms are expected to be
negligible in quantum Hall fluids. One might therefore ask why,
when studying quantum Hall fluids, one should worry about the
dependence of the partition function $\,Z_\Lambda(a\mini,\mini
w)\,$ on the \SU\ connection $\,w$, thereby taking into account
Zeeman and spin-orbit interaction effects?

To answer this question, we first argue that it is of principal,
theoretical interest to know how to incorporate the {\em spin
degrees of freedom$\,$} in a consistent way into the description of
two-dimensional electronic systems. 

Second, as first pointed out  by Halperin~(1983), in $\,GaAs$ (for
example) the $\,g$-factor of the electron is $\,{1 \over 4}\,$ 
of the value in empty space, and the effective mass, $\,m$, of the
electron is about $\,{7\over 100}\,$ of the mass, $\,m_o$, in the
vacuum. Thus, in $\,GaAs\,$, the Zeeman energies are only
approximately $\,{1 \over 60}\,$ of the cyclotron energy (i.e., the
splitting between Landau levels). Furthermore, they are of the same
magnitude as the quasi-particle energies of the fractional quantum
Hall states in magnetic fields of the order of $\,$10$\,$T. One
expects therefore that, at some values of the filling factor
$\,\nu$, the ground state of the system will contain electrons
with reversed spins. Experimental evidence that {\em
spin-unpolarized$\,$} quantum Hall fluids exist has been given
in the works cited in {\bf (D5)} in Subsect.~\ref{ssQH}. We
emphasize that unpolarized (or partially polarized) quantum Hall
fluids can arise in two fundamentally different ways: either
through the presence of two (or more) {\em independent$\mini$}, but
oppositely polarized bands, or through the formation of {\em
spin-singlet$\,$} bands; see Subsect.~VI.C in Fr\"ohlich and Studer
(1993b) and Fr\"ohlich and Thiran~(1994).

Third, we will show in Sects.\,\ref{sAC} and~\ref{sClQHFs} how one
can infer, from the form of $\,Z_\Lambda(a\mini, \mini w)$, {\em
universal$\,$} properties of quantum Hall fluids that will lead to a
classification of such systems in terms of ``universality
classes''. 

Fourth, in Subsect.~\ref{ssLR}, we sketch  how one can
derive  the linear response theory of quantum Hall fluids from
$\,Z_\Lambda(a\mini,\mini w)\,$ describing, among other effects, a 
{\em quantum Hall effect for spin currents$\mini$}. For a
discussion of possible Hall systems where this effect might be
tested experimentally, see Subsect.~VII.A in Fr\"ohlich and
Studer (1993b).

We define the {\em electric charge~-} and {\em current
densities$\mini$}, $\,j^0(x)\,$ and $\,\vec{j}(x)$, by

\bea
j^0(x) & := & \psi^\ast(x)\psi(x) \ ,  \nonumber  \\
j^k(x) & := & -{i\hbar \over 2mc}\, g^{kl}(x)
\left[ (\D{l}\psi)^\ast(x)\psi(x) - \psi^\ast(x)(\D{l}\psi)(x) \right]
\ , 
                             \label{C1}
\eea

\noi
and the {\em spin~-} and {\em spin current densities$\mini$},
$\,s^0_A(x)\,$ and $\,\vec{s}_A(x)$, by

\bea
s_A^0(x) & := & \psi^\ast(x)\,L_A^{(s)}\,\psi(x) \ ,  \nonumber  \\
s_A^k(x) & := & -{i\hbar \over 2mc}\, g^{kl}(x)
\left[ (\D{l}\psi)^\ast(x)\,L_A^{(s)}\,\psi(x) - \psi^\ast(x)
\,L_A^{(s)}\,(\D{l}\psi)(x) \right] \ , 
                             \label{C2}
\eea

\noi
where $\,(L_1^{(s)},\mini L_2^{(s)},\mini L_3^{(s)})\,$ are the
three generators of the spin-$s$ representation of $su(2)$; see
(\ref{DD2}). Similarly, one defines currents associated with
internal symmetries. The electric current is conserved (i.e., the
continuity equation holds; see (\ref{ED0})), but the spin
current is, in general, {\em not$\,$} conserved, because it couples
to a non-abelian gauge potential. It is, however, {\em covariantly
conserved}$\,$; see (\ref{CSC}) below.

It is straightforward to infer from (\ref{Pa1}), (\ref{Cact1}),
(\ref{C1}) and (\ref{C2}) that the {\em connected, time-ordered
current Green functions$\,$} of the system are given, {\em at
non-co\-in\-ci\-ding arguments$\mini$}, by

\bea
\lefteqn{\left< T \Biggl[\, \prod_{i=1}^n j^{\mu_i}(x_i)\,
\prod_{j=1}^m s_{A_j}^{\nu_j}(y_j)\, \Biggr]
\right>^{\!\mbox{\scriptsize con}}_{\!a,\,w}}  \nonumber  \\ & = &
i^{\,n+m} \prod_{i=1}^n {\delta \over \delta a_{\mu_i}(x_i)}\,
\prod_{j=1}^m {\delta \over \delta w_{\nu_j A_j}(y_j)}\: \ln
Z_\Lambda(a\mini,\mini w) \ , 
                         \label{Pa2}
\eea

\noi
where $\,\El\, (\,\cdot\,)\, \Er^{\mbox{\scriptsize con}}_{a,\,w}\,$
denotes the connected expectation functional of the system in an
external gauge field configuration, $\,(a\mini ,\mini w)$, (with
``ground state asymptotic conditions'', as $\,tÊ\rightarrow \pm
\infty$, to be specified), and $\,T\,$ indicates time-ordering. At
{\em coinciding arguments$\mini$}, Eq.~(\ref{Pa2}) is modified by
{\em Schwinger terms$\mini$}, (but their precise form will not be of
importance in our analysis, and therefore we do not display them).

As in Sect.~\ref{sSLaC}, the {\em effective action$\,$} of the
system is defined by

\be
\SLeff(a\mini ,\mini w) \::=\: {\hbar \over i}\, \ln
Z_\Lambda(a\mini,\mini w) \ .
                          \label{Pa2A}
\ee

\noi
The idea is to try to calculate the ``leading terms'' in $\,
\SLeff(a\mini ,\mini w)\,$ which, via (\ref{Pa2}), will provide us with
information on the current Green functions. By leading terms we mean
those terms which dominate at {\em large-distance scales$\,$} and
{\em low frequencies$\mini$}. The calculation of the leading terms
in $\,\SLeff(a\mini ,\mini w)\,$ might appear to be an intractable
problem. Actually, making a single assumption on the excitation
spectrum of the system, {\em incompressibility$\mini$}, and using
the $\USo$ {\em gauge-invariance$\,$} of non-relativistic quantum
mechanics, they can be found explicitly.

Let $\,\chi\,$ be a real-valued function and $\,R\,$ an \SU-valued
function on space-time. Consider the gauge transformations in
Eq.~(\ref{gauget}), i.e.,

\bea
a & \mapsto & {}^\chi a \ , \hspace{1cm} \mbox{with} \hspace{1cm}
{}^\chi a_\mu(x)\::=\: a_\mu(x) + \partial_\mu \chi(x)\
,\hspace{6mm}
                          \label{Pa3}
\eea

\mam

\bea
w & \mapsto & {}^R w \ , \hspace{1cm} \mbox{with} \nonumber  \\
{}^R w_\mu(x) & := & \Us\!(R(x))\,w_\mu(x)\,\Us\!(R(x))^\ast +
\Us\!(R(x))\,\partial_\mu \Us\!(R(x))^\ast\ .\hspace{.3cm}
                          \label{Pa31}
\eea

\noi
Changing integration variables,

\bea
\psi(x) & \mapsto & {}^{\chi,\,R}\psi(x) \::=\:
e^{-i\chi(x)}\,\Us\!(R(x)) \,\psi(x) \ ,
                          \label{Pa4}
\eea

\noi
in the functional integral (\ref{Pa1}), and using the gauge
invariance of $\,\SL(\psi^\ast,\mini\psi\mini;\mini a\mini ,\mini
w)\,$ under the transformations (\ref{Pa3})--(\ref{Pa4}) and the
fact that the Jacobian of (\ref{Pa4}) is unity, we find the {\em
Ward identity}

\be
\SLeff({}^\chi a\mini ,\mini{}^R w) \:=\: \SLeff(a\mini ,\mini w) \ , 
                          \label{Pa5}
\ee

\noi
for all $\,\chi\,$ and $\,R\mini$. For a system of finitely many
particles in a bounded region, $\,\Omega$, of space, (\ref{Pa5})
can be proven rigorously (Fr\"ohlich and Studer, 1992b). The
identity is stable under passing to limits, for $\,\chi$'s$\,$ and
$\,\partial_\mu R$'s$\,$ of compact support.

By differentiating (\ref{Pa5}) in $\,\chi\,$ or $\,R\,$ and setting
$\,\chi=0, \ R= {\bf 1}$, we find, using (\ref{Pa2}) (for
$n+m=1$), that

\be
\sgi\: \partial_\mu\! \left( \sg \El j^\mu(x) \Er_{a,\,w} \right) \:=\:
0 \ , 
                            \label{CEC}
\ee

\vspace{.3cm}
\mam
\vspace{.3cm}

\be
\sgi\:\left(\D{\mu}\Bigl(\mmini\sg\, \El s^\mu(x) \Er_{a,\,w}\Bigr)
\right)_{\!A}  \:=\: 0 \ , \hspace{.5cm} A=1,\,2,\,3\ ,
                            \label{CSC0}
\ee

\vspace{.1cm}
\miem

\be
\sgi\:\partial_\mu\! \left( \sg\, \El s_A^\mu(x) \Er_{a,\,w} \right)
- 2\,\eps{ABC}\,w_{\mu B}(x)\, \El s_C^\mu(x) \Er_{a,\,w} \:=\: 0 \ ,
                            \label{CSC}
\ee

\noi
for arbitrary $\,a\,$ and $\,w$. These infinitesimal Ward identities
play an important role in determining the general form of
$\,\SLeff(a\mini ,\mini w)$. They can be generalized, in an obvious
way, to systems with internal symmetries.

We now proceed to determine the form of $\,\SLeff(a\mini ,\mini w)$
in the {\em scaling limit}. We need to consider {\em ever
larger$\,$} systems and {\em ever slower$\,$} variations in time.
Let $\,1\leq \tha < \infty\,$ be a scale parameter as in
Sect.\,\ref{sSLaC}. We set

\bea
g_{ij}(x) & := & g^\kt_{ij}(x) \:=\: \gamma_{ij}\! \left({x \over
\tha} \right)   \ ,  \hspace{.5cm}  \mbox{and}  \nonumber  \\
\Lambda & := & \Lambda^\kt \:=\: \tha\, \Lambda_o \ ,
                            \label{Pa5A}
\eea

\noi
where $\,\gamma_{ij}(x)\,$ is a fixed metric on $\,M\,$ (e.g.,
$\,\gamma_{ij}(x)=\delta_{ij}$), and $\,\Lambda_o \subset N
=\BB{R} \times M \,$ is a fixed space-time cylinder;

\be
x \::=\:(x^0=c\mini t,\, \vec{x}\,)\:=\: \tha\, (\xi^0,\,
\vec{\xi}\,)\:=\:\tha\,\xi \ , \hspace{.5cm} \xi \in \Lambda_o \ ;
                            \label{Scal}
\ee

\mhm

\be
\dd{x^\mu}\:=\:\tha^{-1}\,\dd{\xi^\mu} \ .
\ee

We propose to study the reaction of the system to a small change
in the ex\-ter\-nal gauge potentials $\,a\,$ and  $\,w$. We
choose {\em fixed background potentials$\mini$}, $\,a_c\,$ and
$\,w_c$, defined on all of space-time $\,N$, and set

\be
a_\mu^\kt(x) \::=\: a_{c,\mu}(x) + \tha^{-1}\,\tilde{a}_\mu\! \left({x
\over \tha} \right) \ ,
                          \label{Pa6}
\ee

\mam

\be
w_\mu^\kt(x) \::=\: w_{c,\mu}(x) + \tha^{-1}\,\tilde{w}_\mu\! \left({x
\over \tha} \right) \ ,
                          \label{Pa7}
\ee

\noi
where $\,\tilde{a}_\mu(\xi)\,$ and $\,\tilde{w}_\mu(\xi)\,$ are
{\em fixed}$\,$ functions defined on $\,\Lambda_o$. If $\,m\,$ is
the effective mass of the particles and $\,\mu\,$ is the quantity
determining their magnetic moment (see (\ref{mamo})) in physical
$\,(t,\mini\vec{x}\mini)$-coordinates, then the mass $\,m^\kt\,$
and the strength of the magnetic moment $\,\mu^\kt\,$ in {\em
rescaled coordinates$\mini$}, $\,(\tau={\xi^0 \over c},\mini
\vec{\xi}\,)$, are given by

\be
m^\kt \:=\: \tha\,m \ ,\hspace{.5cm}  \mbox{and} \hspace{.5cm}
\mu^\kt \:=\: \tha^{-1}\mini \mu \ ,
                           \label{Scal2}
\ee

\noi
as follows from Eqs.~(\ref{Cact1}), (\ref{rho1}), and (\ref{rho2}),
(i.e., in the {\em rescaled$\,$} systems, the particles are heavy
and have small magnetic moment. Moreover, the range of the two-body
potential, in the {\em rescaled$\,$} systems, becomes shorter and
shorter, as the scale parameter $\,\tha\,$ becomes large).

One {\em basic assumption} underlying our analysis is that
$\,\StLeff\,$ is {\em four times continuously (Fr\'{e}chet)
differentiable$\,$} in $\,\tilde{a}_\mu^\kt(x)\mmini :=
\tha^{-1}\mini\tilde{a}_\mu\! \left({x \over \tha} \right)\,$ and
$\,\tilde{w}_\mu^\kt(x)\mmini := \tha^{-1}\mini\tilde{w}_\mu\!
\left({x \over \tha} \right)\,$ at $\,\tilde{a}_\mu^\kt(x) = 0 =
\tilde{w}_\mu^\kt(x)$, for a {\em suitable choice of background
potentials$\mini$}, $\,a_{c,\mu}(x)\,$ and $\,w_{c,\mu}(x) =
\omega_{c, \mu}(x)+\rho_{c,\mu}(x)$, and for
$\,\tilde{a}_\mu(\xi)\,$ and
$\,\tilde{w}_\mu(\xi)=\tilde{\omega}_\mu(\xi)+\tilde{\rho}_\mu
(\xi)\,$ constrained to {\em suitable function spaces$\mini$},
$\,{\cal A}\,$ and $\,{\cal W}$, {\em of perturbation
potentials$\mini$}, to be specified later. We may then
(functionally) expand $\,\StLeff\,$ to third order in
$\,\tilde{a}^\kt (x)\,$ and $\,\tilde{w}^\kt(x)$, with a fourth
order remainder term. Among the terms thus generated we shall
retain only the {\em leading terms$\,$} in $\,\tha$, namely those
{\em scaling with a non-negative power of} $\,\tha\,$ which are
commonly called {\em relevant}$\,$ and {\em marginal}$\,$ terms.
The coefficients of these terms will be denoted by $\,\Ssaw$, a functional
that we call the {\em scaling limit of the effective
action$\mini$}.

Using identity (\ref{Pa2}) and Eq.~(\ref{Pa2A}) to find the Taylor
coefficients of $\,\StLeff$, plugging (\ref{Pa6}) and (\ref{Pa7})
into the resulting expression, and passing to $\,(\xi^0,\,
\vec{\xi}\,)$-co\-ord\-in\-ates, we find that the coefficient of the term
of $\,n^{\rm\mini th}$ order in $\,\tilde{a}(\xi)\,$ and of
$\,m^{\rm\mini th}$ order in $\,\tilde{w}(\xi)\,$ in $\,\StLeff\,$
is given by a distribution

\be
\varphi^{(\tha)\,\mu_1 \dots\mini \mu_n,\,\nu_1 \dots\,
\nu_m}_{\hspace{14.6mm} A_1\dots\mini A_m}(a_c,\mini w_c\mini;\mini
\xi_1,\dots,\mini\xi_n,\mini \eta_1,\dots,\mini\eta_m)  
\: =: \:
\varphi^{(\tha)\,\underline{\mu},\,\underline{\nu}}_{\hspace{7.1mm}
\underline{A}} (a_c,\mini w_c\mini;\mini\underline{\xi}, \mini
\underline{\eta}) \hspace{.4cm}
                            \label{Pa80}
\ee

\noi
which, at non-coinciding arguments, is given by

\be
{(-i)^{n+m+1}\,\hbar\over n!\,m!}\, \left< T \Biggl[\, \prod_{i=1}^n
\left(\tha^2\, j^{\mu_i}(\tha\xi_i)\right) \, \prod_{j=1}^m
\left(\tha^2\, s_{A_j}^{\nu_j}(\tha\eta_j)\right) \, \Biggr]
\right>^{\! \mbox{\scriptsize con}}_{\!a_c,\,w_c} \ ,
                            \label{Pa8}
\ee

\noi
in accordance with the circumstance that, in $\,2+1\,$ space-time
dimensions, the scaling dimension of currents is $2\mini$! [Note,
e.g., that $\,\int_{t\mini=\mini\mbox{\scriptsize const.}}\mini
j^0(c\mini t,\,\vec{x}\,)\: d^{\,2} \vec{x}\;$ is a {\em
di\-men\-sion\-less$\,$} number, a charge.]

We now formulate our {\em basic assumption of
incompressibility$\,$}: We assume that, for {\em certain choices of
the background potentials $\,a_c\,$ and $\,w_c$}, the excitation
spectrum of the system above its ground-state energy is such that
(in the bulk) the {\em connected Green functions of its
currents}$\,$ have ``{\em good}$\,$'' {\em cluster properties$\,$}
(better than in a metal or in a system with Goldstone bosons). More
precisely, we 
assume that the dis\-tri\-bu\-tions in~(\ref{Pa80}) exhibit the
following behaviour, for $\,\tha \rightarrow \infty\mini$:

\bea
\varphi^{(\tha)\,\underline{\mu},\,\underline{\nu}}_{\hspace{7.1mm}
\underline{A}} (a_c,\mini w_c\mini;\mini\underline{\xi}, \mini
\underline{\eta})
& = &
\sum_{\Delta=0}^N\tha^{-D_{\mmini\Delta}}\,\varphi^{\hspace{.8mm}
\underline{\mu}, \,\underline{\nu}}_{\hspace{-.8mm}\Delta
\hspace{2mm} \underline{A}} (a_c,\mini w_c\mini;\mini
\underline{\xi}, \mini \underline{\eta})   \nonumber  \\
& + & \BT^{\underline{\mu}, \,\underline{\nu}}_{\hspace{3mm}
\underline{A}}(a_c,\mini w_c\mini;\mini \underline{\xi}, \mini
\underline{\eta})\: +\: o\!\left(\tha^{-D_{\mmini N}} \right) \ , 
                            \label{Pa81}
\eea

\noi
where $\,\varphi^{\hspace{.8mm}\underline{\mu},\,
\underline{\nu}}_{\hspace{-.85mm}\Delta\hspace{2mm} \underline{A}}
(a_c,\mini w_c\mini;\mini\underline{\xi}, \mini \underline{\eta})\,$
are {\em local distributions$\mini$}, i.e., sums of products of derivatives
of $\delta$-functions, and the {\em scaling dimensions$\mini$},
$\,D_{\mmini \Delta}$, are given by

\be
D_{\mmini\Delta}\: :=\: -2(n+m)+3(n+m-1)+\Delta\: =\: n+m-3+\Delta\;
\in \BB{Z} \ ,
                            \label{Pa82}
\ee

\noi
with $\,\Delta\,$ the number of derivatives present in the
corresponding local distribution. The upper limit, $\,N\,$, in the
sum on the r.h.s.\ of Eq.~(\ref{Pa81}) is chosen such that $\,D_N
\geq 0\mini$. Finally, $\,\BT^{\underline{\mu},
\,\underline{\nu}}_{\hspace{3mm} \underline{A}}(a_c,\mini
w_c\mini;\mini \underline{\xi}, \mini \underline{\eta})\,$ are
distributions ({\em not$\,$} necessarily local ones) that are
completely localized  on the space-time boundary $\,\dLo\,$ of the
rescaled system in $\,\Lambda_o\mini$. These terms will not be
discussed in this section; they form the subject matter of
Sect.~\ref{sAC}. [For a different way of formulating the
incompressibility of a system, see the discussion on the quantum
Hall effect in Subsect.~\ref{ssQH}.]

Our incompressibility assumption is by no means a mild or minor
assumption. It tends to be a hard analytical problem of many-body
theory to show that, for a concrete system, it is satisfied. [For
some recent ideas about how to establish it for quantum Hall fluids
at certain filling factors, see the references given after {\bf
(D5)} in Subsect.~\ref{ssQH}.] What we propose to do here is to use
it to calculate the general form of $\,\Ssaw$, the scaling
limit of the effective action of the system, thereby elucidating
the {\em universal$\,$} properties of two-dimensional,
incompressible quantum fluids. We only sketch some ideas; for
details, see Fr\"ohlich and Studer~(1992b), and Appendix~A in
Fr\"ohlich and Studer~(1993b).

The calculation is based on the following four principles:
\vspace*{4mm}

\begin{list}
{\bf (P\arabic{lico})}{\usecounter{lico}\setlength{\labelsep}{.4cm}
\setlength{\labelwidth}{1cm}\setlength{\leftmargin}{1.4cm}
\setlength{\rightmargin}{.7cm}}
\setcounter{lico}{0}
\item {\em Incompressibility}$\,$: For all $\,n\,$ and $\,m$, with
$\,2\leq n+m\leq 4$, the distributions $\,\varphi^{(\tha)\,
\underline{\mu},\,\underline{\nu}}_{\hspace{7.1mm} \underline{A}}
(a_c,\mini w_c\mini;\mini\underline{\xi}, \mini \underline{\eta})
\,$ ``converge'' (in the bulk), as $\,\tha \rightarrow \infty$, to
local distributions, as specified in Eq.~(\ref{Pa81}).

\item $\USo$ {\em gauge invariance}$\mini$: Ward identities
(\ref{Pa5}) to (\ref{CSC}).

\item {\em Only relevant and marginal terms are kept in}
$\,\Ssaw$.

\item {\em Extra symmetries of the system$\mini$}, e.g., for $\,w_{c,\mu
A}(x) = \delta_{A3}\,w_{c,\mu 3}(x)\,$ (and hence, by~(\ref{rho2}),
for $\,a_{c,\mu}(x)\,$ such that $\,E_{c,3}(x)=0\mini$), global
rotations around the 3-axis in spin space are a continuous, global
symmetry of the system with an associated {\em conserved$\,$}
Noether current, $s^\mu_3(x)\mini$; or translation invariance in
the scaling limit $(\tha \rightarrow \infty),\,\dots \:$, {\em are
exploited to reduce the number of terms$\mini$}. 
\end{list}

{}From {\bf (P1)} and Eqs.~(\ref{Pa6}) and (\ref{Pa7}) it immediately
follows that all terms contributing to $\,\StLeff\,$ of order
$\,4\,$ or higher in $\,\tilde{a}\xx\,$ and  $\,\tilde{w}\xx\,$ are
{\em irrelevant$\mini$}, i.e., they scale like $\,\tha^{-D}$, with
$\,D>0$. In particular, a fourth-order remainder term does {\em
not$\,$} contribute to $\,\Ssaw$, (principle {\bf (P3)}). We 
present the final result of our analysis in the special case of a
system which is incompressible for a choice of $\,w_{c}$ satisfying

\be
w_{c,\mu A}(x)\:=\:\delta_{A3}\,w_{c,\mu 3}(x)\ ,
                           \label{back}
\ee

\noi
or, in view of Eqs.~(\ref{w1})--(\ref{w3}), (\ref{rho1}) and
(\ref{rho2}), for a background electromagnetic field $\,({\bf
E}_c(x),\,{\bf B}_c(x))\,$ with

\be
{\bf E}_c(x)\:=\:(E_{c,1}(x),\mini E_{c,2}(x),\mini 0) \ ,
\hspace{.5cm}  {\bf B}_c(x)\:=\:(0,\mini 0,\mini B_c(x)) \ ,
                           \label{back11}
\ee

\noi
and possibly for some affine spin connection, $\,\omega$, of the
following form (see (\ref{w2}), (\ref{M5}), (\ref{M8}), and
(\ref{w4}), as well as the remarks about the physical relevance of
$\,\omega\,$ at the end of Subsect.~\ref{ssDG} and
after~(\ref{rho3})):

\be
\left( \omega^{A}_{\hspace{.4em} B\mu}(x) \right) = \left( 
\begin{array}{ccc}
 0  &  \omega_{\mu}(x)  & \; 0   \\
-\,\omega_{\mu}(x)  &  0  & \; 0   \\
 0  &  0  & \; 0 
\end{array} \right) \ ,
                           \label{back1}
\ee

\noi
relative to some orthonormal frames $(e^1(x),\, e^2(x),\,
e^3(x))$. [It is natural to work in an \SU\ gauge which respects our
convention of choosing $\,e^3(x) \,$ perpendicular to the cotangent
plane $\,T_{\vec{x}}^\ast(M)\,$ at $\,\vec{x}\in M$, for all times
$\,t\mini$; see the discussion preceding (\ref{M1}). E.g., if
$\,M\,$ were the $(x,\mini y)$-plane in $\,\BBn{E\mini}{3}$, we would
choose $\,e^3(x)\,$ to coincide with $\,dz$. Hence the choice of the
electromagnetic {\em background$\,$} field specified by
(\ref{back11}) corresponds to an electric field, ${\bf E}_c$,
which is tangential to the sample and to a magnetic field, ${\bf
B}_c$, which is perpendicular to it.] In this situation, the {\em
scaling limit of the effective action$\,$} is given by

\samepage{
\bea
-{1 \over \hbar}\,\Ssaw & = & \iLo \jc\,\at{\mu}\,dv \;+\; \iLo
\mm\, \wt{\mu 3}\,dv   
\nonumber   \\
& &\hspace{-2cm}+\;\, \sum_{A=1}^2 \iLo \tau_1^{\mu\nu}\,\wt{\mu
A}\, \wt{\nu A}\, dv\;+ \sum_{A,B=1}^2 \iLo 
\tau_2^{\mu\nu}\,\eps{AB}\, \wt{\mu A}\, \wt{\nu B}\,dv   
\nonumber  \\
& &\hspace{-2cm}+\hspace{3mm} {k \over 4\pi} \iLo
\mbox{tr}\Bigl( w \wedge\! \da w + {2 \over 3}\, w \wedge\! w
\wedge\! w \Bigr)  \rule[-4mm]{0mm}{11.3mm}   
\nonumber  \\
& &\hspace{-2cm}+\hspace{3mm} {\sigma \over 4\pi}\iLo \at{} \wedge\!
\da \at{} \;+\; {\chi_s \over 2\pi}\iLo \at{} \wedge\! \da
\wt{3} \;+\; {\sigma_s \over 4\pi} \iLo \wt{3} \wedge\! \da \wt{3}
\rule[-5mm]{0mm}{12.5mm} 
\nonumber   \\
& &\hspace{-2cm}+ \sum_{A,B,C=1}^3 \iLo
\eta_{ABC}^{\mu\nu\rho}\,\wt{\mu A}\, \wt{\nu B}\,\wt{\rho C}\, dv 
\; +\; \BT(a|_\dLo\mini,\mini w|_\dLo) \ ,
                          \label{Seff}
\eea
}

\noi
where $\,\jc(\xi)\,$ is an electric~- and $\,\mm(\xi)\,$ a magnetic current
circulating in the system when $\,\at{}(\xi)= 0 = 
\wt{}(\xi);\ \tau_1^{\mu\nu}(\xi)\,$ is a function symmetric in
$\,\mu\,$ and $\,\nu$, while $\,\tau_2^{\mu\nu}(\xi)\,$ is
antisymmetric in $\,\mu\,$ and $\,\nu$; the function
$\,\eta_{ABC}^{\mu\nu\rho}(\xi)\,$ is symmetric under interchanges
of $(\mu A),\ (\nu B)$ and $(\rho C)$ and vanishes if two or more
of the indices $A,\,B,\,C$ are equal to $\,3$; $\,dv :=
\sqrt{\gamma(\xi)}\, d^{\,3}\xi$, where $\,\gamma(\xi) :=
\mbox{det}\,(\gamma_{ij}(\xi))$ is the volume element on the
space-time cylinder $\,\Lambda_o$, see (\ref{Pa5A}); $\,\sigma,\ 
\chi_s,\  \sigma_s\mini$, and $\,k\,$ are real constants; $w(\xi)\,$
is the total \SU\ connection given by

\be
w(\xi) \::=\: w_c^\kt(\xi) + \wt{}(\xi)\ , \hspace{.5cm} \mbox{with}
\hspace{.5cm} w_c^\kt(\xi)\::=\: \tha\,w_c(\tha\xi)\ ,
                         \label{wtot}
\ee

\noi
see (\ref{Pa7}); and $\,\BT(a|_\dLo\mini,\mini w|_\dLo)\,$ denotes
boundary terms only depending on the gauge potentials
$\,a|_{\partial \Lambda_o},\mini w|_{\partial \Lambda_o}$
restricted to the boundary, $\,\partial \Lambda_o$, of the
space-time cylinder $\,\Lambda_o$. They will be discussed in
Sect.~\ref{sAC}. Moreover, on the r.h.s.\ of (\ref{Seff}), we are
using the notation:

\bea
\at{} & = & \sum_{\mu=0}^2 \at{\mu}(\xi) \,d\xi^\mu \ , \hspace{1cm}
\da \at{}\;=\;\sum_{\mu,\nu=0}^2 \left(\partial_\mu
\at{\nu}\right)\! (\xi) \,d\xi^\mu \wedge\! d\xi^\nu \ , 
\nonumber  \\  \wt{3} & = & \sum_{\mu=0}^2 \wt{\mu 3}(\xi)
\,d\xi^\mu \ , \hspace{1cm} \wt{}\; =\; i\sum_{\mu=0}^2 \sum_{A=1}^3
\wt{\mu A}(\xi) \, L^{(s)}_A\, d\xi^\mu \ ,   
\eea

\noi
where $\,\partial_\mu = \dd{\xi^\mu}\,$ whenever we are working in
rescaled $\xi$-coordinates. Finally, we note that the terms
in~(\ref{Seff}) are ordered according to their scaling dimensions.

In Fr\"ohlich and Studer~(1993b), we have used results on chiral
\uh\ - and \suh\ current algebras to determine the possible values of
the constants $\,\sigma,\  \chi_s,\  \sigma_s\mini$, and $\,k$; see
Sect.~\ref{sAC} below.

Here we wish to point out that the functions $\,\jc,\, \mm,\,
\tau_1^{\mu\nu},\, \tau_2^{\mu\nu}\mini$, and
$\,\eta_{ABC}^{\mu\nu\rho}\,$ are {\em not$\,$} all independent,
but are constrained by the infinitesimal Ward identities
(\ref{CEC}) and (\ref{CSC}): By (\ref{Pa2}) and (\ref{Pa2A})

\bea
\El j^\mu(\xi) \Er_{a^\kt,\,w^\kt} & = & -\hi\,{\delta \Ssaw \over
\delta\mini\at{\mu}(\xi)} + \cdots\ ,
                       \label{LR1}
\eea

\mam

\bea
\El s_A^\mu(\xi) \Er_{a^\kt,\,w^\kt} & = &-\hi\, {\delta \Ssaw \over
\delta\wt{\mu A}(\xi)} + \cdots\ .
                       \label{LR2}
\eea

\noi
The dots stand for contributions from irrelevant terms in the
effective action. We calculate the r.h.s.\ of these equations by
using (\ref{Seff}) and plug the result into Eqs.~(\ref{CEC}) and
(\ref{CSC}). As a result we obtain the following {\em constraints$\,$}
(Fr\"ohlich and Studer, 1992b):

$$
\hspace{-51.6mm} {\bf (i)} \hspace{40mm}
\sggi\: \partial_\mu\! \left( \sgg\,j_c^\mu \right)\!(\xi)\; =\; 0
\ .
$$

$$
\hspace{-51.4mm} {\bf (ii)} \hspace{40mm}
\sggi\: \partial_\mu\! \left( \sgg\,m_3^\mu \right)\!(\xi)\; =\; 0 \
.  
$$

\bea
\lefteqn{\hspace{-27.9mm} {\bf (iii)} \hspace{10mm}
\sum_{B=1}^2 \eps{AB} \left[\mm(\xi)-2\,\tau_1^{0\mu}(\xi)\,
w^\kt_{c,03}(\xi) \right]\wt{\mu B}(\xi) }   \nonumber  \\
 &  & \mbox{} + 2\,w^\kt_{c,03}(\xi) \sum_{j=1}^2
\tau_2^{0j}(\xi)\,\wt{jA}(\xi)\; =\; 0 \ , \hspace{.5cm} A=1,\,2 \ .
\nonumber
\eea

\vspace{-.9cm}
\bea
\lefteqn{\hspace{-14.7mm} {\bf (iv)} \hspace{10mm}
\sggi\; \partial_\mu\! \left[ \sgg\,\tau_1^{\mu\nu}\, \wt{\nu A} +
\sgg\sum_{B=1}^2\eps{AB}\, \tau_2^{\mu\nu}\, \wt{\nu B}
\right]\!(\xi)\; = }  \nonumber   \\
 &  & \mbox{} - 2 \left[\sum_{B=1}^2 \eps{AB}\,
\tau_1^{\mu\nu}(\xi)\, \wt{\nu B}(\xi) - \tau_2^{\mu\nu}(\xi)\,
\wt{\nu A}(\xi) \right]\wt{\mu 3}(\xi)  \nonumber  \\
 &  & \mbox{} - 3 \sum_{B=1}^2 \eps{AB} \sum_{C,D=1}^3
\eta_{BCD}^{0\nu\rho}(\xi)\, w^\kt_{c,03}(\xi)\, \wt{\nu C}(\xi)\,
\wt{\rho D}(\xi)\ , \hspace{.5cm} A=1,\,2 \ .
                         \label{constr} 
\eea

\noi
Constraints {\bf (i)} and {\bf (ii)} just express the conservation
of the currents $\,\jc\,$ and $\,\mm\,$ when $\, \at{} = 0 =
\wt{}\mini$.

If we impose constraints {\bf (iii)} and {\bf (iv)}, for {\em
arbitrary$\,$} smooth perturbation potentials $\,\wt{}$, then
it follows that

\be
m_3^\mu(\xi) \:=\: \tau_1^{\mu\nu}(\xi) \:=\:  \tau_2^{\mu\nu}(\xi)
\:=\: 0 \ , \hspace{.5cm} \mbox{for all}\;
\mu,\mini\nu=0,\mini 1,\mini 2\ ;
\ee

\noi
in particular, the system {\em cannot$\,$} be {\em magnetized}
$\,(m_3^0=0)\,$ and cannot support persistent spin currents.
This may seem rather strange, because we would expect that if
$\,\rho_{c,03}=-{g\mini\mu \over 2\hbar c}B_c\,$ (see (\ref{rho1})),
for some large background magnetic field $\,{\bf
B}_c=(0,\mini 0,\mini B_c)$, then the system would be magnetized in
the $\mini 3$-direction. What has gone wrong? The point is that the
assumed properties, that $\,\StLeff\,$ is four times continuously
differentiable in $\,\at{}\,$ and $\,\wt{}\,$ and that the system
remains incompressible in an {\em arbitrary$\,$} function-space
neighbourhood of $\,(a_c,\,w_c)\,$ of sufficiently small diameter,
must fail for magnetized systems! The reason is that an {\em
arbitrarily small}$\,$ perturbation field $\,\wt{}\,$ which
oscillates rapidly in time can destroy the incompressibility of the
system, and hence our estimate on the fourth order remainder term
in the Taylor expansion of $\,\StLeff\,$ breaks down!

We thus assume, for example, that, for a {\em time-independent$\,$}
background field $\,w_c\,$ satisfying~(\ref{back}), the system
remains incompressible and $\,\StLeff\,$ is four times continuously
differentiable in $\,(\at{},\,\wt{})\,$, provided $(\at{}\, ,\,
\wt{})$ belongs to a set of ${\cal A} \times {\cal W}$ given by 

\bea
{\cal A} & := & \{\:\at{}\;\:|\;\: \at{\mu}(\tau,\mini\vec{\xi}\,) =
\eta(\tau)\, f_\mu(\vec{\xi}\,)\ , \ f_\mu(\vec{\xi}\,) \in {\cal
S}\:\} \ ,  \nonumber   \\    
{\cal W} & := & \{\:\wt{}\;\,|\;\: \wt{\mu A}(\tau,\mini\vec{\xi}\,)
= \eta(\tau)\,g_{\mu A}(\vec{\xi}\,)\ , \ g_{\mu A}(\vec{\xi}\,)
\in {\cal S} \:  \}\ , 
                        \label{constr0}
\eea

\noi
where $\,\eta(\tau)\,$ describes an adiabatic process of turning
on and off the perturbations: $\,\eta(\tau)=1$, for
$\,\tau\in[\tau_1,\mini\tau_2]$, some finite interval in
(rescaled) time, and $\,\eta(\tau)=0$, for $\,\tau \ll \tau_1\,$ or
$\,\tau \gg \tau_2$, while smoothly interpolating in between; and
$\,\cal S\,$ is some Schwartz space neighbourhood of $\,0\mini$.
Then constraints {\bf (iii)} and {\bf (iv)} imply that

\be
\ti{00} \:=\: {m^0_3(\xi) \over 2\mini\wcc} \ , \hspace{.5cm}
\ti{0i} \:=\: \ti{ij} \:=\: 0 \ , \hspace{.5cm} \tii{\mu\nu} \:=\:
0 \ ,
                        \label{constr1}
\ee

\mam

\be
\eta_{AA3}^{000}(\xi)\:=\:-{\ti{00} \over 3\mini\wcc} \ ,
\hspace{.5cm} \mbox{for}\; A=1,\mini 2\ , \hspace{.5cm} \mbox{all
other }\, \eta_{ABC}^{\mu\nu\rho}(\xi)\, \mbox{ vanish.}
                        \label{constr2}
\ee

\noi
Hence, $\,(m^\mu_3)=(m^0_3,\,{\bf 0})$. Under somewhat more
restrictive assumptions on $\,\cal W$, imposing, for example,
relations of the form (\ref{rho1}) and (\ref{rho2}) on
$\,\wt{}=\tilde{\omega} + \tilde{\rho}\,$ which couple $\,\wt{}\,$
and $\,\at{}$, a non-zero spin current $\,{\bf
m}_3=(m^1_3,\,m^2_3)\,$ is possible, too. For a more detailed
discussion, see Fr\"ohlich and Studer~(1992b).

A corollary of our derivation of $\,\Ssaw$, using gauge
invariance and incompressibility, is the {\em Goldstone
theorem$\,$} (Goldstone, 1961; Goldstone, Salam, and Weinberg,
1962): Recalling that $\,w_{c,03}= -{g\mini\mu \over 2\hbar
c}\,B_c$, if $\,\omega_0=0\mini$ (see (\ref{rho1}) and
(\ref{back1})), and denoting by $\,{\cal M}:={g\mini\mu \over
2}\,m^0_3\,$ the {\em magnetization$\,$} in the background field
$\,B_c\mini$ (see~(\ref{LRS1}) below), one finds,
by~(\ref{constr1}), that, for an incompressible system, the
following identity must hold:

\ben
{\cal M}(\xi)\:=\: -{2\mini\hbar c\over g^2\mu^2}\,\tau_1^{00}(\xi)\,
B_c(\xi) \ . 
\een

\noi
Hence, with $\,|\tau_1^{00}|<\infty\,$, one finds that, if $\,{\cal
M}\,$ does {\em not$\,$} tend to $\,0$, as the background magnetic
field, $\,B_c$, tends to $\,0$, then the system {\em cannot$\,$} be
incompressible at $\,B_c=0$, i.e., there are gapless extended modes,
the {\em Goldstone bosons$\mini$}, coupled to the ground state by
the spin current (Fr\"ohlich and Studer, 1992b). We note that our
proof also works for systems with continuous non-abelian internal
symmetries. An analysis of the form of the effective action and of the
spectrum of low-energy modes in systems with broken continuous
symmetries and Goldstone bosons has been presented in Leutwyler
(1993), using ideas very similar to those described in this section.

\bigskip
{\bf Remark}$\mini$.$\;$ In Eq.~(\ref{Seff}) for the scaling
limit of the effective action $\,\Ssaw\,$ of a two-dimensional,
incompressible quantum fluid we have, as explained above, retained
only {\em relevant$\,$} and {\em marginal terms$\mini$}, i.e., terms
scaling as $\,\tha^{-D}$, for $\,\tha \rightarrow \infty$, with
$\,D\leq -1\,$ and $\,D=0$, respectively. Although 
we are mainly interested in the physics corresponding to the
relevant and marginal terms in the effective action $\,\StLeff\,$
we display, here, the most important {\em subleading order
terms$\mini$}. These are the unique terms that are of second order
in the perturbation potentials $\,\tilde{a}\,$ and $\,\tilde{w}\,$
and of scaling dimension $\,D=1$, the so-called ``Maxwell terms''.
Added to the r.h.s.\ of Eq.~(\ref{Seff}) they take the form

\bea
&  & \hspace{-6.3mm} -\, {1\over 4}\: \biggl[\; \sum_{i=1}^2\: 
\gp{i} \iLo \fft{0i}^{\,2}\,dv + \go \iLo \fft{12}^{\,2} \,dv 
\nonumber  \\  
&  & +\; \sum_{i=1}^2\: \lp{i} \iLo \mbox{tr}\,
[\,\kk{0i}^2\,]\,dv + \lo \iLo \mbox{tr}\,[\,\kk{12}^2\,]
\,dv \;\biggr] \ ,
                         \label{MAX}
\eea

\noi
where $\,\fft{\mu\nu} := \partial_\mu\at{\nu} - \partial_\nu
\at{\mu}\,$ is the \U\ curvature (or field strength), and likewise
$\,\kk{\mu\nu} = \partial_\mu w_{\nu} - \partial_\nu w_{\mu} +
[w_\mu,\,w_\nu]$, where $\,w\,$ is given by (\ref{wtot}), is the
\SU\ curvature. Moreover, $\,\gp{i}, \ \go, \ \lp{i}\,$ and
$\,\lo\,$ are constants of dimension of a length. Rotation invariance
in the scaling limit would imply that $\,\gp{1} = \gp{2} =:
g_\Vert^{\mbox{}}\,$ and $\,\lp{1}=\lp{2} =: l_\Vert^{\mbox{}}\,$. A
discussion of the consequences of the \U\ curvature terms can
be found in Fr\"ohlich and Studer~(1992b). For an application of the
\SU\ curvature terms to a spin-pairing mechanism, see the end of
Subsect.~\ref{ssQP}.


\subsection{Linear Response Theory and Current Sum Rules}
\label{ssLR}

Next, we discuss the {\em linear response equations$\,$}
(\ref{LR1}) and (\ref{LR2}) that follow from our (universal)
expression (\ref{Seff}) for the scaling limit $\,\Ssaw\,$ of
the effective action of systems characterized by the conditions
(\ref{back}) and (\ref{constr0})--(\ref{constr2}). It is a simple
exercise to verify that

\bea
\sggx\: \El j^\mu\xx \Er_{a,\,w} & = & \sggx\:j_c^\mu\xx + {\sigma
\over 2\pi}\, \ops{\mu\nu\rho} \left( \partial_\nu\at{\rho}
\right)\!\xx  \nonumber  \\
 & & \mbox{} + {\chi_s \over 2\pi}\, \ops{\mu\nu\rho} \left(
\partial_\nu \wt{\rho 3} \right)\!\xx + \cdots \ ,
                       \label{LRE}
\eea

\mam

\bea
\sggx\: \El s_A^\mu\xx \Er_{a,\,w} & = &
\sggx\:\delta_{A3}\,\delta^\mu_0\,m_3^0\xx + \delta_{A3}\, {\chi_s
\over 2\pi}\, \ops{\mu\nu\rho} \left( \partial_\nu \at{\rho}
\right)\!\xx   \nonumber  \\
 & & \mbox{} + \delta_{A3}\, {\sigma_s \over 2\pi}\, \ops{\mu\nu\rho}
\left( \partial_\nu \wt{\rho 3} \right)\!\xx  \nonumber  \\
 & & \mbox{} - {k \over \pi}\, \ops{\mu\nu\rho} \Bigl[
\left( \partial_\nu w_{\rho A} \right)\!\xx - \eps{ABC}\, w_{\nu
B}\xx w_{\rho C}\xx \Bigr]  \nonumber  \\
 & & \mbox{} + \sggx\: 2(1-\delta_{A3})\,\delta^\mu_0\, \ti{00}\,
\wt{0A}\xx + \cdots \ ,
                            \label{LRS}
\eea

\noi
where the dots stand for terms coming from irrelevant terms in the
effective action, or from terms of second order in $\,\wt{}\mini$
(e.g. a term proportional to $\,\eta^{000}_{ABC}\,$) which are of
little interest in linear response theory. Furthermore, we recall
that $\,w_{\mu A}=w^\kt_{c,\mu A} + \wt{\mu A}\mini$;
see~(\ref{wtot}).

In order to understand the physical contents of these equtions, we
should recall the physical meaning of the connections
$\,a\,$ and $\,w\,$ elucidated in Subsect.~\ref{ssSS}$\mini$:
{}From Eqs.\ (\ref{a}), (\ref{T11}), and (\ref{cori}) we know that

\be
\am{j}\:=\: -{q \over \hbar c} A_j(x) - {m \over \hbar}f_j(x) \ ,
                            \label{Ta1}
\ee

\noi
where $\,\bA\,$ is the electromagnetic vector potential, $\,q\,$
is the charge and $\,m\,$ the effective mass of the particles in the
quantum fluid, (for electrons, we have $\,q=-e\,$), and $\,{\bf
f}\,$ is a divergence free velocity field generating some
incompressible superfluid flow. Furthermore, by (\ref{a}),

\be
\am{0}\:=\: {q \over \hbar c} \Phi(x) \ ,
                            \label{Ta2}
\ee

\noi
where $\,\Phi\,$ is the electrostatic potential.

In our study of two-dimensional, incompressible
quantum fluids on a surface $\,M\,$ embedded in
$\,\BBn{E\mini}{3}$, it is natural to choose an \SU\ gauge with the
property that $\,e^3(t,\mini\vec{x})$ is orthogonal to the
cotangent space of $\,M\,$ at $\,\vec{x}$, for all times $\,t\mini$;
see~(\ref{back})--(\ref{back1}) and the discussion at the beginning
of Subsect.~\ref{ssDG}. Then, by~(\ref{back1}), a possible affine
\SU\ connection, $\,\omega^{(s)}_{\mu}$, has the form

\be
\oms{\mu}\:=\:i\,\omega_\mu(x)\,L^{(s)}_3 \ .
                           \label{oo}
\ee

\noi
It then follows from (\ref{w1})--(\ref{w3}), (\ref{rho1}),
(\ref{T11}), and (\ref{wh1}) that

\be
w_{0A}(x)\:=\: -{g\mini\mu \over 2\hbar c}B_A(x)+ \delta_{A3}\,
[\,\Omega(x) + \omega_0(x)\,] \ ,
\ee

\noi
where, by (\ref{Rot0}), $\,{\bf\Omega}(x) = (0,\mini 0,\mini
\Omega(x)) = \halb\, curl \:{\bf f}(x)\,$ with respect to the
orthonormal frame $(e^A(x))_{A=1}^3\,$ at $\,x$, and the magnetic
moment of the particles is given by $g\mu \vec{L}^{(s)}$, (for electrons,
we have $\,\mu=-\mu_B\mini$; see (\ref{mamo})). Moreover, by
(\ref{w1})--(\ref{w3}), (\ref{rho2}), (\ref{T7}), and (\ref{oo}),

\be
w_{jA}(x)\:=\: \left( -{g\mini\mu \over 2 \hbar c} + {q \over 4mc^2}
\right) \sum_{B=1}^{3} \eps{kAB}(x)\, E_B(x) + \delta_{A3}\,
[\,\omega_j(x)+ \cdots\,] \ , 
\ee

\noi
where the dots correspond to terms proportional to derivatives of
$\,\Omega(t',\mini\vec{x}\mini ),\ t'\leq t$, (and are generated by
the \SU\ gauge transformation $\,\Us(R)\,$ with $\,R\,$ defined in
(\ref{Rot})).

Finally, we define, {\em in physical units$\mini$}, the {\em charge density
(operator)$\,$} by

\be
\rho\xx \::=\: q\:\sggx\:j^0\xx \ ,
                           \label{ec1}
\ee

\noi
the {\em electric current density$\,$} by

\be
\JJ \::=\: qc\:\sggx\:j^i\xx \ ,
                           \label{ec2}
\ee

\noi
the {\em spin density$\,$} by

\be
S_{\, A}^{\; o}\;(\xi) \::=\: {\hbar \over 2}\,\sggx\:s_A^0\xx \ ,
\ee

\noi
and the {\em spin current density$\,$} by

\be
S_{\, A}^{\; i}\;(\xi) \::=\: {\hbar c \over 2}\,\sggx\:s_A^i\xx \ .
\ee

Then, for the $(\mu=0)$-component, Eq.~(\ref{LRE}) reads

\bea
\El \rho\xx \Er_{a,\,w} & = & \rho_c\xx - {\sH \over c}\,\Bt -
\sigma\, {2qm \over h}\,\Ot  \nonumber  \\
 & & \mbox{} - \chi_s \left[\,{qg\mu \over 4hc}\,\vec{\nabla} \cdot
\Etv - {q \over 2\pi}\,\R \,\right] + \cdots \ ,
                       \label{LRE1}
\eea

\noi
where the {\em Hall conductivity (for the electric current)}, $\sH$,
is defined by

\be
\sH \::=\: {q^2 \over h}\,\sigma \ ,
                          \label{sH}
\ee

\noi
$\Etv:=(\tilde{E}_1\xx,\mini \tilde{E}_2\xx)\,$,
$\,\vec{\nabla} = (\partial_1,\mini\partial_2) :=(\dd{\xi^1},\mini
\dd{\xi^2})$,  $\,\R:=curl\;\vec{\omega}\xx$ is the scalar curvature
of $\,M\,$ at $\,\xi$, and the dots stand for contributions from
irrelevant terms. It will turn out that

\be
\chi_\bot\::=\:-{qg\mu \over 2hc}\, \chi_s
                       \label{MS}
\ee

\noi
is the {\em magnetic susceptibility$\,$} of the system in the
$\mini 3$-direction normal to the surface. In Eq.~(\ref{LRE1}) and
the following formulas the tildes $\;\tilde{}\;$ indicate
contributions from the perturbation potentials $\,\at{}\,$ and
$\,\wt{}$; (we have absorbed the affine spin connection $\,\omega\,$
into $\,\wt{}$, but without decorating it with a $\;\tilde{}\;$).
Next, one verifies that

\bea
\El \JJ \Er_{a,\,w} & = & \JJc - \sH\,\ops{ij}\Et_j\xx  + \sigma\,
{qm \over h}\,\ops{ij}\dd{\tau}\ft_j\xx     \nonumber  \\
 & & \mbox{} - \chi_s \left[\,{qg\mu \over 2h}\, \ops{ij}
\partial_j \Bt - {qc \over 2\pi}\,\ops{ij} \partial_j \Ot
\,\right]    \nonumber  \\
 & & \mbox{} + \chi_s \left[\,{qg\mu \over 4hc}\, \dd{\tau} \Et^i\xx
- {q \over 2\pi}\,\ops{ij} \dd{\tau}\lambda_j\xx\, \right]
+ \cdots \ ,
                       \label{LRE2}
\eea

\noi
where $\,\tau:=\xi^0/c\,$ is the rescaled time variable.

{}From Eq.~(\ref{LRS}) we find, for example, that for $\,\mu=0\,$ and
$\,A=3\,$ (i.e., for the spin density along the $\mini 3$-direction
in the spin~- or (co)tangent space)

\bea
{g\mu \over \hbar}\,\El S_{\,3}^{\,0}\;(\xi) \Er_{a,\,w} & = & {\cal
M}_c\xx + \sHs \left[\,{g\mu \over 2\hbar c}\, \vec{\nabla}
\cdot \Etv - 2\,\R \,\right] + k\,{g^2\mu^2 \over 4hc}\,
\vec{\nabla} \cdot \Etvc  \nonumber  \\
 & & \mbox{} + \chi_\bot \left[\,\Bt - {hc \over g\mu\, \pi}\,\Ot\,
\right] + \cdots \ ,
                       \label{LRS1}
\eea

\noi
where $\,{\cal M}_c\,$ is the {\em magnetization} of the system in
the background field $\,(a_c,\mini w_c)\,$ given by
(\ref{back})--(\ref{back1}), $\,\chi_\bot\,$ is the magnetic
susceptibility at $\,(a_c,\mini w_c)\,$ defined in (\ref{MS}), and

\be
\sHs \::=\: {g\mu \over 4\pi}\,k - {g\mu \over
8\pi}\,\sigma_s
                       \label{LRS11}
\ee

\noi
is the {\em Hall conductivity for the spin current$\mini$}. As
Eqs.~(\ref{LRS1}) and~(\ref{Seff}) show, $\,\sHs\,$ is a {\em
pseudoscalar$\mini$}.
Note that (when $\tilde{f}_j = \frac{\partial}{\partial\tau} \;
\tilde{E}_j \;=\;\frac{\partial}{\partial\tau}\;\lambda_j\;=\;0$)
equations (6.51), (6.52) and (6.50) imply the {\it Hall law} (4.13)! 
 Next, for $\,\mu=i=1,\,2\,$ and $\,A=3\,$
(i.e., for the spin current density in the $\mini i$-direction in
the surface $\,M\,$ and polarized along the $\mini 3$-direction in
the spin~- or (co)tangent space)

\bea
\El S_{\, 3}^{\, i}\;(\xi)\Er_{a,\,w} & = & \sHs\, \left[\, \ops{ij} \partial_j
\Bt - {hc \over g\mu\, \pi}\, \ops{ij} \partial_j \Ot - {1 \over
2c} \dd{\tau}\Et^i\xx + {h \over g\mu\,\pi}\, \dd{\tau}\lambda^i\xx\,
\right]   \nonumber    \\
 & & \hspace{-7.6mm}\mbox{} + k\, {g\mu \over 4\pi}\, \ops{ij}
\partial_j B_c\xx + \chi_\bot \left[\, {\hbar c \over g\mu}\,
\ops{ij} \Et_j\xx - {m\hbar c \over qg\mu}\, \dd{\tau}\ft^i\xx\,
\right] + \cdots \ ,
                       \label{LRS2}
\eea

\noi
where the dots stand for terms proportional to $\,\omega_0\xx\,$
and further irrelevant and higher-order terms. Similar equations
hold for the remaining $su(2)$ components of $\,\El S_{\, A}^{\,\mu}
\Er_{a,\,w}$, but we refrain from displaying them explicitly and
refer the reader to the discussion in Fr\"ohlich and Studer~(1992b).

We encourage the reader to notice how neatly our formulas summarize
the laws of the {\em Hall effect$\mini$}, including effects due to {\em
tidal forces$\,$} coming from {\em (superfluid) flow$\,$} and
effects due to {\em curvature$\mini$}. [We believe that the tidal
terms might be relevant in the study of transitions between certain
plateaux of $\sigma_H$ in very pure samples.]

\bigskip
Let us comment on the relation of our definition of the {\em
Hall conductivity} $\,\sH = {e^2 \over h}\sigma\,$ as the
coefficint of a Chern-Simons term, $\,{\sigma \over 4\pi}\iLo \at{}
\wedge\! \da \at{}\,$ in the effective action $\,\Ssaw$, see
(\ref{Seff}), of an incompressible quantum Hall fluid to the more
conventional definition via the {\em Kubo formula$\,$} (see,
e.g., Fradkin, 1991). It follows easily from Eqs.~(\ref{Pa2}),
(\ref{Pa2A}), and (\ref{Seff}) that $\,\sigma\,$ appears in the
following {\em current sum rules$\,$}: For every choice of a
permutation $(\mu\,\nu\,\rho)$ of $\,(0\,1\,2)$,

\be
\sigma \:=\: 2\pi i\; \mbox{sign}(\mu\,\nu\,\rho) \int (y-x)^\mu\,
\El\, T[\, j^\nu(x)\,j^\rho(y)]\, \Er_{a,\,w}^{\mbox{\scriptsize
con}}\; d^{\,3}y \ .
                            \label{sum}
\ee

\noi
These are {\em three$\,$} equations for one and the same quantity
$\,\sigma$. The equation for $\,(\mu\,\nu\,\rho)=(0\,1\,2)\,$ is

\be
\sigma \:=\: 2\pi i \int (s-t)\, \El\, T[\, j^1(t,\mini\vec{x}\,)
\,j^2(s,\mini\vec{y}\,)]\, \Er_{a,\,w}^{\mbox{\scriptsize con}}\;
ds\,d^{\,2}\vec{y} \ ,  
\ee

\noi
which is just the Kubo formula (in ``mathematical'' units);
compare, e.g., to  Fradkin~(1991). The other two equations are an
automatic consequence of \U\ gauge invariance.

Thouless and coworkers (Thouless {\em et al.}, 1982; Niu and
Thouless, 1984, 1987; Kohmoto, 1985), and followers (Avron,
Seiler, and Simon, 1983; Avron and Seiler, 1985; Dana, Avron, and
Zak, 1985; Avron, Seiler, and Yaffe, 1987; Kunz, 1987), have derived
from the Kubo formula that

\be
\sigma \:=\: {1\over n_o}\,c_1 \ ,
                               \label{deg}
\ee

\noi
where $\,n_o\,$ is the ground-state degeneracy and $\,c_1\,$ is the
{\em first Chern number$\,$} of a vector bundle over a two-torus of
magnetic fluxes $\,(\Phi_1,\mini\Phi_2)$. Thus, $\,c_1\,$ is an
integer. Does our formulation ``know''
that $\,n_o\,$ is the degeneracy of the ground state? Yes it does!
This follows, e.g., from the material in Sects.\,\ref{sAC}
and~\ref{sClQHFs}, and has been noted in Wen~(1989, 1990a), and
Wen and Niu~(1990).

Bellissard~(1988a, 1988b) and Avron, Seiler, and Simon~(1990, 1994)
have also given a definition of $\,\sigma\,$ as an {\em
index$\mini$}. Their definition is equivalent to ours, too, and the
proof follows from the material in Sects.\,\ref{sAC}
and~\ref{sClQHFs}$\mini$; see also Sect.~6 in Fr\"ohlich and
Kerler~(1991).

Finally, we note that from Eqs.~(\ref{Pa2}), (\ref{Pa2A}), and
(\ref{Seff}) it also follows that $\,\sHs={g\mini\mu_B \over 8\pi}
\sigma_s\,$ (for $\,k=0$, i.e., fully spin-polarized quantum Hall
fluids) is given by a Kubo formula involving spin currents,

\be
\sigma_s \:=\: 2\pi i\; \mbox{sign}(\mu\,\nu\,\rho) \int (y-x)^\mu\,
\El\, T[ s_3^\nu(x)\,s_3^\rho(y)]\, \Er_{a,\,w}^{\mbox{\scriptsize
con}}\; d^{\,3}y \ , 
\ee

\noi
and it can then be shown to be proportional to a first Chern number
of a vector bundle over a two-dimensional torus of electric charges
per unit length $\,(Q_1,\mini Q_2)$. We refer the reader
to Fr\"ohlich and Studer~(1992b) for a more systematic study of current sum rules and
``proofs''. Here we merely give a last example by expressing the
{\em magnetic susceptibility} $\,\chi_\bot=-{qg\mini\mu \over
2hc}\chi_s\,$ in the form of a (mixed) sum rule,

\be
\chi_s \:=\: 2\pi i\;\mbox{sign}(\mu\,\nu\,\rho) \int (y-x)^\mu\, \El\,
T[\,j^\nu(x)\, s_3^\rho(y)]\, \Er_{a,\,w}^{\mbox{\scriptsize
con}}\; d^{\,3}y \ .  
\ee


\subsection{Quasi-Particle Excitations and a Spin-Singlet Electron
Pairing Mechanism}  
\label{ssQP}

In this subsection, we present a first analysis of quasi-particle
excitations above the ground state in a two-dimensional,
incompressible quantum fluid, whose scaling limit of the effective
action is given by the action $\,\Ssaw\,$ presented in
Eq.~(\ref{Seff}). A systematic, general analysis of quasi-particle
excitations in quantum Hall fluids is given in
Sect.~\ref{sClQHFs}. 

At the end of this subsection, we describe a natural mechanism for
spin-singlet pairing of electrons that are moving in some two- or
three-dimensional magnetic background.

\subsubsection{Laughlin Vortices and Fractional Statistics}

For simplicity, we begin our analysis by considering a flat,
two-dimensional system of charged fermions with vanishing magnetic
moment $\,(\mu=0)$, so that the \SU\ connection $\,w\,$ vanishes
identically in an appropriate \SU\ gauge; (the local frames
$\,(e^1(x),\mini e^2(x),\mini e^3(x))\,$ are chosen to be
time-independent, so that there is no tidal Zeeman term; see
Sect.\,\ref{sGI}). We suppose that, in a small neighbourhood of a
suitably chosen background potential $\,a_c\,$ (typically $\,
a_{c,0}=0\,,\ b_c=\partial_1 a_{c,2} -\partial_2\mini a_{c,1} =
\const\,$ and of suitable magnitude), the system is incompressible.
Then the action in the scaling limit is given by

\be
-\hi\,\Ssa \:=\: \iLo j^\mu_c\,\at{\mu}\, dv + {\sigma \over 4\pi}
\iLo \at{} \wedge\! \da\at{}\ ,  
\ee

\noi
up to boundary terms. The first term on the r.h.s.\ is unimportant in
the following discussion, and we set $\,j^\mu_c = 0$.

Let us produce a ``{\em Laughlin vortex$\,$}'' (Laughlin, 1983a,
1983b, 1990) in this system by turning on a (perturbing) magnetic
field $\,\bt = \partial_1 \at{2} - \partial_2 \at{1}\,$ in a small
disc. [Actually, $\,\bt\,$ could be a vorticity field of a
superfluid flow if, instead of an electronic quantum Hall fluid, we
consider a superfluid film. We shall, however, use ``magnetic
language'' in the following discussion.] {}From our discussion of the
Aharonov-Bohm effect in Subsect.\,\ref{ssAB} we know that this
excitation only disturbs the system locally, and thus may have a
finite energy difference from the ground-state energy, if

\be
{1 \over 2\pi} \int \bt\mini (t,\mini\vec{\xi}\,)\, d^{\,2}\vec{\xi}
\:=\: n\ , \hwh n\in \BB{Z}\ . 
                             \label{Q0}
\ee

\noi
By Eq.~(\ref{LRE1})  [see also Eq.~(\ref{ED3})], we
have that

\be
\El j^0\xx \Er_{a,\,w}\:=\: {\sigma \over 2\pi}\,\bt\mini\xx \ ,
\ee

\noi
and hence the charge of this excitation (in units where $e=1$ and
with the background charge normalized to 0) is given by

\be
Q\:=\:\int \El j^0(t,\mini\vec{\xi}\,) \Er_{a,\,w}\,d^{\,2}\vec{\xi} \:=\:
\sigma\,n \ .
                             \label{Q1}
\ee

\noi
If $\,\sigma\,$ is not an integer then $\,Q\,$ will be
{\em fractional$\mini$}, in general. Now consider two such
excitations localized in two disjoint small disks and interchange
them (adiabatically) along some path oriented anticlockwise.
According to Subsect.\,\ref{ssAB}, the Aharonov-Bohm phase picked up
in this process is given by

\be
e^{i\pi \mini \tha} \::=\: e^{i\pi\mini Q\mini n} \:=\:
e^{i\pi\mini \sigma \mini n^2}\ ,
                             \label{Q2}
\ee

\noi
where we have normalized the {\em statistical phase} $\,\tha\,$ such
that $\,\tha\equiv 1\ (\mbox{mod }2)\,$ corresponds to {\em Fermi
statistics$\mini$}, $\,\tha\equiv 0\ (\mbox{mod }2)\,$ corresponds to
{\em bosons$\mini$}, and $\,\tha \not\equiv 0,\mini 1\ (\mbox{mod }2)\,$ to
{\em anyons$\,$} (Leinaas and Myrheim, 1977; Goldin, Menikoff, and
Sharp, 1980, 1981, 1983; Wilczek, 1982a, 1982b; for a review, see
Fr\"ohlich, 1990). Thus Laughlin vortices are anyons, unless
$\,\sigma\,n^2\,$ is an integer.

Among the excitations that one can produce in this fashion there
should be the particles constituting the system, namely electrons
(or holes). Let us suppose that the state of the system is fully
spin-polarized, (as is the case for filling factors $\,\nu
\stackrel{\rm e.g.}{=} {1 \over 3},\,{1 \over 5}\,$ in electronic
quantum Hall fluids). Supposing that a magnetic flux $\,n_o\,$ (in
units where the elementary flux quantum $\,{hc \over e} = 1\mini$) produces
a state of $\,N\,$ electrons, we infer, from (\ref{Q1}), that

\be
\sigma \:=\: {N \over n_o} \ .
                          \label{sigm}
\ee

\noi
If $\,N\,$ is {\em odd$\,$} this state is composed of $\,N\,$
fermions and hence describes an excitation with Fermi statistics, so
that by (\ref{Q2}) 

\be
e^{i\pi\mini N\mini n_o}\:=\:-1Ê\ .
                             \label{Q3}
\ee

\noi
Thus $\,n_o\,$ must be {\em odd$\mini$}, too. In fact, one may show that
if $\,N\,$ and $\,n_o\,$ have no common divisor then $\,n_o\,$ is
odd. In particular, for $\,N=1$, we conclude that

\be
\sigma \:=\: {1 \over n_o}\ ,\hspace{.5cm} \mbox{with $\,n_o\,$ odd} \ .
\ee

\noi
This is the famous {\em odd-denominator rule$\,$} (see, e.g., Tao
and Wu, 1985). An excitation associated to a magnetic flux (or
vorticity) 1  then has fractional
charge $\,Q={1\over n_o}\,$  and is an anyon, for $\,n_o>1$.

Note that the vector potential $\,\at{}$ created by a point-like
excitation of charge $\,Q\,$ located at $\,\vxx = \vec{0}\,$ is
given by

\be
\at{i}(t,\mini\vxx) \:=\: -{Q\over \sigma}\,\sum_{j=1}^2 \eps{ij}\,
{\xi^j \over \mbox{$\left|\mini\vxx\right|$}^2} \ , \hspace{.5cm}
i=1,\,2 \ ,  \ee

\noi
as follows from (\ref{Q0}) and (\ref{Q1}), for $\,\El j^0(t,\mini
\vxx) \Er_{a,\,w}= Q\,\delta(\vxx)$. This is the
``$U(1)$ Knizhnik-Zamolodchikov connection''.

\subsubsection{Spinon Quantum Mechanics}

Next, we consider an ``in vitro'' system, namely a ``{\em chiral
spin liquid$\,$}''. [It is not entirely clear that such systems
exist in nature, but they might appear as subsystems of superfluid
$\mbox{}^3\mmini He$ layers.] A chiral spin liquid is a system of
neutral particles of spin $\,s_o>0\,$ and with non-zero magnetic
moment (i.e., $\,\mu \neq 0\,$) having a spin-singlet ground state
for some constant magnetic field, ${\bf B}_c$. It is assumed, here,
to be incompressible and to exhibit breaking of parity (reflections
in lines) and time reversal, but no spontaneous magnetization. In
our formalism, the scaling limit ot the effective action of such a
system is given by

\be
-\hi\,\Ssw \:=\: {k \over 4\pi} \iLo \mbox{tr}\left(w \wedge\! \da w +
{2 \over 3}\, w \wedge\! w \wedge\! w \right) \ ,
                             \label{spi0} 
\ee

\noi
up to boundary terms. Under reflections in lines, $\,w_i\,$
transforms as a vector, $\,w_0\,$ as a pseudoscalar, and $\,k\,$ as
a pseudoscalar. Let us consider an excitation created by turning on
an $\,SU(2)$ gauge field $\,w\,$ with curvature (or field strength)
$\mini h$ given by

\bea
h\xx & := & \da w\xx + w\xx \wedge\! w\xx   \nonumber  \\
 & = & {i\over 2} \sum_{\mu,\nu=0}^2 \sum_{A=1}^3 h_{\mu\nu A}\xx\,
\LsA{s_o}{A}\,d\mini\xi^\muÊ\wedge\! d\mini\xi^\nu \ ,   
\eea

\noi
where $\,h_{\mu\nu A}=\partial_\mu w_{\nu A}-\partial_\nu w_{\mu
A}-2\,\eps{ABC}\,w_{\mu B}\mini w_{\nu C}$; (see also the discussion
following (\ref{MAX})). For example, we may choose $\,h\,$ to be
given by

\be
\bfh_{0i}\xx \::=\: (h_{0i\,1}\xx,\mini h_{0i\,2}\xx,\mini
h_{0i\,3}\xx) \:=\: {\bf 0} \ , \hah  \bfh_{12}\xx \:=\: -\bfn\,
h_o(\vxx) \ , 
\ee

\noi
where $\,\bfn\,$ is some unit vector in $\,\BBn{R}{3}$ and
$\,h_o\,$ is a time-independent function. By Eq.\
(\ref{LRS}), the spin density of this excitation is given by

\be
\El \bfs^0\xx \Er_{w} \::=\: \El\mini (s_1^0\xx,\mini s_2^0\xx,
\mini s_3^0\xx) \mini\Er_{w} \:=\: -{k \over \pi}\mini
\bfh_{12}(\vxx) \:=\: \bfn\, {k \over \pi}\mini h_o(\vxx)   \ , 
                          \label{spi1}
\ee

\noi
so that the expectation value of its total spin operator, $\,{\bf
S} := {\hbar\over 2}\mini {\bf L}^{(s)}\,$ (in the spin-$s\,$
representation; see~(\ref{spin})), is given by

\be
{\hbar\over 2}\mini \El {\bf L}^{(s)} \Er_{w} \:=\: \El\mini {\bf S}
\mini\Er_{w} \:=\: \bfn\,{k\mini \hbar \over 2\pi}Ê\int h_o(\vxx)\,
d^{\,2}\vxx \ .
\ee

\noi
Such an excitation is commonly called a ``{\em spinon$\,$}''.
Quantum-mechanically, spin is quantized: $\,{\bf S}\cdot{\bf
S}=s\mini (s+1)\mini \hbar^2$, with $\,s\!\in\halb\mini \BB{Z}\,$.
Consider a spinon of spin $\mini s\mini$ {\em localized$\,$} at the
point $\,\vxx=\vxxo_1\,$. Then Eq.\ (\ref{spi1}) says that
$\,\bfh_{12}\,$ has to solve the equation

\be
\El {\bf L}^{(s)} \Er_{w}\, \delta(\vxx-\vxxo_1) \:=\: - {k \over
\pi}\, \bfh_{12}(\vxx) \ .
                             \label{spi2}
\ee

\noi
An \SU\ connection $\,w = i \sum_{\mu} \bfw_{\mu} \cdot
{\bf L}^{(s)}\, d\mini\xi^\mu\,$ for the field strength $\,h\,$
satisfying (\ref{spi2}), with $\,\bfh_{0i} = 0$, is given by

\be
\bfw_0\xx\:=\:{\bf 0} \ , \hah \bfw_j\xx \:=\: {1 \over 2\mini k}\, \El {\bf
L}^{(s)} \Er_{w}\,\sum_{l=1}^2\eps{jl}\, {\xi^l-\xi_1^l \over
\mbox{$\left|\mini \vxx-\vxxo_1 \right|$}^2} \ ,\hspace{.5cm}
j=1,\mini 2 \ .
                             \label{spi3} 
\ee

Suppose we now create a second spinon of spin $\mini s'\mini$ moving
in the background gauge field $\,w\,$ excited by the first spinon.
Its dynamics is coupled to $\,w\,$ through the covariant derivatives
(see Eqs.~(\ref{DD1}) and~(\ref{DD2}))

\be
\D{\mu}\:=\: \partial_\mu + i\,\bfw_\mu\xx \cdot {\bf L}^{(s')} \ ,
                             \label{spi4}
\ee

\noi
with $\,\bfw\,$ as in (\ref{spi3}). Let us imagine that it makes
sense to do ``two-spinon quantum mechanics'' on a Hilbert space
$\,{\cal H}^{(s)} \otens {\cal H}^{(s')}$,
with

\ben
{\cal H}^{(s)} \:=\: {\cal D}^{(s)} \otens L^2(M,\,dv) \ , 
\een

\noi
where $\,{\cal D}^{(s)}$ carries the spin-$s\,$ representation of 
\SU. By (\ref{spi3}) and (\ref{spi4}), the covariant derivatives on
$\,{\cal H}^{(s)} \otens {\cal H}^{(s')}\,$ are then given by

\be
\D{0}^1\:=\:\dd{\xi^0_1}\ , \hspace{.5cm} \D{j}^1\:=\:\dd{\xi^j_1}+{i
\over 2\mini k}\sum_{l=1}^2\eps{jl}\, {\xi_1^l-\xi_2^l \over
\mbox{$\left|\mini \vxxo_1-\vxxo_2 \right|$}^2} \sum_{A=1}^3
\LsA{s}{A} \otens \LsA{s'}{A}\ , \nonumber 
\ee

\mam

\be
\D{0}^2\:=\:\dd{\xi^0_2}\ , \hspace{.5cm} \D{j}^2\:=\:\dd{\xi^j_2}+{i
\over 2\mini k}\sum_{l=1}^2\eps{jl}\, {\xi_2^l-\xi_1^l \over
\mbox{$\left|\mini \vxxo_1-\vxxo_2 \right|$}^2} \sum_{A=1}^3
\LsA{s}{A} \otens \LsA{s'}{A}\ .
                             \label{spi5}
\ee

\noi
These are the covariant derivatives associated with the celebrated
{\em Knizhnik-Zamo\-lodchikov connection$\,$} (Knizhnik and
Zamolodchikov, 1984; Tsuchiya and Kanie, 1987). For the 
``two-spinon quantum mechanics'' with parallel transport given by
(\ref{spi5}) to be consistent with unitarity, it is necessary that

\be
k\:=\:\pm (\kappa+2)\ , \hspace{.5cm} \kappa=1,\,2,\,\dots\ .
\ee

\noi
This follows from results in Belavin, Polyakov, and
Zamolodchikov~(1984), Knizhnik and Zamolodchikov~(1984),
 and Tsuchiya and Kanie~(1987). Recalling what
we have said in Subsect.~\ref{ssAC} about the Aharonov-Casher
effect, we observe that the ``phase factor'' arising in the
parallel transport of a quantum-mechanical spinon in the field
excited by a classical spinon with spin orthogonal to the plane of
the system is an Aharonov-Casher phase factor.

Consider an exchange of the positions of two quantum-mechanical,
point-like spinons along some anticlockwise oriented path. Then the
``Aharonov-Casher phase factor'' multiplying the wave function is
given by a matrix

\ben
R^{(\kappa)}_{s\mini s'}\; : \;{\cal D}^{(s)} \otens {\cal
D}^{(s')}\; \rightarrow \;  {\cal D}^{(s')} \otens {\cal D}^{(s)} 
\ ,  
\een

\noi
which is the {\em braid matrix$\,$} for exchanging a chiral vertex
of spin $\,s\,$ with a chiral vertex of spin $\,s'\,$ in the {\em
chiral Wess-Zumino-Novikov-Witten model$\,$} (Belavin, Polyakov, and
Zamolodchikov, 1984; Knizhnik and Zamolodchikov, 1984; Tsuchiya
and Kanie, 1987; see also Gawedzki, 1990) at level $\,\kappa$. It is
given by

\be
R^{(\kappa)}_{s\mini s'}\:=\: T\:\pi_s \otens \pi_{s'}\mini ({\cal
R}^{(\kappa)}) \ ,
                             \label{spi6} 
\ee

\noi
where $\,{\cal R}^{(\kappa)}\,$ is the universal $\,R$-matrix of the
quantum group $\,U_q(sl_2)$, with $\,q=\exp \left( {i\,\pi \over
\kappa+2} \right)$, and $\,T\,$ is the flip (transposition of
factors). All this can be extended to ``$n$-spinon quantum
mechanics''. The matrices $\,(R^{(\kappa)}_{s\mini s'})\,$ determine
an exotic quantum statistics described by non-abelian (for
$\,\kappa>1$, and $\,s,\mini s'< {\kappa \over 2}\mini$)
representations of the braid groups (more precisely, the groupoids
of coloured braids) which is commonly called {\em non-abelian braid
statistics$\,$} (Fr\"ohlich (1988), Fredenhagen, Rehren, and Schroer,
1989; Fr\"ohlich 
and Gabbiani, 1990).  We note that $\,s\,$ and
$\,s'\,$ are forced to be $\leq {\kappa \over 2}\mini$, i.e., there
are no spinons of spin $>{\kappa \over 2}\mini$. One might call
this phenomenon ``{\em spin screening$\,$}''. If the particles of
spin $\,s_o\,$ constituting the chiral spin liquid appear as spinon
excitations above the ground state and $\,s_o\,$ is half-integer
then

\ben
\kappa \:\geq\: 2\, s_o \ .
\een

\noi
One can argue that the
statistics of these particles must be {\em abelian$\,$} braid
statistics, i.e., they are anyons. In fact, it then follows that
they are semions $\,(\tha={1\over 2})$. Now, for a given level
$\,\kappa$, the matrices $\,(R^{(\kappa)}_{s\mini s}) \,$ define an
abelian representation of the braid groups if, and only if,
$\,2\mini s=\kappa$. Thus it follows that for a chiral spin liquid
made of particles of spin $\,s_o\,$

\be
\kappa \:=\: 2\, s_o \ .
                             \label{spi7}
\ee

\noi
Any spinon-excitation of spin $\,0<s<s_o\,$ then exhibits {\em
non-abelian$\,$} braid statistics!

The reader may feel that our ``derivation'' of ``spinon quantum
mechanics'' from the effective action $\,\Ssw\,$ given in
(\ref{spi0}) is based on idealizations -- see (\ref{spi2}) -- and
jumps in the logics -- reasoning between (\ref{spi4}) and
(\ref{spi5}) -- that might be doubtful.
Actually, it turns out that our conclusions concerning spinon
statistics, in particular Eqs.~(\ref{spi6}) and (\ref{spi7}), are
perfectly correct. This follows from an analysis that we
have presented in Sect.~VI in Fr\"ohlich and Studer~(1993b).

In order to understand quantum Hall fluids with {\em
  spin-singlet$\,$} 
ground state, one must glue the Laughlin vortices described in
(\ref{Q0})--(\ref{Q3}) to the spinons discussed above. One checks
that for $\,\sigma={2\over n_o},\ n_o\,$ odd, and
$\,\kappa=2\mini s_o=1$, a Laughlin vortex of vorticity $\,n=-{n_o
\over 2}\,$(!) glued to a spinon of spin $\,s=\halb\,$ is an
excitation of charge $\,Q=-1$, spin $\halb$ and {\em Fermi
statistics$\mini$}; see Fr\"ohlich and Kerler~(1991), and
Sect.~VI in Fr\"ohlich and Studer~(1993b). These are the
properties of an electron. In a quantum Hall fluid ({\em
without$\,$} any {\em very exotic$\,$} internal symmetries) one does
{\em not$\,$} find any finite-energy excitations with non-abelian braid
statistics. However, if one could manufacture a quantum Hall fluid
made of charge carriers of spin $\,s_o={3 \over 2},\,{5 \over
2},\,\dots\ \,$, with a spin-singlet ground state it would display
excitations with non-abelian braid statistics (Fr\"ohlich, Kerler, and
Marchetti, 1992). It may appear difficult to build such a system, in practice. 
 
It may be worthwhile emphasizing that in quantum Hall fluids with
non-vanish\-ing magnetic susceptibility (spin-polarized Hall fluids)
the fractional statistics of Laughlin vortices always appears as a
consequence of a {\em combination$\,$} of the Aharonov-Bohm~- {\em
and$\,$} the Aharonov-Chasher effect; (but notice that, for
spin-polarized quantum Hall fluids, the Aharonov-Casher phase
factors are automatically abelian). This is a consequence of the
fact that electrons have a non-vanishing magnetic moment and follows
form Eq.~(\ref{Seff}).

\subsubsection{A Spin-Singlet Electron Pairing Mechanism}

In this paragraph, we describe a mechanism for spin-singlet pairing
of dopant  electrons (or holes) moving in an antiferromagnetic~- or a
resonating valence-bond (RVB) background. Our mechanism may be
related to the ``spin-bag mechanism'' (Schrieffer, Wen, and Zhang,
1988). For definiteness, we consider two-dimensional systems, but
our  arguments can easily be extended to systems in three dimensions.

The magnetic properties of the system are described by an order 
parameter, $\phi$, transforming under the adjoint representation
of \SUs. 
Integrating out the order parameter $\,\phi$, at a fixed temperature 
$\,T>0$, we obtain the free energy, $\,F_T(w) = - kT\,\ln Z_T(w)$,
as a functional of the gauge field $\,w$.

For an antiferromagnet or a system with an RVB ground state, described, 
e.g., by a Landau-Ginzburg type Lagrangian in which $\,\phi\,$ is 
coupled minimally to $\,w$, the free energy $\,F_T(w)\,$ is expected to
be smooth in $\,w\,$ in a neighbourhood of $\,w=0$. This is in
contrast to the behaviour of $\,F_T(w)\,$ in a system with {\em 
ferromagnetic$\,$} ordering. In a ferromagnetic system, the order 
parameter is the spin density which is the time-component of the spin
current density. The spin current density is the variable conjugate to
the \SU\ gauge field $\,w\mini$; see Eq.~(\ref{Pa2}). Thus if the 
expectation value of the total spin operator in an equilibrium state 
of the system at some temperature $\,T\,$ is non-zero then the free 
energy $\,F_T(w)\,$ must have a cusp-like singularity at $\,w=0$.
But in systems with an antiferromagnetic~- or RVB magnetic
structure,  the order parameter $\,\phi\,$ is {\em not$\,$} given
by a component of some current density. Viewing \SUs$\,$ as an
internal symmetry group of the  system, $\,\phi\,$ turns out to be
a {\em scalar$\,$} field transforming  under the adjoint
representation of the internal symmetry group. In this situation, it
is consistent to assume that $\,F_T(w)\,$ is quadratic in $\,w\,$
at $\,w=0$.

We shall assume that, in the {\em scaling limit$\mini$}, the system does
not break parity~- and time-reversal invariance and is rotation~-
and translation invariant. It then follows from the assumed
smoothness of $\,F_T(w)\,$ near $\,w=0$, from \SU\ diamagnetism ,
i.e., $F_T (w)$ is increasing in $w$, and
from the invariance  of $\,F_T(w)\,$ under time-independent
\SU\ gauge transformations that  $\,F_T(w)\,$ is given by

\bea
F_T(w) & = & -\, {1\over 4}\: \biggl[\;{1\over l(T)}\mini \int
\mbox{tr}\,[\,w_0^2(\vxx)\,] \:d^{\,2}\vxx \:+\: {1\over l^\prime(T)}
\, \sum_{i=1}^2\: \int \mbox{tr}\, [\,(w_i^T)^2(\vxx)\,] \:d^{\,2}
\vxx \nonumber  \\
& &\hspace{-1.8cm} -\, {1\over 4}\: \biggl[\;\lpp(T)\,
\sum_{i=1}^2\: \int  \mbox{tr}\, [\,\kk{0i}^2(\vxx)\,]
\:d^{\,2}\vxx \:+\: \lo(T)\! \int  \mbox{tr}\,[\,
\kk{12}^2(\vxx)\,] \:d^{\,2}\vxx \;\biggr] \:+\:  \cdots \ .
                            \label{PP1}
\eea
\noi
Here $w\,$ is a time-independent, external \SU\ gauge field,
and our  conventions have been chosen such that the traces
in~(\ref{PP1}) are negative;  see Eq.~(\ref{DD2}). The third and
the fourth term on the r.h.s.\  of (\ref{PP1}) are the ``Maxwell
terms'' discussed in~(\ref{MAX});  (note that, because of
rotation~- and translation invariance, we have  $\,\lp{1}=\lp{2} =:
\lpp\,$ and $\,g_{ij}(\vxx)=\delta_{ij}\mini$,  in (\ref{DD2})). In
the first and second term on the r.h.s.\  of~(\ref{PP1}), $\,l\,$
and $\,l^\prime\,$ are constants of dimension  $\,cm$, and the
``transversal'' gauge field $\,w_i^T$, $\,i=1, \mini 2$, is defined
by $\,w^T_i := \left[\,\delta^j_i - \D{i}\, \Delta_{cov}^{-1}\,{\cal
D}^{\mini j} \right] \!w_j \mini$, $\,$with $\,\Delta_{cov} :=
\D{j}\,{\cal D}^{\mini j}$, where $\,\D{j}\,$ has been defined
in~(\ref{CSC0}) and~(\ref{CSC}). Note that the first term is
invariant under time-independent \SU\ gauge
transformations~(\ref{Pa31}), and the second term respects  the
\SU\ Ward identity~(\ref{CSC0}), (i.e., it is invariant under infinitesimal 
(time-independent) \SU\ gauge transformations), but it is {\em
non-local$\mini$}. In fact, this term is the non-abelian analogue
of the term we have encountered in the free energy~(\ref{FE}) of a
superconductor. Its presence mirrors the fact that we do {\em
not$\,$} require the system to be incompressible. \SU\ diamagnetism
implies that $\,\lpp,\mini \lo,\mini  l$, and $\,l^\prime\,$ are
non-negative. [For a system with an RVB~- or VBS ground state, the
non-local term is expected to be absent.]  Finally, the dots
in~(\ref{PP1}) stand for terms of higher order in  $\,w\,$ or
involving higher derivatives acting on $\,w$. 

Given the free energy~(\ref{PP1}), we can study how the system
responds to turning on an (external) \SU\ gauge field $\,w$.
Choosing some $\,w\,$ with the property that $\,\bfw_j=0,\
\,j=1,\mini 2$, we find, similarly to~(\ref{LRS}) and~(\ref{spi1}),
that

\be
\El \bfs^0\xx \Er_{w} \:=\: {\delta F_T(w) \over
\delta\mini\bfw_0\xx} \:=\: -\mini \lpp\, \Delta \bfw_0\xx + {1\over
l}\,\bfw_0\xx + \cdots  \ ,
                              \label{PP2}
\ee

\noi
where $\,\Delta\,$ is the two-dimensional Laplacian.

In order to create a spin-$\halb\mini$ excitation (a dopant electron) 
localized at $\,\vxx=\vxxo_1$, the three \SU\ gauge field components
of  $\,\bfw_0\,$ have to solve Eq.~(\ref{PP2}) for a l.h.s.\ of the
form

\be
\El \bfs^0\xx \Er_{w} \:=\: {2\over \hbar}\,\El {\bf S} \Er_{w}\,
\delta (\vxx- \vxxo_1) \:=:\: \bfsig_1 \, \delta (\vxx-
\vxxo_1) \ .
                              \label{PP3}
\ee

\noi
If $\,\lpp\,$ and $\,l\,$ are {\em positive$\,$} constants,
the solution of~(\ref{PP2}) and~(\ref{PP3}) is given by

\be
\bfw_0\xx \:=\: {\bfsig_1 \over \lpp}\, K \Bigl(\mini
{1\over \ell}\mmini\left|\mini\vxx-\vxxo_1
\right| \Bigr) \ , \hah \bfw_j\xx \:=\: {\bf 0}\ ,
\hspace{.5cm}j=1,\mini 2\ , 
                              \label{PP4}
\ee

\noi
where $\,\ell= \sqrt{\lpp\, l}\mini$, and $\,K\,$ is a function
with the following asymptotic behaviour:

\be
K(x) \:=\: \sqrt{{1\over x}}\, e^{-x} \left( \alpha + {\cal
  O}(\mbox{${1\over x}$}) \right)\ ,\hspace{.5cm}
\mbox{if $\: x\gg 1$} \ ,
                              \label{PP5}
\ee

\noi
with $\,\alpha\,$ a positive constant.

We now consider a second dopant electron moving in the background gauge field 
$\,w\,$ excited by the first one; see~(\ref{PP4}). Recalling the form 
of the coupling in~(\ref{spi4}), we expect the motion of the second 
electron to be subject to a force resulting from a ``Zeeman term'' 
given by $\,2\mini c\,\bfw_0\cdot{\bf S}_2$, where $\,{\bf S}_2\,$ 
denotes the spin operator of the second electron. Classically, we 
find a contribution to the energy of the two-electron system of
the form

\be
E_{12}\:=\: \hbar\mini c\, {\bfsig_1\cdot \bfsig_2 \over \lpp}\, K
\Bigl(\mini {1\over \ell}\mmini \left| \mini \vxxo_2-\vxxo_1 \right| 
\Bigr) \ ,
                              \label{PP6}
\ee

\noi
where $\,\bfsig_2\,$ is the expectation value of the spin
operator, $\mini {\bf S}_2$, of the second electron which we assume to
be localized at $\,\vxx=\vxxo_2\mini$. A similar expression 
to~(\ref{PP6}) is obtained in a ``more symmetric'' treatment: One 
solves Eq.~(\ref{PP2}) for several dopant electrons localized at points 
$\,\vxxo_1,\dots,\mini \vxxo_n\,$ and considers the interaction
term in  the free energy $\,F_T(w)\,$ corresponding to the resulting
\SU\ gauge  field $\,w$. 

The form of the energy $\,E_{12}\,$ in~(\ref{PP6}) suggests that a 
term of the form

\ben
J(\vxxo_1-\vxxo_2)\;{\bf S}_{\vxxo_1}\!\cdot {\bf S}_{\vxxo_2} \ ,
\een

\noi
must be included in the Hamiltonian of the dopant system, where 
$\,J(\vxx)\,$ is some positive function with $\,J(\vxx)\sim 
e^{-|\vxx|/\ell}$, for $\,|\vxxo\mini|\rightarrow \infty\mini$.

To summarize, when we consider two dopant electrons (or holes) moving in a
magnetic background characterized by the free 
energy~(\ref{PP1}) then Eq.~(\ref{PP6}) implies that, as a result of
the collective response of the background, the two electrons
experience a mutual {\em attraction$\,$} if their spins are ``{\em
antiparallel$\,$}'' and a mutual {\em repulsion$\,$} if their spins
are ``{\em parallel$\,$}''. This interaction could result in a 
superconducting state for spin-singlet pairs of dopant electrons (or
holes), as studied in Part II. 



\section{Anomaly Cancellation and Algebras of Chiral Edge Currents
in Two$\,$-Dimensional, Incompressible Quantum Fluids}
\label{sAC}

The purpose of this section is to make a first step in discussing
the origin of the quantization of the constants $\,\sigma,\,\chi_s,\,
\sigma_s\mini$, and $\,k\,$ which appear as the coefficients of the
Chern-Simons terms in the scaling limit of the effective action
of two-dimensional, incompressible quantum fluids; see (\ref{Seff}).
{}From the linear response equations displayed in (\ref{LRE1}) through
(\ref{LRS2}) we recall that these constants completely determine the
response (on large-distance scales and at low frequencies) of such
quantum fluids when perturbed by small external electromagnetic,
``tidal'', and geometric fields. In particular, they specify the Hall
effect for the electric~- and for the spin current. The rationality
of the constants $\,\sigma,\ \chi_s,\ \sigma_s\mini$, and $\,k\,$
follows from a {\em consistency analysis} of the theory presented
hitherto. This analysis can be based on a study of the
so far mysterious boundary terms, ``B.T.'', on the r.h.s.\ of
Eq.~(\ref{Seff}), and it has been carried out in full detail in
Fr\"ohlich and Studer (1993b). [A complementary approach in which
the bulk terms are further exploited is given in Sect.~\ref{sClQHFs}
below.] By the requirement of {\em anomaly cancellation} we find,
among the boundary terms, gauge-anomalous contributions which turn
out to be the generating functionals of the connected time-ordered
Green functions of chiral current operators which generate \uh\ - and
\suh\ current (Kac-Moody) algebras. Basic, physical requirements on
the spectrum of (finite-energy) excitations in incompressible
quantum fluids, together with elements of the representation theory
of current algebras, enter this consistency analysis. The analysis
naturally leads to a  list of possible values of quantum numbers (such
as (fractional) charges and (fractional) statistical phases) of
physical excitations that one expects to find in such quantum
fluids. 

For the sake of concreteness, we restrict our attention to
two-dimensional, incompressible quantum fluids composed of {\em
electrons} (or {\em holes}). For a different example of a physical
system where we can apply similar ideas, see the discussion of
superfluid $\,\mbox{}^3\mmini He$-A/B-interfaces with broken
parity~- and time-reversal invariance that we have presented in
Subsect.~VII.B in Fr\"ohlich and Studer~(1993b).

In the first part of this section, we review the physics at the
boundary of incompressible quantum Hall fluids (for short, QH fluids)
by following some basic ideas of Halperin~(1982). Extending these
ideas by making use of some facts concerning chiral \uh\ current
algebra and introducing the idea of anomaly cancellation, we present
a very natural explanation of the {\em integer quantum Hall
effect$\mini$}. How to generalize this explanation to cover the
{\em fractional quantum Hall effect\,} is sketched in the second
part of this section. More details on the classification of QH
fluids based on a general analysis of \uh\ - and \suh\ current
algebras describing the boundary excitations of a Hall sample can
be found in Subsects.~VI.B and VI.C in Fr\"ohlich and Studer~(1993b).

Independent work about current algebras in incompressible quantum
Hall fluids that bears resemblance with ours (Fr\"ohlich and Kerler,
1991; Fr\"ohlich and Zee, 1991; Fr\"ohlich and Studer, 1992b,
1992c, 1993a, 1993b) has been carried out by Wen and collaborators
(Wen, 1990a--1990c, 1991a, 1991b; Block and Wen, 1990a,
1990b; Wen and Niu, 1990). Additional work vaguely or closely
related to Wen's and ours can be found in B\"uttiker (1988),
Beenakker (1990), MacDonald (1990), Stone (1991a, 1991b), Balatsky
and Fradkin (1991), and Balatsky and Stone~(1991).


\subsection{Integer Quantum Hall Effect and Edge Currents}
\label{ssIQ}

We consider a system of electrons confined to a two-dimensional
domain $\,\Omega$ in the $\mini(x,\mini y)$-plane. We choose
$\,\Omega\,$ to be an annulus and denote by $\,\partial
\Omega\,$ the boundary of $\,\Omega$. In our example, $\,\partial
\Omega\,$ consists of two connected components, $C_1\,$ and $\,C_2$,
which are circles of radius $R_1$ and $R_2$, respectively. We imagine
that there is a (uniform) external magnetic field,  $\,{\bfÊB}_c
=(0,\mini 0,\mini B_c)\,$ (with vector potential $\,{\bf A}_c$,
i.e.,$\ B_c=curl\,{\bf A}_c$), perpendicular to the plane of the
sample. Note that the magnetic field $\bBc$ breaks time-reversal~-
and parity (reflections-in-lines) invariance.

In Subsect.~\ref{ssQH}, we have mentioned that if the Hall
conductivity of the system is on a plateau the longitudinal
resistance $\,R_L\,$ vanishes and the system is {\em
dissipationless$\mini$}, or {\em incompressible$\mini$}. In this
subsection, we intend to study the physics at the boundary of the
sample. [If $\,R_L\,$ is non-vanishing the physics of the
system is complicated, and simple concepts of universality fail to
capture the basic properties of the system.]

Classically, in the absence of an external electric field, there
are no currents in the system. Quantum-mechanically, however, the
picture is different as has been emphasized by Halperin~(1982).
In the absence of an external electric field, currents supported by
the system are localized within approximately one cyclotron radius
of the boundary $\,\partial \Omega$, and they are expected to
persist even in the presence of a moderate amount of disorder
in the sample. Because of the presence of the external magnetic field
$\bBc$ the {\em edge currents} are {\em chiral$\mini$}, i.e., the
electrons drifting in the field $\bBc$ can only move in one
direction along the boundary components $\,C_1\,$ and $\,C_2\,$ of
$\,\partial \Omega$. We can choose the orientation of the annulus
$\,\Omega$, and therefore of $\,C_1\,$ and $\,C_2$, such that the
chirality of the edge current localized near $\,C_i\,$ is given by
the orientation of $\,C_i,\ i=1,\,2$.

In order to be more explicit, we temporarily assume that there is
no disorder in the system, and the electrons are moving in a
confining one-body potential, $\,V$, which is constant in the bulk
and rises steeply at the boundaries of the sample, i.e., at
$|\vec{x}\mini|\approx R_1,\mini R_2$. Furthermore, we assume that
electron-electron interactions are turned off ({\em independent
electron approximation}), so that the many-electron states of the
system can be constructed by filling up one-electron states,
$\,\psi_{\uparrow/ \downarrow}$, in accordance with the Pauli
principle. The one-particle wave function
$\,\psi_{\uparrow/\downarrow}\,$ describes a two-dimensional,
charged {\em scalar} fermion (i.e., a fermion with a fixed spin
polarization ``up'' $(\uparrow)$ or ``down'' $(\downarrow)$). It
satisfies the Schr\"odinger equation

\be
i\hbar \dd{t}\psi_{\uparrow/\downarrow}(t,\mini \vec{x})\:=\: \left[-
{\hbar^2 \over 2m^\ast}\left(\bgrec{\nabla}+i{e \hbar \over
c}\bgrec{A}_c(\vec{x}) \right)^2 + V_{\uparrow/\downarrow}(\vec{x})
\right] \psi_{\uparrow/\downarrow}(t,\mini\vec{x}) \ ,
                          \label{Schr}
\ee

\vspace{2mm}
\mwm

\ben
V_{\uparrow/\downarrow}(\vec{x}) \::=\: V(\vec{x}) \pm {g\mini\mu_B
\over 2}B_c \ . 
\een

\noi
The second term in $\,V_{\uparrow/\downarrow}\,$ is the Zeeman
energy ($\mu_B > 0$) whose sign depends on whether the spin of the
electron is parallel ($\uparrow$) or antiparallel ($\downarrow$) to
the magnetic field $\bBc$. We assume that the effective electron
mass, $\,m^\ast$, is less then the vacuum mass, $\,m_o$, so that all
Landau bands in the combined spectrum of the two Hamiltonians in
Eq.~(\ref{Schr}) have different energies. Hence, each Landau band of
one-electron states is {\em fully spin-polarized$\mini$}.

Because of the rotational symmetry of the annular domain $\,\Omega$,
the eigenvalues, $\,m \in \BB{Z}\mini$, of the z-component, $L_z$, of
the orbital angular momentum operator are good quantum numbers for
the one-electron states. For a given Landau band, a one-electron
state with magnetic quantum number $\,m\,$ is localized, in the
radial direction, within about one cyclotron radius, $r_c=\left(
{\hbar c \over e B_c}\right)^{\!1/2}$, from some mean radius $\,r_m$,
with $\,r_m=r_c \left({m \over \pi}\right)^{\!1/2}$, provided that
$\,R_1<r_m<R_2$, and $|R_{\mini i}-r_m| \gg r_c$, for $\,i=1,\mini
2$. In the presence of a confining one-body potential,
$\,V(|\vec{x}\mini|)$, the energy, $\,\cal E\,$ (in units of
$\,\hbar \omega_c$, where $\,\omega_c= {e B_c \over m^\ast c}\,$ is
the cyclotron frequency), of one-electron states as a function of
the angular momentum $\,m\mini$ (i.e., of the square of their mean
radius $\,r_m$) is approximately constant in the bulk and
rises at the edges; see Halperin (1982).

We define quantities $\,m_i\,$ by setting $\,r_{m_i} \approx
R_{\mini i},\ i=1,\,2$. Furthermore, the magnetic quantum numbers
$\,m^F_{i,\nu}\mini,\ i=1,\,2$, are determined by requiring that all
one-particle levels up to the Fermi energy, $\,E^F$, be filled in
the Landau band indexed by $\,\nu=(n,\mini s)$, where $\,n \in
\BB{N}_{\mini o}$, and $\,s=\:\uparrow$ or $\downarrow\mini$. [Note
that the incompressibility assumption on the system is reflected by
the requirement that, in the bulk region (i.e., for $\,m_1 \ll m \ll
m_2$) the Fermi energy $\,E^F\,$ lies between two Landau bands.]
States corresponding to values $\,m\,$ well between $\,m_1\,$ and
$\,m_2\,$ carry no net electric current. However, states with
$\,m\,$ below, but close, to $\,m_1\mini$, or above $\,m_2\,$ carry
{\em gapless$\mini$}, chiral currents localized within approximately
one cyclotron radius of the boundaries $\,C_1\,$ and $\,C_2$,
respectively (Halperin, 1982).

To summarize, we find that the gapless  excitations near
the Fermi surface of a given filled Landau band are {\em charged,
chiral, scalar fermions$\,$} propagating along the boundary of the
sample. In order to describe the dynamics of these chiral fermions
(chiral Luttinger model) in more detail, we introduce the
one-dimensional momenta

\be
p_{i,\nu} \::=\: {m - m^F_{i,\nu} \over R_{\mini i}}\, \hbar\ ,
\hspace{.5cm} i=1,\mini 2\ ,\hah \nu=(n,\mini s) \in \BB{N}_{\mini o}
\times \{ \uparrow,\downarrow \} \ ,
                             \label{odm}
\ee

\noi
and we redefine the energy of the one-electron states relative to the
Fermi surface, i.e., we write $\,E:={\cal E}-E^F$. Let us set
$\,R_{\mini i}=\theta\mini r_i$, with $\,r_i\,$ fixed, $i=1,\mini 2,\
0<\theta<\infty$. We scale space- {\em and$\,$} time-coordinates
as in (\ref{Scal}), i.e., $(t,\mini \vec{x}\mini)=\theta\mini
(\tau,\mini \vec{\xi}\,)$, where $\,(\tau, \mini \vec{\xi}\,)\,$
belongs to a fixed space-time domain $\,\Lambda_o=\BB{R} \times
\Omega_o\mini$. Then, as $\,\theta\,$ grows, we are interested in
that part of the spectrum of the edge excitations,
$\,E=E(p_{i,\nu})$, which belongs to an ever smaller interval
around $\,p_{i,\nu}=0\mini$. Note that, by Eq.~(\ref{odm}),
$\,p_{i,\nu}\,$ scales with $\,\theta^{-1}$. Thus, for the rescaled
systems in $\Omega_o$, the spectra of the edge excitations
associated with a given Landau band converge towards the {\em
linear\mini} energy-momentum dispersion law of a {\em massless,
chiral}, ``{\em relativistic\mini}'' {\em Fermi field$\,$}
propagating along a circle of radius $\,r_i,\ i=1,\mini 2\mini$.

Before we turn to the description of relativistic Fermi fields
we note that, in order to observe the quantum Hall effect
experimentally, it is necessary to perturb the  system.
This can be achieved, for example, by applying a small voltage
between the inner and outer edge of the annular sample, thereby
changing the chemical potentials of the electrons at the two edges.
More generally, we shall couple the  system to an
additional, external electromagnetic vector potential, $A$, where
$A=A_{\mbox{\scriptsize tot}}-A_c$ is a small perturbation, and
$\,A_c\,$ is the vector potential corresponding to ${\bf
B}_c$. Our next step is thus to recall the description of a
massless, chiral, relativistic Fermi field circulating along one
component, $\,C_{\!o}$, of the boundary of the (rescaled) system and
coupled to $\,\Ag$, (the restriction of $\,A\,$ to $\,\Gamma_{\!o}
=\BB{R}\times C_{\!o} \subset \partial \Lambda_o$).

It is convenient to
use ``light-cone'' coordinates on the $(1+1)$-dimensional boundary
space-time $\,\Gamma_{\!o}\mini$. We set

\be
u_\pm \::=\: {1 \over \sqrt 2} (\zeta^0 \pm \zeta^1) \:=\: {1
\over \sqrt 2} (v\tau \pm {L \over 2\pi} \eta) \ , 
                        \label{lcc}
\ee

\noi
where $\,\eta \in [0,\,2\pi)\,$ is an angular variable along
$\,C_{\!o}$ (whose length is given by $\,L$), $\,\tau\,$ is (rescaled)
time, and the constant $\,v\,$ physically corresponds to the
propagation speed of {\em charged waves at the edge of the
sample$\mini$}. [The value of the velocity $\,v\,$ does not
matter in the following.]  We write

\be
\Ag \:=\: A_+(u)\,du_+ + A_-(u)\,du_- \ ,
                          \label{Ano1}
\ee

\mwm

\ben
A_\pm(u) \:=\: {1 \over \sqrt 2}\, \Bigl( A_0|_{\Gamma_{\!o}}(\zeta)
\pm A_1|_{\Gamma_{\!o}}(\zeta) \Bigr) |_{\zeta=\zeta(u)} \ , 
\een

\noi
with $\,u:=(u_+,\mini u_-)\,$ and $\,\zeta:=(\zeta^0, \mini
\zeta^1)$. The (1+1)-dimensional d'Alembertian,
$\,\Box:=\partial_0^2-\partial_1^2 = {1 \over v^2} {\partial^2
\over \partial \tau^2} - \left({2\pi \over
L}\right)^2\!\!{\partial^2 \over \partial \eta^2}$, $\,$is given in
``light-cone'' coordinates by

\be
\Box \:=\: 2\,\partial_+ \partial_- \ ,
\ee

\noi
where $\,\partial_\pm := \dd{u_\pm}\mini$.

In $\,1+1\,$ dimensions, a relativistic Fermi field (fermion, for
short) is described by a two-component Dirac (i.e., complex)
spinor, $\,\psi$. We choose the chiral representation of Dirac
matrices:

\be
\gamma^0Ê\:=\: \sigma_1 \ , \hspace{.5cm} \gamma^1 \:=\:
-i\sigma_2 \ , \hah \gamma_5 \::=\: \gamma^0\gamma^1 \:=\:
\sigma_3 \ , 
                          \label{gam}
\ee

\noi
where $\,\sigma_1,\,\sigma_2,\,\sigma_3\,$ are the standard Pauli
matrices. Moreover, we define by $\,\overline{\psi}:=\psi^\ast
\gamma^0$ the conjugate of the Dirac spinor $\,\psi\,$ and identify
its left-/right-handed component with

\be
\psi\tlr(\zeta) \::=\: \halb (1 \mp \gamma_5)\psi(\zeta) \ .
                          \label{Ano2}
\ee

Our aim is to write down an action for a massless, chiral
fermion, $\,\psi\tlr$, coupled to the vector potential $\,\Ag$, and
to calculate an effective gauge field action, $\,\tGLR$, by
integrating out the chiral fermion degrees of freedom located at
the edge of the sample. Naively, one might try to
start with an action of the form

\be
\mbox{``\ }S\tlr(\overline{\psi}\tlr,\, \psi\tlr \hspace{.5mm};\,
\Ag) \:=\: i \int_{\Gamma_{\!o}} \overline{\psi}\tlr(\zeta)\,\dir\,
\psi\tlr(\zeta)\: d^{\,2}\zeta \mbox{\ ''}\ ,
                       \label{Cac}
\ee

\noi
where the {\em Dirac operator$\mini$}, $\,\dir$, is defined by

\be
\dir \::=\: \gamma^\mu \left( \partial_\mu + i {e \over \hbar c}
A_\mu|_{\Gamma_{\!o}}(\zeta) \right) \:=\: \partial\slashi + i
{e \over \hbar c} A\slashi|_{\Gamma_{\!o}}(\zeta) \ . 
\ee

\noi
However, it is {\em not$\,$} possible to compute the effective action
of a massless, chiral fermion coupled to $\,\Ag\,$ by a fermionic
(Berezin) path integral based on the action~(\ref{Cac}). Put
differently, it is {\em not$\,$} possible to define a 
determinant of the Dirac operator $\,\dir\,$ {\em restricted$\,$} to
the subspace of either only left- or only right-handed field modes.
This is because of the simple fact that the Dirac operator
$\,\dir\,$ maps left- to right-handed modes and vice versa, i.e.,
the chirality subspaces are not invariant under the action of
$\,\dir$; see, e.g., Alvarez-Gaum\'e and Ginsparg~(1984). Using
(\ref{Ano1}) and (\ref{Ano2}), we can rewrite the standard action of
a massless, two-component Dirac field $\,\psi$ on $\,\Gamma_{\!o}$
in terms of its components $\,\psi_L\,$ and $\,\psi_R$. We find that

\bea
\lefteqn{i \int_{\Gamma_{\!o}} \overline{\psi}(\zeta)\,\dir\,
\psi(\zeta) \:d^{\,2}\zeta}    \nonumber  \\
& = & i\sqrt{2} \int_{\Gamma_{\!o}} \left[ \psi^\ast_L
\left(\partial_- + i {e \over \hbar c} A_- \right)\psi_L +
\psi^\ast_R \left(\partial_+ + i {e \over \hbar c} A_+ \right)\psi_R
\right]\!(\zeta) \: d^{\,2}\zeta \ . 
                           \label{tdf}
\eea

\noi
[The equation of motion for $\,\psi\tlr\,$ following from (\ref{tdf})
reads $\,\left(\partial_\mp + i {e \over \hbar c} A_\mp \right)
\psi\tlr = 0$. Hence, in $\,1+1\,$ dimensions the left-/right-handed
modes are actually left-/right-moving excitations, provided $\,A_\mp
= 0\,$.] By inspecting the coupling structure of $\,A_\mp\,$ to
$\,\psi\tlr\,$ in Eq.~(\ref{tdf}), we are led to the following
expression for the {\em effective gauge field action of a massless,
chiral (left-/right-moving) relativistic Fermi field
coupled to the external vector potential} $\,\Ag$:

\bea
\ih\mini\Gamma\!_{L/R}(\Ag) & := & \biggl[\, \ln\, \det \dir
\biggr]_{A_\pm=0} + {a \over 4\pi} \ehc\! \int_{\Gamma_{\!o}}
A_+(u)A_-(u)\, d^{\,2}u  \nonumber   \\ 
& = & \ln\, \det \left[ \partial\slashi + i {e \over \hbar c}
A\slashi|_{\Gamma_{\!o}}\, \halb \left( 1 \mp \gamma_5 \right)
\right]   \ , 
                           \label{rcf}
\eea

\noi
see Jackiw (1985), and Jackiw and Rajaraman (1985). In
Eq.~(\ref{rcf}), the  determinant of the Dirac operator $\,\dir\,$
is calculated on the full mode-space of a $(1+1)$-di\-men\-sional,
two-component Dirac field, and its evaluation goes back to
Schwinger (1962). In the second term, $\mini a\,$ is an {\em
arbitrary$\,$} real constant. This term stands for a finite
renormalization ambiguity and is related to the fact that one cannot
invoke \U\ gauge invariance as a guiding principle in the
calculation of {\em chiral$\,$} effective actions (Jackiw and
Rajaraman, 1985; Leutwyler, 1986 and references therein). Chiral
effective actions are {\em anomalous$\mini$}, a fact that we will
exploit shortly! We set $\,a=1$, which is a particularly convenient
choice for the subsequent discussion (see also Jackiw, 1985), and
find that

\be
\GLR \:=\: {1 \over 4\pi} \ehc\! \int_{\Gamma_{\!o}}  \biggl[A_+(u)\,
A_-(u) - 2 A_\mp(u)\, {\partial^2_\pm \over \Box}\mini A_\mp(u)
\biggr]\, d^{\,2}u \ .
                           \label{rcf2}
\ee

Let us define the left-/right-handed current (operator),
$\,j^\mu\tlr$, by

\be
j^\mu\tlr(\zeta) \::=\;\; :\overline{\psi}(\zeta)\,\gamma^\mu \halb
\left( 1\mp \gamma_5 \right)\,\psi(\zeta):  \ ,
                           \label{jLR}
\ee

\noi
where $\;:\hspace{2.5mm} :\;$ indicates normal ordering. Then,
using Eqs.~(\ref{tdf}) and~(\ref{rcf}), we observe that the effective
action $\,\tGLR|_{A_\pm=0}\,$ is the {\em generating functional$\,$}
for time-ordered, connected Green functions of $\,j^\mu\tlr$. Both
currents, $\,j^\mu_L\,$ and $\,j^\mu_R$, independently generate a
{\em chiral \uh\ current algebra$\mini$}. This will be discussed in
more detail in the next subsection. We note that the choice of the
{\em left}- or {\em right$\,$}-handed current, $\,j^\mu_L\,$ or
$\,j^\mu_R$, depends on the physics of the given system, namely, on
the sign of the external magnetic field and on whether the
physical edge currents are carried by {\em electrons$\,$} or {\em
holes$\mini$}.

Next, we exploit the fact that effective actions for {\em chiral$\,$}
fermions are breaking \U\ gauge invariance, i.e., that they are {\em
anomalous}$\,$! Performing a \U\ gauge transformation, $\,A \mapsto
A + \da\chi$, the explicit expression for the chiral effective
action $\,\tGLR\,$ given by (\ref{rcf2}) implies that

\bea
\lefteqn{\hi \Gamma\!_{L/R}([A+\da\chi]|_{\Gamma_{\!o}})\; =\; \GLR} 
\nonumber \\ &  & \pm {1 \over 4\pi} \ehc\! \int_{\Gamma_{\!o}} 
\biggl[A_+(u)\, \partial_-\chi(u) -  A_-(u)\, \partial_+\chi(u)
\biggr]\, d^{\,2}u  \ . 
                           \label{Ano0} 
\eea

\noi
Thus, the {\em chiral anomaly}$\,$ of the effective action produced
by the quantum-mechanical degrees of freedom located at the edge of
the sample takes the form

\bea
\lefteqn{\hspace{-3cm}\pm\mini {1 \over 4\pi}\ehc\!
\int_{\Gamma_{\!o}} \biggl[ A_+(u) \,\partial_-\chi(u) -
A_-(u)\,\partial_+\chi(u) \biggr]\, d^{\,2}u}  \nonumber   \\   
& = & \pm\mini {1\over 4\pi}\ehc\!\int_{\Gamma_{\!o}} \da\chi
\wedge\! A  \ .
                           \label{Ano} 
\eea

Remember that a Landau band gives rise to an algebra of chiral edge
currents at {\em each$\,$} connected component, $\mini C_{\!o}$, of
the boundary, $\,\partial\Omega_o$, of the sample. Hence the total
chiral effective boundary action resulting from a given Landau band
is obtained by simply adding up the contributions of the form
(\ref{rcf2}) for the different connected components of $\,\partial
\Omega_o$. We recall our assumption that, in the bulk, the Fermi
energy, $\,E^F$, lies well {\em between} two Landau bands
(incompressibility) and that there are $\,N=0,\mini 1,\mini
2,\ldots\,$ {\em filled$\,$} Landau bands {\em below$\mini$} the
Fermi energy, each of fixed spin-polarization (i.e., all the
electrons can be treated as scalar fermions). Then, in the scaling
limit, all the quantum-mechanical degrees of freedom localized near
the boundary $\,\partial \Omega_o\,$ of the sample together give
rise to the following boundary contribution, $\,\zeta^{L/R}_\dLo
(A|_\dLo)$, to the {\em total partition function$\mini$},
$\,Z_{\Lambda_o}(A)$, of the system:

\be
\zeta^{L/R}_\dLo(A|_\dLo) \:=\: \prod_{j=0}^N\; \prod_{\Gamma_{\!o}
\subset \partial\Lambda_o} \exp\!\left[ \ih
\Gamma\!_{L/R}(\Ag)\right] \ .
                       \label{cBT}
\ee 

{}From Eqs.~\cfrs{Ano0}{Ano} it follows that the {\em total chiral
anomaly$\,$} of the boundary contribution (\ref{cBT}) is given by

\be
\pm {N \over 4\pi} \ehc\! \int_{\dLo} \da\chi \wedge\! A \ .
                       \label{tAno}
\ee

We are now in a position to describe the basic idea of {\em anomaly
cancellation}$\,$: In Sects.\,\ref{sPE} and~\ref{sGI}, we have seen
that non-relativistic quantum mechanics is \U\ gauge invariant which
means that the total partition function,
$\,Z_{\Lambda_o}(A)$, of the system {\em is} \U\ gauge
invariant! In other words, the total chiral anomaly (\ref{tAno}) due
to the {\em degrees of freedom localized at the boundary$\,$} has to
be {\em cancelled$\,$} by the one of an anomalous term in the total
effective action resulting from the {\em degrees of freedom in the
bulk$\,$} of the system. The term with the required anomaly under
\U\ gauge transformations is the (by now familiar) {\em abelian
Chern-Simons$\,$} term,

\be
\mp {N \over 4\pi} \ehc\! \int_{\Lambda_o} A \wedge\! \da A \ .
                       \label{tAno2}
\ee

\noi
Thus, to leading order in the scale parameter $\mini\theta$,
the total partition function, $\,Z_{\Lambda_o}(A)$, of a
(non-interacting) quantum Hall fluid with
$\,N=0,\mini 1,\mini 2,\ldots\,$ filled Landau bands coupled to a
small perturbing vector potential $\,A\,$ takes the form

\be
Z_{\Lambda_o}(A) \:=\: \zeta^{L/R}_\dLo(A|_\dLo)\: \exp\!\left[ \mp\, i{N
\over 4\pi} \ehc\! \int_{\Lambda_o} A \wedge\! \da A + \mbox{G.I.}
\mini\right] \ ,
                       \label{tAno3}
\ee
 
\noi
where ``G.I.'' stands for possible \U\ gauge-invariant terms.
As we have seen in Eqs.~(\ref{ED11}) and (\ref{ED14}), it is the
Chern-Simons term (\ref{tAno2}) in the total (bulk plus boundary)
effective action which reproduces the basic response equations,
(\ref{Hall}) and (\ref{ED3}), of the quantum Hall effect. In
particular, comparing Eqs.~(\ref{ED11}) and (\ref{tAno3}), we
identify the coefficient of the Chern-Simons term (\ref{tAno2}) with
the Hall conductivity $\,\sH\,$ of the system, i.e., 

\be
\sH \:=\: \pm N\, {e^2 \over h} \ , \hspace{.5cm}
N=0,\mini 1,\mini 2,\ldots \ . 
\ee
 
\noi
These considerations yield a natural description of the physics
underlying the {\em integer quantum Hall effect$\mini$}, provided
the system consists of non-interacting electrons.

One can argue that
the picture of chiral edge excitations given above, and hence the
form of the anomaly, still hold when the system is perturbed by a
small amount of disorder. A {\em chiral$\,$} Luttinger liquid
perturbed by a weak random potential will {\em not$\,$} exhibit
Anderson localization, because there is no interference between left-
and right-moving waves! This is in contrast to what happens in the
bulk where a moderate amount of disorder leads to plenty of localized
states; (we then require that the Fermi energy, $\,E^F$, lie in a
region of localized states, in order to conclude incompressibility of
the fluid). In fact, we recall that Anderson localization in the
bulk is crucial in order for the (integer) quantum Hall effect to be
observable experimentally (Halperin, 1983; Morandi, 1988; Prange,
1990; and references therein). More precisely, the width of a
Hall plateau depends on the amount of disorder in the system -- which
determines the density of localized states -- and, in the
thermodynamic limit, would tend to $\,0\,$ as the strength of the
disorder tends to $\,0\mini$.

Taking electron-electron interactions into account, the
{\em form$\,$} of the ano\-ma\-ly will not change, because it is
universal. However, the value of the Hall conductivity $\,\sH\,$
can and will change.

In the following subsection, we discuss spin-polarized,
two-dimensional, incompressible systems of electrons which do not
necessarily have integer filling factors $\mini\nu\,$, due to {\em
electron-electron interactions$\,$}; see~(\ref{fifa1}). 

However, in contrast to the logic of the present subsection, we
shall start from the {\em universal}$\,$ form, $\,\Ssaw$, given in
Eq.~(\ref{Seff}), that the effective action for the {\em bulk$\,$}
degrees of freedom takes in the scaling limit. [We recall that
expression~(\ref{Seff}) of $\,\Ssaw\,$ takes the spin degrees of
freedom into account and that it does {\em not\,} exclude any
effects of electron-electron interactions that respect $\USo$ gauge
invariance and are compatible with incompressibility of the quantum
fluid!] We will identify those terms in $\,\Ssaw\,$ which exhibit an
anomalous behaviour under \U\ and \SU\ gauge transformations not
vanishing at the boundary $\,\partial \Lambda_o$. The idea of {\em
anomaly cancellation$\,$} then will lead us to the study of the
dynamics of degrees of freedom at the {\em boundary$\,$} of the
system compensating the gauge non-invariance of the bulk terms in
$\Ssaw$. Similarly as above, we shall find charge~- {\em
and$\,$} spin carrying chiral edge currents forming \uh\ - and
\suh\ current (Kac-Moody) algebras, respectively. The
representation theory of these current algebras  then captures
the {\em universal features$\,$} of systems exhibiting a {\em
fractional quantum Hall effect for the electric and for the spin
current$\mini$}. 

In the following subsection, we shall illustrate this logic by
treating the simplest situation of fully spin-polarized (or scalar)
quantum Hall fluids. For a complete implementation of the above
logic, see Subsects.~VI.B and VI.C in Fr\"ohlich and
Studer~(1993b). There properties of ``chiral boundary systems''
($\!$\uh\ current algebras) associated with fully spin-polarized
quantum Hall fluids are discussed in detail, and the additional
features arising for systems exhibiting spin~- ($\!$\suh\ current
algebra) and possibly internal symmetries ($\mmini$\gh\ current
algebra associated to some compact Lie group $\,G\mini$) are
explained.


\subsection{Edge Excitations in Spin-Polarized Quantum Hall Fluids}
\label{ssEE}

In this subsection, we consider {\em interacting$\mini$},
{spin-polarized, two-dimensional, incompressible quan\-tum Hall
fluids$\mini$}, where ``spin-polarized'' means that the spin degrees
of freedom are ``frozen''. Moreover, we shall neglect the magnetic
moments of the electrons. [For a treatment of spin-polarized Hall
fluids including the effects of the magnetic moments of the
electrons, see Subsect.~VI.C in Fr\"ohlich and Studer~(1933b).] For
such systems, the {\em universal$\,$} form of the scaling limit
$\,\Ssa\,$ of the effective action is thus obtained by discarding
all terms depending on the \SU\ gauge fields $\,w\,$ or
$\,\tilde{w}\,$ in expression (\ref{Seff}). We find that

\be
-{1 \over \hbar}\mini\Ssa \:=\:  \iLo
\jc(\xi)\,\at{\mu}(\xi)\,d^{\,3}\xi + {\sigma \over 4\pi}\iLo \at{}
\wedge\! \da \at{}\, +\: \BT(\at{}|_\dLo) \ ,
                          \label{Ssa}
\ee

\noi
where $\,\BT(\at{}|_\dLo)\,$ stands for boundary terms only
depending on the restriction of the perturbing, external vector
potential $\,\at{}\,$ to the boundary of the system. [Note that we
have returned to working in ``mathematical units''; see,
e.g.,~(\ref{a}). Moreover, we consider a euclidean background metric,
$\,\gamma_{ij}=\delta_{ij}$; see (\ref{Pa5A}).] So far, the
coefficient $\,\sigma\,$ of the Chern-Simons term on the r.h.s.\ of
Eq.~(\ref{Ssa}) can be an arbitrary real constant. 

The particular situation where $\,\sigma\,$ takes {\em integer$\,$}
values has been identified in the previous subsection as describing
the {\em integer quantum Hall effect$\mini$}. The 
main purpose of this subsection is to understand the basic aspects
of the more general situation of the {\em fractional$\,$}
quantization of the values of the constant $\,\sigma\,$ arising
in systems where electron-electron interactions cannot be neglected.
We recall that, by Eq.~(\ref{sH}), $\,\sigma\,$ determines the value
of the Hall conductivity, $R_H^{-1} = \sigma \mini{e^2 \over h}$, which
encodes the linear response properties of incompressible
quantum Hall fluids, provided we neglect the spin degrees of
freedom; see Eqs.~(\ref{LRE1}) and (\ref{LRE2}). In this
subsection, we develop a picture of universal aspects of the
{\em fractional quantum Hall effect$\,$} in spin-polarized,
two-dimensional electronic systems by focusing on the physics at
their edges.

We follow  the ideas described  in Subsect.~\ref{ssIQ}: In a
first step, we make use of  {\em anomaly
cancellation} in the direction opposite to the one explained in
Subsect.~\ref{ssIQ}, i.e., we start from the bulk terms. As
explained in Fr\"ohlich and Studer~(1992b, 1993a), $\,\Ssa\,$ must
be invariant under \U\ gauge transformations

\be
\at{} \:\mapsto\: {}^\chi\at{} \::=\: \at{} + \da \chi\ ,
                      \label{Ugt}
\ee

\noi
in spite of the fact that $\,\at{}\,$ is the vector potential
of a {\em perturbation\mini} of the electromagnetic field (i.e.,
$\,\at{}=a-a_c\,$ is the {\em difference$\,$} of two connection
1-forms)! For a trivial \U\ bundle, always realized for our
choice of space-time domains $\,\BB{R} \times \Omega_o$, the
space of connections (vector potentials) is a real {\em
vector space$\mini$}. In particular, any gauge transformation of a
sum of connections can be rewritten as the result of gauge
transforming each summand in the sum {\em separately$\mini$}. This is
in contrast to the situation encountered for 
non-abelian gauge fields. Performing a gauge transformation
(\ref{Ugt}) on $\,\at{}$ with $\,\chi\,$ non-vanishing at $\,\dLo$,
we find from (\ref{Ssa}) that

\bea
{1 \over \hbar}\,S^\ast_{\Lambda_o}(\tilde{a}+\da \chi) & = & {1
\over \hbar}\,\Ssa + \idLo (\ast j_c)\, \chi    \nonumber  \\
& &\hspace{-1.5cm} \mbox{} + {\sigma \over 4\pi}\idLo \da \chi
\wedge\! \at{} -\: \BT([\at{}+\da \chi]|_\dLo)\, +
\:\BT(\at{}|_\dLo) \ ,
                          \label{Ssadc}
\eea

\noi
where $\,\ast j_c = \halb \sum_{\mu,\,\nu}\,
\varepsilon_{\mu\nu\sigma}\,j_c^\sigma\,(\xi) \,\da\xi^\mu
\wedge\! \da\xi^\nu\,$ is the (Hodge) dual of the  current $\,j_c$. 

Note that $\,S^\ast_{\Lambda_o} (\tilde{a} + \da \chi)\,$ would be
equal to $\,\Ssa$, and hence we could set $\,\BT(\at{}|_\dLo) = 0\,$
(up to gauge-invariant terms at $\,\dLo$), if

\be
\ast\! j_c|_\dLo \:=\: {\sigma \over 4\pi}\,\da \at{}|_\dLo \ ,
                           \label{jca}
\ee

\noi
since $\,\idLo \da \chi \wedge\! \at{} = -\idLo \chi\,\da\at{}$.
However, $\,j_c\,$ is an arbitrary current in the Hall
fluid when $\,a=a_c\,$ (i.e., $\,\at{}=0$), and $\,\at{}\,$ is the
potential of a small but {\em arbitrary$\mini$}, external perturbation.
Therefore the relation (\ref{jca}) cannot hold in general.

Experimentally, for a Hall fluid in a heterostructure or MOSFET, the
boundary $\,\dLo$ of the sample is such that there is no leakage
of electric charge through $\,\dLo\,$, which means that the normal
component of $\,\jc\,$ at $\,\dLo\,$ has to vanish, or equivalently,

\be
\ast\! j_c|_\dLo \:=\: 0 \ .
\ee

\noi
In this case, the second term on the r.h.s.\ of (\ref{Ssadc})
{\em vanishes$\mini$}, and requiring the total effective action
$\,\Ssa\,$ to be \U\ gauge-invariant implies the following
functional equation for the boundary terms:

\be
\BT([\at{}+\da \chi]|_\dLo) - \:\BT(\at{}|_\dLo) \:=\:  {\sigma \over
4\pi} \idLo \da \chi \wedge\! \at{} \ , 
                         \label{BT}
\ee

\noi
which must hold for arbitrary $\,\at{}\,$ and $\,\chi$. 

{}From the discussion in the previous subsection we can immediately
infer the solution to Eq.~(\ref{BT}). Introducing ``light-cone''
coordinates on each connected component, $\,\Gamma_{\!o}$, of the
$\,(1+1)$-dimensional boundary space-time $\,\dLo\,$ (see
Eq.~(\ref{lcc})), we have, as in Eq.~(\ref{Ano1}), that

\bea
\at{}|_{\Gamma_{\!o}} & = &{e \over \hbar c} A_+(u)\,du_+ + {e
\over \hbar c} A_-(u)\,du_-   \nonumber  \\
& =: &  \alp(u)\,du_+ + \alm(u)\,du_-  \ .
                            \label{ano1}
\eea

\noi
{}From Eqs.~(\ref{Ano0}) and (\ref{Ano}) it follows that the
solution to Eq.~(\ref{BT}) is given by

\be
\BT(\at{}|_\dLo) \:=\: \sigma_L\! \sum_{\Gamma_{\!o} \subset \partial
\Lambda_o} \frac{1}{\hbar}\Gamma_L (\alpha)\, +\, \sigma_R\!
\sum_{\Gamma_{\!o} \subset \partial \Lambda_o} \frac{1}{\hbar}
\Gamma_R (\alpha) \, +\:\GI(\alpha) \ ,
                            \label{ano2}
\ee

\vspace{5mm}
\mwim

\be
\sigma \:=\: \sigma_L - \sigma_R \ ,
                            \label{ano21}
\ee

\noi
where $\Gamma_{L/R} (\alpha)$ is as in Eq.~(\ref{rcf2}), $\,\sigma_L\,$
and $\,\sigma_R\,$ are non-negative constants, and ``$\GI$''
stands for manifestly gauge-invariant terms only depending on
$\tilde{a}|_{\partial\Lambda_o}$. We recall from Eq.~(\ref{Ano0}) that
the  
contributions to the total anomaly of the two terms $\Gamma_L(\alpha)$
and $\Gamma_R(\alpha)$ are of opposite sign which explains the sign
in~(\ref{ano21}). We do not need to discuss the terms in ``$\GI$''
any further and hence omit them in the following discussion. 

In Subsect.~\ref{ssIQ}, we have seen that, for $\,\sigma_{L/R}=N\,$
a (positive) {\em integer}, there is a straightforward
interpretation of the boundary terms in Eq.~(\ref{ano2}) in terms of
$\,N\,$ bands of {\em non-interacting$\mini$}, chiral
(left-/right-moving) fermions propagating along the different
connected components of the sample boundary $\,\partial\Omega_o$. In
order to understand the more general, {\em fractional$\,$}
quantization of the values of the constant $\,\sigma$, we have to
generalize this physical picture. For this purpose, we use
bosonization techniques always available in two space-time
dimensions; see Sect.\,\ref{sSLaC}. In this subsection, we derive an
expression for $\Gamma_{L/R} (\alpha)$ (see~(\ref{rcf})) in terms of one chiral
Bose field. In a second step, this bosonic expression would have to
be generalized to several bands of excitations
at the boundary. For details on this second step, see Sect.~VI.B in
Fr\"ohlich and Studer~(1993b).

In the following we assume, for simplicity, that the topology
of the sample $\,\Omega_o\,$ is that of a disk, i.e., the boundary
$\,\partial\Omega_o\,$ consists of but {\em one$\,$} connected
component $\,C_{\!o}$.

Let us first suppose that the external gauge field $\alpha
:=\at{}|_\dLo$ is set to $\,0$. In Eq.~(\ref{jLR}) we have
introduced the currents 

\be
j^\mu\tlr(\zeta) \:=\:\halb\, \bigl[\, j^\mu(\zeta) \mp j_5^\mu(\zeta)
\bigr]\ ,
                            \label{jj5}
\ee

\noi
of a massless Dirac spinor $\,\psi$. Recalling the equation of
motion satisfied by the left-/right-handed component of $\,\psi\,$
(see (\ref{Ano2}), (\ref{gam}) and the remark after (\ref{tdf})) we
see that $\,j^\mu\,$ and $\,j_5^\mu\,$ are conserved currents, i.e.,

\be
\partial_\mu j^\mu(\zeta) \:=\: 0 \:=\: \partial_\mu j_5^\mu(\zeta) \ .
                             \label{cc}
\ee

\noi
The general solution to (\ref{cc}) is

\be
j^\mu(\zeta) \:=\: \sqrt{2}\,\ops{\mu\nu}\partial_\nu \phi(\zeta) \ ,
\hah j_5^\mu(\zeta)\:=\: \sqrt{2}\,\ops{\mu\nu}\partial_\nu \phi_5(\zeta) \ , 
\ee

\noi
where $\,\phi\,$ and $\,\phi_5\,$ are scalar fields, and $\,\ops{01}
= -\ops{10}=1$. However, in $\,1+1\,$ dimensions, it follows
form (\ref{jLR}) and (\ref{gam}) that $\,j_5^0 = j^1\,$ and
$\,j_5^1=-j^0$. Thus, $\,j_5^\mu = -\sqrt{2}\, \partial^\mu \phi$,
and (\ref{cc}) implies that

\be
\partial_\mu\partial^\mu \phi(\zeta) = \Box\phi(\zeta) \:=\: 0 \ ,
                           \label{ff}
\ee

\noi
i.e., $\,\phi\,$ is a {\em free, massless, relativistic Bose
field$\mini$}. Any solution to Eq.~(\ref{ff}) has the form (see (\ref{lcc}))

\be
\phi(\zeta(u))\:=\: \phi_L(u_+) + \phi_R(u_-) \ .
                           \label{pLR}
\ee

\noi
Moreover, since $\,j^\mu\,$ and $\,j^\mu_5\,$ are currents
propagating along the boundary $\,\partial\Omega_o$, $\,\phi\,$ has
to satisfy the periodicity conditions

\be
\partial_\pm \phi(\zeta^0,\,\zeta^1+L) \:=\:
\partial_\pm \phi(\zeta^0,\,\zeta^1) \ . 
\ee

\noi
In terms of the chiral components $\phi_{L}$ and $\phi_{R}$ of the
Bose field $\phi$, we can define the {\em chiral currents$\mini$}
$J_{L}$ and $J_{R}\mini$,

\bea
{2\pi\over\ell} J_L(u_+) & := & \partial_+\phi_L(u_+)\: =\:
j^0_L(\zeta(u)) \ , \hah    \nonumber  \\ 
{2\pi\over\ell} J_R(u_-)  & := & \!-\partial_-\phi_R(u_-)\: =\:
j^0_R(\zeta(u)) \ ,
                           \label{JLR} 
\eea

\noi
where $\,\ell={L\over \sqrt{2}}$. Clearly $\,
\partial_\mp J_{L/R}=0\mini$. We emphasize that
Eqs.~(\ref{jj5})--(\ref{JLR}) hold at the level of {\em
quantized$\,$} fields. They are at the origin of {\em abelian
bosonization$\,$} in two space-time dimensions. 

The currents $\,J_L\,$ and $\,J_R\,$ both generate a chiral
\uh\ current algebra as follows: We can decompose the current
$\,J_L\,$ (and similarly $\,J_R\,$) into its Fourier modes:

\be
J_L(u_+)\: =\: {\ell\over 2\pi}\mini\partial_+\phi_L(u_+)\: =\:
{1\over\kappa}\,p + {1\over\sqrt{\kappa}} \sum_{n\neq 0}
\alpha_n\,e^{-2\pi i\mini n\mini u_+/\ell} \ ,
                              \label{Jff}
\ee

\noi
where $\,\kappa\,$ is a positive normalization constant whose
meaning will become apparent shortly. Integrating
relation~(\ref{JLR}) we find that

\be
\phi_L(u_+) \:=\: q + {2\pi\over\ell\kappa}\,p\,u_+ + {i \over
\sqrt{\kappa}} \sum_{n\neq 0} {1\over n}\,\alpha_n\,e^{-2\pi
i\mini n\mini u_+/\ell} \ ,
                               \label{ff1}
\ee

\noi
where  $\,q\,$ is a real integration constant. Since $\,\phi_L\,$
is a real quantum field, the operator $\,\alpha_{-n}\,$ is the
adjoint of $\,\alpha_n$, i.e., $\,\alpha_{-n} = {(\alpha_n)}^{\dag}
,\ n\in\BB{Z}\,\backslash\{0\}$. Moreover, the Fourier coefficients
in (\ref{ff1}) are subject to the following commutation relations

\be
[q,\,p]\:=\:i \ ,  \hah [\alpha_m,\,\alpha_n]\:=\:m\,\delta_{m,-n}\ ,
\hspace{.25cm} m,\,n\in \BB{Z}\,\backslash\{0\} \ .
                             \label{uKM}
\ee

\noi
Eq.~(\ref{ff1}) and the relations~(\ref{uKM}) define 
\uh\ {\em current (Kac-Moody) algebra} (at level~$\kappa$); see,
e.g., Goddard and Olive (1986), Buchholz, Mack, and Todorov (1990), 
Ginsparg (1990), and also Frenkel (1981) and Kac (1983). The vacuum
(charge-$0$) sector of this quantum field theory is the Fock
space built from the vacuum state, $\,|0\Er$, which is defined by

\be
\alpha_n\mini |0\Er \:=\: 0\ , \hspace{.25cm} n>0 \ , \hah p\mini |0\Er
\:=\: 0 \ ,
\ee

\noi
by applying polynomials in the creation operators $\,\alpha_{-n}$,
$\, n=1,2,\ldots$, to $\,|0\Er$.

These constructions are well-known in string~- and conformal
field theory. We describe them below. First, however, we
derive an expression, in terms of Bose fields, for the boundary term
``$\BT$'', given in Eq.~(\ref{ano2}).

In the path integral formulation, a free, massless Bose field
$\,\phi$, propagating along the boundary $\,\partial\Omega_o\,$ of
the sample, is described by a gaussian action

\be
\hi\mini S(\phi) \::=\: {\kappa \over 4\pi} \idLo
\dmm\phi(u)\,\dpp\phi(u) \du \ ,
                             \label{Gauss}
\ee 

\noi
where $\du:=\da u_-\wedge\! \da u_+\,$ is the (oriented) space-time
measure on $\,\dLo$, and $\,\kappa\,$ is the same normalization
constant that appeared already in Eqs.~(\ref{Jff}) and (\ref{ff1}). If
we consider but one chiral component of the 
field $\,\phi\,$ we have to supplement the action (\ref{Gauss}) by a
chirality constraint

\be
\dmm\phi(u)\:=\:0 \ , \hoh \dpp\phi(u)\:=\:0 \ .
                             \label{cco}
\ee

Next, we couple $\,\phi\,$ to an external gauge potential
$\,\alpha = \at{}|_\dLo$. We present a formal argument allowing us to
identify the correct way of coupling $\,\phi\,$ to $\,\alpha$ and
providing a path integral derivation of the fermion-boson
equivalence in $\,1+1\,$ dimensions. The state sum (a divergent
constant) for a free, chiral (left-moving), massless Bose field reads

\be
Z_L\::=\:\int \D{}\phi \,\exp\!\left[\ih S(\phi)\right]
\,\delta\!\left(\dmm\phi\right)  \ , 
                             \label{ff2}  
\ee

\noi
where $\,S(\phi)\,$ is given by (\ref{Gauss}). Since the field
$\,\phi\,$ is an angle variable, we may shift it according to
$\,\phi \mapsto \mbox{}^\chi\phi := \phi + {1 \over
\kappa}\mini\cdL\mini$. Using the fact that the integration
measure is gauge invariant, i.e.,
$\,\D{}\,\mbox{}^\chi\phi=\D{}\phi$, we find, after partial
integration, that

\bea
Z_L & = & \exp\!\left[{i \over \hbar\mini\kappa^2}\, S(\chi)\right] 
\nonumber  \\
& & \hspace{-2.2cm}\mbox{} \times \int \D{}\phi \,\exp\!\left[\ih
S(\phi) + {i \over 2\pi\hbar} \idLo\dpp\phi(u)\,\dmm\chi(u) \du\right]
\,\delta\!\left(\dmm\phi + {1\over\kappa}\, \dmm\cdL \right) \,.
                           \label{ano3}
\eea

\noi
Let us choose $\,\chi\,$ such that $\,\dmm\cdL = -Q\,\alm$, where 
$\,Q\,$ is some constant. Then Eq.~(\ref{ano3})
and expression~(\ref{rcf2}) for the action functional $\,
\Gamma\!_{L} (\alpha)$ imply that the following identity holds

\be
\exp\!\left[{i\mini Q^2\over\hbar\mini\kappa}\,\Gamma\!_{L}(\alpha)
\right] \:=\: {1 \over Z_L} \int \D{}\phi \,\exp\!\left[\ih
\SWZ(\phi\hspace{.4mm}; \mini \alpha) \right]
\,\delta\!\left(\dmm\phi - {Q\over\kappa}\,\alm \right) \ , 
                           \label{ano4}
\ee

\mwm

\bea
\hi \SWZ(\phi\hspace{.4mm};\mini\alpha) & := & {\kappa \over 4\pi}
\idLo \dmm\phi(u) \,\dpp\phi(u) \du   \nonumber   \\
& - & {Q \over 2\pi} \idLo \Bigl[\dpp\phi(u) \,\alm(u) -
\left(\dmm\phi - {Q\over\kappa}\,\alm \right)\hspace{-1.1mm}(u)
\,\alp(u) \Bigr]\du  \nonumber  \\ 
& + & {Q^2 \over 4\pi\,\kappa} \idLo \alm(u)\,\alp(u)\du \ ,
                           \label{SWZ}
\eea

\noi
with $\,\alp\,$ as in~(\ref{ano1}). We note that the
$\,\alp$-term in the second line of~(\ref{SWZ}) vanishes when 
the constraint in~(\ref{ano4}) is imposed. It has been added for symmetry
reasons discussed below. Moreover, the third term on
the r.h.s.\ of~(\ref{SWZ}) is independent of $\,\phi$. It has been
added in order for the l.h.s. of~(\ref{ano4}) to coincide with the
expression given in~(\ref{rcf2}).

The field theory with action~(\ref{SWZ}) is known as the {\em gauged,
abelian Wess\--Zumino\--Novikov\--Witten (WZNW) model$\,$}; see,
e.g., Gawedzki~(1990) \ (and 
Floreanini and Jackiw~(1987), Sonnenschein~(1988), and
Harada~(1990)). The {\em chirality constraint$\,$}
in Eq.~(\ref{ano4}),

\be
\dmm\phi(u) - {Q\over\kappa}\, \alm(u) \:=\: 0\ ,
                            \label{cco1}
\ee

\noi
is invariant under gauge transformations

\bea
\phi(u) & \mapsto & \mbox{}^\chi\phi(u) \::=\: \phi(u) + {Q \over
\kappa}\,\cdL(u) \ ,  \nonumber  \\
\alpha & \mapsto & {}^\chi\alpha \::=\: \alpha + \da\cdL \ .
                            \label{gtp}
\eea

\noi
By Eqs.~(\ref{Ano0}) and (\ref{ano4}), the theory specified by
the r.h.s.\ of~(\ref{ano4}), with an action as given in
Eq.~(\ref{SWZ}), gives rise, under a gauge
transformation~(\ref{gtp}), to the anomaly

\be
{\sigma \over 4\pi} \idLo \da\chi \wedge\! \at{}\ , \hwh \sigma \:=\:
{Q^2 \over \kappa} \ . 
                            \label{ano5}
\ee

\noi
It is clear from (\ref{SWZ}) that, physically, $\,Q\,$ specifies the
charge of the left-moving component of the Bose field $\,\phi\,$ in
units where the elementary charge $- e = 1$; see (\ref{ano1})
and~(\ref{JLR}). If $\,{Q^2\over \kappa}\,$ equals $\,1$, then
Eq.~(\ref{ano4}) can be used to prove the equivalence of the theory
of one chiral fermion, given in~(\ref{rcf}), to the theory of one
chiral boson, given in~(\ref{SWZ}).

Replacing $L$ by $R$ on the l.h.s.\ of (\ref{ano4})
corresponds to interchanging $+$ and $-$, and replacing $Q$ by
$-Q$, and $\kappa$ by $-\kappa$ on the r.h.s.\ of (\ref{ano4}).
[Interchanging $+$ and $-$, the measure $\du\,$ goes over
into $\,-\!\du$. In order for this symmetry property to become
evident, we have included, on the r.h.s.\ of (\ref{SWZ}), the
$\,\alp$-term that vanishes upon imposing the constraint
(\ref{cco1}).] Clearly the resulting anomaly has the opposite
sign of the one given in (\ref{ano5}). Physically, this symmetry
property of (\ref{ano4}) corresponds to replacing {\em
electrons$\,$} $(L)\,$ by {\em holes$\,$} $(R)\,$ as the elementary
charge carriers for the edge currents at the boundary of the
sample. It reflects the fact that, for a given external magnetic
field $\,{\bf B}_c$, electrons and holes circulate in opposite
directions. {}From this it follows that we can continue our discussion
by only considering the left-moving excitations along the boundary
$\,\partial\Omega_o$, the corresponding equations for the
right-moving ones following by applying this symmetry.

\medskip
There is yet another way of deriving expression~(\ref{SWZ}) that
should be mentioned: We start from the Chern-Simons term
$\,\exp\!\left[-{i\,\sigma \over 4\pi} \iLo \at{} \wedge\! \da
\at{}\right]\mini$ and perform a gauge transformation, $\,\at{}
\mapsto \mbox{}^\varphi\at{} := \at{} + \da \varphi$. Then,
integrating the gauge-transformed Chern-Simons functional,
$\,\exp\!\left[-{i\,\sigma \over 4\pi} \iLo\! \mbox{}^\varphi\at{}
\wedge\! \da\mbox{}^\varphi \at{}\right]\mini$, over all gauge
transformations $\,\varphi\,$ satisfying the boundary condition,
$\,\dmm\varphi - \alm = 0$, where $\,\alpha = \at{}|_\dLo$, we
reproduce the same expressions as in Eqs.~(\ref{ano4}) and
(\ref{SWZ}) by setting $\,\varphi={\kappa \over Q}\phi$. This
procedure can also be applied in the non-abelian situation; see
Fr\"ohlich and Studer~(1993b).

\medskip
So far, the constants $\kappa$ and $Q$ are arbitrary real numbers.
Next, we sketch how to find natural constraints on these two
constants by making use of the representation theory of \uh\ current
algebra and some basic requirements on the spectrum of physical
excitations in a Hall fluid. As a consequence of these constraints
we find that $\,\sigma\,$ has to take {\em rational$\,$} values!
 
The basic objects in the representation theory of chiral \uh\ current
algebra are the chiral vertex (Weyl) operators which play the role of
Clebsch-Gordan operators. We recall some basic properties of these
operators. It is convenient (see (\ref{Jff}) and (\ref{ff1})) to
introduce the coordinates 

\be
z \::=\: e^{\mini 2\pi i\mini u_+/\ell} \ , \hspace{1cm} \bar{z}
\::=\: e^{\mini 2\pi i\mini u_-/\ell} \ .
                           \label{zco}
\ee

In terms of the left-moving Bose field $\,\phi_L\,$ given
in~(\ref{ff1}), we define the {\em chiral vertex operators}

\be
V_n(z) \::=\; \; :e^{i\mini n\mini \phi_L(z)}:  \ , \hwh n \in \BB{R}
\ ,
                            \label{VO}
\ee

\noi
where $\;:\hspace{2.5mm} :\;$ denotes normal ordering (moving
all $\,\alpha_n\,$ with $\,n>0\,$ to the right of $\,\alpha_m\,$ with
$\,m<0$, and $\,p\,$ to the right of $\,q$). Applying the
commutation relations~(\ref{uKM}), one obtains the basic exchange
(or Weyl) relations of chiral vertex operators,

\be
V_n(z)\,V_m(w) \:=\: e^{\pm i\pi {n\mini m \over \kappa}}\,
V_m(w)\,V_n(z) \ , \hspace{.5cm} z\neq w \ , 
                            \label{sta}
\ee

\noi
where the sign, $+$ or $-$, depends on the sign of $\,(\arg z-\arg
w)\,$ relative to a fixed choice of a reference direction. 

In the presence of an external gauge field $\,\alpha\,$, the {\em
charge operator$\,$}  $\,\cal{Q}$ is defined by

\be
{\cal Q} \::=\: \oint_{|z|\mini=\mini{\rm const.}} \left[J_L(z) -
{Q\over \kappa}\,\alpha_z(z) \right]\, {dz \over 2\pi i\mini z} \ ,
                              \label{cha1} 
\ee

\noi
where $\,\alpha_z = \alp \ddd{u_+}{z}$. This operator
is manifestly gauge invariant. Recalling Eqs.~(\ref{Jff}) and
(\ref{uKM}), we find that

\be
[{\cal Q},\,V_n(z)] \:=\: {n\over\kappa}\,V_n(z) \ .
                              \label{cha}
\ee 

Thus, the chiral vertex operator $\,V_n(z)\,$ creates a left-moving
excitation of charge $\,q={n\over\kappa}\,$. 
It follows from~(\ref{cha1}) that, in order to change the charge of
the system by an amount $\,{n\over\kappa}$, the magnetic flux penetrating the
system has to be changed by an amount $\,n\,$ (in units where 
$\,-{hc\over e} = 1$).

Next, we attempt to identify those chiral vertex
operators~(\ref{VO}) that when applied to the ground state create
states with properties that are {\em consistent$\,$} with two basic assumptions
about the physics of two-dimensional, incompressible quantum Hall
fluids whose elementary charge carriers are (spin-polarized) electrons.
\vspace*{4mm}

\begin{list}
{\bf (a\arabic{lico})}{\usecounter{lico}\setlength{\labelsep}{.4cm}
\setlength{\labelwidth}{1cm}\setlength{\leftmargin}{1.4cm}
\setlength{\rightmargin}{.7cm}}
\setcounter{lico}{0}

\item The system describes finite-energy excitations with the quantum numbers
of a {\em (scalar) electron$\mini$}, i.e., with {\em charge\mini}
$1$  and obeying {\em Fermi statistics$\mini$}.

\item The quantum-mechanical state vector describing an {\em
arbitrary\mini} physical state of the system is {\em single-valued\,}
in the position coordinates of all those (finite-energy) excitations
that are composed of {\em electrons\mini} and/or {\em holes}.
\end{list}

Exploiting  formulae~\cfrs{sta}{cha} in the light of the two
assumptions {\bf (a1)} and {\bf (a2)}, we find that $\,Q=\pm 1\,$ and
$\,\kappa=2\mini p+1$, 
with $\,p=0,1,2,\ldots\ $; see Fr\"ohlich and Studer (1993b). Hence,
by~\cfr{ano5}, we find that $\,\sigma={1\over{2\mini p+1}}$, and the
{\em Hall conductivity} $R_H^{-1}=\sigma\mini {e^2\over h}\,$ is a {\em
rational$\,$} multiple of $\,{e^2\over h}$. Our analysis thus 
provides a description of the celebrated Laughlin fluids (Laughlin,
1983a, 1983b) in an ``edge-current picture''. In particular, in the
simplest example of $\,p=0$, we are describing an integer quantum
Hall fluid with $\,\sigma=\pm 1$, and the discussion above coincides
with the one given in Subsect.\,\ref{ssIQ}.

In order to describe quantum Hall fluids with more general, rational
values of the Hall conductivity (see Fig.~4.1), the ideas presented
above must be  generalized 
to ``multi-channel chiral boundary systems''; see Fr\"ohlich and
Studer~(1993b). This generalization leads naturally to the notion of
a ``{\em quantum Hall lattice\mini}''. In Subsects.\,\ref{ssDic}
and~\ref{ssBI}, we review the defining
properties of a quantum Hall lattice in a ``bulk picture'' of
incompressible quantum Hall fluids. This complementary picture has
been shown to be equivalent to the above ``edge-current picture''
(Fr\"ohlich and Thiran, 1994). Equipped with the notion of a quantum
Hall lattice, we shall describe, in Subsects.\,\ref{ssBI}--\ref{ssDis},
the main classification results of incompressible quantum Hall
fluids that have been derived in Fr\"ohlich, Studer, and
Thiran~(1995), and Fr\"ohlich, Kerler {\em et al.}~(1995). For
additional ideas worth thinking about, see Cappelli {\em et al.}~(1995). 

Our results can be interpreted as a ``{\em gap labelling
theorem$\,$}'': Assuming {\em in\-com\-press\-i\-bi\-li\-ty$\,$} of
a two-dimensional quantum Hall fluid, we prove that its Hall
conductivity $R_H^{-1}$ has to belong to a certain set of  {\em
  rational$\,$} multiples  of
$\,{e^2\over h}$. Conversely, if $R_H^{-1}$ is {\em not$\,$} a
rational multiple of $\,{e^2\over h}$, the corresponding
two-dimensional electron system {\em cannot$\,$} be
incompressible, i.e., there {\em cannot$\,$} be a positive energy
gap $\delta$ above the ground-state energy in the spectrum of the
many-body Hamiltonian of this system; (see also Subsect.\,\ref{ssQH}).



\section{Classification of Incompressible Quantum Hall Fluids}
\label{sClQHFs}

In this section, we review  a classification of {\em incompressible
(dissipation-free) quantum Hall fluids\mini} (or, for short, {\em
QH fluids\mini}) in terms of integral lattices and arithmetical
invariants thereof. Our classification enables us to
characterize the plateau values of the Hall conductivity $\sH$
in the interval $\,(0,1]\,$ (in units where $\,e^2/h=1$)
corresponding to ``stable'' incompressible quantum Hall fluids.
Theoretical results are carefully compared to experimental data, and
various predictions and experimental implications of our
theory are discussed.


\subsection{QH Fluids and QH Lattices: Basic Concepts}
\label{ssQHFQHL}

Our analysis of the fractional quantum Hall effect is based on a
precise notion of incompressible (dissipation-free) quantum Hall
fluids\,--\,see assumptions {\bf (A1)} through {\bf (A4)}
below\,--\,from which  their main features can be derived; see
Fr\"ohlich and Studer (1992b, 1992c, 1993b), Fr\"ohlich and Thiran
(1994), Fr\"ohlich, Studer, and Thiran (1995), and Fr\"ohlich,
Kerler {\em et al.}~(1995). Our mathematical characterization, in
terms of ``QH lattices'', of (universality classes of) QH fluids
enables us to enumerate and classify QH fluids. In this subsection,
we review the defining properties of QH lattices.

We recall from Subsect.~\ref{ssQH} that the fractional QH effect is
observed in two-dimensional gases of electrons at temperatures
$\mini T \approx 0\,K\mini$ subject to a very nearly constant
magnetic field, ${\bf B}_{c}$, transversal to the plane of the
system. In Subsect.~\ref{ssSL}, we have characterized a QH fluid by
the property that connected Green functions of the electric charge -
and current densities have cluster decomposition properties stronger
than those encountered in a metal or in a system where the electric
charge - and
current densities couple the ground state of the system to a Goldstone
boson (as in a London superconductor). One can then show (see
Subsect.~\ref{ssLR}) that the {\em longitudinal resistance}, $\RL$,
of a QH fluid {\em vanishes}, which can also be used as a {\em
definition} of an (incompressible) QH fluid.

The longitudinal resistance $\RL$ is a complicated function of the
filling factor $\nu$ (see~(\ref{fifa1})), and it is a difficult
problem of many-body theory to predict where $\RL$ vanishes as a
function of $\nu$. We do not solve this problem in this section.
Instead, we show that {\em if\,} $\RL$ {\em vanishes\mini} then the
{\em Hall conductivity} $\RH^{-1}$ necessarily belongs to a
certain set of {\em rational multiples} of $e^2/h$. Furthermore, given
such a value of $R_H^{-1}$, we
can determine the possible types of quasi-particles, i.e., the
different ``Laughlin vortices'' of the system, their electric
charges and their statistical phases.

Next, we describe the basic assumptions and physical principles
underlying our analysis of QH fluids.
\vspace*{4mm}

\begin{list}
{\bf (A\arabic{lico})}{\usecounter{lico}\setlength{\labelsep}{.4cm}
\setlength{\labelwidth}{1cm}\setlength{\leftmargin}{1.4cm}
\setlength{\rightmargin}{.7cm}}
\setcounter{lico}{0}

\item The temperature $T$ of the system is close to $0\,K$. For
an (incompressible) QH fluid at $T=0\,K$, the {\em total electric
charge} is a good quantum number to label physical states of the
system describing finite-energy excitations above the ground
state. The charge of the ground state of the system is normalized to
be zero; see Sect.~5, Eqs.~(5.11) - (5.12), and Fr\"ohlich,
G\"otschmann and Marchetti (1995a).

\item In the regime of very short wave vectors and low frequencies,
the {\em scaling limit}, the total electric current density is the
sum of $\,N=1,2,3,\ldots$ {\em separately conserved\,} \ui\ current
densities, describing electron and/or hole transport in $N$ separate
``channels'' distinguished by conserved quantum numbers. [For a
finite, but macroscopic sample, this assumption implies that there
are $N$ approximately separately conserved chiral \ui\ edge currents
circulating
around the boundary of the system.] In our analysis, we regard $N$ as
a free parameter.\,--\,Physically, $N$ turns out to depend on the
filling factor $\nu$ and other parameters characterizing the system.

\item In the units chosen in this section where $\,h=-e=1$, the
electric charge of an {\em electron/hole} is $+1/-1$. Any local
finite-energy
excitation (quasi-particle) above the ground state of the system
with {\em integer\mini} total electric charge $q_{el}$ satisfies {\em
Fermi-Dirac statistics\mini} if  $q_{el}$ is {\em odd}, and {\em
Bose-Einstein statistics\mini} if $q_{el}$ is {\em even}.

\item The quantum-mechanical state vector describing an {\em
arbitrary\mini} physical state of an (incompressible) QH fluid is
{\em single-valued\,} in the position coordinates of all those
local excitations that are composed of {\em electrons\mini}
and/or {\em holes}.
\end{list}

\smallskip

The basic contention advanced in Fr\"ohlich and Studer
(1992b, 1992c, 1993b) and Fr\"ohlich and Thiran (1994) (see also
Sect.~\ref{sSL}) is that if a QH fluid is interpreted as a
two-dimensional system of electrons with vanishing longitudinal
resistance $\RL$ satisfying assumptions {\bf (A1)} through  {\bf
(A4)} above then, in the scaling limit, its quantum-mechanical
description is completely coded into a {\em quantum Hall lattice (QH
lattice)}. A QH lattice $(\G,\Q)$ consists of an odd, integral
lattice $\G$ and an integer-valued, linear functional $\Q$ on $\G$.
In general, the metric on $\G$ need not be positive definite. The
number of positive eigenvalues of the metric corresponds,
physically, to the number of edge currents of one chirality, the
number of negative eigenvalues correponds to the number of egde
currents of opposite chirality. In this section, we present results
mainly on QH fluids with edge currents of just {\em one\mini}
chirality. These QH fluids correspond to QH lattices $(\G,\Q)$ where
$\G$ is a {\em euclidian\mini} lattice, i.e., a lattice with a {\em
positive-definite\mini} metric. These special QH lattices are called
{\em chiral QH lattices (CQHLs)}.

Our study of {\em chiral\,} QH lattices is motivated by a physical
hypothesis expressing a ``{\em chiral factorization\,}'' property of
QH fluids.
\vspace*{0mm}

\begin{list}
{\bf (A\arabic{lico})}{\usecounter{lico}\setlength{\labelsep}{.4cm}
\setlength{\labelwidth}{1cm}\setlength{\leftmargin}{1.4cm}
\setlength{\rightmargin}{.7cm}}
\setcounter{lico}{4}

\item The fundamental charge carriers of a QH fluid are
electrons and/or holes. We assume that, in the scaling limit, the
dynamics of electron-rich subfluids of a QH fluid is {\em
independent\mini} of the dynamics of hole-rich subfluids, and, up to
charge conjugation, the theoretical analysis of an electron-rich
subfluid is {\em identical\,} to that of a hole-rich subfluid.
\end{list}

A discussion of the ``working hypothesis''~{\bf^^ca(A5)}, including
some proposals for its experimental testing, is given in
Fr\"ohlich, Studer, and Thiran (1995). There, it is also shown that
all ``hierarchy fluids'' of the Haldane-Halperin (Haldane, 1983;
Halperin, 1984) and Jain-Goldman (1992) scheme, respectively,
satisfy our assumptions {\bf (A1--4)} and (a slightly weaker form of)
assumption {\bf (A5)}.

\subsubsection{Chiral Quantum Hall Lattices (CQHLs)}

Let $V$ be an $N$-dimensional, real vector space equipped with a {\em
positive-definite} inner product, $\la.\,,.\ra\!$. In $V$ we choose
a basis $\{ {\bf e}_{1},\ldots,{\bf e}_{N}  \}$ such that

\be
K_{ij} \:=\: K_{ji} \::=\:\: \sca{{\bf e}_{i}\mini}{{\bf e}_{j}}
\,\in \Z\ , \hfa i,j = 1, \ldots,N\ ,
\label{K}
\ee

\noi
i.e., {\em all\mini} matrix elements of $(K_{ij})$ are {\em
  integers}. The basis
$\{ \avecd{{\bf e}} \}$ generates an {\em integral, euclidean} (i.e.,
positive-definite) {\em lattice}, $\G$, defined by

\be
\G \::=\: \{\, \q = \sum_{i=1}^{N} q^i \e_i \,|\  q^i \in \Z\mini,
\mbox{ for all } i=1,\ldots,N \, \}\ .
\label{G}
\ee

\noi
The matrix $K$ in~\cfr{K} is called the {\em Gram matrix\mini} of
the basis $\{ \avecd{{\bf e}} \}$ generating the lattice $\G$. Let
$\{ \avecu{\es} \}$ be the basis of $V$ {\em dual\,} to $\{
\avecd{\e} \}$, i.e., the basis satisfying

\be
\sca{\es^{i}}{\e_{j}} \:=\: \delta_j^i\ , \hfa i,j = 1,\ldots,N\ .
\ee

\noi
Then,

\be
\es^i \:=\: \sum_{j=1}^N ( K^{-1} )^{ij} \e_j\ ,
\label{epsi}
\ee

\noi
where $K^{-1}$ is the inverse of the matrix $K = ( K_{ij} )$, with $K_{ij}$
given in~\cfr{K}, and $( K^{-1} )^{ij} =\; \sca{\es^{i}}{\es^{j}}$.
The basis $\{ \avecu{\es} \}$ generates the {\em dual lattice},
$\Gs$, defined by

\bea
\Gs & := & \{ \, \n \in V \,|\ \sca{\n\,}{\q} \in \Z\mini, \mbox{ for
all } \q \in\!\G \,\}  \nonumber \\
& = & \{\, \n = \sum_{i=1}^{N} n_i\, \es^i \,|\  n_i
\in \Z\mini, \mbox{ for all } i=1,\ldots,N \, \}\ ,
\label{Gs}
\eea

\noi
and, by~\cfr{epsi}, $\Gs$ contains $\G$. Denoting by $\,\Del := \det
K \in\!\Z\,$ the determinant of the matrix $K$ ($\det K > 0\,$ for
euclidean lattices), the matrix elements $( K^{-1}
)^{ij}$ corresponding to the
scalar products between basis vectors in $\Gs$ are, in general (for
$\Del \neq 1$), {\em rational\,} numbers; since, by {\em Kramer's
rule}, we have

\be
( K^{-1} )^{ij}=( \widetilde{K} )^{ij}/\Del\ ,
\label{Kram}
\ee

\noi
with $\widetilde{K}$ the integer-valued matrix of cofactors of the
matrix $K$.

The (equivalence) classes of elements in $\Gs$ modulo $\G\mini$ form
an abelian group, $\Gs / \, \G$. There are $\,\Del = | \Gs/\,\G|\,$
distinct classes; $\Del$ is often referred to as the lattice
{\em discriminant}. An integral lattice $\G$ is said to be {\em
odd\,} if it contains a vector $\q$ such that $\,\sca{\q\,}{\q}$ is
an {\em odd\,} integer. Thus $\G$ is odd if, and only if, $\,K_{ii}
=\: \sca{\e_{i}\mini}{\e_{i}}$ is odd for at least one $i$.
(Otherwise, $\G$ is said to be {\em even}.)

Given a set of integers, $\avecd{n}$, we denote by
$\,\gcd(\avecd{n})\,$ their {\em greatest common divisor}. A vector
$\,\n = \sum_{i=1}^N n_i\, \es^i \in\!\Gs\,$ is called {\em
primitive} (or ``visible'') if, and only if,

\be \gcd(\avecd{n}) \:=\: \gcd (
\sca{\n\,}{\e_1},\ldots,\sca{\n\,}{\e_N} ) \:=\: 1\ .
\label{primi}
\ee

\noi
In geometrical terms, a vector $\,\n \in\!\Gs\,$ is primitive if,
and only if, the line segment joining the origin to the point
$\,\n\,$ is free of any lattice point. In particular, one can always
take a primitive vector as the first vector of a lattice basis.

A lattice $\G$ is said to be {\em decomposable\mini} if it can be
written as an orthogonal direct sum of sublattices,

\be
\G \:=\: \bigoplus_{j=1}^r \Gamma_j \ , \hfs r \geq 2 \ ,
\label{Gd}
\ee

\noi
with the property that $\,\sca{\q^{(i)}}{\q^{(j)}}\:= 0$, for
arbitrary vectors  $\q^{(i)}$ in $\Gamma_i$  and  $\q^{(j)}$ in
$\Gamma_j$, and for arbitrary $i \neq j$. Otherwise, $\G$ is said to
be {\em indecomposable}.

If $\,m\,$ and $\,n\,$ are two integers we shall write $\,m \equiv n
\bmod p\,$ if, and only if, $\,m-n\,$ is an integer multiple of $p$.

We are now prepared to define what we mean by a {\em chiral
QH lattice (CQHL)}:
\rule[-4mm]{0mm}{5mm}

{\bf Definition~\ref{ssQHFQHL}.1.} {\em A CQHL is a pair $(\G,\Q)$
where $\G$ is an odd, integral euclidean lattice, and $\Q$ is a
primitive vector in the dual lattice $\,\Gs$ with the property that

\be
\sca{\Q\,}{\q} \:\equiv\: \sca{\q\,}{\q}\:\bmod\ 2\ ,\hfa\,\q \in\!
\G\ ,
\label{parity}
\ee

\noi
i.e., $\Q$ is an\,} odd {\em vector in the dual lattice $\,\Gs$.}
\rule[-4mm]{0mm}{5mm}

Let $( \G , \Q )$ be a CQHL for which $\,\G = \bigoplus_{j=1}^r
\Gamma_j \,$ is  decomposable. Then $\,\Gs = \bigoplus_{j=1}^r
\Gamma_j^\ast \,$ is the associated decomposition of the dual
lattice. We say that $(\G,\Q)$ is {\em proper\mini} if, in the
decomposition of $\Q$

\be
\Q \:=\: \sum_{j=1}^r \Q^{(j)}\ , \hwh \Q^{(j)} \::=\:
\Q\!\mid_{\Gamma_j^\ast}\;\in\Gamma_j^\ast \ ,
\label{Qd}
\ee

\noi
corresponding to the decomposition of $\,\Gs$, every $\mini\Q^{(j)}$
is {\em non-zero}. In this section we discard improper CQHLs; see the
discussion at the end of Subsect.\,\ref{ssDic}.

\rule[-4mm]{0mm}{5mm}

{\bf Definition~\ref{ssQHFQHL}.2.} {\em A CQHL is called\,}
primitive {\em if, in the decomposition~\cfr{Qd} of $\,\Q$, the
component $\,\Q^{(j)}$ is a\,} primitive {\em vector in
$\Gamma_j^\ast$, for all $\,j = 1,\ldots, r$.}
\rule[-4mm]{0mm}{5mm}

We note that any {\em indecomposable} CQHL (i.e., one with $r=1$
in~\cfrs{Gd}{Qd}) is proper and primitive.


\subsection{A Dictionary Between the Physics of QH Fluids and the
Mathematics of QH Lattices}
\label{ssDic}

In this subsection, we provide a precise dictionary between
mathematical properties of QH lattices and physical properties of QH
fluids. Such a  dictionary has been presented
in Fr\"ohlich and Studer (1993a, 1993b), and Fr\"ohlich and Thiran
(1994). Here we recall its main contents and significance. The
starting point is the idea to describe the physics of a QH fluid in
the scaling limit in terms of an effective field theory of its
conserved current densities; see Sect.~\ref{sSLaC}. Since a QH fluid
has a strictly positive mobility gap $\delta$, the scaling limit of
the effective theory of its conserved current densities must be a
{\em topological field theory} (Fr\"ohlich and Zee, 1991;
Fr\"ohlich and Studer, 1992b). The presence of a non-zero external
magnetic field transversal to the plane to which the electrons  are
confined implies that the quantum dynamics of the
system {\em violates} the symmetries of parity
(reflections-in-lines) and time reversal. Thus the topological field
theory will {\em not\,} be parity - and time-reversal invariant.

In $2 + 1$ space-time dimensions, the continuity equation

\be
{\partial\over \partial t}\, j^0 + \vec{\bigtriangledown}\cdot\vec{j}
\:=\: 0\ , \ee

\noi
obeyed by a conserved current density $\,j^\mu=(j^0,\vec{j}\,)$,
implies that $j^{\mu}$ is the curl of a vector potential, i.e.,
$j^{\mu} = \varepsilon^{\mu \nu \lambda} \, \partial_{\nu}
b_{\lambda}$, for a vector potential $\,b =
(b_0,\vec{b}\,)\,$ that is unique up to the gradient of a scalar
field (gauge invariance!); see Sect.~\ref{sSLaC}.

Let us consider a QH fluid which, in the scaling limit, has $N$
independent, conserved current densities, $\,j^{\mu\mini
1},\ldots,j^{\mu\mini N}$. The electric current density,
$J_{el}^\mu$, is then a linear combination, $J_{el}^\mu=\sum_{i=1}^N
Q_i\, j^{\mu\mini i}$, of these current densities. When formulated in
terms of the vector potentials $b^1,\cdots,b^N$ of these
conserved current densities, the topological field theory describing
the physics of the {QH fluid} in the scaling limit can only be a {\em
pure, abelian Chern-Simons theory\mini} in the fields
$b^1,\cdots,b^N$, because it
must violate parity - and time-reversal invariance.  The bulk
action is given by

\be
S( {\bf b}) \:=\: \frac{1}{4 \pi}
\int C_{ij} \, b_{\mu}^i\, \partial_{\nu} b_{\lambda}^j \,
\varepsilon^{\mu \nu \lambda}\,dt\diix \:=:\: \frac{1}{4 \pi} \int
\sca{{\bf b}_{\mu}\,}{ \partial_{\nu} {\bf b}_{\lambda}}
\!\varepsilon^{\mu \nu \lambda}\,dt\diix\ ,
\ee

\noi
where $\,C_{ij}= C_{ji}\,$ is some non-degenerate quadratic form
(metric) on \BBn{R}{N}.

Physical states of a pure, abelian Chern-Simons theory describe
static $N$-tuples of ``charges'' localized in bounded disks of
space. Each $N$-tuple, $(\avecu{q})$, of ``charges'' localized in
some disk $D$ of space is an $N$-tuple of eigenvalues of the
operators

\be
\int\limits_D j^{0i} (\vec{x},t)\,\diix \:=\:
\oint\limits_{\partial D}  \vec{b}^{\mini i} (\vec{x},t)\cdot
d\vec{x} \ ,\hsp i = 1,\ldots,N\ .
\ee

\noi
The equations
of motion of pure, abelian Chern-Simons theory, with the currents
$j^{\mu i}$  minimally coupled to $N$ external gauge fields $a_{\mu
i}$, $i=1,\ldots,N$, read

\be
\varepsilon_{\mu \nu \rho} j^{\rho\mini i} \:=\: \partial_{\mu}
b_{\nu}^i - \partial_{\nu} b_{\mu}^i \:=\:  (C^{-1})^{ij} \left(
\partial_{\mu} a_{\nu j} - \partial_{\nu} a_{\mu j} \right)
\:=:\:  (C^{-1})^{ij} f_{\mu \nu j} \ ,
\label{CSeq}
\ee

\noi
These equations imply that, to an $N$-tuple of charges,
$(\avecu{q})$, there corresponds an $N$-tuple of ``fluxes'',
$(\avecd{\varphi})$, with $\,\varphi_j = \int_D f_{12j}\,\diix\,$,
related to $q^i$ through

\be
q^i \:=\: (C^{-1})^{ij} \varphi_j\ ,\hsp  i=1,\ldots,N\ .
\label{qphi}
\ee

Consider a state of Chern-Simons theory describing two excitations
with {\em identical\,} charges $(\avecu{q})$ localized in
disjoint, congruent disks of space, $D_1$ and $D_2$. {}From the
theory of the {\em Aharonov-Bohm effect\,} we know that if the
positions of the two disks are exchanged adiabatically along
counter-clockwise oriented paths, the state vector only changes by a
phase factor, $\exp i\pi\theta$, given by

\be
\exp {i\pi\theta} \:=\: \exp{i\pi\sca{\q\,}{\q}} \ ,
\label{staph1}
\ee

\vwv

\be
\theta \:=\: \sum_{i=1}^N \varphi_i \, q^i
\:=\: \sum_{i=1}^N (C^{-1})^{ij} \varphi_i\, \varphi_j
\:=\: \sum_{i=1}^N C_{ij}\, q^i q^j
\:=:\: \sca{\q\,}{\q} \ .
\label{staph2}
\ee

\noi
It is well known  that $\,\exp i\pi\!\sca{\q\,}{\q}$ has the meaning
of a {\em statistical phase} of the excitation corresponding to the
charges $(\avecu{q})$.  The $N$-dimensional vector $\q\,$
introduced here is equivalently defined through its components
$(\avecu{q})$, or, through~\cfr{qphi}, in
terms of its dual (w.r.t.~the metric $C_{ij}$) components
$(\avecd{\varphi})$, the ``fluxes''. An excitation labelled by
$\q\,$ with $\,\mbox{$\sca{\q\,}{\q}$} \equiv 1 \bmod 2\,$ is a
fermion, while if  $\,\sca{\q\,}{\q} \equiv 0 \bmod 2\,$ it
represents a boson.

Next, we consider a state describing two excitations with two
{\em different\mini} charge vectors ${\bf q}_{(1)}$ and ${\bf
q}_{(2)}$ localized in disjoint disks $D_1$ and $D_2$. Imagine that
$D_2$ is transported around $D_1$ adiabatically, along a
counter-clockwise oriented loop. Then the theory of the
Aharonov-Bohm effect teaches us that the state vector only changes
by a phase factor given by

\be
\exp {2 i\pi\sca{{\bf q}_{(1)}\mini}{{\bf q}_{(2)}}}\ ,
\label{monodr}
\ee

\vwv

\be
\sca{{\bf q}_{(1)}\mini}{{\bf q}_{(2)}}
\::=\: \sum_{i=1}^N C_{ij}\, q_{(1)}^i q_{(2)}^{\mini j}
\:=\:\sum_{i=1}^N \varphi_{(1)i} \, q_{(2)}^i
\:=\: \sum_{i=1}^N (C^{-1})^{ij} \varphi_{(1)i}\,\varphi_{(2)j}
\ee

Since the charges, $q^i$, of physical states\,--\,i.e., the
eigenvalues of the operators $\int_D j^{0\mini i}
(\vec{x},t)\,\diix,\ i=1,\ldots,N$\,--\,are
{\em additive\mini} quantum numbers, the set of vectors
$\mini\q\mini$ of physical excitations of a QH fluid form a {\em
lattice}, $\Gp$, whose dimension can be taken to be $N$. Expressing
the electric current density $J_{el}^\mu$ as a linear combination,
$J_{el}^\mu=\sum_{i=1}^N Q_i\,j^{\mu\mini i}$, of the current
densities $\,j^{\mu\mini 1},\ldots,j^{\mu\mini N}$, the {\em total
electric charge}, $q_{el}(\q)$, of an excitation
of the system with charge vector $\q$ localized in some disk $D$ is
found to be given by

\be
q_{el}(\q) \:=\: \sum_{i=1}^N Q_i\,q^i \:=:\; \sca{\Q\,}{\q}\ ,
\label{qelq}
\ee

\noi
where the electric charge assignments $Q_i$ of each current have
been collected in an $N$-dimensional vector $\Q$ henceforth
referred to as the ``charge vector''. Note that, according to the
above definition, the $Q_i$ are the {\em dual\,} components of the
charge vector $\Q$. The electric charge $q_{el}(\q)$ is an
eigenvalue of the electric charge operator

\be
\int J_{el}^0 (\vec{x},t)\,\diix \:=\:
\int \left( \sum_{i=1}^N Q_i \, j^{0\mini i}(\vec{x},t)
\right)\diix\ .
\ee

We define $\G$ as the set of all physical excitations $\q$ with {\em
integral\,} electric charge, i.e.,

\be
\G \::=\: \{\, \q \in \Gp \,|\ \sca{\Q\,}{\q}=q_{el}(\q) \in \Z
\,\}\ .
\ee

\noi
Clearly $\G$ is a sublattice of $\,\Gp$. If $\,\G$ has at least one
excitation $\q$ of electric charge $\,q_{el}(\q) = 1\,$ then the
dimension of $\,\G$ is equal to $N$. Assumption~{\bf (A3)} entails
that arbitrary vectors $\q_{(i)}$ in $\G$ must  have integer
statistical phases, $\,\sca{{\bf q}_{(i)}\mini}{{\bf q}_{(i)}}$,
related to their electric charges by the congruence $\,q_{el}({\bf
q}_{(i)}) \equiv\:\sca{{\bf q}_{(i)}}{{\bf q}_{(i)}} \bmod\
2\mini$. Writing the scalar product $\,\sca{{\bf
q}_{(1)}\mini}{{\bf q}_{(2)}}$ between two arbitrary charge
vectors in $\G$ as

\ben
\sca{{\bf q}_{(1)}\mini}{{\bf q}_{(2)}} \:=\: \frac{1}{2}\left(
\sca{{\bf q}_{(1)}+{\bf q}_{(2)}\mini}{{\bf q}_{(1)}+{\bf q}_{(2)}}
- \,\sca{{\bf q}_{(1)}\mini}{{\bf q}_{(1)}} - \,\sca{{\bf
q}_{(2)}\mini}{{\bf q}_{(2)}} \right) \ ,
\een

\noi
we see that this is always an integer and conclude that $\G$ is
an {\em integral\,} lattice. Since there is at least one
electron-like excitation $\q$ in $\G$ with $\,q_{el}(\q)=1$, {\bf
(A3)} also constrains the lattice $\G$ to be {\em odd\,}, because
$\,\sca{\q\,}{\q}  \equiv q_{el}(\q) \equiv 1 \bmod 2\mini$. For
an arbitrary lattice basis $\{ \q_{(1)},\ldots,\q_{(N)}
\}$ of $\,\G$, all the scalar products $\,Q_i=\:
\sca{\Q\,}{\q_{(i)}},\ i=1,\ldots,N$, are integers, and hence the
vector $\Q$ belongs to the {\em dual}, $\Gs$, of $\,\G$;
(see~\cfr{Gs}). The charge vector $\Q$ must necessarily be {\em
primitive}\,--\,i.e., $\gcd(\avecd{Q}\, )=1$\,--\,in order for an
excitation of electric charge $1$ to exist.

Every vector $\q$ of the sublattice $\G$ of $\,\Gp$ can now be
consistently interpreted as labelling a physical excitation describing a
(multi-)electron and/or (multi-)hole configuration (depending on
the integral value of the electric charge). Note that, starting {}from
an electron-like
excitation $\q_{(1)}$ in $\G$, one can construct a  basis of  $\G$
consisting of electron vectors $\q_{(i)}$, with
$\,q_{el}(\q_{(i)})=1$, for $\,i=1,\ldots,N$. We call such a
basis a ``symmetric'' lattice basis.

The next step consists in finding restrictions on the lattice $\Gp$
by making use of our last assumption {\bf (A4)}. Consider, for each
$\,i=1,\ldots,N$, a physical state of the system describing an
excitation with ``charges'' $\q^\prime$ localized in a disk $D_1$
and an excitation corresponding to an electron with vector
$\q_{(i)}$ localized in a disk $D_2$ disjoint {}from $D_1$. Then we
derive {}from assumption {\bf (A4)} and~\cfr{monodr} that

\be
\exp 2\mini \pi i \sca{\q^\prime}{\q_{(i)}} \;=\: 1 \ ,
\ee

\noi
i.e., $\sca{\q^\prime}{\q_{(i)}}$ must be an {\em integer}, for
all $\,i=1,\ldots,N$. {}From this it follows that $\q^\prime$
belongs to the dual lattice $\Gs$. Thus, vectors $\q^\prime$ of
{\em arbitrary\mini} physical excitations of a QH fluid described
by the lattice $\G$ and the charge vector $\Q$ must all belong to
the lattice $\Gs$ dual to $\G$. We then have the following
inclusions between the various lattices introduced so far:

\be
\Gs \:\supseteq\: \Gp \:\supseteq\: \G\ .
\ee

To discover how the data $(\G,\Q)$ predict the value of the Hall
conductivity $\sH$, we consider the effect of turning on a
perturbing external magnetic field $\,{\bf B}={\bf B}_{total} - {\bf
B}_{c}\,$ localized in a small disk $D$ inside the system and
creating a total magnetic flux $\Phi$, with

\be
\Phi \:=\: \int\limits_D B (\vec{x})\,\diix \ ,
\ee

\noi
where $B\,$ is the component of $\,{\bf B}$  perpendicular to the
plane of the system. Since the total electric current density
$J_{el}^\mu$ is given by $\,J_{el}^\mu=\sum_{i=1}^N Q_i\,j^{\mu\mini
i}$, the ``flux'' $\varphi_i$ created by $\,{\bf B}$ is given by

\be
\varphi_i \:=\: Q_i \, \Phi \ ,
\label{fluxB}
\ee

\noi
for $\,i=1,\ldots,N$. This follows {}from the way in which, in
Chern-Simons theory, the currents $j^{\mu \mini i}$ are coupled
to an external gauge field. Let $\,A = (A_0,\vec{A}\,)\,$ denote
the electromagnetic vector potential of the external magnetic field
$\,{\bf B}$. Then from the fact that the total electric current
$J_{el}^\mu$ couples to the vector potential $A_\mu$ through the
term $\,J_{el}^\mu\,A_\mu$, it follows that the currents $j^{\mu\mini
i}$ are minimally coupled to the field $\,Q_i\mini A_\mu\mini$. Thus
the gauge fields $\,a_{\mu\mini i}\,$ appearing on the r.h.s.\
of~\cfr{CSeq} are given by $\,a_{\mu\mini i} = Q_i\mini A_\mu\mini$
which implies~\cfr{fluxB}; see Fr\"ohlich and Studer
(1992b, 1992c), Fr\"ohlich and Thiran (1994), and
Subsect.~\ref{ssEE}. Since the charges $\,q^i = \int_D  j^{0\mini i}
(\vec{x},t)\,\diix = \oint_{\partial D}\vec{b}^i (\vec{x},t)\cdot
d\vec{x}$, corresponding to the fluxes $\,\varphi_k\mini$, are equal
to $\,(C^{-1})^{ik}\, \varphi_k\mini$, by the equations of
motion~\cfr{CSeq} of Chern-Simons theory, we conclude that the {\em
total excess electric charge} created by the excess magnetic flux
$\Phi\mini$ is given by

\bea
q_{el}(\Phi) &=& \sum_{i=1}^N
Q_i \, q^i \:=\: \sum_{i,k=1}^N Q_i \, (C^{-1})^{ik}\,\varphi_k
\nonumber \\
&=& \left( \sum_{i,k=1}^N Q_i \,(C^{-1})^{ik}\,Q_k \, \right) \Phi
\:=:\:\: \sca{\Q\,}{\Q}\!\Phi\ .
\label{chargePhi}
\eea

\noi
Comparing this equation to the equations describing the
electrodynamics of a QH fluid which have been described
in Subsect.~\ref{ssQH}, in particular, to Eq.~\cfr{ED3},

\ben
J_{el}^0 \:=\: \sH \, B \ ,
\een

\noi
we find that the coefficient of $\mini\Phi$ on the r.h.s.\
of~\cfr{chargePhi} is the {\em Hall conductivity}, $\sH$, of the QH
fluid. Thus we arrive at the fundamental equation

\be
\sH \:=\:\: \sca{\Q\,}{\Q}\ .
\label{sHdic}
\ee

Since, as shown above, $\Q$ belongs to the dual lattice $\Gs$, we
have that

\ben
\Q \:=\: \sum_{i=1}^{N} Q_i\, \es^i\ ,
\een

\noi
with $Q_i\in\!\Z\mini$, for $\,i=1,\ldots,N$, where $\{ \avecu{\es}
\}$ is the basis of $\,\Gs$ dual to a basis $\{ \avecd{\e} \}$ of
the {\em integral\,} lattice $\G$ with Gram matrix $\,K_{ij}
=\:\sca{{\bf e}_{i}\mini}{{\bf e}_{j}} \in\!\Z\mini$. We could
choose, for example, a symmetric basis of electron-like excitations,
i.e., $\e_i =\q_{(i)}$, for all $\,i=1, \ldots,N$. In this
situation the electric charge requirements fix the coefficients
$\,Q_i=1$, for all $\,i=1,\ldots,N$, since we must have that
$\,q_{el}(\e_i)=\:\sca{\Q\,}{\e_i}=Q_i=1\mini$. But any other choice
of basis is admissible as well. The all important consequence of
$\mini\Q\mini$ being a vector in the dual of $\,\G$ is that its
squared length $\,\sca{\Q\,}{\Q}$ is a {\em rational number}. To see
this, we return to the discussion after~\cfr{Gs}. We have that
$\,\sca{\es^i}{\es^j}=( K^{-1} )^{ij}=( \widetilde{K} )^{ij}/\Del$,
where $\,\Del = \det K\,$ (the lattice discriminant) is an integer,
and $\,(\widetilde{K})^{ij}\,$ are integers. It follows that

\be
\sca{\Q\,}{\Q} \:=\; \sum_{i,j=1}^N Q_i\, Q_j\sca{\es^i}{\es^j}
\;=\: \left( \sum_{i,j=1}^N Q_i\,( \widetilde{K} )^{ij}\, Q_j
\right)\,\Del^{-1}
\label{Qrat}
\ee

\noi
is a rational number whose denominator is a {\em divisor} of
$\Del\mini$; (in general there will be non-trivial common divisors of
$\Del$ and of the numerator of the expression on the r.h.s.\
of~\cfr{Qrat}). We conclude that the Hall conductivity $\sH$ of a QH
fluid satisfying assumptions {\bf (A1)} through {\bf (A4)} is
necessarily a {\em rational number}.

The analysis just completed shows that, in the scaling limit, the
physics of a QH fluid is described by a pair $(\G,\Q)$ of an
integral, odd lattice $\G$ and a primitive vector $\Q$ in the dual
lattice, with $\,\sca{\Q\,}{\q} \equiv\:\sca{\q\,}{\q} \bmod\ 2$,
for all vectors $\q$ in $\G$, i.e., by data that we have termed {\em
QH lattice\mini} in Subsect.\,\ref{ssQHFQHL}.

In the following, we mainly specialize to {\em chiral\,} QH lattices
(CQHLs) which are QH lattices where the lattice $\G$ is
{\em euclidean}. They describe (universality classes of) QH fluids
with edge currents of {\em definite chirality}, the basic charge
carriers being, say, electrons. The theory of an electron-rich QH
fluid differs {}from that of a hole-rich QH fluid only in relative
minus signs in all equations relating charges to fluxes and
involving the electric charge of the basic charge carriers. One
simply reverses the sign of the charge vector $\Q$ and the metric
$C:\ \Q \rightarrow -\mini\Q\,$ and $\,C_{ij} \rightarrow -\mini
C_{ij}$,  i.e., $\sca{.\,}{.} \rightarrow -\sca{.\,}{.}\!$. A more
general type of QH fluids consisting of electron- {\em and\,}
hole-rich subsystems is therefore described by {\em two\mini} CQHLs,
$(\G_e,\Q_e)\, \mbox{ and } \, (\G_h,\Q_h)$, with

\be
\sH \:=\:\: \sca{\Q_e\mini}{\Q_e\mini} - \:\sca{\Q_h}{\Q_h}\ .
\ee

\noi
In such QH fluids, the subsystems described by $(\G_e,\Q_e)$ and
by $(\G_h,\Q_h)$ are independent of each other; (in particular,
left- and right-moving edge excitations are independent of each
other). The mathematical properties of $(\G_e,\Q_e)$ and of
$(\G_h,\Q_h)$ are analogous. It is therefore sufficient to study
the properties of $(\G_e,\Q_e)$, say, and we shall omit the
subscript ``$e$'' henceforth. Note that this ``factorized''
situation implements assumption~{\bf (A5)} of Subsect.\,\ref{ssQHFQHL}.

As in~\cfrs{Gd}{Qd}, let $\,\G = \bigoplus_{j=1}^r \Gamma_j \,$ be
the decomposition of the lattice $\G$ into an orthogonal direct sum
of indecomposable sublattices $\G_j$, and let $\,\Q = \sum_{j=1}^r
\Q^{(j)}$, with $\,\Q^{(j)} \in \Gamma_j^\ast$, be the associated
decomposition of the charge vector $\Q$. Then, by~\cfr{sHdic},

\be
\sH \:=\:\: \sca{\Q\,}{\Q} \:=\: \sum_{j=1}^r
\sca{\Q^{(j)}}{\Q^{(j)}} \:=\: \sum_{j=1}^r \sH^{(j)}
\ee

\noi
is the corresponding decomposition of the Hall conductivity (or
{\em Hall fraction}) as a sum of Hall fractions of subfluids
described by the pairs $(\G_j,\Q^{(j)}),\ j=1,\ldots,r$. Let us
imagine that, for some $j$, $\Q^{(j)}=0$. Then $\sH^{(j)}=0$, and
the subfluid corresponding to $(\G_j,\Q^{(j)})$ does not have any
interesting electric properties. For the purpose of describing
electric properties of QH fluids and classifying the possible values
of the Hall fraction $\sH$ of QH fluids, subfluids described by
$(\G_j,\Q^{(j)})$, with $\Q^{(j)}=0$, can therefore be discarded.
Thus we may henceforth assume that $\Q^{(j)}\neq 0$, for all
$\,j=1,\ldots,r$, i.e., we may limit our analysis to {\em
proper\mini} CQHLs; see~\cfr{Qd}. [However, if there are QH fluids
with spin-charge separation, subfluids $(\G_j,\Q^{(j)})$ with
$\Q^{(j)}=0$ will appear, and spin currents could receive
contributions {}from such neutral subfluids; see Subsect.~VI.C in
Fr\"ohlich and Studer (1993b).]

Next, suppose that, for some $j$ with $1\leq j \leq r$, $\,\Q^{(j)}$
is an integer multiple of a {\em primitive} vector $\,\Q_\ast^{(j)}
\in\!\G_j^\ast$, i.e., $\,\Q^{(j)}= \nu_j \, \Q_\ast^{(j)}$, with
$\nu_j \geq 2\mini$; see~\cfr{primi}. Then, for {\em any} $\,\q\in\!
\G_j,\ \,\sca{\Q^{(j)}}{\q\mini}\,$ is an {\em integer multiple} of
$\,\nu_j$. This would mean that the electric charge of an {\em
arbitrary\mini} quasi-particle of the subfluid described by
$(\G_j,\Q^{(j)})$ would be an integer multiple of $\nu_j$ (in units
where $-e=1$), i.e., only bound states of electrons and holes of
electric charge $\,n\mini\nu_j,\  n \in\!\Z\mini$, would appear as
quasi-particles of such a subfluid. There appear to exist QH fluids
where this situation arises (e.g. ``hierarchy QH fluids'', see
Appendix~E in Fr\"ohlich, Studer, and Thiran (1995), or films of
superfluid $\mbox{He}^3$, see Subsect.~VII.B in Fr\"ohlich and
Studer (1993b)). But for simplicity, we shall assume henceforth
that $\,\nu_j=1$, i.e., that $\Q^{(j)}$ is a {\em primitive} vector
in $\Gs_j$, for each $\,j=1,\ldots,r$. In other words, we limit our
analysis of CQHLs to what we have called {\em primitive\mini} CQHLs;
see Def.\,\ref{ssQHFQHL}.2 at the end of the previous subsection.


\subsection{Basic Invariants of Chiral QH Lattices (CQHLs) and their
Physical Interpretations}
\label{ssBI}

Let $(\Gamma, \Q)$ be a {\em chiral\,} QH lattice (CQHL), i.e., a QH
lattice describing (a universality class of) QH fluids with edge
currents of a {\em definite chirality}. In this subsection, we
describe elementary mathematical properties of $(\Gamma, \Q)$. They
are formulated in terms of {\em invariants\mini} of $(\Gamma,
\Q)$. Numerical invariants of a CQHL $(\Gamma, \Q)$ are numbers
which only depend on its {\em intrinsic\mini} properties. They are
{\em independent\,} of the choice of a basis in $\Gamma$ and of a
``reshuffling'' of electric charge assignments corresponding to a
transformation of $\Q$ by an orthogonal symmetry of $\,\G$.

Choosing a basis, $\{\ve{1},\ldots,\ve{N}\}$, in $\mini\G$,
the CQHL is specified by the (integral) Gram matrix $\,K_{ij}\!
=\:<\!\ve{i},\ve{j}\!>,\ i,j=1,\ldots,N$, and by the row vector
$\,\Qv=(Q_1\mini,\ldots,Q_N)\,$ which specifies the components of
the charge vector $\mini\Q$ in the dual basis
$\{\ved{1},\ldots,\ved{N}\}$ of $\,\Gs$, i.e., $\Q=\sum_{j=1}^N
Q_j\,\ved{j}$; see Eqs.~(\ref{K}) through~(\ref{Gs}).
Choosing a different basis in $\mini\G$, the resulting pair
$(K^\prime,\Qv^\prime)$ is related to the pair $(K,\Qv)$ by

\be
K^\prime\:=\:S^T\,K\,S\ , \hah \Qv^\prime\:=\Qv\,S\ ,
\label{equiv}
\ee

\noi
where $S$ is an element of $GL(N,\BB{Z}\mini)$, the group of
integral, non-degenerate $N\times N$-matrices. Note that, for
$S^{-1}$ to be an element of the group, the determinant of any
element $\mini S\mini$ of $GL(N,\BB{Z}\mini)$ has to be $\mini\pm
1\mini$.

A concise
presentation of the numerical invariants of a CQHL, \QHL, is given
by the {\em symbol}

\be
\raisebox{-3.6mm}{$\scriptstyle N$} {\left( \sH = \ndH
\right)}^g_\lambda\;[\mini\lm,\lM] \ ,
\label{symb}
\ee

\noi
where the invariants summarized in the symbol have the following
mathematical definitions and physical interpretations:
\rule[-4mm]{0mm}{5mm}

{\bf (1)} $N:=\dim\G=\mbox{rank}\,\G$, the {\em lattice dimension}
$N$, is equal to the number of separately
conserved \ui\ current densities of the corresponding QH fluid, (in
the scaling limit).
Although no rigorous results are known, a mean field theory of the
FQHE suggests that  $\mini N\mini$ depends on the filling factor
$\mini\nu\mini$ and on the density and
strength of impurities (disorder) in the system. We expect that the
absolute upper bound, $\mini N_\ast$, on the dimension $\mini N\mini$ of
physically realizable CQHLs tends to $\mini\infty$, as the density and
strength of impurities tend to $\mini 0\mini$; see Fr\"ohlich,
Kerler {\em et al.}~(1995).
\rule[-4mm]{0mm}{5mm}

{\bf (2)} By~(\ref{sHdic}) and~(\ref{equiv}), the {\em Hall
conductivity} (or {\em Hall fraction}) $\sH$ is clearly a CQHL
invariant: $\sH =\:<\!\Q\,,\Q\!> \:= \Qv\cdot\Ki \Qv^T$.
By~(\ref{Kram}) and the definition of $\Qv$, it is a positive
{\em rational\,} number.
\rule[-4mm]{0mm}{5mm}

{\bf (3)} Writing $\,\sH=n_H/d_H$, with $\,\gcd(n_H,d_H)=1$, the
important invariant of the lattice given by its {\em discriminant}
$\Del$ can be written as

\be
\Del\::=\:\det K \:=\: l\,d_H\ ,
\label{disc}
\ee

\noi
where the invariant $\mini l\,$ is called the {\em level\,} of
the CQHL \QHL; see~(\ref{Kram}).
\rule[-4mm]{0mm}{5mm}

{\bf (4)} The level $\mini l\,$ satisfies an interesting
factorization property, namely $\,l=g\mini\lb$, where $\mini
g\mini$ is defined by $\,g:=\gcd(Q^1, \ldots,Q^N)$, with
$\,Q^j:=\Del<\!\Q\,,\ved{j}\!>,\ j=1,\ldots,N$, and $\{\ved{1},
\ldots,\ved{N}\}$ {\em any} dual basis of $\mini\Gs$. Thus,
by~(\ref{disc}), the discriminant is given by $\,\Del=g\mini\lb\mini
\dH\mini$. The invariant $\lb$ is called the {\em charge
parameter}, and its physical significance derives from the following
fact: one can prove (Fr\"ohlich and Thiran, 1994) that, in our units
where $\,e\!=\!-1$, the smallest possible (fractional) electric
charge of a quasi-particle of the corresponding QH fluid is given by

\be
e^\ast\::=\: \min_{\vn\in\mini\mbox{$\scriptstyle \Gamma
\raisebox{1.3mm}{$\scriptscriptstyle\ast$}$}\mimini,\,
<\Q\mini,\mini\vn>\mini \neq\mini 0} | <\!\Q\,,\vn\!> |
\:=\: {1 \over \lambda\mini d_H} \ .
\label{emin}
\ee

{\bf (5)} Next, we define the {\em
relative-angular-momentum invariants} $\,\lm$ and $\lM\,$. Since
$\Q$ is a primitive vector in $\mini\Gs$ (see~(\ref{primi})) there is
a basis of $\mini\G,\ \{\vq_1,\ldots,\vq_N\}$, such that
$\,<\!\Q\,,\vq_i\!>\:=\!1,\ i=1,\ldots,N$. The set of all such
``symmetric'' bases is denoted by $\BQ$. Then, for any CQHL \QHL,
we define the invariants

\be
\Lm \::=\: \min_{\vq\in\G\mimini,\,<\Q\mini,\mini\vq>\mini =\mini 1}
\mini <\!\vq,\vq\!>\ ,
\label{Lmin}
\ee

\vav

\be
\LM \::=\: \min_{\{\vq_1,\ldots,\vq_N \}\in \BQs}
\mini \left( \max_{1\leq i\leq N} <\!\vq_i,\vq_i\!>\right)\ .
\label{Lmax}
\ee

In the situation where \QHL\, is a {\em primitive, decomposable}
CQHL with decomposition $\,(\G,\Q)=\oplus_{j=1}^k(\G_j,\Q^{(j)})\,$
(see Definition 8.1.2) it is natural to refine the
definitions~(\ref{Lmin}) and~(\ref{Lmax}) as follows:

\be
\lm\QHL \::=\: \min_{1\leq j\leq k} \Lm(\G_j,\Q^{(j)}) \:\geq\:
\Lm\QHL\ ,
\label{lmin}
\ee

\vav

\be
\lM\QHL \::=\: \max_{1\leq j\leq k} \LM(\G_j,\Q^{(j)}) \:\geq\:
\LM\QHL\ .
\label{lmax}
\ee

We note that, because ${\bf Q}$ is an odd vector in $\Gamma^\ast$
(see~(\ref{parity})), the
relative-angular-momentum invariants~(\ref{Lmin})
through~(\ref{lmax}) are positive, {\em odd\,} integers satisfying

\be
\Lm \:\leq\: \LM \ ,\hah \lm \:\leq\: \lM\ .
\label{mlM}
\ee

Exploiting the Chern-Simons description of QH fluids, it
has been argued in Fr\"ohlich and Thiran (1994), and Fr\"ohlich,
Kerler {\em et al.}~(1995) that, for an {\em
elementary\mini} chiral QH fluid corresponding to the {\em
indecomposable} CQHL \QHL, $\,\lm=\Lm\,$ is the smallest
possible relative angular momentum of a state of two electrons excited above
the ground state of the fluid. The physical significance  of the
quantity $\mini\lM\mini$ as well as its role in the classification
of CQHLs will be expounded in
Subsect.\,\ref{ssGT}; (see also Fr\"ohlich, Studer, and Thiran
(1995), and Fr\"ohlich, Kerler {\em et al.}~(1995)).

The elementary invariants defined in {\bf (1--\mini 4)}, above, are
well-defined for general QH lattices; see Fr\"ohlich and Thiran
(1994).

\bigskip
{\bf Examples.} We conclude this subsection by considering some
examples of QH fluids in the language of QH lattices, thereby
illustrating the usefulness of the above invariants.
\rule[-4mm]{0mm}{5mm}

{\bf (a)} {\em QH fluids with $\,\sH=N,\ N=1,2,\ldots\,$, in the
non-interacting electron approximation.} These integer QH
fluids correspond to the (self-dual) simple euclidean lattices in N
dimensions: $\G=\Gp=\Gs=\BBn{Z}{N}= \BB{Z}\,\oplus\cdots
\oplus\,\BB{Z}\mini$. Here, $N$ is the number of separately
conserved edge currents (Halperin, 1982) or filled Landau levels.
Denoting by $\mini\ve{i}$ the generator of the $i$th summand,
$i=1,\ldots,N$, we have $\,K_{ij}=\:<\!\ve{i},\ve{j}\!>\:
=\delta_{ij}$. By the primitivity condition on the charge vector
$\Q$, we have $\,\Q= \ve{1}+\cdots + \ve{N}$, and $\,\sH
=\:<\!\Q\,,\Q\!>\: =1+\cdots+1=N$. We note that, by the self-duality
of $\BBn{Z}{N}$, there are no fractionally charged excitations with
fractional statistics (``anyons'') in these fluids.

The symbols characterizing these integer QH fluids read

\be
\thesymbol \;=\: \sub{N}{(N)}^1\hspace{-1.7mm}\sub{1}\;[\mini 1,1]
\ ,\hsp N=1,2,\ldots  \ .
\ee

\noi
Note that, by the decomposability of these CQHLs, we can write
$\,\subi{N}{(N)}^1_1 =\subi{1}{(1)}^1_1\oplus\cdots
\oplus\subi{1}{(1)}^1_1\mini$, in accordance with the physical
picture of $N$ independent, filled Landau levels.
\rule[-4mm]{0mm}{5mm}

{\bf (b)} {\em The Laughlin fluids\,} (Laughlin, 1983a, 1983b) {\em
at $\,\sH=1/m\,$, where $\,m=2\mini p+1,\  p=(0),1,2,\ldots\,$.} Here
$\G=\sqrt{m}\,\BB{Z}\,$, which is the one-dimensional lattice
generated by $\mini\veb\mini$ with squared length
$\,<\!\veb\,,\veb\!>\:=m$. The dual lattice $\,\Gs=(1/\sqrt{m})
\,\BB{Z}\,$ which is generated by \mbox{$\,\veed=\veb/m$}. The
charge vector, being primitive in $\mini\Gs$, takes the form
$\,\Q=\veed$, and thus $\,\sH= \mbox{$<\!\Q\,,\Q\!>$} =1/m$.
The quasi-particles are labelled by $\,n\veed\in\!\Gp =\Gs,\
 n\in\!\BB{Z}\mini$. By~(\ref{qelq}), they have fractional
electric charges $\,\qel(n)=\mbox{$<\!\Q\,,n\veed\!>$}=n/m$,
and, by Eqs.~\cfrs{staph1}{staph2}, they have fractional
statistical phases $\,\theta(n)\equiv
\mbox{$<\! n\veed,n\veed\!>$}\equiv \mbox{$n^2/m$}
\pmod{2}$.

In this example, the knowledge of the electric
charges $\mini\qel\mini$ of the excitations uniquely determines their
statistical phases $\mini\theta$. Such a charge-statistics
relation is a property of many interesting higher-dimensional QH
lattices; see Thm.\,\ref{ssGT}.5 below. However, a relation of this
kind does {\em not\,} hold for arbitrary QH lattices.

The symbols of the CQHLs describing the Laughlin fluids are
\vspace{-5mm}

\be
\thesymbol \;=\: \QHLsymbol{1}{1\over 2\mini p+1}{1}{1} \;[\mini
2\mini p+1,\mini 2\mini p+1\mini]\ ,\hsp p=1,2,\ldots  \ .
\ee

{\bf (c)} For each $\,p=1,2,\ldots\,$, there is the series of Hall
fractions $\,\sH=N/(2\mini pN\!+\!1)\,$ with $\,N=1,2,\ldots\ $.
{}From the data presented in Fig.\,\ref{sKE}.1, it is clear that many
of the experimentally most prominent Hall fractions belong to these
series (or to the charge-conjugated partner series of the one with
$\,p=1$; see Subsect.\,\ref{ssDis}). These fractions
also figure prominently in Jain's work (Jain, 1989, 1990,
1992)\,--\,the Jain-Goldman (1992) hierarchy
scheme\,--, and we refer to Thm.\,\ref{ssGT}.8 below where, from a
classification point of view, the uniqueness of the associated CQHLs
is proven. The above series of Hall fractions can be obtained from
the following series of {\em indecomposable} CQHLs: the data pairs
$(K,\Qv)$ which determine these CQHLs are given, in some bases that
we call ``normal'', by

\be
K \:=\,
\left.\left(\renewcommand{\arraystretch}{.8}
\begin{array}{c|ccccc}
2\mini p+1 & -1 & 0  & \cdot & \cdot & 0\rule[-1.8mm]{0mm}{4mm} \\
\hline
-1 & 2 & -1 & 0 & \cdot & 0\rule[0mm]{0mm}{4.5mm} \\
0 & -1 & 2 & -1 & \cdot & \cdot \\
\cdot & 0 & -1 & \cdot & \cdot & 0 \\
\cdot & \cdot & \cdot & \cdot & \cdot & -1 \\
0 & 0 & \cdot & 0 & -1 & 2
\end{array} \right)\  \right\}\mbox{\scriptsize\em N}
\ ,\hah
\Qv \:=\:
(\mini\underbrace{1,0,\ldots,0}_{\mbox{\scriptsize\em N}}\,) \ ,
\label{promKQ}
\ee

\noi
and the associated symbols read

\be
\thesymbol \;=\: \QHLsymbol{N}{N\over 2\mini pN+1}{1}{1} \;[\mini
2\mini p+1,\mini 2\mini p+1\mini]\ ,\hsp p,N=1,2,\ldots  \ .
\label{promsymb}
\ee

\noi
Note that the $(N\!-\!1)$-dimensional submatrix in the lower right
of $K$ is the Cartan matrix of the simple Lie algebra $\,A_{N-1}
= su(N),\ N=2,3,\ldots\ $. For $\,N=1$, we
recognize in~(\ref{promKQ}) and~(\ref{promsymb}) the expressions
corresponding to the Laughlin fluids; see example~{\bf (b)}. In
connection with the QH effect, the matrices in~(\ref{promKQ}) first
appeared in Read (1990). Combining results of Read (1990)
and Fr\"ohlich and Zee (1991) (see Appendix~E in Fr\"ohlich, Studer,
and Thiran (1995)) the CQHLs specified by~(\ref{promKQ}) can be seen
to correspond to the ``basic Jain states'' (Jain, 1989, 1990, 1992)
at $\,\sH=\mbox{$N/(2\mini pN\!+\!1)$}\mini$. Moreover, it has been
shown in Fr\"ohlich and Zee (1991) that the QH fluids corresponding
to~(\ref{promKQ}) exhibit large symmetries, namely \suhN\ current
algebras at level 1. In
Subsect.\,\ref{ssMS}, we show that the above CQHLs belong to an
interesting class of CQHLs with ``large'' symmetries, so-called
``maximally symmetric'' CQHLs.

By extending definitions~(\ref{promKQ})
and~(\ref{promsymb}) to $\,p=0$, the {\em composite} integer QH
fluids of example~{\bf (a)} can also be included as special cases
of~{\bf (c)}. (The Chern-Simons mean field theory of the FQHE offers
an explanation of this fact.)


\subsection{General Theorems and Classification Results for~CQHLs}
\label{ssGT}

The purpose of this subsection is to review general facts and
classification results for CQHLs. We summarize, in the form of eight
theorems, results that have been presented in Fr\"ohlich and Thiran
(1994), Fr\"ohlich, Studer, and Thiran (1994, 1995), and Fr\"ohlich,
Kerler {\em et al.}~(1995) where more details can be found. We
sketch some of the proofs and discuss  phenomenological implications
of the theorems.

The first two theorems are based on {\em CQHL inequalities\mini}
that establish relations between some of the numerical
invariants introduced in Subsect.\,\ref{ssBI}\mini.
\rule[-4mm]{0mm}{5mm}

{\bf Theorem~\ref{ssGT}.1.} {\em The set of (proper) CQHLs
\QHL\, with rank $\,N\!\leq\! N_\ast\,$ and
relative-angular-momentum invariant $\,\lM\leq \Ls$, where
$\mini N_\ast$ and $\,\Ls$ are two arbitrary, finite integers, is\mini} finite.
\rule[-4mm]{0mm}{5mm}

This theorem implies that the set of Hall fractions $\sH$ that can
be realized by CQHLs satisfying the above bounds on $N$ and
$\mini\lM$ is finite. However, the number of
possible fractions is growing superexponentially in $N_\ast$
and $\Ls$; (e.g., for $\,N_\ast=2\,$ and $\,\Ls=3\,$, there are
$\mini 10\mini$ CQHLs, while for $\,N_\ast=3\,$ and $\,\Ls=5$, one
finds already more than $\mini 250\mini$ CQHLs). Fortunately, in
the physically relevant situation where one also has a natural upper
bound, $\sigma_\ast$, on the Hall fractions to be considered, the
number of CQHLs satisfying this bound and the ones in
Thm.\,\ref{ssGT}.1 is drastically reduced!

The basic tools in proving Thm.\,\ref{ssGT}.1 are {\em Hadamard's
inequality\mini} for positive-definite quadratic forms (see,
e.g., Siegel, 1989), which implies that

\be
\lb\mini g\, \dH \:=\: \Del \:=\: \det K \:\leq\: \lM^{\,N}\ ,
\label{Dbound}
\ee

\noi
and the fact that (Fr\"ohlich, Kerler and Thiran, 1995)

\be
\lb\mini g\, \nH \:\leq \: C(N)\,\lM^{\,N-1} \ ,
\ee

\noi
where $C(N)$ is a constant depending on the lattice dimension
$N\,$ (e.g., for two-dimensional CQHLs, one finds that
$\,C(2)=4$). \rule{2.5mm}{2.5mm}

Recall that $N$ is the number of separately conserved \ui\ current
densities in a QH fluid (in the scaling limit). Increasing the amount of
disorder (the density or strength of impurities) in the
system is expected to reduce $\mini N_\ast\mini$ because of
``channel-mixing''  effects.

Theoretical arguments and numerical simulations (Meissner,
1992,1993) suggest that the relative-angular-momentum invariant
$\lM$ obeys a universal upper bound $\,\lM\leq\Ls\,$, with $\,\Ls=
7\,$ or $9$. This can be understood as follows: If
$\mini\lM\mini$ were larger than $7$ or $9$, say, the density of
electrons in the ground state of a (pure) QH fluid would be so small
that it would be energetically more favourable for the electrons to
form a Wigner crystal, thereby destroying the incompressibility of
the system (Lam and Girvin, 1984; Levesque, Weiss, and MacDonald,
1984). This remark shows that the following basic CQHL inequality is of
interest.
\rule[-4mm]{0mm}{5mm}

{\bf Theorem~\ref{ssGT}.2.} {\em For a CQHL \,\QHL, the Hall
fraction $\sH$ and the relative-angular-momentum invariants $\Lm,\
\lm$, and $\,\lM$ satisfy}

\be
{1\over \sH} \:\leq\: \Lm \:\leq\: \lm \:\leq\: \lM\ .
\label{sHbound}
\ee

This theorem is a direct consequence of the {\em Cauchy-Schwarz
inequality\mini} (in the real vector space $\,V\supset \G\mini$),
$<\!\Q\,,\vq \!>^2\,\leq\:<\!\vq,\vq\!>\,<\!\Q\,,\Q\!>$, and the fact
that, for any vector $\,\vq\in\!\G$, with $\,\qel(\vq)\!=
\:<\!\Q\,,\vq\!>\:\neq 0$, we have that $\,<\!\Q\,,\vq\!>^2\,\geq
1$. \rule{2.5mm}{2.5mm}

If we suppose that chiral QH fluids satisfy a universal
bound $\,\lM \leq \Ls\,$ then~(\ref{sHbound}) tells us that CQHLs
with $\,\sH\!<\!1/\Ls\,$ cannot be realized. The data in
Fig.\,\ref{sKE}.1 of Subsect.\,\ref{ssQH}
are consistent with a choice of $\,\Ls =7$.
\rule[-4mm]{0mm}{5mm}

Given these observations on the quantities $N$ and $\mini\lM$, one
is led to the following heuristic principle:
\rule[-4mm]{0mm}{5mm}

{\bf Stability Principle.} {\em A QH fluid described (in the
scaling limit) by a CQHL $(\G,\Q)$ is the\mini} more stable,
{\em the\mini} smaller {\em the value of the invariant $\,\lM$
and, given the value of $\,\lM$, the\mini} smaller  {\em the
dimension $N$ of $\,\G$.}
\rule[-4mm]{0mm}{5mm}

This  principle has been discussed in
detail in Fr\"ohlich, Studer, and Thiran (1995), and in Fr\"ohlich,
Kerler {\em et al.}~(1995). It is also exemplified by the results
summarized in Fig.\,\ref{sClQHFs}.1 below.
\rule[-4mm]{0mm}{5mm}

{\bf Theorem~\ref{ssGT}.3.} {\em Let \,\QHL\, be a CQHL with\,} even
{\em Hall denominator $\dH$. Then the charge parameter $\lb$ has to
be} even, {\em too.}
\rule[-4mm]{0mm}{5mm}

For a proof of this theorem, we first define the vector $\,\vv:=
\lb\mini\dH\mini\Q\in\!\Gs$. Then, for all dual vectors $\,\vn=
\sum_{j=1}^N n_j\, \ved{j} \in\!\Gs$, we find that
\mbox{$\,<\!\vv,\vn\!>\: = \!\lb\mini\dH <\!\Q\,,\vn\!>\: =$}
\mbox{$\sum_{j=1}^N (\Del <\!\Q\,,\ved{j} \!>\mmini /\mini g)\, n_j =
\sum_{j=1}^N (Q^j/g)\,n_j \in \!\BB{Z}\,$}, by using $\,\Del=
g\mini\lb\mini\dH\,$ and the definition of $\mini g\mini$; see points
{\bf (3)} and {\bf (4)} in Subsect.\,\ref{ssBI}\mini. Thus $\vv$ is
actually an element of $(\Gs)^\ast \simeq\G$ and hence, by the
oddness of $\Q$ (see~(\ref{parity})), the congruence
$<\!\Q,\vv\!>\equiv <\!\vv,\vv\!>$ (mod~2) holds.
{}From the definitions of $\sH$ and $\vv$ it follows that the
l.h.s.\ of the congruence equals $\,\lb\mini \nH\,$ and the r.h.s.
equals $\,\lb^2\mini\dH\mini\nH\mini$, i.e.,
$\,\lb\mini\nH\equiv\lb^2\mini\dH\mini\nH\mE\! \pmod{2}$. Finally,
since for $\dH$ even $\nH$ is odd (see point {\bf (3)} in
Subsect.\,\ref{ssBI}), the latter congruence implies that $\lambda$ is
even.
\rule{2.5mm}{2.5mm}

The phenomenologically most interesting implication of
Thm.\,\ref{ssGT}.3 is that, in QH fluids with an even Hall
denominator $\dH$, there exist quasi-particle
excitations with ``fractional'' fractional
charges, i.e., since $\,\lb=2,4,\ldots\,$,

\be
e^\ast \:=\: {1 \over \lambda\mini d_H} \:\leq\: {1 \over 2\mini
d_H}\ .
\label{even}
\ee

\noi
It would be interesting to test this model-independent prediction
experimentally for even-denominator QH fluids at $\,\sH\!=\!1/2\,$
and $5/2$: We predict that
$\,e^\ast\!\leq\mmini 1/4$ (in units where $\,e=\mimini-1$)\mini!
\rule[-4mm]{0mm}{5mm}

{\bf Theorem~\ref{ssGT}.4.} {\em At every Hall fraction $\,\sH=1/m$,
$m$ odd, there is a} unique {\em indecomposable CQHL with the
property that its level $l=\lb\mini g=1$. This CQHL is
one-dimensional and corresponds to the Laughlin fluid at
$\,\sH=1/m$. Moreover, any CQHL with $\,\sH=1/m$, $m$ odd, and
$\,N\geq 2$ has a charge parameter $\,\lb\geq 2$.}
\rule[-4mm]{0mm}{5mm}

A proof of this theorem has been given in Subsect.\,7.5 in
Fr\"ohlich and Thiran (1994).

The CQHLs corresponding to the Laughlin fluids have been described
explicitly in example~{\bf (b)} at the end of
Subsect.\,\ref{ssBI}\mini. By an argument similar
to the one in~(\ref{even}), the last statement in Thm.\,\ref{ssGT}.4
has implications that are, in principle, observable! In
Subsect.\,\ref{ssDis}, an example illustrating this point is
discussed in connection with possible phase transitions at $\,\sH=1$.
\rule[-4mm]{0mm}{5mm}

An interesting subclass of CQHLs is formed by CQHLs with level
$\,l=1$, i.e., their lattice discriminant $\mini\Del\mini$ equals the
Hall denominator $\mini\dH\mini$.  Indecomposable CQHLs with level
$\,l=1$, and thus $\,\lb=g=1$, have been classified for $\,\dH\leq
25\,$ and $N$ below some fairly high
$\mini N_c(\sH)$, typically around $10\mini$; see Fr\"ohlich and
Thiran (1994), and Fr\"ohlich, Studer, and Thiran (1994).

This partial classification has been achieved by combining the recent
lattice-clas\-si\-fi\-ca\-tion results of Conway and
Sloane (1988a) with a systematic investigation of the possible
charge vectors $\Q$ in the dual of all the classified (odd,
integral, euclidean) lattices. For the latter search, one makes use
of the following fact: from the Cauchy-Schwarz inequality and the
defining relation $\,\sH\!=\:<\!\Q\,,\Q\!> \mini$, one infers that,
for a CQHL $(\Gamma$,{\bf Q}),  the dual components $\mini Q_j$ of the charge
vector $\,\Q=\sum_{j=1}^N Q_j\,\ved{j}\,$ are constrained by

\be
Q_j^2 \:\leq\: \sH\, \lM\ , \hfa j=1,\ldots, N\ .
\label{Qbound}
\ee

\noi
Thus, focussing attention on CQHLs with $\,\lM\leq\Ls\,$ and
$\,\sH\leq\sigma_\ast$, Eq.~(\ref{Qbound}) implies that the search
for all possible charge vectors $\Q$ in the dual of a given lattice
$\mini\G$ is a {\em finite} problem. \rule{2.5mm}{2.5mm}

Next, we summarize a few general properties of CQHLs with level
$\,l=1\mini$:
\rule[-4mm]{0mm}{5mm}

{\bf Theorem~\ref{ssGT}.5.} {\em Let \,\QHL\, be a (proper) CQHL with
level $\,l=\lb\mini g=1$. Then}

\hspace*{5.3mm}{\bf (i)} {\em $\dH$ is} odd, {\em and} $\,\Gs/\G
\simeq \BB{Z}_{\mini\dH}\,$;

\ssp{\bf (ii)} {\em in order to realize a Hall fraction $\sH$ with
$\nH$ even (odd), $N\mini$ has to be even

\sspp (odd, and $\,N\equiv\nH$ (mod $4$)\mini);}

\hspace*{2.6mm}{\bf (iii)} {\em for quasi-particles labelled by
$\,\vn\in\!\Gs\!$, a charge-statistics relation holds:

\sspp if $\,\qel(\vn)=\varepsilon/\dH\,$ then $\,\theta(\vn)
\equiv (\nH)^{-1}\,\varepsilon^2/\dH\!$ (mod $2$).}
\rule[-4mm]{0mm}{5mm}

In the last statement of this theorem, the number
$(\nH)^{-1}$ is defined as follows: if $\nH$ is odd, then
$\,\nH\mini(\nH)^{-1} \equiv 1$ (mod $2 \dH$), and if $\nH$ is even,
then $\,(\nH)^{-1}:=2\mini(2\nH)^{-1}+\dH\mini$, with $\,2\nH\mini
(2\nH)^{-1} \equiv 1$ (mod $\dH$). A proof of this theorem can be
found in Sect.\,5 in Fr\"ohlich and Thiran (1994).

\subsubsection{Shift Maps and their Implications}

In the remaining part of this subsection, we study ``structurally
similar'' chiral QH fluids. At the level of CQHLs, ``structural''
relationships are realized by particular maps, called {\em shift
maps}. {}From a classification point of view, shift maps
allow\,--\,under suitable conditions\,--\,to immediately carry over
classification results for CQHLs with Hall fractions in a certain
interval to corresponding results for other intervals.
Phenomenologically  interesting implications of structural
relationships are outlined in Subsect.\,\ref{ssDis}\mini.

First, we divide the interval $(0,\infty)$ of possible Hall
fractions $\mini\sH\mini$ into a sequence of suitable subintervals,
``{\em windows\,}'', $\Sigp^\pm$ defined by

\ben
\Sigp^+ \::=\: \{\, \sH\ |\ {1\over 2\mini p+1}\leq \sH <{1\over
2\mini p}\,\}\ ,\hsp p=1,2,\ldots\ ,
\een

\vav

\be
\Sigp^- \::=\: \{\, \sH\ |\ {1\over 2\mini p}\leq \sH <{1\over
2\mini p-1}\,\}\ ,\hsp p=1,2,\ldots\ .
\label{Sig}
\ee

\noi
The ``$+$'' superscripts are
chosen because these subintervals contain the ``first
main series'' of Hall fractions, $\,\sH=N/(2\mini pN\!+\! 1),\
N=1,2,\ldots\ $. Similarly, the ``$-$'' superscripts for the
``complementary'' windows indicate that these windows contain the
``second main series'' of Hall fractions, $\,\sH=N/(2\mini p N
\!-\!1),\ N=2,3,\ldots\ $. Moreover, we denote by $\Sigma_0^+$ the
interval $\,[\mini 1\mini,\infty)$, and by $\Sigp$ the union of the
two complementary subintervals $\mini\Sigp^+$ and $\mini\Sigp^-$,
i.e., $\Sigp:=\Sigp^+ \cup\Sigp^-,\ p=1,2,\ldots\ $.

Second, we define a class of CQHLs that will figure prominently in
the sequel.
\rule[-4mm]{0mm}{5mm}

{\bf Definition~\ref{ssGT}.1.} {\em A primitive CQHL \QHL\, (see
Def.\,\ref{ssQHFQHL}.2) with Hall fraction $\,\sH\in\Sigp\,$ is
called\,} \Lmini\ {\em if, and only if, $\,\lM$ takes the smallest
possible value consistent with~(\ref{sHbound}), namely $\,\lM=2\mini
p+1,\ p=1,2,\ldots\ $.}
\rule[-4mm]{0mm}{5mm}

\noi
By (\ref{lmin})--(\ref{mlM}), \Lmini\ CQHLs satisfy $\,\Lm=\lm=
\LM=\lM=2\mini p+1$. General, powerful implications that follow from
$L$-minimality are summarized below in
Thms.\,\ref{ssGT}.6--\ref{ssGT}.8; for proofs, see Fr\"ohlich, Kerler
{\em et al.}~(1995).
\rule[-4mm]{0mm}{5mm}

{\bf Theorem~\ref{ssGT}.6.} {\em For $p=1,2,\ldots\,$, let \,\QHL\,
be a (proper) CQHL with $\,\sH\in\Sigp\,$ and $\,\LM=2\mini p+1$.
Then \,\QHL\, is primitive and\,} \Lmini, {\em i.e., we also have that
$\,\Lm=\lM =2\mini p+1$. Moreover, \,\QHL\, is\,}
indecomposable {\em if $\,\sH<2/3\mini$.}
\rule[-4mm]{0mm}{5mm}

The bound $\,\sH<2/3\,$ for indecomposability is
sharp. As a matter of fact, at $\,\sH=2/3\mini$, there is an
\Lmini $\ (\lM\mimini=3)$ {\em composite} CQHL. It is given by the
direct sum of two Laughlin fluids at $\,\sH=1/3\mini$; see
example~{\bf (b)} in Subsect.\,\ref{ssBI}\mini.

Next,  ``shift maps'' are defined.
\rule[-4mm]{0mm}{5mm}

{\bf Definition~\ref{ssGT}.2.} {\em Shift maps, denoted by
$\mini\Sp\mini,\ p=1,2,\ldots\,$, are maps between (proper) CQHLs of
equal dimensions, $\Sp\mini:\, \QHL \mapsto (\G^\prime,\Q^\prime)$.
Starting from an arbitrary basis $\{\ve{1},\ldots,\ve{N}\}$ of
\,\QHL, the image $(\G^\prime, \Q^\prime)$ is uniquely specified by
constructing a basis $\{\ve{1}^\prime,\ldots,\ve{N}^\prime \}$ and
a charge vector $\mini\Q^\prime$ that satisfy the conditions}

\ben
K_{ij}^\prime \:=\; <\!\ve{i}^\prime,\ve{j}^\prime\!> \;=\;
<\!\ve{i},\ve{j}\!> +\, 2\mini p <\!\Q\,,\ve{i}\!>\,
<\!\Q\,,\ve{j}\!> \;=\: K_{ij} + 2\mini p\, \qel(\ve{i}) \mini
\qel(\ve{j})\ ,
\een

\vav

\be
Q_i^\prime \:=\; <\!\Q^\prime,\ve{i}^\prime\!> \;=\;
<\!\Q\,,\ve{i}\!> \;=\: Q_i\ , \hfa i,j=1,\ldots,N\ .
\label{shiftdef}
\ee

\noi
Although the conditions in~(\ref{shiftdef}) are
formulated w.r.t.\ given bases, they specify the image $(\G^\prime,
\Q^\prime)$ uniquely, since different choices of bases and charge
vectors in~(\ref{shiftdef}) simply lead to data pairs $(K^\prime,
\Qvp)$ for $(\G^\prime, \Q^\prime)$ which are related by the
equivalence transformations~(\ref{equiv}).

Denoting by $\,\G_0\subset\G\,$ the {\em neutral sublattice} of a
CQHL \QHL, i.e.,

\be
\G_0 \::=\: \{\, \vq\in\G\ |\ <\!\Q\,,\vq\!>\:=\qel(\vq)=0 \,\}\ ,
\label{neutraldef}
\ee

\noi
it is straightforward to show that shift maps leave neutral
sublattices invariant,

\be
\G_0^\prime \:=\: \G_0\ .
\label{neutral}
\ee

\noi
As will be explained in more detail in Subsect.\,\ref{ssMS},
Eq.~(\ref{neutral}) implies that (in the scaling limit) the
corresponding chiral QH fluids exhibit the {\em same symmetries}.
This fact is the mathematical basis for calling two chiral QH
fluids {\em structurally similar}.

What is the action of the shift map $\,\Sp\mini:\, \QHL \mapsto
(\G^\prime,\Q^\prime)$, for $\,p=1,2,\ldots\,$, on the space of
invariants introduced in Subsect.\,\ref{ssBI}\mini?

{\bf (i)} The discriminant $\Del^\prime$ of the (odd, integral,
euclidean) lattice $\mini\G^\prime$ is given by

\be
\Del^\prime \:=\: \Del\,(1+2\mini p\,\sH)\ .
\ee

{\bf (ii)} The Hall conductivity changes according to

\be
{1\over \sH^\prime} \:=\: {1\over \sH} + 2\mini p\ ,
\label{sHprime}
\ee

\noi
which corresponds to the ``$D$-operation'' in the
Jain-Goldman (1992) hierarchy scheme; see also Jain (1989, 1990,
1992) and Fr\"ohlich and Zee (1991). Note that Eq.~(\ref{sHprime})
implies that any CQHL which is the image under a shift map
$\mini\Sp\mini,\ p=1,2,\ldots\,$, necessarily has a Hall fraction
strictly below $1/(2\mini p)\mini$.

{\bf (iii)} The level $\mini l$, $g$, and the charge parameter $\lb$
are all invariant under the action of a shift map $\mini\Sp\mini$.

We summarize {\bf (i)--(iii)} by giving a succinct representation of
the action of the shift map $\mini\Sp\mini$ at the level of the CQHL
symbol,

\be
\QHLsymbol{N}{\sH=\ndH}{g}{\lb}
\;\:\stackrel{\Sp}{\longmapsto}\;
\QHLsymbol{N}{\sH^\prime = {{n_H\rule[-.8mm]{0mm}{2mm}} \over
d_H+2\mini p\,n_H}}{g}{\lb}\ ,\ p=1,2,\ldots\ .
\label{shift}
\ee

{\bf (iv)} The name ``shift map'' for $\Sp$ is motivated by the fact
that the relative-angular-momentum invariants $\Lm$ and $\LM$ are
simply shifted by $2\mini p$,

\be
\Lm^\prime \:=\: \Lm + 2\mini p\ , \hah \LM^\prime \:=\: \LM +
2\mini p\ .
\label{LL}
\ee

\noi
Unfortunately, for the invariants $\mini\lm$ and
$\mini\lM$ of {\em generic} primitive CQHLs, there does {\em not,
in general}, hold a transformation rule similarly simple
to~(\ref{LL})!\,--\,However, for {\em indecomposable}
CQHLs, the identities $\,\lm = \Lm\,$ and $\,\lM = \LM\,$ hold.

{}From the definitions above one sees that the shift maps $\Sp$
are invertible on the set of (proper) CQHLs with Hall fractions
$\,\sH\leq 1/(2\mini p),\ p=1,2,\ldots\
$.\,--\,From~(\ref{shiftdef}) it follows that $\,\Sp^{-1} =
{\cal S}_{-p}$.\,--\,The preimages of these CQHLs are readily seen
to be (proper) CQHLs. The set of (proper) CQHLs is closed under the
action of the maps $\Sp$ and their inverses (when defined).

However, the maps $\Sp$ and their inverses do {\em not\mini}
necessarily preserve the decomposability properties of
CQHLs.\,--\,E.g., composite CQHLs can be mapped into indecomposable
ones, as illustrated in Thm.\,\ref{ssGT}.8 below.\,--\,Moreover, the
maps $\Sp$ and their inverses do {\em not}, in general, preserve the
primitivity property we have imposed on
(composite) CQHLs; see Def.\,\ref{ssQHFQHL}.2.\,--\,For an
example of a primitive CQHL with a preimage that is non-primitive,
see Sect.\,4 in~Fr\"ohlich, Kerler {\em et al.}~(1995).\,--\,From
these remarks and definitions~(\ref{lmin}) and~(\ref{lmax}) of
the invariants $\mini\lm$ and $\mini\lM$ it is clear that the
transformation properties of these invariants under shift maps are
not as straightforward as the ones in~(\ref{LL}).

The main objective of the present section is the classification of
primitive CQHLs. Although this set is not closed under the action of
shift maps and their inverses, it is remarkable that a subset of the
primitive CQHLs, the class of {\em\Lmini\,} CQHLs (see
Def.\,\ref{ssGT}.1) {\em is closed\,} under the action of shift maps
and their inverses. This is the key to powerful classification
results described presently.

It is convenient to partition the class of \Lmini\ CQHLs into the
following subsets:

\be
\Hp^\pm \::=\: \{\, \QHL\ |\ \sH\in\Sigp^\pm\mini,\mbox{ \Lmini,
i.e., primitive and }\,\lm=\lM=2\mini p+1\,\}\ ,
\ee

\noi
where $\,p=(0),1,2,\ldots\,$, in accordance with the definition of
the windows $\Sigp^\pm$ given in~(\ref{Sig}).

The next two theorems show that, on the one hand, the sets
$\,\Hp:=\Hp^+\cup\Hp^-\,$ are structurally similar for different
$\mini p$'s, while, on the other hand, there is an essential
{\em structural asymmetry\mini} between the sets $\Hp^+$ and
$\Hp^-\mmini$, for a given $\mini p$.
\rule[-4mm]{0mm}{5mm}

{\bf Theorem~\ref{ssGT}.7 (Bijection Theorem).} {\em The sets $\Hp$ of
\mini\Lmini\ CQHLs with $\,\sH\in\!\Sigp\mini$, for
$\,p=2,3,\ldots\,$, are in\mini} one-to-one correspondence {\em with
the set $\Hi$. The corresponding bijections are realized by the
shift maps $\,{\cal S}_{p-1} \mini : \, \Hi \rightarrow \Hp\mini$.}
\rule[-4mm]{0mm}{5mm}

The proof of this theorem rests on Thm.\,\ref{ssGT}.6
above, and it should be emphasized that chirality and $L$-minimality
are crucial for the theorem to hold; see Fr\"ohlich, Kerler {\em et
al.}~(1995). The bijection theorem (Thm.\,\ref{ssGT}.7) implies that,
for the classification of \Lmini\ CQHLs, we can restrict our
analysis to the lattices with Hall fractions $\sH$ in the
``fundamental domain'' $\,\Sigi:=[\mini 1/3\mini, 1)\mini$!
\rule[-4mm]{0mm}{5mm}

In Fr\"ohlich, Kerler {\em et al.}~(1995), the set $\mini{\cal
H}_0^+$ of \Lmini\ CQHLs with $\,\sH\in\![\mini 1\mini,\infty)\,$ has
been constructed. Applying the shift map ${\cal S}_1$ to it, we
obtain the set $\Hi^+$ of \Lmini\ CQHLs in the window
$\,\Sigi^+:=[\mini 1/3\mini,1/2)\subset \Sigi\mini$. Hence, by the
bijection theorem, all the sets $\Hp^+,\ p\geq 1$, are known. In
fact, we have the following result.
\rule[-4mm]{0mm}{5mm}

{\bf Theorem~\ref{ssGT}.8 (Uniqueness Theorem).} {\em  For each
$\,p=0,1,2,\ldots\,$, the set $\mini\Hp^+$ of \mini\Lmini\ CQHLs
with $\,\sH\in\!\Sigp^+\,$ coincides with the (infinite) series,
$N=1,2,\ldots\,$, of indecomposable, $N$-dimensional, maximally
symmetric CQHLs with $SU(N)$ symmetry of $\mini N$-ality $1$,
meaning that the elementary charge\mini-1 fermions (electrons)
described by these CQHLs transform under the fundamental
representation of $SU(N)$. For a given $p$, the corresponding
symbols read}

\be
\QHLsymbol{N}{\sH={N\over 2\mini pN+1}}{1}{1} \;[\mini \lm=\lM=
2\mini p+1\mini]\ ,\hsp N=1,2,\ldots  \ .
\label{symbHp+}
\ee

The maximally symmetric CQHLs of this theorem have been described
explicitly in example~{\bf (c)} at the end of
Subsect.\,\ref{ssBI}\mini. Each Hall fraction appearing
in~(\ref{symbHp+}) is realized by a unique \Lmini CQHL with
$SU(N)$ symmetry (of $N$-ality 1). In the notation of the next
subsection (see~(\ref{dataLoG}) below), the sets $\Hp^+$ can be
written as

\be
\Hp^+ \:=\: \{ \: (\mini2\mini p+1\,|\,^1\mA A_{N-1})\ |\
N=1,2,\ldots\: \}\ .
\label{Hp+}
\ee

In Thm.~\ref{ssGT}.6, it has been stated that all CQHLs
in~(\ref{Hp+}) with $\,p>0\,$ are {\em indecomposable}.
Furthermore, since their level $l$ equals unity, Thm.\,\ref{ssGT}.5
states that a {\em charge-statistics relation} holds for the
quasi-particle excitations of the corresponding QH fluids.

A first comparison of the Hall fractions appearing in~\cfr{symbHp+}
with the experimental data presented in Fig.\,\ref{sKE}.1 of
Subsect.\,\ref{ssQH} reveals an impressive agreement between the Hall
fractions predicted by \Lmini\ CQHLs and the experimentally observed
values in the windows $\Sigp^+,\ p=1,2$, and $3\mini$. Note that
CQHLs with higher dimensions and/or higher values of $\,\lM$ are
associated with less stable QH fluids, which is in accordance with
our stability principle advocated at the beginning of this
subsection. In the windows, $\Sigp^+,\ p=1,2$, and $3\mini$, there
is {\it only one} Hall fraction, $4/11$, for which there are some
experimental indications (however, only very weak ones!) that {\it cannot}
be realized by an \Lmini\ CQHL.


\subsection{Maximally Symmetric CQHLs}
\label{ssMS}

Maximally symmetric CQHLs form the most natural generalizations of
the ``elementary'' $A\mini$- (or $su(N)$-) fluids that appeared in
the uniqueness theorem (Thm.\,\ref{ssGT}.8) of the last subsection
and which have been shown to encompass the Laughlin fluids as well
as the ``basic'' Jain fluids. Before we can give a precise
definition of the class of maximally symmetric CQHLs, we need to
investigate a general geometrical feature of CQHLs, namely their
``Witt sublattices''. We use technical language and then translate
our definitions into  explicit statements at the level of the data
pairs $(K,\Qv)$ associated with CQHLs; for the definition of these
pairs, see the beginning of Subsect.\,\ref{ssBI}\mini.

Let \QHL\, be a CQHL. Then the {\em Witt sublattice}, $\mini\Gw
\subset\G$, is defined to be the sublattice of $\mini\G$ generated by
all vectors of length squared $1$ and $2$. The general theory of
integral, euclidean lattices (Conway and Sloane, 1988a, 1988b) tells
us that $\mini\Gw$ is of the form

\be
\Gw \:=\: \G_A \oplus \G_D \oplus \G_E \oplus I_l\ ,
\label{Witt}
\ee

\noi
where $I_l$ denotes the (self-dual) simple euclidean lattice in
$\mini l\mini$ dimensions and $\,\G_A,\,\G_D$, and $\mini\G_E$ are
direct sums of root lattices of the simple Lie algebras $\mini
A_{m-1}=su(m), \,D_{m+2} =so(2m+4),\ m=2,3,\ldots$, and $E_m,\
m=6,7,8$, respectively. The subscripts in the symbols $\mini
A_n,\,D_n$, and $E_n$ indicate the {\em ranks} of these algebras. We
note that all the root lattices of these Lie algebras are generated
by vectors of only one length, namely of length squared $2$. [In the
mathematical literature, the $A\mini$-, $D\mini$-, and $E\mini$-Lie
algebras are called simply-laced.] For a general reference on Lie
algebras and their representation theory, see Slansky (1981).

Denoting by $\OO$ the orthogonal complement of $\mini\Gw$ in $\mini\G$,
whose dimension satisfies $\,\dim \OO = N-\dim\Gw \geq 1$, the
sublattice $\mini\Gw\oplus\OO$ is  called the {\em Kneser
shape\mini} of $\mini\G$, and one has the following embeddings of
lattices:

\be
\Gw\oplus\OO \:\subseteq\: \G \:\subseteq\: \Gs \:\subseteq\:
\Gws\oplus\OO^\ast\ ,
\label{Kneser}
\ee

\noi
where $\,{}^\ast\,$ denotes the dual of a lattice, as defined
in~(\ref{Gs}).

It can be shown (Fr\"ohlich and Thiran, 1994) that, for {\em
indecomposable\mini} CQHLs \QHL, $\Gw$ does {\em not\,} contain any
$I_l$ and $\mini\G_{E_8}$ sublattices. In the following, we will
concentrate on indecomposable CQHLs, or, correspondingly, on
``elementary'' chiral QH fluids.
\rule[-4mm]{0mm}{5mm}

{\bf Theorem~\ref{ssMS}.1.} {\em Let \QHL\, be an indecomposable
CQHL with $\,\sH<2\,$. Then $\Q$ is\,} orthogonal {\em to $\mini\Gw$,
i.e., $\Q\in\!\OO^\ast$, and $\,\Gw\subseteq\G_0\,$ where $\mini\G_0$ is
the neutral sublattice of \QHL. Moreover, if $\,\Gw\neq \emptyset$
all the inclusions in~(\ref{Kneser}) are} proper.
\rule[-4mm]{0mm}{5mm}

For a proof of this theorem and more details on the constructions
above\,--\,which constitute the basis of the complete
classification program of (general) CQHLs\,--,
see Sect.\,6 in Fr\"ohlich and Thiran (1994).

Thm.\,\ref{ssMS}.1 has an interesting corollary concerning the
{\em symmetry properties} of the chiral QH fluid corresponding to
\QHL: To every point in $\mini\G$ there corresponds a
vertex operator of the algebra of edge currents (compare with
Subsect.\,\ref{ssEE}). Let $\mini\GG$ denote the Lie algebra\,--\, a
direct sum of simple algebras $\mini A_n, \,D_n$, and
$E_{6,7}$\,--\,whose root lattice is given by the Witt sublattice
$\mini\Gw$ of \QHL. It is not hard to show (Fr\"ohlich and Zee,
1991; Fr\"ohlich and Thiran, 1994) that the algebra generated by the
vertex operators corresponding to the Witt sublattice, $\Gw$, of
$\mini\G$ and the neutral $u(1)$ currents is the enveloping algebra
of the Kac-Moody current algebra $\mini\GGh$  at level 1 (denoted
$\mini\GGh_1$).

The (infinite-dimensional) {\em symmetry algebra} $\mini\GGh_1$
canonically contains the (finite dimensional) Lie algebra $\mini\GG$
that consists of global symmetry generators. Thus the
Lie group $G$ corresponding to $\mini\GG$ is the {\em group of global
symmetries\mini} of the QH fluid. This implies that,
given $m\mini$ electrons\,--\,fermionic quasi-particles with
charge $1$ and labelled by, say, $\,\vq_1, \ldots,
\vq_m\in\!\G\subset\Gs$\,--, they transform under particular unitary
irreducible representations (irreps.) of $\mini G$. These unitary
irreps.\ are specified as follows: Let

\be
\vq_{\mini i} \:=\: \vq_{\mini i,W} + \vq_{\mini i}^\prime\,, \hwh
\vq_{\mini i,W}\in\Gws\,, \hah \vq_{\mini i}^\prime
\in\OO^\ast\ ,
\label{weights}
\ee

\noi
be the decomposition, according to~(\ref{Kneser}), of the $i\mini$th
electron vector, $i=1,\ldots,m$. Then we may write $\,\vq_{\mini
i,W}=\vom_i+ {\bf r}_i\mini$, with $\,{\bf r}_i\in\!\Gw$ a {\em root
vector}, and with $\,\vom_i \in\!\Gws\,$ an {\em elementary
weight\mini}, i.e., a smallest length representative of the cosets
(or ``congruence classes'' in Lie group terminology) in the quotient
$\mini\Gws/\Gw$ (see, e.g., Slansky, 1981). Furthermore, by the
general representation theory of Lie - and Kac-Moody algebras (see,
e.g., Goddard and Olive, 1986), the elementary weight $\vom_i$
determines uniquely a unitary irrep., $\pi_{\vom_i}$, of $\mini G$
according to which the one-electron state labelled by $\vq_{\mini
i}$ transforms, $i=1,\ldots,m$.

{}From the general results about lattices given in Conway and Sloane
(1988a), it follows that all
the elementary weights $\,\vom_i\in\!\Gw^\ast\,$ that can appear
in~(\ref{weights}) are such that the corresponding irreps.\ of
$\mini\GG$ {\em can be extended\,} to unitary highest-weight
representations of $\mini\GGh$ at level $\mini 1$. For a discussion
of the latter point, see, e.g., Subsect.\,3.4 in Goddard and Olive
(1986). We call these elementary weights ``{\em
admissible\mini}'' weights, and the ones that can occur for the
simple algebras $\mini A_n,\,D_n$, and $E_{6,7}$ have been given
explicitly in Appendix~A in Fr\"ohlich, Studer, and Thiran (1995).

One can show (Fr\"ohlich and Thiran, 1994) that if $\,\dim\OO=1\,$
and $\,\G_0=\Gw\,$ then all one-electron states transform under the
{\it same} unitary irrep.\ $\pi_{\vom}$ of $\mini G$, i.e., $\vq_{\mini
1,W}\equiv\cdots \equiv\vq_{\mini m,W}\equiv \vom\,$ mod $\mini\Gw$.
\rule[-4mm]{0mm}{5mm}

These results motivate the following definition of
{\em maximally symmetric CQHLs}.
\rule[-4mm]{0mm}{5mm}

{\bf Definition~\ref{ssMS}.1.} {\em A (proper) CQHL \QHL\, is
called\,} maximally symmetric {\em iff $\,\dim \OO=1\,$
and $\,\G_0=\Gw\,$, i.e., the neutral sublattice of \mini\QHL\, and
its Witt sublattice coincide. Furthermore, denoting by $\mini\GG$
the Lie algebra associated with the root lattice $\mini\Gw$, the
one-electron states described by \mini\QHL\, are required to
transform under a unitary irrep.\ of the Lie algebra $\,\GG$
which can be extended to a unitary highest-weight representation of
$\,\GGh_1\mini$.}
\rule[-4mm]{0mm}{5mm}

Maximally symmetric CQHLs \QHL\, are specified by the data

\be
(\, L\:|\; {}^{\vom} \Gw \,)\ ,
\label{dataLoG}
\ee

\noi
where $L\,$ is an odd, positive integer, $\,\Gw = \G_A \oplus
\G_D \oplus \G_{E\neq 8}\,$ is the Witt sublattice of \QHL, and
$\,\vom\in\!\Gws/\Gw\,$ is an admissible weight labelling an irrep.\ of
the Lie algebra $\mini\GG$ associated with $\mini\Gw$. The possible weights $\vom$
are further restricted by the value of $L$, namely
$\,<\!\vom,\vom\!>\;< L\mini$; see~(\ref{Lgrom}) below.

If the Witt sublattice is a direct sum of simple root
lattices, $\Gw=\Gwi{1}\oplus\cdots\oplus\Gwi{k},\ k\geq 2$, then
the associated Lie algebra $\mini\GG$ is semi-simple with decomposition
$\,\GG=\GG_1\oplus\cdots\oplus\GG_k$, and, correspondingly, the
admissible weight reads $\,\vom=\vom_1 +\cdots+\vom_k\,$
where $\,\vom_i\in\!\Gwi{i}^\ast/\Gwi{i},\ i=1,\ldots,k$. In order to
get an {\em indecomposable\mini} lattice, every projection $\vom_i$
must represent a non-trivial coset in $\mini\Gwi{i}^\ast/\Gwi{i}$.
This can also be shown to be sufficient; see Fr\"ohlich, Kerler, and
Thiran (1995). In the sequel, we always assume that admissible
weights $\vom$ fulfill this requirement. Hence {\em all\,} the
maximally symmetric CQHLs mentioned in this section are {\em
indecomposable\mini}.

Equivalently to~(\ref{dataLoG}), we can also specify maximally
symmetric CQHLs \QHL\, by their corresponding data pair $(K,\Qv)$,
once a basis has been chosen in $\mini\G$; see the beginning of
Subsect.\,\ref{ssBI}\mini. Relative to a ``normal'' basis
$\{\vq,\ve{1},\ldots,\ve{{N-1}}\}$ of $\mini\G$, \QHL\, is specified by

\be
K\:=\,
\left.\left(\renewcommand{\arraystretch}{1.7}
\begin{tabular}{c|c}
$L$ & $\omv\rule[-3.5mm]{0mm}{5mm} $  \\
\hline
$\omv^T$ & $\,C(\Gw)\rule[-3.5mm]{0mm}{5mm} $
\end{tabular} \right)\  \right\}\mbox{\scriptsize N}
\ ,\hah
\Qv \:=\:
(\mini\underbrace{1,0,\ldots,0}_{\mbox{\scriptsize N}}\,) \ ,
\label{dataKQ}
\ee

\noi
where $\,L=\; <\!\vq\,,\vq\!>\,$ is the same odd integer as
in~(\ref{dataLoG}), $C(\Gw)$ is the Gram matrix of the basis
$\{\ve{1},\ldots,\ve{{N-1}}\}$ of $\mini\Gw$\,--\,in a convenient
normal basis, it equals the {\em Cartan matrix} of the Lie algebra
$\mini\GG$ associated with $\mini\Gw$\,--, and, finally,
$\,\omv=(\omega_1,\ldots,\omega_{N-1})\,$ is the vector of the dual
components of $\vom$ which are given by $\,\omega_j
=\:<\!\vom,\ve{j}\!>\mini,\ j=1,\ldots,N-1$. According to the
decomposition~(\ref{Kneser}), the basis vector $\vq$ can be written
as $\,\vq ={\sH}^{-1} \Q + \vom \mini$.

If $\mini\Gw$ is a direct sum of simple root lattices then
$\,C(\Gw)=C(\Gwi{1})\oplus\cdots\oplus C(\Gwi{k})\,$ is a
block-diagonal matrix, and $\,\omv=\omvi{1}+\cdots+\omvi{k} \mini$.
An example of data pairs~(\ref{dataLoG}) and~(\ref{dataKQ}) has been
given by~(\ref{Hp+}) and~(\ref{promKQ}), respectively. The explicit
forms of the Cartan matrices for the simple algebras $\mini A_n,
\,D_n$, and $E_{6,7}$, and of the dual vectors $\omv$ for the
admissible weights $\vom$ have been given in Appendix~A of
Fr\"ohlich, Studer, and Thiran (1995).

We denote by $\Del(\Gw)$ the discriminant of the Witt
sublattice, $\Gw$, of $\mini\G$, i.e., $\,\Del(\Gw) := \det C(\Gw)
=| \Gws/\Gw\mini |$. {}From~(\ref{dataKQ}), it immediately follows that
for maximally symmetric CQHLs

\bea
\Del\:=\: \det K & = & \Del(\Gw)\:[\mini L-\,\omv\cdot
C(\Gw)^{-1}\omv^T\mini] \nonumber \\
 & = & \Del(\Gw)\:[\mini L\,- <\!\vom,\vom\!>\mini ]\ssp
(\mini>0\mini)\ ,
\label{Lgrom}
\eea

\vav

\be
\sH \:=\;\, <\!\Q\,,\Q\!> \;=\: \Qv\cdot\Ki\Qv^T \:=\:
{\Del(\Gw) \over \Del} \:>\: {1\over L}\ .
\label{sHms}
\ee

\noi
These two equations are basic for proving the following theorem.
\rule[-4mm]{0mm}{5mm}

{\bf Theorem~\ref{ssMS}.2.} {\em The symbol of a maximally symmetric
CQHL \,\QHL\, specified by~(\ref{dataLoG}) or, equivalently,
by~(\ref{dataKQ}), takes the form}

\be
\QHLsymbol{1\mini+\,\ranks\Gw\mini=\mini N}{\sH={1\over
L\,-<\!\vom,\vom\!>}}{g\mini=\mini \Del(\Gw)/\hws}
{\hspace{-17.8mm}\lb\mini=\mini\hws/n_H} \hsp .
\label{symbms}
\ee

\noi
{\em where $\hw$ is the order of the elementary weight $\vom$ in
$\mini\Gws/\Gw\mini$. Furthermore, for the
relative-angular-momentum invariants $\,\lM$ and $\,\lm$, the
equalities $\,\lm=\lM=L\,$ hold.}
\rule[-4mm]{0mm}{5mm}

If $\mini\Gw$ is a direct sum of simple root lattices,
$\Gw=\Gwi{1} \oplus\cdots\oplus\Gwi{k},\ k\geq 2$, and $\,\vom=
\vom_1 +\cdots+\vom_k$, as above, then the following
identities hold:

\bea
\rank\Gw &=& \sum_{i=1}^k \rank\Gwi{i}\ , \nonumber \\
<\!\vom,\vom\!> &=& \sum_{i=1}^k <\!\vom_i,\vom_i\!>\ , \nonumber \\
\Del(\Gw) \:=\: \det C(\Gw) &=& \prod_{i=1}^k \det C(\Gwi{i})\ ,
\nonumber
\eea

\vav

\bea
\hw &=& \lcm(\mini h_{\vom_1},\ldots,h_{\vom_{\mimini k}})\ ,
\label{dsum}
\eea

\noi
where the {\em least common multiple\mini} (lcm) of two
integers $\mini a\mini$ and $\mini b\mini$ is defined by
$\,\lcm(a,b)\!:= a\mini b/\gcd(a,b)$, and similarly for more than
two integers.

For the simple Lie algebras $\mini A_{m-1}=su(m), \,D_{m+2}
=so(2m+4),\ m=2,3,\ldots\,$, and $E_{6,7}$, all the ranks and
determinants of their Cartan matrices, as well as all the lengths
squared and orders of their admissible weights can be found in
Appendix~A of Fr\"ohlich, Studer, and Thiran (1995).

\subsubsection{Classification}

Exploiting the results of Thm.\,\ref{ssMS}.2 and the identities
in~(\ref{dsum}), it is possible to list all maximally symmetric
CQHLs which have a fixed value of $L$ and whose Hall fractions $\sH$
($>1/L\mini$; see~(\ref{sHms})) belong to a given interval. In
Fr\"ohlich, Studer, and Thiran (1995), {\em all\,} maximally
symmetric CQHLs with $\,\lm=\lM=L=3\,$ and $\,\sH<1\,$ have been
listed. They are organized in four infinite one-para\-meter, one
infinite two-para\-meter, and six finite series of CQHLs. For a
physically relevant subset of Hall fractions (with $\,\dH\!<\!21\,$
and odd), the resulting CQHLs are indicated in
Fig.\,\ref{sClQHFs}.1 below.

The most powerful implication of this classification result is
obtained by combining it with the discussion about the shift maps in
the second part of Subsect.\,\ref{ssGT}: In Fr\"ohlich,
Studer, and Thiran (1995), we have shown that this leads to {\em
the complete classification of \mini\Lmini, maximally symmetric
CQHLs}. For some alternative ideas on the classification of QH fluids,
see also Cappelli et al.~(1995).

Some of our main insights will be summarized in the following
subsection; see, in particular, Fig.\,\ref{sClQHFs}.1\mini.
For a detailed discussion of this classification result and comparison
with experimental data, however, we refer to the
second part of Sect.\,5 in Fr\"ohlich, Studer, and Thiran (1995).


\subsection{Summary and Physical Implications of the Classification
Results}
\label{ssDis}

The purpose of this subsection is to summarize phenomenological
implications of the theoretical results presented in the previous
subsections. We confront these results
with the experimental data presented in Subsect.\,\ref{ssQH}; see, in
particular, Fig.\,\ref{sKE}.1 which summarizes data on
single-layer QH systems. So far, we have focused our attention mainly
on {\em chiral\,} QH lattices with Hall fractions $\,\sH\leq
1\mini$. We recall that such lattices describe QH fluids that have
only electrons (or holes) as fundamental charge carriers and whose
edge currents are of {\em  definite chirality\mini}. When restricting
considerations to CQHLs, we do {\em not\,}
make use of the idea of ``charge conjugation'', i.e.,
$\sH=1-\sH^{\mini\prime}$, in the discussion of QH fluids with
$\,1/2<\sH<1\mini$. In this interval, charge conjugation is usually
invoked in other approaches to the fractional QH effect; see Haldane
(1983), Halperin (1984), Jain (1989, 1990, 1992), and Jain and
Goldman (1992). There is no difficulty, within our framework, to go
beyond the chirality assumption and to consider the general
classification problem of mixed-chirality QH lattices including,
e.g., those corresponding to the QH fluids proposed by the
charge-conjugation picture; see Fr\"ohlich, Studer, and Thiran
(1995). The {\em general\,} problem, however, is exorbitantly
involved, and conclusions lack the simplicity and transparency of
the results obtained when restricting the analysis to the class of
{\em chiral\,} QH lattices. Experimentally, it would be interesting
to test the chirality assumption by direct edge-current measurements,
e.g., of the type reported in Ashoori {\em et al.}~(1992); (which,
by the way, are compatible with a {\em purely chiral\,} structure
for the QH fluid at $\,\sH=2/3$ that has been studied there).

\subsubsection{General Structuring Results}

\noi
$\bullet\ $ {\em The Inequality $\,\lM\geq\sH^{-1}$.} The first
important result of our analysis is that, for an {\em arbitrary\,}
CQHL $(\G,\Q)$, the associated invariants $\sH$ and $\,\lM$ satisfy
the fundamental inequality $\,\lM \geq \sH^{-1}$;
see~\cfr{sHbound}. This result implies a first {\em organizing
principle\,} for CQHLs with $\,0<\sH\leq 1\mini$. It offers a
natural splitting of the interval $\mini(0,1]\mini$ into successive
windows, $\Sigp$, defined by $\,1/(2p+1)\leq \sH < 1/(2p-1),\
p=1,2,\ldots\ $. The relative-angular-momentum invariant $\,\lM$
characterizing an arbitrary CQHL with $\,\sH\in\!\Sigp\,$ is thus
greater or equal to $\,2\mini p + 1,\ p=1,2,\ldots\ $. Adopting the
(heuristic) stability principle of Subsect.\,\ref{ssGT}, CQHLs in
successive windows $\Sigp$, for {\em increasing\mini} values of $\mini
p\mini$, are expected to describe QH fluids of {\em decreasing\mini}
stability.

Combining the fundamental inequality above with a (universal) upper
bound $\Ls$ on the relative-angular-momentum invariant
$\,\lM$, we conclude that, physically, {\em no} chiral QH fluid can
form with a Hall conductivity  $\,\sH < 1/\Ls$. Theoretical and
numerical arguments (see the remarks preceding Thm.\,\ref{ssGT}.2)
as well as the present-day experimental data (see Fig.\,\ref{sKE}.1
in Subsect.\,\ref{ssQH}) suggest that $\,\Ls=7$. In the sequel, we
shall impose this bound.

Thus, for the classification of {\em physically relevant\,} CQHLs,
we have to consider only the first three windows, $\Sigma_1,\
\Sigma_2$, and $\Sigma_3$, because only there CQHLs with $\,\lM\leq
7\,$ can be found. In particular, in the third window $\Sigma_{3}$,
any physically relevant CQHL $(\G,\Q)$ {\em must\,} have
$\,\lM = 7\,$, meaning that $(\G,\Q)$ has to be {\em
\Lmini\mini}, i.e., all relative-angular-momentum invariants take
their smallest possible value; see Def.\,\ref{ssGT}.1.
\rule[-4mm]{0mm}{5mm}

\noi
$\bullet\ $ {\em Uniqueness Theorem (Thm.\,\ref{ssGT}.8).} The
uniqueness theorem of the previous subsection classifies all {\em
\Lmini\,} CQHLs in the left halfs, $\Sigp^+$, of the windows
$\Sigp$, i.e., in the subintervals $\,1/(2\mini p+1) \leq
\sH<1/(2\mini p),\ p=1,2,\ldots\ $. A unique, $N$-dimensional,
\Lmini\ CQHL is found at every Hall fraction $\,\sH=N/(2\mini
pN+1),\ p,N=1,2,\ldots\ $. Given this uniqueness theorem and the
bound $\,\Ls=7$, we find the following remarkable result: The
classification problem of {\em physically relevant\,} CQHLs can be
{\em completely\,} solved in the small subwindow $\Sigma_3^+ :=
[1/7,1/6)$. In $\Sigma_3^+$, the only fractions at which such
lattices can be found are $\,\sH=N/(6N+1),\ N=1,2,\ldots\,$, and the
electrons of the corresponding unique, chiral QH fluids carry an
$SU(N)$-symmetry.

In Thm.\,\ref{ssGT}.1 we have mentioned that any set of CQHLs
satisfying upper bounds on their invariants $\,\lM$ and $N$ is
finite. It is interesting to note that, {\em
independently\mini} of any upper bound, $\Ns$, on the dimension $N$,
every {\em closed\,} subinterval in $\Sigma_3^+$ contains only a
{\em finite\mini} number of physically relevant $(\lM\leq 7)$ CQHLs.
Presently we do not know to which extent this is also true in other
(sub)windows. The relevance of this remark derives {}from the fact
that it is rather difficult to find satisfactory bounds $\Ns$ on the
dimension $N$ of physically relevant QH lattices; see the remarks
preceding Thm.\,\ref{ssGT}.2. Thus, by-passing the need for a bound
on the dimension $N\mini$ represents  progress in the theory.

So far, the only indication of a QH fluid in $\Sigma_3^+$ has been
found at $\,\sH= 1/7$, corresponding to the first member of this
series. It coincides with the Laughlin fluid at $\,m=7$; see
example~{\bf (b)} in Subsect.\,\ref{ssBI}\mini. Further probing
(although difficult experimentally) of the subwindow $\Sigma_3^+$
would clearly be interesting! \rule[-4mm]{0mm}{5mm}

\noi
$\bullet\ $ {\em \LMiniy.} Next, we comment on the {\em assumption
of \Lminiy\,} that we impose when trying to classify CQHLs that
correspond to stable physical QH fluids. As discussed above, this
assumption is strictly justified only in the window $\Sigma_{3}$. In
general, it is an implementation of the {\em heuristic stability
principle\mini} described in Subsect.\,\ref{ssGT} stating that
the {\em lower\,} the values of $\,\lM$ and $N$ of a CQHL, the {\em
more stable\,} the corresponding QH fluid. Thus, the justification
of the assumption of \Lminiy\ in the other two windows, $\Sigma_2$
and $\Sigma_1$, is directly connected to the validity of this
stability principle. Thanks to the precise predictions it yields,
experimental tests can be proposed to settle its validity. Such
tests are discussed in detail by Fr\"ohlich, Studer, and
Thiran (1995).

In order to formulate such tests, one has to go beyond the
assumption of \Lminiy\ in classifying CQHLs, which requires much
more work and can be achieved, in full generality, only for small
values of the bounds $\Ns$ and $\Ls$. In Fr\"ohlich, Studer, and
Thiran (1995), we have fully classified all CQHLs with upper bounds
on $\,\lM$ and $N$ given by $\,(\Ls,\Ns)=(7,2)$, $\,(\Ls,\Ns)=
(5,3)$, and $\,(\Ls,\Ns)=(3,4)$. The general problem is
bound to be very complicated, since it involves as an input the
knowledge of the complete classification of integral lattices with
given bounds on the lattice discriminant, and this is a very
intricate mathematical problem (unsolved, in general, for euclidean
lattices, as needed here). With more patience and computing skill
one might extend the above results slightly (at most, however, up to
cases where $\,\Ns=6\,$ or $7$) because, in low dimensions, complete
lists of lattices can be computed (Conway and Sloane, 1988b;
Dickson, 1930); see Sect.\,6\, in Fr\"ohlich, Studer, and Thiran
(1995). Partial results (for small discriminants) derived {}from the
lattice classification in Conway and Sloane (1988b) have already
been given in Table~2 in Fr\"ohlich and Thiran (1994). In addition
to these results, however, we have also classified all \Lmini,
maximally symmetric CQHLs with $\,\sH\leq 1$, ({\em independent\,}
of any upper bound on the dimension $N$) in Fr\"ohlich, Studer, and
Thiran (1995), as mentioned in Subsect.\,\ref{ssMS}.


\begin{figure}[p]

\begin{flushleft}
\begin{picture}(10,8)(1,0)

\put(3.68,0){\noLmin}   
\put(4.21,0){\noLmin}   
\put(4.74,0){\makebox(0,0)[l]{$\hspace*{.4mm}\cdot$}}
  \put(4.74,0){\maxsym}\put(4.74,0){\dimms{9}}  
\put(5.26,0){\chacon}   
\put(5.79,0){\noLmin}   
\put(6.32,0){\maxsym}\put(6.32,0){\dimms{6}}   
\put(6.84,0){\noLmin}   
\put(7.37,0){\lowdim}\put(7.37,0){\dimg{4}}   
\put(7.89,0){\nolodiLmin}\put(7.89,0){\dimms{15}}   
\put(8.42,0){\nolodiLmin}\put(8.42,0){\dimms{19}}   
\put(8.95,0){\nolodiLmin}\put(8.95,0){\dimms{33}}   
\put(9.47,0){\nolodiLmin}\put(9.47,0){\dimms{20}}   

\put(3.53,.8){\noLmin}   
\put(4.12,.8){\noLmin}   
\put(4.71,.8){\makebox(0,0)[l]{$\circ$}}
  \put(4.71,.8){\maxsym}\put(4.71,.8){\dimms{8}}   
\put(5.29,.8){\makebox(0,0)[l]{$\circ$}}
  \put(5.29,.8){\chacon}   
\put(5.88,.8){\makebox(0,0)[l]{$\hspace*{.4mm}\cdot$}}
  \put(5.88,.8){\maxsym}\put(5.88,.8){\dimms{6}}   
\put(6.47,.8){\nolodiLmin}\put(6.47,.8){\dimms{23}}   
\put(7.06,.8){\maxsym}\put(7.06,.8){\dimms{9}}   
\put(7.65,.8){\lowdim}\put(7.65,.8){\dimg{4}}   
\put(8.24,.8){\nolodiLmin}\put(8.24,.8){\dimms{20}}   
\put(8.82,.8){\maxsym}\put(8.82,.8){\dimms{7}}   
\put(9.41,.8){\nolodiLmin}\put(9.41,.8){\dimms{18}}   

\put(4.67,1.6){\makebox(0,0)[l]{$\circ$}}
  \put(4.67,1.6){\maxsym}\put(4.67,1.6){\dimms{7}}   
\put(5.33,1.6){\makebox(0,0)[l]{$\circ$}}
  \put(5.33,1.6){\chacon}   
\put(7.33,1.6){\maxsym}\put(7.33,1.6){\dimms{11}}   
\put(8.67,1.6){\nolodiLmin}\put(8.67,1.6){\dimms{25}}   
\put(9.33,1.6){\maxsym}\put(9.33,1.6){\dimms{8}}   

\put(3.85,2.4){\noLmin}   
\put(4.62,2.4){\makebox(0,0)[l]{$\bullet$}}
  \put(4.62,2.4){\maxsym}\put(4.62,2.4){\dimms{6}}   
\put(5.38,2.4){\makebox(0,0)[l]{$\bullet$}}
  \put(5.38,2.4){\lowdim} \put(5.38,2.4){\chacon}
  \put(5.38,2.4){\dimg{3}}   
\put(6.15,2.4){\makebox(0,0)[l]{$\bullet$}}
  \put(6.15,2.4){\maxsym}\put(6.15,2.4){\dimms{9}}   
\put(6.92,2.4){\makebox(0,0)[l]{$\circ$}}
  \put(6.92,2.4){\maxsym}\put(6.92,2.4){\dimms{9,11}}   
\put(7.69,2.4){\maxsym}\put(7.69,2.4){\dimms{6}}   
\put(8.46,2.4){\lowdim}\put(8.46,2.4){\dimg{4}}   
\put(9.23,2.4){\maxsym}\put(9.23,2.4){\dimms{7,8}}   

\put(3.64,3.2){\makebox(0,0)[l]{$\hspace*{.4mm}\cdot$}}
\put(3.64,3.2){\noLmin}   
\put(4.54,3.2){\makebox(0,0)[l]{$\bullet$}}
  \put(4.54,3.2){\maxsym}\put(4.54,3.2){\dimms{5}}   
\put(5.45,3.2){\makebox(0,0)[l]{$\bullet$}}
  \put(5.45,3.2){\maxsym} \put(5.45,3.2){\chacon}
  \put(5.45,3.2){\dimms{4}}   
\put(6.36,3.2){\makebox(0,0)[l]{$\circ$}}
  \put(6.36,3.2){\maxsym}\put(6.36,3.2){\dimms{7}}   
\put(7.27,3.2){\makebox(0,0)[l]{$\bullet$}}
  \put(7.27,3.2){\maxsym}\put(7.27,3.2){\dimms{11}}
  \put(7.27,3.2){\lowdim}\put(7.27,3.2){\dimg{4}}   
\put(8.18,3.2){\nolodiLmin}\put(8.18,3.2){\dimms{17}}   
\put(9.09,3.2){\lowdim}\put(9.09,3.2){\dimg{4}}   

\put(4.44,4){\makebox(0,0)[l]{$\bullet$}}
  \put(4.44,4){\maxsym}\put(4.44,4){\dimms{4}}   
\put(5.56,4){\makebox(0,0)[l]{$\bullet$}}
  \put(5.56,4){\maxsym}\put(5.56,4){\lowdim}\put(5.56,4){\chacon}
  \put(5.56,4){\dimms{5}}\put(5.56,4){\dimg{4}}   
\put(7.78,4){\maxsym}\put(7.78,4){\dimms{7}}   
\put(8.89,4){\maxsym}\put(8.89,4){\dimms{8,10,11}}   

\put(4.29,4.8){\makebox(0,0)[l]{$\bullet$}}
  \put(4.29,4.8){\maxsym}\put(4.29,4.8){\dimms{3}}   
\put(5.71,4.8){\makebox(0,0)[l]{$\bullet$}}
  \put(5.71,4.8){\maxsym}\put(5.71,4.8){\lowdim}
  \put(5.71,4.8){\chacon}\put(5.71,4.8){\dimms{5,6}}
  \put(5.71,4.8){\dimg{4}}   
\put(7.14,4.8){\makebox(0,0)[l]{$\bullet$}}
  \put(7.14,4.8){\maxsym}
  \put(7.14,4.8){\lowdim \makebox(0,0)[l]{\hspace{1.6mm}
  {\scriptsize\em (B-p\hspace{-.2mm})}}}
  \put(7.14,4.8){\dimms{9,10}}\put(7.14,4.8){\dimg{3,4}}   
\put(8.57,4.8){\maxsym} \put(8.57,4.8){\lowdim}
  \put(8.57,4.8){\dimms{6,7,\ldots}}\put(8.57,4.8){\dimg{4}}  

\put(4,5.6){\makebox(0,0)[l]{$\bullet$}}
  \put(4,5.6){\maxsym \makebox(0,0)[l]{\hspace{1.6mm}
  {\scriptsize\em (B-p\hspace{-.2mm})}}}
  \put(4,5.6){\dimms{2}}   
\put(6,5.6){\makebox(0,0)[l]{$\bullet$}}
  \put(6,5.6){\maxsym} \put(6,5.6){\lowdim}
  \put(6,5.6){\chacon \makebox(0,0)[l]{\hspace{1.6mm}
  {\scriptsize\em B-p}}}
  \put(6,5.6){\dimms{5,6,7}}\put(6,5.6){\dimg{3,4}}   
\put(8,5.6){\makebox(0,0)[l]{$\bullet$}}
  \put(8,5.6){\maxsym} \put(8,5.6){\lowdim}
  \put(8,5.6){\dimms{6,7,\ldots}}\put(8,5.6){\dimg{4}}   

\put(3.33,6.4){\makebox(0,0)[l]{$\bullet$}}
  \put(3.33,6.4){\maxsym}\put(3.33,6.4){\dimms{1}}   
\put(6.67,6.4){\makebox(0,0)[l]{$\bullet$}}
  \put(6.67,6.4){\maxsym}\put(6.67,6.4){\lowdim}
  \put(6.67,6.4){\chacon \makebox(0,0)[l]{\hspace{1.6mm}
  {\scriptsize\em B/n-p}}}\put(6.67,6.4){\dimms{4,5,\ldots}}
  \put(6.67,6.4){\dimms{4,5,\ldots}}\put(6.67,6.4){\dimg{3,4}}  

\put(10,7.2){\makebox(0,0)[l]{$\bullet$}}
  \put(10,7.2){\maxsym}\put(10,7.2){\dimms{1}}   

\put(2.72,7.2){\makebox(0,0)[r]{\small$\dH=1$}}
\put(2.72,6.4){\makebox(0,0)[r]{\small$3$}}
\put(2.72,5.6){\makebox(0,0)[r]{\small$5$}}
\put(2.72,4.8){\makebox(0,0)[r]{\small$7$}}
\put(2.72,4){\makebox(0,0)[r]{\small$9$}}
\put(2.72,3.2){\makebox(0,0)[r]{\small$11$}}
\put(2.72,2.4){\makebox(0,0)[r]{\small$13$}}
\put(2.72,1.6){\makebox(0,0)[r]{\small$15$}}
\put(2.72,.8){\makebox(0,0)[r]{\small$17$}}
\put(2.72,0){\makebox(0,0)[r]{\small$19$}}

\thinlines
\put(2.83,-.45){\hspace{1mm}\line(1,0){7.67}}
\put(2.83,7.6){\hspace{1mm}\line(1,0){7.67}}

\multiput(3.33,7.6)(0,-0.1){11}{\hspace{1mm}\line(0,-1){0.04}}
\multiput(3.33,6)(0,-0.1){64}{\hspace{1mm}\line(0,-1){0.04}}
\put(3.33,-.4){\hspace{1mm}\line(0,-1){0.1}}    
\multiput(5,7.6)(0,-0.2){40}{\hspace{1mm}\line(0,-1){0.1}}
\put(5,-.4){\hspace{1mm}\line(0,-1){0.1}}       
\multiput(10,7.6)(0,-0.1){3}{\hspace{1mm}\line(0,-1){0.04}}
\multiput(10,6.8)(0,-0.1){72}{\hspace{1mm}\line(0,-1){0.04}}
\put(10,-.4){\hspace{1mm}\line(0,-1){0.1}}      

\thicklines
\put(3.33,-.54){\hspace{1mm}\makebox(0,0)[t]{$1\over 3$}}
\put(3.33,-.41){\hspace{1mm}\line(1,0){1.67}}
\put(5,-.54){\hspace{1mm}\makebox(0,0)[t]{$1\over 2$}}
\put(4.26,-1.5){\mbox{$\renewcommand{\arraystretch}{.7} \ba
\uparrow
  \rule[-2mm]{0mm}{5mm} \\ \mbox{\small\em Fermi liquid} \\
  \mbox{\small\em behaviour}\ea$}}
\put(10,-.6){\hspace{1mm}\makebox(0,0)[t]{$1$}}
\put(10,-.41){\hspace{1mm}\line(1,0){.5}}
\put(10.4,-.66){\hspace{1mm}\makebox(0,0)[t]{$\sH$}}
\put(8.41,-1){\mbox{$\underbrace{\hspace*{30mm}}_{
  \renewcommand{\arraystretch}{.7} \ba \mbox{\small\em domain of}^^ca\\
  \mbox{\small\em attraction} \\
  \mbox{\small\em of $\,\sH=1$}\ea}$}}

\put(3.33,7.56){\hspace{1mm}\line(1,0){1.67}}
\put(4.17,7.75){\hspace{1.2mm}\makebox(0,0)[b]{$\Sigma_1^+$}}
\put(7.5,7.75){\hspace{1.2mm}\makebox(0,0)[b]{$\Sigma_1^-$}}
\put(10,7.56){\hspace{1mm}\line(1,0){.5}}

\end{picture}
\end{flushleft}

\vspace{31mm}
\noi
{\bf Figure \ref{sClQHFs}.1. }{\em Compilation of \Lmini\ $\,(\mini\lM\!=
\!3\mini)$ chiral quantum Hall lattices (CQHLs) with
odd-denominator Hall fractions $\mini\sH$ in the interval
$\,1/3\leq\sH\leq 1\mini$.}\rule[-4mm]{0mm}{5mm}

\noi
{\small The experimental status of the Hall fractions displayed
here is indicated, for single-layer systems, by
``$\,\bullet\,,\circ\,$'', and ``$\,\cdot\,$'', as in Fig.\,4.1.
Superposed on the interval  $\,1/3\leq\sH\leq 1$ of that figure
is a list of different \Lmini\ CQHLs:
``\raisebox{1.5mm}{\hspace{2.3mm}\maxsym\hspace{3mm}}'' indicates
maximally symmetric, \Lmini\ CQHLs of dimension $\mini
N\!\leq\!11\mini$ (where the corresponding dimensions are given
below the symbols);
``\raisebox{1.4mm}{\hspace{1.6mm}\lowdim\hspace{2.5mm}}''
indicates generic, indecomposable, \Lmini\ CQHLs of low
dimension, $N\!\leq\!4\mini$ (the respective dimensions are given
above the symbols). For fractions decorated with
``\raisebox{1.3mm}{\hspace{1.5mm}\nolodiLmin\,}'', there are no
low-dimensional ($N\!\leq\!4\mini$), \Lmini\ CQHLs. However, there
are maximally symmetric ones in ``high'' dimensions (with the
lowest such dimension indicated below the symbols). At fractions
with ``\raisebox{1.3mm}{\,\mini\noLmin\,}'', there are neither
low-dimensional, \mbox{\Lmini} CQHLs, nor maximally symmetric ones
in ``higher'' dimensions. In addition,
``\raisebox{1.5mm}{\hspace{2.5mm}\chacon\hspace{3.1mm}}'' stands
for non-chiral QH lattices that are ``charge-conjugated'' to the
maximally symmetric, \Lmini\ CQHLs in $\Sigi^+$.}

\end{figure}

The upshot of all the results in Fr\"ohlich, Studer, and Thiran
(1995) is that {\em the assumption of \Lminiy\ for physically
relevant CQHLs is well supported by the presently available
experimental data}. Hence, in the sequel, we adopt it as our basic
working hypothesis.
\rule[-4mm]{0mm}{5mm}

\noi
$\bullet\ $ {\em Bijection Theorem (Thm.\,\ref{ssGT}.7).} An
important theorem described in the previous subsection,
the bijection theorem (for a proof, see Fr\"ohlich, Kerler {\em et
al.}, 1995), asserts that there are one-to-one correspondences
between the sets of \Lmini\ CQHLs in the different windows $\Sigp,\
p=1,2,\ldots\ $. Although this theorem does not classify CQHLs, it
is a powerful {\em structuring device\,} for \Lmini\ CQHLs with
$\,\sH\leq 1\mini$. In particular, it reduces the classification of
\Lmini\ CQHLs with Hall fractions in the entire interval
$\,0<\sH<1\,$ to the classification of such lattices in the
``fundamental domain'' $\,\Sigma_1:=[1/3,1)$! For a pictorial
summary of results for this fundamental domain, see
Fig.\,\ref{sClQHFs}.1.


\subsubsection{Theoretical Implications versus Experimental Data}

In the remaining part of this subection, we shall discuss explicit
consequences of the uniqueness and bijection theorems (Fr\"ohlich,
Kerler {\em et al.}, 1995), complement them with classification
results given in Fr\"ohlich, Studer, and Thiran (1995), and confront
our conclusions with the experimental data summarized in
Fig.\,\ref{sKE}.1 in Subsect.\,\ref{ssQH}.
\rule[-4mm]{0mm}{5mm}

\noi
$\bullet\ $ {\em The Subwindows $\Sigp^+$.} Adopting the hypothesis
of \Lminiy, the uniqueness theorem tells us that, in the subwindows
$\Sigma_{p}^{+}$, {\em no} (chiral) QH fluids can be found with Hall
conductivities $\,\sH\neq N/(2\mini p N + 1),\ N,p=1,2,\ldots\ $.
Taking a look at Fig.\,\ref{sKE}.1, remarkable agreement between this
theoretical prediction and experimental data is found: QH fluids have
been observed at $\,N/(2N+1),\ N=1, \ldots, 9$, in
$\,\Sigma_{1}^{+}$; at $\,N/(4N+1),\ N=1,2$, and $3$, in
$\,\Sigma_{2}^{+}$; and, as already mentioned, at just one value of
$\,N/(6N+1)$, namely $\,N=1$, in $\,\Sigma_{3}^{+}$. As we have
mentioned in Subsect.\,\ref{ssBI}, one may recognize these Hall
fractions as the ones of the ``basic Jain states'' (Jain, 1989, 1990,
1992). One can show (Fr\"ohlich, Studer, and Thiran,
1995) that, at the above fractions, the proposals of the hierarchy
schemes (Haldane, 1983; Halperin, 1984; Jain and Goldman, 1992) and
of our \Lmini-CQHL scheme coincide. The additional insight our
approach offers is that all these proposals have a {\em unique\,}
status as \Lmini, chiral QH fluids!

A closer inspection of Fig.\,\ref{sKE}.1 shows that, in the
subwindows $\Sigp^+$, there seems to be {\em only one\mini} fraction,
$\sH=4/11$, at which a weak signal of a QH fluid has been reported,
and which does {\em not\,} belong to the set of fractions described
by the uniqueness theorem. The corresponding experimental data
(reported only once) are somewhat controversial; see Hu (1991).
Theoretically, a QH fluid at $\,\sH=4/11\,$ is predicted by the
Haldane-Halperin and the Jain-Goldman hierarchy scheme at {\em
low(!)} ``level'' $2$ and $3$, respectively. These two proposals can
be shown (Fr\"ohlich, Studer, and Thiran, 1995) to belong to one
and the {\em same} universality class of QH fluids described by a
{\em non-\Lmini}, two-dimensional (primitive) CQHL which, in some
sense, provides the ``simplest'' example of a non-\Lmini\ CQHL. This
fraction marks thus an interesting plateau value where further
experiments might challenge the hierarchy schemes and/or our working
hypothesis of \Lminiy.

It has been pointed out in the literature (Hu, 1991) that the
{\em absence\mini} in the data of Fig.\,\ref{sKE}.1 of a QH
fluid at $\,\sH=5/13\,$ is quite remarkable. Indeed, this fraction is
conspicuous by its absence from the list of observed Hall fractions
(in single-layer systems) with denominator $\,\dH=13$, which are
$\,\sH=3/13,\ 4/13,\ 6/13,\ 7/13,\ 8/13$, and $9/13$. Theoretically,
the Haldane-Halperin and the Jain-Goldman hierarchy schemes predict a
QH fluid with $\,\sH=5/13\,$ at {\em low(!)} ``level'' $3$ and $2$,
respectively. These two proposals correspond to a {\em non-chiral\,}
QH lattice. We note that, in addition, there is an (inequivalent)
{\em chiral}, but {\em non-\Lmini\,} QH lattice in three dimensions
with $\,\sH=5/13$; see Fr\"ohlich, Studer, and Thiran (1995). This
fraction is thus another interesting plateau value where the
hierarchy schemes and/or the \Lminiy\ assumption can be tested
further.

By a similar reasoning process, in the first subwindow $\Sigma_1^+$,
all fractions in the open intervals $\,N/(2N+1)<\sH< (N+1)/(2N+3),\
N=1,2,\ldots\,$, are interesting plateau values for testing the
\Lminiy\ assumption. To be explicit, we do {\em not\,} expect stable
QH fluids to form at $\,\sH=4/11(!),\ 5/13(!),\ 6/17,\ 7/19,\
8/21,\ldots\,$ in $\,(1/3\mini,\mini 2/5)$, and at $\,\sH=7/17,\
8/19,\ldots\,$ in $\,(2/5\mini,\mini 3/7)$. Note that, with the help
of the shift maps ${\cal S}_1$ and ${\cal S}_2$ (see~\cfr{shift}),
these predictions can be translated into predictions in the
subwindows $\,\Sigma_2^+$ and $\,\Sigma_3^+$.
\rule[-4mm]{0mm}{5mm}

\noi
$\bullet\ $ {\em The Subwindows $\Sigp^-,\ p=1,2,\ldots\ $.} In the
``complementary'' subwindows, $\Sigp^-$, defined by
$\,1/(2\mini)p\leq\sH<1/(2\mini p-1),\ p=1,2,\ldots\,$, we do
{\em not\,} have a complete classification of \Lmini\ CQHLs.
Nevertheless, we can make interesting observations for these
subwindows by exploiting the bijection theorem of the previous
subsection and the (partial) classification results given in
Fr\"ohlich, Studer, and Thiran (1995); see
\mbox{Fig.\,\ref{sClQHFs}.1\mini.}

First, we note that the experimental data in $\,\Sigma_1^-\!:=
[1/2,1)\,$ (see Fig.\,\ref{sKE}.1) can hardly be interpreted as a
complete ``mirror image'' of the data in the
interval $\,(0\mini,\mini 1/2]$, as one would expect if charge
conjugation were at work {\em in general\mini}. Second, comparing,
the data in the two complementary subwindows $\,\Sigma_1^-$ and
$\,\Sigma_1^+$, we find, besides the prominent series of fractions
$\,\sH=n/(2n-1),\ n=2,\ldots,9$, ``mirroring'' the unique fractions
in $\Sigma_1^+$ (i.e., $\sH=1-\sH^{\mini\prime}$), data points at
$\,\sH=4/5,\ 5/7,\ 7/11,\ 8/11,\ 8/13,\ 9/13$, and possibly at
$10/17$. This is a first {\em experimental\,} indication that the
sets of QH fluids appearing in the complementary subwindows
$\Sigp^+$ and $\Sigp^-,\ p=1,2,\ldots\,$, are ``{\em structurally
distinct\,}'', as our theory predicts.

We may ask to which extent the experimental data in
Fig.\,\ref{sKE}.1 also support the one-to-one correspondences
predicted by the bijection theorem between QH fluids in the
different subwindows $\Sigp^-$. We can act ``formally'' with the
shift maps $\,{\cal S}_{p-1},\ p=2\,$ and $3$, of the bijection
theorem on the fractions $\mini\sH\mini$ given in $\,\Sigma_1^-$ of
Fig.\,\ref{sKE}.1, for example, $\,{\cal S}_{p-1}\mini:\mini n/(2n-1)
\mapsto n/(2\mini p\mini n-1)$; see~\cfr{shift}. The resulting
fractions $\sH$ that we obtain in the two subwindows $\,\Sigma_2^-$
and $\,\Sigma_3^-$ are fully {\em consistent\,} with the experimental
data given in Fig.\,\ref{sKE}.1. Experimentally observed are the
fractions $\,\sH=n/(4n-1),\ n=2,3,4$, and $\,\sH=4/13\,$ (very
weakly) in $\,\Sigma_2^- = [1/4,1/3)$, and only one fraction in
$\,\Sigma_3^- = [1/6,1/5)$, namely $\,\sH=n/(6n-1)$, with $\,n=2$.

If the QH fluids corresponding to points in $\Sigma_1^-$ were
described by
\Lmini\ CQHLs then, by the logic of the bijection theorem, we would
predict the formation of (chiral) QH fluids at $\,\sH=4/13(!),\
5/17,\ 5/19,\ldots\,$ in $\Sigma_2^-$, and at $\,\sH=3/17,\
4/21,\ldots\,$ in $\Sigma_3^-$. These are thus interesting
plateau-values for experimentation.

What do we know {\em explicitly\mini} about \Lmini\ CQHLs in the
subwindows $\Sigp^-$? As mentioned above, the analysis presented
in Fr\"ohlich, Studer, and Thiran (1995) contains, in particular, a
complete classification of all {\em low-dimensional\,}
$(N\leq\Ns=4)$ and of all maximally symmetric, \Lmini\ CQHLs with
$\,\sH\leq 1\mini$.
\rule[-4mm]{0mm}{5mm}

\noi
$\bullet\ $ {\em Summary of Classification Results  for the
  Fundamental Subdomain $\Sigma_1^-$.} The upshot of our analysis (see
Fig.\ref{sClQHFs}.1) is that, in $\Sigma_1^-$, natural proposals for
QH fluids at the fractions of the series $\,\sH=n/(2n-1),\
n=2,3,\ldots\,$, are provided by the {\em charge-conjugation
picture}, meaning that the corresponding QH fluids are {\em
composite}. They consist of an electron-rich subfluid with a partial
Hall fraction $\,\sigma_{(1)}=1$, and of a hole-rich subfluid
corresponding to an \Lmini\ CQHL of the $su(N)$-series in
$\Sigma_1^+$ with partial Hall fraction $\,\sigma_{(2)} =-N/(2N+1)$,
where $\,N=n-1\mini$. This is, however, {\em not\,} the full story!
It is a {\em structural property\,} of the subwindows $\,\Sigp^-$
that, at a given Hall fraction $\sH$, one typ\-i\-cally finds {\em
more than one\mini} \Lmini\ CQHL realizing that fraction. This is
much in contrast to the {\em unique\mini} realization of the
fractions $\,\sH= N/(2N+1)\,$ in the complementary subwindow
$\Sigma_1^+$.

For example, at the {\em finite\mini} series of fractions
$\,\sH=n/(2n-1),\ n=2,\ldots,7$, one finds maximally symmetric,
\Lmini\ CQHLs in dimensions $\,N=10-n\,$ which are based on the root
lattices of the exceptional Lie algebras $E_{\mini 9-n}$, similarly
to the way in which the $su(N)$-QH lattices in $\Sigma_1^+$ are
based on the root lattices of $su(N)$; see
Subsect.\,\ref{ssMS}\mini. While the last two members of this finite
series, $\sH=6/11$ and $7/13$, are realized by unique,
low-dimensional ($N=4$ and $3$, respectively), \Lmini\ CQHLs, the
higher dimensional members of this ``$E$-series'' of CQHLs contain
several embedded \Lmini, chiral ``{\em QH (sub)lattices\mini}'' of
lower dimensions. All these QH (sub)lattices (although not
necessarily maximally symmetric) represent possible proposals for QH
fluids at the corresponding Hall fractions $\sH$.

Let us define precisely what we mean by a ``QH lattice embedding'':
\rule[-4mm]{0mm}{5mm}

{\bf Definition~\ref{ssDis}.1.} {\em A QH lattice $(\G^\prime,
\Q^\prime \in\! \G^{\prime\mini\ast})$ is embedded into another QH
lattice $(\G,\Q\in\!\Gs)$ if (i) both QH lattices correspond to the same
Hall fraction, i.e., $\sH^\prime = \;<\!\Q^\prime,\Q^\prime\!> \;=\;
<\!\Q\,,\Q\!> \;= \sH\mini$, (ii) $\G^\prime$ is a sublattice of
$\,\G$, and (iii) the two charge vectors $\Q^\prime$ and $\Q$ are
compatible in the sense that all multi-electron/hole states
described by $(\G^\prime,\Q^\prime)$ remain physical states when
viewed (via the lattice embedding $\,\G^\prime\subset\G$) as states
described by $(\G,\Q)$. In particular, all the electric charges stay
the same, i.e., $<\!\Q^\prime,\vq^\prime\!> \;=\;
<\!\Q\,,\vq^\prime\!> \mini$, for all $\,\vq^\prime \in\!\G^\prime
\subset\G$.}
\rule[-4mm]{0mm}{5mm}

At the level of symbols (see~(\ref{symb})), we write such embeddings
as

\be
\thesymbolpr\:\emb\thesymbol\ .
\label{embed}
\ee

\noi
Note that, as an immediate consequence of definition~(\ref{lmin}),
$\,\lm^{\mini\prime}\geq\lm\mini$.

Physically, a QH fluid described by the QH lattice $(\G^\prime,
\Q^\prime)$ embedded into another lattice \QHL\, is
characterized by a {\em restricted\,} set of possible
multi-electron/hole excitations, as compared
to the corresponding set of the fluid associated with the lattice
\QHL. Furthermore, since the neutral sublattice
(see~(\ref{neutraldef})) of $(\G^\prime,\Q^\prime)$ is a sublattice
of the neutral sublattice of \QHL, the embedded fluid exhibits a
(global) symmetry group $G^{\mini\prime}$ (see~(\ref{weights}))
which is a {\em subgroup} of $\mini G$, the symmetry group exhibited
by the fluid associated with \QHL. Thus, in this precise sense, {\em
the embedded fluid exhibits a more restricted symmetry than the one
it embeds into.} Put differently, {\em passing from a QH fluid to an
embedded subfluid corresponds to a ``reduction or breaking of
symmetries''.} [As a mathematical aside, we remark that the study of
embeddings of maximally symmetric CQHLs into one another is
equivalent to the study of regular conformal embeddings of level-$1$
Kac-Moody algebras and the respective branching rules. For recent
results on the latter subject, see, e.g., Schellekens and Warner
(1986), Bais and Bouwknegt (1987), and Kac and Sanielevici (1988).]
Experimentally, symmetry breaking might be realized in {\em phase
transitions\mini} that are driven, at a given Hall fraction, by
varying external control parameters. Hence, it is most interesting
to see at which fractions in $\mini\Sigi^-$ such ``{\em
structural\,}'' {\em phase transitions\mini} can be expected within
our framework.

As an example (see Appendix~D in Fr\"ohlich, Studer, and Thiran
(1995)), one finds $13$ and $5$ QH sublattices embedded into the
QH lattice of the $E$-series at $\,\sH=2/3\ (E_7)\,$ and $3/5\
(E_6)$, respectively. The {\em composite\,} \Lmini\ CQHL at
$\sH=2/3$ which consists of two Laughlin subfluids with partial Hall
fraction $\,\sigma_{(i)}=1/3,\ i=1,2$, is the lowest-dimensional
($N=2$) QH sublattice of the $E_7$-QH lattice at $\,\sH=2/3$. All
the other QH sublattices at $\,\sH=2/3\,$ are {\em indecomposable}.
(Recall that we have mentioned in Thm.\,\ref{ssGT}.6 that all \Lmini\
CQHLs with $\sH<2/3$ are indecomposable.) In
$\,\Sigma_1^-=[1/2\mini,\mini 1)$, complex embedding patterns of
\Lmini\ CQHLs are found at the fractions $\,\sH=2/3,\ 3/5,\ 4/5,\
4/7,\ 5/7,\ 5/9$, and at the even-denominator fraction
$\,\sH=1/2\mini$! These fractions are interesting in the light of the
data in Fig.\,\ref{sKE}.1, where phase transitions are indicated at
$\,\sH=2/3,\ 3/5$, and possibly at $5/7$, driven by an added
in-plain component of the external magnetic field (Clark {\em et
al.}, 1990; Engel {\em et al.}, 1992; Sajoto {\em et al.}, 1990),
and at $\,\sH=2/3$, driven by changing the density of charge
carriers in the system (Eisenstein, Stormer {\em et al.}, 1990b);
see also the data reported in Suen {\em et al.}~(1994) on phase
transitions in wide-single-quantum-well systems!

[Clearly, by the uniqueness of the \Lmini\ CQHLs
in the subwindows $\Sigp^+$ and our heuristic stability principle,
we do {\em not\,} expect structural phase transitions there. So,
what about a possible indication of a magnetic field driven phase
transition at $\,\sH=2/5$? As a matter of fact, in Fr\"ohlich and
Studer (1993b) (see also Fr\"ohlich, Studer, and Thiran (1995)) we
have argued that $\,\sH=2/5\,$ is the most likely plateau value
where we may expect a phase transition from a {\em spin-polarized\,}
to a {\em spin-singlet\,} QH fluid. While, structurally, the two
phases are described by {\em one and the same\mini} \Lmini\ CQHL,
the phase transition corresponds to a change from an {\em internal\,}
$SU(2)$-symmetry to a {\em spatial\,} $SU(2)_{spin}$-symmetry.]

We complete our short review of results derived in Fr\"ohlich,
Studer, and Thiran (1995)\,--\,see Fig.\,\ref{sClQHFs}.1\,--\,by
mentioning that to {\em all\,} data points in $\,\Sigma_1^-\,$
(including the fraction $\,\sH=1/2$) one can associate {\em at least
one\mini} \Lmini\ CQHL that is either generic (without special
symmetry properties) and low-dimensional $(N\leq 4)$, maximally
symmetric with dimension $\,N\leq 9$ (based on the root lattice of a
simple or semi-simple Lie algebra), or charge-conjugated to an
$su(N)$-lattice in $\Sigma_1^+$. Within these three subclasses of
CQHLs, predictions of {\em new\mini} QH fluids are made at
$\,\sH=6/7,\ 10/13,\ 10/17(!),\ 13/17,\ 10/19,\ 12/19,\
14/19,\ldots\,$, and at the {\em even-denominator} fractions
$\,\sH=3/4$ and $5/8$. The CQHLs that yield even-denominator Hall
fractions have a structure that can naturally be interpreted as
describing {\em double-layer/component} QH {\em systems}. Furthermore, staying
within these three subclasses, we do {\em not\,} expect {\em
stable(!)} QH fluids to form at $\,\sH=9/11,\ 11/17,\ 14/17,\
13/19$, and $\mini 15/19\mini$ in $\Sigma_1$, where we have omitted
fractions with $\dH\geq 21\,$ and within the ``domain of
attraction'' of the most stable Laughlin fluid at $\,\sH=1$. {\em
None\mini} of these fractions has been observed experimentally!
These predictions are rather different from those of the standard
hierarchy schemes (Haldane, 1983; Halperin, 1984; Jain and Goldman,
1992); for further discussions, see Fr\"ohlich, Studer, and Thiran
(1995).
\rule[-4mm]{0mm}{5mm}

\noi
$\bullet\ $ {\em Phase Transitions at\,} $\mini\sH\!=\!1\mini$. An
interesting question to ask is whether one should expect to observe
{\em phase transitions at} $\,\sH\!=\!1\mini$. The unique \Lmini\
($\lM\!=\!1$) CQHL at $\,\sH=1\,$ is the one-dimensional ``Laughlin
lattice'' with $m\!=\!1$; see example~{\bf (b)} in
Subsect.\,\ref{ssBI}\mini. Thus, any other CQHL realizing this
fraction necessarily has to be {\em non-\Lmini\,} ($\lM\!\geq\! 3$),
a fact that suggests a {\em markedly reduced stability\mini} for the
corresponding fluids, as compared to the (\Lmini) Laughlin fluid!
Moreover, by Thm.\,\ref{ssGT}.4, we know that any other
indecomposable CQHL at this fraction exhibits a charge parameter
$\mini\lb$ strictly larger than $1$. By an argument similar to the
one in~(\ref{even}), this leads to the prediction of {\em
fractional\,} charges in these fluids! For the purpose of
illustration, we give the lowest-dimensional examples of such CQHL's
classified in Appendix~C of Fr\"ohlich, Studer, and Thiran (1995).
Using the symbol introduced in~\cfr{symb}, one finds the following
embeddings of non-\Lmini\ CQHLs at $\,\sH=1$:

\be
\subi{2}{(1)}^4_2\fsp[3,3]
\emb
\left\{\renewcommand{\arraystretch}{1.2}
\ba
 \subi{3}{(1)}^6_2\fsp[3,3]\rule[-3.8mm]{0mm}{5mm}
\\
 \subi{3}{(1)}^8_2\fsp[3,3]
\ea\right\}
\emb
\subi{5}{(1)}^8_2\fsp[3,3]
\emb\ldots\;.
\label{embed1}
\ee
\noi
This ``chain of embeddings'', with the corresponding
possibilities of structural phase transitions, is particularly
interesting in the light of the recent experimental data given
in Murphy {\em et al.}~(1994). There, evidence for a phase transition
between different QH fluids at $\,\sH=1\,$ has been reported. The
phase transition appears to be driven by an in-plane magnetic field,
$\Bcpara$, and is observed in {\em double-layer\mini} QH systems.
This matches our theoretical predictions: In~(\ref{embed1}), e.g., the first two CQHLs, (the
lattice with symbol $\,\subi{2}{(1)}^4_2\,$ and the one with symbol
$\,\subi{3}{(1)}^6_2$), both are natural candidates for describing
double-layer/component QH fluids. The first one can be shown to
exhibit a discrete $\BB{Z}_{\mini 2}$ layer symmetry, while the
second one can be seen to exhibit a continuous $\,A_1=su(2)\,$
layer symmetry; see Fr\"ohlich, Studer, and Thiran (1995).
Furthermore, since for all lattices in~(\ref{embed1}) the charge
parameter $\mini\lb$ equals $2$, we would expect, as mentioned
above, that quasi-particles with fractional charge $1/2$ exist in  the
corresponding QH fluids. An
experimental investigation of this prediction would seem to be
revealing and is encouraged!
\rule[-4mm]{0mm}{5mm}

\noi
$\bullet\ $ {\em Concluding Remarks.} By the bijection theorem, {\em
all\,} the statements about \Lmini\ {\em chiral\,} QH lattices in
$\,\Sigma_1^-:=[1/2,1)\,$ have their precise analogues in the
shifted subwindows $\,\Sigp^-:=[1/(2\mini p),1/(2\mini p-1)\mini)$,
for $p=2,3,\ldots\ $. For example, interpreting the phase
transitions observed at $\,\sH=2/3,\ 3/5$, and possibly $5/7$ in
$\,\Sigma_1^-$ as {\em structural phase transitions}, we predict
analogous transitions at the Hall fractions $\,\sH=2/7,\ 3/11,\
(5/17)\,$ in $\,\Sigma_2^-$, and at $\,\sH=2/11,\ 3/17,\ (5/27)$ in
$\,\Sigma_3^-$.

When acting with the shift maps $\Sp$ on the
composite, {\em mixed-chirality\,} QH lattices at $\,\sH=n/(2n-1),\
n=2,3,\ldots\,$, corresponding to the charge-conjugation picture,
we obtain composite {\em non-euclidean\,} QH lattices which are
however {\em not\,}
primitive (where primitivity has been defined at the end of
Subsect.\,\ref{ssQHFQHL}\mini). A discussion of non-primitive QH
lattices and the
corresponding QH fluids has been given in Appendix~E in Fr\"ohlich,
Studer, and Thiran (1995). In particular, we have described
composite, non-euclidean (but ``factorizable'') QH lattices
corresponding to the hierarchy fluids in the windows $\,\Sigp^-,\
p=1,2,3$.

For the complementary subwindows $\,\Sigp^+:=[1/(2\mini
p+1),1/(2\mini p)\mini)$, the one-to-one correspondences between
the different sets of \Lmini\ CQHLs are contained in the
classification result of the uniqueness theorem discussed above.

These results may suffice to convince the reader that there is a
significant {\em structural asymmetry\,} between the sets of QH
fluids with Hall conductivities in the two complementary subwindows
$\,\Sigp^+$ and $\,\Sigp^-$, for a given $\,p=1,2,\ldots\ $, while
there is a {\em structural similarity\,} among QH
fluids with conductivities in the ``$+$-windows'' $\,\Sigp^+$ and
among the ones with conductivities in the ``$-$-windows''
$\,\Sigp^-$, for different values of $p$.

After more than ten years since its
discovery (Tsui, Stormer, and Gossard, 1982), the fractional QH
effect in the interval $\,0<\sH\leq 1\,$ is thus still an interesting
field of experimental and theoretical research. In our recent work,
summarized in this section, we have argued that the QH-lattice
approach provides an efficient instrument for describing universal
properties of QH fluids. In our analysis of {\em physically
relevant\,} QH lattices we have assumed chirality and \Lminiy\ as
basic properties. New and refined experimental data in the
neighbourhood of the various plateau-values discussed above would
either support or question these hypotheses, and hence could lead
to further progress in the understanding of this fascinating effect.
The idea, however, that the physics of incompressible QH fluids at
large distance scales and low frequencies can be encoded into pairs of
an integral lattice $\Gamma$ and a primitive vector {\bf Q} in its
dual lattice should be regarded as a firmly established theoretical
result! (Number theory makes its appearance in condensed matter
physics.)







\setcounter{secnumdepth}{0}
\section{References}

\begin{list}
{}{\setlength{\leftmargin}{4mm}\setlength{\rightmargin}{0mm}
\setlength{\parsep}{0cm}\setlength{\listparindent}{-4mm}}

\item\mbox{}
\vspace{-6mm}

Aharonov, Y., and D.~Bohm, 1959, \PR {\bf 115}, 485.

Aharonov, Y., and G.~Carmi, 1973, \FP {\bf 3}, 493.

Aharonov, Y., and A.~Casher, 1984, \PRL {\bf 53}, 319.

Alvarez-Gaum\'e, L., and P.~Ginsparg, 1984, \NPB {\bf 243}, 449.

Alvarez-Gaum\'e, L., and P.~Ginsparg, 1985, \AP {\bf 161}, 423.

Anandan, J., 1989, \PL A {\bf 138}, 347.

Anandan, J., 1990, in {\em Proceedings of the 3rd International
Sympsymposium on the Foundations of Quantum Mechanics}, Tokyo, 1989
(Phys.\ Soc.\ Jpn., Tokyo), p.~98.

Anderson, P.W., 1990, \PRL {\bf 64}, 1839.

Arovas, D., J.R.~Schrieffer, and F.~Wilczek, 1984, \PRL {\bf 53},
722.

Ashoori, R.C., H.L.~Stormer, L.N.~Pfeiffer, K.W.~Baldwin, and
K.W.~West, 1992, \PRB {\bf 45}, 3894.

Avron, J.E., and R.~Seiler, 1985, \PRL {\bf 54}, 259.

Avron, J.E., R.~Seiler, and B.~Simon, 1983, \PRL {\bf 51}, 51.

Avron, J.E., R.~Seiler, and B.~Simon, 1990, \PRL {\bf 65}, 2185.

Avron, J.E., R.~Seiler, and B.~Simon, 1994, \CMP {\bf 159}, 399.

Avron, J.E., R.~Seiler, and L.~Yaffe, 1987, \CMP {\bf 110}, 33.

Bais, F.A., and P.G.~Bouwknegt, 1987, \NPB{\bf 279}, 561.

Balatsky, A.V., and E.~Fradkin, 1991, \PRB {\bf 43}, 10622.

Balatsky, A.V., and M.~Stone, 1991, \PRB {\bf 43}, 8038.

Bargmann, V., L.~Michel, and V.L.~Telegdi, 1959, \PRL {\bf 2}, 435.

Beenakker, C.W.J., 1990, \PRL {\bf 64}, 216.

Belavin, A.A., A.M.~Polyakov, and A.B.~Zamolodchikov, 1984, \NPB
{\bf 241}, 333.

Bell, J.S., and J.M.~Leinaas, 1983, \NPB {\bf 212}, 131.

Bell, J.S., and J.M.~Leinaas, 1987, \NPB {\bf 284}, 488.

Bellissard, J., 1988a, in {\em Localization in Disordered Systems},
Texts in Physics, edited by W.~Weller, and P.~Ziesche (Teubner,
Leipzig).

Bellissard, J., 1988b, in {\em Operator Algebras and Applications},
Vol.~2, Lecture Notes {\bf 136}, edited by D.E.~Evans, and
M.~Takesaki (London Math.\ Soc., Cambridge), p.~49.

Benfatto, G., and G.~Gallavotti, \JSP {\bf 59}, 541.

Bleecker, D., 1981, {\em Gauge Theory and Variational Principles},
Global Analysis, Pure and Applied; No. 1 (Addison-Wesley, Reading,
MA).

Block, B., and X.G.~Wen, 1990a, \PRB {\bf 42}, 8133.

Block, B., and X.G.~Wen, 1990b, \PRB {\bf 42}, 8145.

Buchholz, D., G.~Mack, and I.~Todorov, 1990, in {\em The Algebraic
Theory of Superselection Sectors: Introduction and Recent Results},
edited by D.~Kastler (World Scientific, Singapore), p.~356.

B\"uttiker, M., 1988, \PRB {\bf 38}, 9375.

Cappelli, A., C.A. Trugenberger, and G.R. Zemba, 1993, \NPB {\bf 396},
465; \PL B {\bf 306}, 100.

Cattaneo, A.S., P. Cotta-Ramusino, J. Fr\"ohlich, and M. Martellini,
1995, ``Topological BF theories in 3 and 4 dimensions'', to appear in
J.~Math.~Phys.

Cattaneo, A.S., 1995, ``Teorie topologiche di tipo BF ed invarianti
dei nodi'', PhD thesis, University of Milan. 

Chakraborty, T., and P.~Pietil\"ainen, 1987, \PRL {\bf 59}, 2784.

Chakraborty, T., and P.~Pietil\"ainen, 1988, {\em The Fractional
Quantum Hall Effect: Properties of an Incompressible Quantum Fluid},
Springer Series in Solid State Sciences {\bf 85} (Springer, Berlin).

Chang, A.M., and J.E.~Cunningham, 1989, Solid State Commun.\ {\bf
72}, 652.

Clark, R.G., S.R.~Haynes, J.V. Branch, A.M.~Suckling, P.A.~Wright,
P.M.W.~Oswald, J.J.~Harris, and C.T.~Foxon, 1990, \SurS {\bf
229}, 25.

Clark, R.G., S.R.~Haynes, A.M.~Suckling, J.R.~Mallett,  P.A.~Wright,
J.J.~Harris, and C.T.~Foxon, 1989, \PRL {\bf 62}, 1536.

Clark, R.G., J.R.~Mallett, S.R.~Haynes, J.J.~Harris, and
C.T.~Foxon, 1988, \PRL {\bf 60}, 1747.

Conway, J.H., and N.J.A.~Sloane, 1988a, Proc.\ R.\ Soc.\ Lond.\ A
{\bf 418}, 17.

Conway, J.H., and N.J.A.~Sloane, 1988b, {\em Sphere-Packings,
Lattices and Groups} (Sprin\-ger, New York).

d'Ambrumenil, N., and R.~Morf, 1989, \PRB {\bf 40}, 6108.

Dana, I., J.E.~Avron, and J.~Zak, 1985, J.\ Phys.\ C{\bf 18}, L679.

Deaver, B.S., and W.M.~Fairbank, 1961, \PRL {\bf 7}, 43.

Deser, S., and R.~Jackiw, 1988, \CMP {\bf 118}, 495.

de Gennes, P.G., 1966, {\em Superconductivity of Metals and
Alloys$\,$} (Benjamin, New York).

de Rham, G., 1984, {\em Differentiable Manifolds}, Grundlehren der
mathematischen Wissenschaften 266 (Springer, Berlin/Heidelberg)

Dickson, L.E., 1930, {\em Studies in the Theory of Numbers\mini}
(University of Chicago Press, Chicago), reprinted by Chelsea
Publishing Company (New York, 1957).

Doll, R., and M.~N\"abauer, 1961, \PRL {\bf 7}, 51.

Du, R.R., H.L.~Stormer, D.C.~Tsui, L.N.~Pfeiffer, and K.W.~West,
1993, \PRL {\bf 70}, 2944.

Eguchi, T., P.B.~Gilkey, and A.J.~Hanson, 1980, Phys.\ Rep.\ {\bf
66}, 213.

Eisenstein, J.P., G.S.~Boeblinger, L.N.~Pfeiffer, K.W.~West, and
Song He, 1992, \PRL {\bf 68}, 1383. 

Eisenstein, J.P., H.L.~Stormer, L.N.~Pfeiffer, and K.W.~West,
1989, \PRL {\bf 62}, 1540.

Eisenstein, J.P., H.L.~Stormer, L.N.~Pfeiffer, and K.W.~West,
1990a, \SurS {\bf 229}, 21.

Eisenstein, J.P., H.L.~Stormer, L.N.~Pfeiffer, and K.W.~West,
1990b, \PRB {\bf 41}, 7910.

Eisenstein, J.P., R.L.~Willett, H.L.~Stormer, D.C.~Tsui,
A.C.~Gossard, and J.H.~Eng\-lish, 1988, \PRL {\bf 61}, 997.

Eisenstein, J.P., R.L.~Willett, H.L.~Stormer, L.N.~Pfeiffer, and
K.W.~West, 1990, \SurS {\bf 229}, 31.

Engel, L.W., S.W.~Hwang, T.~Sajoto, D.C.~Tsui, and M.~Shayegan, 1992,
\PRB {\bf 45}, 3418.

Fano, G., F.~Ortolani, and E.~Colombo, 1986, \PRB {\bf 34}, 2670.

Feldman, J., and E. Trubowitz, 1990, \HPA {\bf 63}, 156; {\bf 64},
214. 

Feldman, J., H.~Kn\"orrer, and E.~Trubowitz, 1992, ``Mathematical
Methods of Many-Body Quantum Field Theory,'' Lecture Notes, ETH
Z\"urich.

Feldman, J., J.~Magnen, V.~Rivasseau, and E.~Trubowitz, 1992, \HPA
{\bf 65}, 679. 

Fetter, A.L., and J.D.~Walecka, 1980, {\em Theoretical Mechanics of
Particles and Continua\mini} (McGraw-Hill, New York).

Floreanini, R., and R.~Jackiw, 1987, \PRL {\bf 59}, 1873.

Fradkin, E., 1991, {\em Field Theories of Condensed Matter
Systems}, Frontiers in Physics, Vol.~82 (Addison-Wesley, Redwood
City, CA).

Fredenhagen, K., K.-H.~Rehren, and B.~Schroer, 1989, \CMP {\bf
125}, 201.

Frenkel, I.B., 1981, J.\ Funct.\ Anal.\ {\bf 44}, 259.

Fr\"ohlich, J., 1988, in ``Non-perturbative Quantum Field Theory'',
G. 't Hooft et al. (eds.), New York: Plenum Press 1988.

Fr\"ohlich, J., 1990, in {\em Proceedings of The Gibbs Symposium},
Yale University, 1989, edited by D.G.~Caldi, and G.D.~Mostow (Am.\
Math.\ Soc., Providence, RI), p.~89.

Fr\"ohlich, J., 1992, in {\em First European Congress of
Mathematics}, Paris, 1992, edited by A.~Joseph, F.~Mignot, F.~Murat,
B.~Prum, and R.~Rentschler, Vol.~II, p.~23.

Fr\"ohlich, J., and F.~Gabbiani, 1990, Rev.\ Math.\ Phys.\ {\bf 2},
251.

Fr\"ohlich, J., F.~Gabbiani, and P.-A.~Marchetti, 1990, in {\em The
Algebraic Theory of Superselection Sectors: Introduction and Recent
Results}, edited by D.~Kastler (World Scientific, Singapore),
p.~259. 

Fr\"ohlich, J., R.~G\"otschmann, and P.-A.~Marchetti, 1995a,
J.~Phys.\ A: Math.\ Gen.\ {\bf 28}, 1169.

Fr\"ohlich, J., R.~G\"otschmann, and P.-A.~Marchetti, 1995b,
``The Effective Gauge Field Action of a System of Non-Relativistic
Electrons'', \CMP (in press).

Fr\"ohlich, J., and T.~Kerler, 1991, \NPB {\bf 354}, 369.

Fr\"ohlich, J., T.~Kerler, and P.-A.~Marchetti, 1992, \NPB {\bf
374}, 511.

Fr\"ohlich, J., T.~Kerler, and E.~Thiran, 1995, in preparation.

Fr\"ohlich, J., T.~Kerler, U.M.~Studer, and E.~Thiran, 1995,
``Structuring the Set of Incompressible Quantum Hall Fluids'',
preprint ETH-TH/95-5; see also cond-mat/9505156.

Fr\"ohlich, J., and P.-A.~Marchetti, 1991, \NPB {\bf 356}, 533.

Fr\"ohlich, J., and U.M.~Studer, 1992a, \IJMP B {\bf 6}, 2201.

Fr\"ohlich, J., and U.M.~Studer, 1992b, \CMP {\bf 148}, 553.

Fr\"ohlich, J., and U.M.~Studer, 1992c, in {\em New Symmetry
Principles in Quantum Field Theory}, Carg\`ese Lectures, 1991,
edited by J.~Fr\"ohlich, G. t'Hooft, A. Jaffe, G. Mack, P.K. Mitter,
and R. Stora (Plenum, New York), p.~195.

Fr\"ohlich, J., and U.M.~Studer, 1993a, in {\em Phase Transitions:
Mathematics, Physics, Biology,$\ldots\,$}, Prague, June 1992, edited
by R.~Koteck\'y (World Scientific), p.~85.

Fr\"ohlich, J., and U.M.~Studer, 1993b, Rev.~Mod.~Phys.\ {\bf 65},
733.

Fr\"ohlich, J., U.M.~Studer, and E.~Thiran, 1994, in {\em On Three
Levels: Micro-, Meso-, and Macro-Approaches in Physics}, edited by
M.~Fannes, C.~Maes, and  A.~Verbeure (Plenum, New York), p.~225; see
also cond-mat/9406009.

Fr\"ohlich, J., U.M.~Studer, and E.~Thiran, 1995, ``A Classification
of Quantum Hall Fluids'', J.~Stat.~Phys.\ (in press); see also
cond-mat/9503113.

Fr\"ohlich, J., and E.~Thiran, 1994, J.~Stat.~Phys. {\bf 76}, 209.

Fr\"ohlich, J., and A.~Zee, 1991, \NPB {\bf 364}, 517.

Gawedzki, K., 1990, in {\em Constructive Quantum Field Theory II},
Erice Lectures, 1988, edited by G.~Velo, and A.S.~Wightman (Plenum,
New York), p.~89.

Ginsparg, P., 1990, in {\em Fields, Strings and Critical
Phenomena}, Les Houches, Session XLIX, 1988, edited by E.~Br\'ezin,
and J.~Zinn-Justin (North-Holland, Amsterdam), p.~1.

Girvin, S., and A.H.~MacDonald, 1987, \PRL {\bf 58}, 1252.  

Goddard, P., and D.~Olive, 1986, \IJMP A {\bf 1}, 303.

Goldin, G.A., R.~Menikoff, and D.H.~Sharp, 1980, J.\ Math.\ Phys.\
{\bf 21}, 650.

Goldin, G.A., R.~Menikoff, and D.H.~Sharp, 1981, {\bf 22}, 1664.

Goldin, G.A., R.~Menikoff, and D.H.~Sharp, 1983, \PRL {\bf 51},
2246.

Goldstone, J., 1961, Nuovo Cimento {\bf 19}, 154. 

Goldstone, J., A.~Salam, and S.~Weinberg, 1962, \PR {\bf 127}, 965.

Haldane, F.D.M., 1983, \PRL {\bf 51}, 605. 

Haldane, F.D.M., 1990, in {\em The Quantum Hall Effect}, Second
Edition, Graduate Texts in Contemporary Physics, edited by
R.E.~Prange, and S.M.~Gervin (Springer, New York), p.~303.

Haldane, F.D.M., 1992, \HPA {\bf 65}, 152. 

Halperin, B.I., 1982,  \PRB {\bf 25}, 2185.

Halperin, B.I., 1983, \HPA {\bf 56}, 75.

Halperin, B.I., 1984, \PRL {\bf 52}, 1583; 2390(E).

Harada, K., 1990, \PRL {\bf 64}, 139.

Heidenreich, R., R.~Seiler, and D.A.~Uhlenbrock, 1980, \JSP {\bf
22}, 27.

Hu, C.-R.,1991, \IJMP B {\bf 5}, 1739.

Hwang, S.~W., J.A.~Simmons, D.C.~Tsui, and M.~Shayegan, 1992,
\SurS {\bf 263}, 72.

Jackiw, R., 1985, in {\em Current Algebra and Anomalies}, edited
by  S.B.~Treiman, R.~Jack\-iw, B.~Zumino, and E.~Witten (World
Scientific, Singapore), p.~211.

Jackiw, R., and R.~Rajaraman, 1985, \PRL {\bf 54}, 1219; 2060(E).

Jackson, J.D., 1975, {\em Classical Electrodynamics}, Second
Edition (Wiley, New York).

Jaffe, A., and C.~Taubes, 1980, {\em Vortices and Monopoles:
Structure of Static Gauge Theories}, Progress in Physics; 2
(Birkh\"auser, Boston).

Jain, J.K., 1989, \PRL {\bf 63}, 199.

Jain, J.K., 1990, \PRB {\bf 41}, 7653. 

Jain, J.K., 1992, Adv.\ Phys.\ {\bf 41}, 105.

Jain, J.K., and V.J.~Goldman, 1992, \PRB {\bf 45}, 1255.

Kac, V.G., 1983, {\em Infinite Dimensional Lie Algebra$\,$}
(Birkh\"auser, Boston).

Kac, V.G., and M.N.~Sanielevici, 1988, \PRD {\bf 37}, 2231.

Katanaev, M.O., and I.V.~Volovich, 1992, \AP {\bf 216}, 1.

Kay, B.S., and U.M.~Studer, 1991, \CMP {\bf 139}, 103.

Kerman, A.K., and N.~Onishi, 1981, Nucl.\ Phys.\ A {\bf 361}, 179.

Kleinert, H., 1989, {\em Gauge Fields in Condensed Matter},
Vol.~I,~II (World Scientific, Singapore).

Knizhnik, V.G., and A.B.~Zamolodchikov, 1984, \NPB {\bf 247}, 83.

Kohmoto, M., 1985, \AP {\bf 160}, 343.

Kohno, T., 1987, Ann.~de l'Inst.\ Fourier de l'Univ.\ de Grenoble
{\bf 37}, 139.

Kohno, T., 1988, in {\em Braids}, Contemporary Mathematics, Vol.~78,
edited by J.S.~Birman, and A.~Libgober (Am.\ Math.\ Soc.,
Providence, RI), p.~339.

Kunz, H., 1987, \CMP {\bf 112}, 121.

Lam, P.K., and S.M.~Girvin, 1984, \PRB {\bf 30}, 473.

Landau, L.D., and E.M.~Lifshitz, 1960, {\em Electrodynamics of
Continuous Media}, Course of Theoretical Physics, Vol.~8 (Pergamon,
Oxford).

Laughlin, R.B., 1981, \PRB {\bf 23}, 5632.

Laughlin, R.B., 1983a, \PRL {\bf 50}, 1395.

Laughlin, R.B., 1983b, \PRB {\bf 27}, 3383.

Laughlin, R.B., 1984, \SurS {\bf 141}, 11.

Laughlin, R.B., 1990, in {\em The Quantum Hall Effect}, Second
Edition, Graduate Texts in Contemporary Physics, edited by
R.E.~Prange, and S.M.~Gervin (Springer, New York), p.~233.

Lee, D.H., and S.C.~Zhang, 1991, \PRL {\bf 66}, 1220.

Leinaas, J.M., and J.~Myrheim, 1977, Nuovo Cimento B {\bf 37}, 1. 

Leutwyler, H., 1986, \HPA {\bf 59}, 201.

Leutwyler, H., 1993, ``Nonrelativistic Effective Lagrangians'', to
appear in \NPB. 

Levesque, D., J.J.~Weiss, and A.H.~MacDonald, 1984, \PRB {\bf 30},
1056.

Lieb, E., D.~Mattis, 1965, J.~Math.~Phys.~{\bf 6}, 304.

Luther, A., 1979, \PRB {\bf 19}, 320.

Luther, A., I.~Peshel, 1975, \PRB {\bf 12}, 3908.

MacDonald, A.H., 1990, \PRL {\bf 64}, 220.

Meissner, G., 1992, in {\em From Phase Transitions to Chaos: Topics
in Modern Statistical Physics}, edited G.~Gy\"orgyi, I.~Kondor,
L.~Sasv\'ari, and T.~Te\'el (World Scientific, Singapore), p.~145.

Meissner, G., 1993, Physica B {\bf 184}, 66.

Morandi, G., 1988, {\em Quantum Hall Effect$\,$} (Bibliopolis,
Napoli).

Murphy, S.Q., J.P.~Eisenstein, G.S.~Boebinger, L.N.~Pfeiffer, and
K.W.~West, 1994, \PRL {\bf 72}, 728.

Negele, J.W., and H.~Orland, 1987, {\em Quantum Many-Particle
Systems}, Frontiers in Physics, Vol.~68 (Addison-Wesley, New York).

Niu, Q., and D.J.~Thouless, 1984, J.\ Phys.\ A {\bf 9}, 30.
   
Niu, Q., and D.J.~Thouless, 1987, \PRB {\bf 35}, 2188.

Pines, D., and P.~Nozi\`eres, 1989, {\em The Theory of Quantum
Liquids$\,$} (Addison-Wesley, Redwood City, CA).

Prange, R.E., 1990, in {\em The Quantum Hall Effect}, Second
Edition, Graduate Texts in Contemporary Physics, edited by
R.E.~Prange, and S.M.~Gervin (Springer, New York), p.~69.

Prange, R.E., and S.M.~Gervin, 1990, Eds., {\em The Quantum Hall
Effect}, Second Edition, Graduate Texts in Contemporary Physics
(Springer, New York).

Read, N., 1989, \PRL {\bf 62}, 86.

Read, N., 1990, \PRL {\bf 65}, 1502.

Rezayi, E.H., 1987, \PRB {\bf 36}, 5454.

Sajoto, T., Y.W.~Suen, L.W.~Engel, M.B.~Santos, and M.~Shayegan,
1990, \PRB {\bf 41}, 8449.

Sakurai, J.J., 1980, \PRD {\bf 21}, 2993.

Salomaa, M.M., and G.E.~Volovik, 1989, J.\ Low Temp.\ Phys.\ {\bf
75}, 209.

Schellekens, A.N., and N.P.~Warner, 1986, \PRD {\bf 34}, 3092. 

Schrieffer, J.R., 1964,\,{\em Theory of Superconductivity\mini}
(Benjamin-Cummings, Menlo Park).

Schrieffer, J.R., X.-G.~Wen, and S.-C.~Zhang, 1988, \PRL {\bf 60},
944.

Schmutzer, E., and J.~Pleba\'nski, 1977, Fortsch.\ Physik {\bf 25},
37.

Schwinger, J., 1962, \PR {\bf 128}, 2425.

Semon, M.D., 1982, \FP {\bf 12}, 49.

Shankar, R., 1991, Physica A {\bf 177}, 530.

Siegel, C.L., 1989, {\em Lectures on the Geometry of Numbers\mini}
(Springer-Verlag, Berlin, Heidelberg).

Simmons, J.A., H.P.~Wei, L.W.~Engel, D.C.~Tsui, and M.~Shayegan,
1989, \PRL {\bf 63}, 1731.

Slansky, R., 1981, Phys.~Reports {\bf 79}, 1.

Sonnenschein, J., 1988, \NPB {\bf 309}, 752.

Stone, M., 1991a, \IJMP B {\bf 5}, 509.

Stone, M., 1991b, \AP {\bf 207}, 38.

Stone, M., 1992, Ed., {\em Quantum Hall Effect\,} (World Scientific,
Singapore).

Stormer, H.L., 1992, Physica B {\bf 177}, 401.

Suen, Y.W., L.W.~Engel, M.B.~Santos, M.~Shayegan, and D.C.~Tsui,
1992, \PRL {\bf 68}, 1379.

Suen, Y.W., H.C.~Manoharan, X.~Ying, M.B.~Santos, and M.~Shayegan,
1994, \SurS {\bf 305}, 13.

't Hooft, G., 1978, \NPB {\bf 138}, 1.

't Hooft, G., 1988, \CMP {\bf 117}, 685.

Tao, R., and Y.-S.~Wu, 1985, \PRB {\bf 31}, 6859.

Thomas, L.T., 1927, Philos.\ Mag.\ {\bf 3}, 1.

Thouless, D.J., M.~Kohmoto, M.P.~Nightingale, and M.~den Nijs,
1982, \PRL {\bf 49}, 405.

Tilley, D.R., and J.~Tilley, 1986, {\em Superfluidity and
Superconductivity$\,$}, Second Edition, Graduate Student Series in
Physics (Adam Hilger, Bristol).

Trugman, S., and S.~Kivelson, 1985, \PRB {\bf 26}, 3682.

Tsakadze, J.S., and S.J.~Tsakadze, 1980, J.\ Low Temp.\ Phys.\ {\bf
39}, 649.

Tsuchiya, A., and Y.~Kanie, 1987, Lett.\ Math.\ Phys.\ {\bf 13}, 303.

Tsui, D.C., H.L.~Stormer, and A.C.~Gossard, 1982, \PRB {\bf 48},
1559.

Tsui, D.C., 1990, Physica B {\bf 164}, 59.

Vinen, W.F., 1961, Proc.\ R.\ Soc.\ London, Ser.\ A\ {\bf 260}, 218.

von Klitzing, K., G.~Dorda, and M.~Pepper, 1980, \PRL {\bf 45}, 494.

Wegner, F., 1971, J.~Math.~Phys.\ {\bf 12}, 2259.

Wen, X.G., 1989, \PRB {\bf 40}, 7387.

Wen, X.G., 1990a, \IJMP B {\bf 2}, 239.

Wen, X.G., 1990b, \PRL {\bf 64}, 2206. 

Wen, X.G., 1990c, \PRB {\bf 41}, 12 838.

Wen, X.G., 1991a, \PRL {\bf 66}, 802.

Wen, X.G., 1991b, \PRB {\bf 43}, 11025. 

Wen, X.G., and Q.~Niu, 1990, \PRB {\bf 41}, 9377.

Wen, X.G., and A.~Zee, 1992, \PRB {\bf 46}, 2290.

Werner, S.A., J.L.~Staudenmann, and R.~Colella, 1979, \PRL {\bf
42}, 1103.

Weyl, H., 1918, Sitzber.\ Preuss.\ Akad.\ Wiss.\ 465.

Weyl, H., 1928, {\em Gruppentheorie und Quantenmechanik$\,$}
(Hirzel, Leipzig); Reprinted in English, 1949, {\em The Theory of
Groups and Quantum Mechanics$\,$} (Dover, New York).

Wilczek, F., 1982a, \PRL {\bf 48}, 1144.

Wilczek, F., 1982b, \PRL {\bf 49}, 957.

Wilczek, F., 1990, Ed., {\em Fractional Statistics and Anyon
Superconductivity$\,$} (World Scientific, Singapore).

Willett, R.L., J.P.~Eisenstein, H.L.~Stormer, D.C.~Tsui,
A.C.~Gossard, and J.H.~Eng\-lish, 1987, \PRL {\bf 59}, 1776.

Yoshioka, D., 1986, J.\ Phys.\ Soc.\ Jpn.\ {\bf 55}, 885.

Zhang, S.C., 1992, Int.\ J.\ Mod.\ Phys.\ B {\bf 6}, 25.

Zhang, S.C., T.H.~Hansson, and S.~Kivelson, 1989, \PRL {\bf 62},
82; 980(E). 

Zimmerman, J.E., and J.E.~Mercereau, 1965, \PRL {\bf 14}, 887.

\end{list}



\end{document}